\documentclass[10pt,twoside,openright]{book}
\pdfoutput=1
\usepackage[utf8]{inputenc}
\usepackage[danish,english]{babel}
\usepackage[sc]{mathpazo}
\usepackage[T1]{fontenc}
\usepackage{xcolor}
\definecolor{aaublue}{RGB}{0,0,0}
\usepackage{graphicx}
\usepackage{caption}
\usepackage{array,booktabs}
\usepackage{tabularx}
\usepackage{multirow}
\usepackage{framed}
\usepackage{tikz,pgfplots}
\usetikzlibrary{shapes.multipart}
\usepackage{amsmath}
\usepackage{amssymb}
\usepackage{amsfonts}
\usepackage[framed,amsmath,amsthm,thmmarks]{ntheorem}

\usepackage[
  paperwidth=17cm, 
  paperheight=24cm, 
  outer=2.5cm, 
  inner=2.5cm, 
  top=2.5cm, 
  bottom=2.5cm 
  ]{geometry}

\usepackage{titlesec}
\titleformat*{\section}{\normalfont\Large\bfseries\color{aaublue}}
\titleformat*{\subsection}{\normalfont\large\bfseries\color{aaublue}}
\titleformat*{\subsubsection}{\normalfont\normalsize\bfseries\color{aaublue}}
\addto\captionsenglish{
}

\usepackage{fancyhdr}
\usepackage{calc}
\usepackage{mparhack}

\usepackage[sectionbib]{chapterbib}
\usepackage{cite}

\usepackage[nottoc,section]{tocbibind}
\usepackage{appendix}
\usepackage{lastpage}
\usepackage[
  colorinlistoftodos, 
  textwidth=2cm, 
  textsize=scriptsize, 
  ]{todonotes}
\usepackage{hyperref}
\hypersetup{%
	pdfpagelabels=true,%
	plainpages=false,%
	pdfauthor={Author},%
	pdftitle={Title},%
	pdfsubject={Subject},%
	bookmarksnumbered=true,%
	colorlinks=true,%
	citecolor=aaublue,%
	filecolor=aaublue,%
	linkcolor=aaublue,
	urlcolor=aaublue,%
	pdfstartview=FitH%
}

\usepackage[export]{adjustbox}
\usepackage[printonlyused]{acronym}
\usepackage{enumitem}
\usepackage{cleveref}
\usepackage{stackengine}
\usepackage{chngcntr}
\usepackage{changepage}
\usepackage{multicol}
\usepackage{makecell}

\theoremheaderfont{\normalfont\bfseries}
\theorembodyfont{\normalfont}
\theoremstyle{break}

\newshadedtheorem{exa}{Example}[section]

\newcommand{\papertitlepage}[5]{
  \chapter{#1}\label{#2}
  \chaptermark{}
  \vspace{2cm}
  \begin{center}
    \large #3
  \end{center}
  \vspace{3cm}
  \begin{center}
  \normalsize #4
  \end{center}
  \vspace*{\fill}
  \newpage\thispagestyle{empty}
  \vspace*{\fill}
    #5\par
    \noindent{\em The layout has been revised.}
  \vspace*{\fill}
  \cleardoublepage
}

\newenvironment{abstract}{\section*{Abstract}\it}{}

\newcommand{\IEEEPARstart}[2]{#1#2}

\makeatletter
\let\cite\relax
\DeclareRobustCommand{\cite}{%
	\let\new@cite@pre\@gobble
	\@ifnextchar[\new@cite{\@citex[]}}
\def\new@cite[#1]{\@ifnextchar[{\new@citea{#1}}{\@citex[#1]}}
\def\new@citea#1{\def\new@cite@pre{#1}\@citex}
\def\@cite#1#2{[{\new@cite@pre\space#1\if\relax\detokenize{#2}\relax\else, #2\fi}]}
\makeatother

\usepackage{tocloft}
\iftrue
\makeatletter
\def\cleardoublepage{\clearpage%
	\if@twoside
	\ifodd\c@page\else
	\vspace*{\fill}
	\hfill
	\begin{center}
		This page intentionally left blank.
	\end{center}
	\vspace{\fill}
	\thispagestyle{empty}
	\newpage
	\if@twocolumn\hbox{}\newpage\fi
	\fi
	\fi
}
\makeatother

\makeatletter
\def\@endpart{\vfil\newpage}
\makeatother

\fi

\begin{document}

\frontmatter
\pagestyle{empty} 
\pagenumbering{roman} 

%
%
%
%
\pdfbookmark[0]{Front page}{label:frontpage}%
\begin{titlepage}
  \addtolength{\hoffset}{0.5\evensidemargin-0.5\oddsidemargin} 
  \noindent%
  \begin{tabular}{@{}p{\textwidth}@{}}
    \toprule[2pt]
    \midrule
    \vspace{0.2cm}
    \begin{center}
    \huge{\textbf{
      Single-Microphone Speech Enhancement and Separation \\ Using Deep Learning
    }}
    \end{center}
    \vspace{0.2cm}\\
    \midrule
    \toprule[2pt]
  \end{tabular}
  \vspace{4 cm}
  \begin{center}
    {\large
      PhD Thesis
    }\\
    \vspace{0.2cm}
    {\Large
      Morten Kolbæk
    }
  \end{center}
  \vfill
  \begin{center}
  2018
  \end{center}
\end{titlepage}
\clearpage

\thispagestyle{empty}
\noindent
{\small
\begin{tabularx}{\textwidth}{@{}lX}
    Thesis submitted: & August 31, 2018\\
    \\
    PhD Supervisor: & Professor Jesper Jensen\\
                    & Aalborg University, Denmark\\
    \\                 
    Assistant PhD Supervisor: & Professor Zheng-Hua Tan\\
                   & Aalborg University, Denmark\\
    \\
    PhD Committee: & Associate Professor Thomas Arildsen (chairman)\\
    			   & Aalborg University, Denmark\\
    			   \\
    
                   & Professor Reinhold H{\"a}b-Umbach\\
                   & Paderborn University, Germany\\
                   \\
                   
                   & Professor John H. L. Hansen\\
                   & The University of Texas at Dallas, USA\\
                   \\
                   
    PhD Series:    & Technical Faculty of IT and Design, Aalborg University\\
    			   \\
    			   
    Department:    & Department of Electronic Systems\\
\end{tabularx}
\strut\vfill
\noindent
%
\strut\vfill
\noindent
\begin{tabularx}{\textwidth}{@{}lX}
    ISSN: & xxxx-xxxx\\
    ISBN: & xxx-xx-xxxx-xxx-x\\
\end{tabularx}
\strut\vfill
\noindent Published by:\newline
Aalborg University Press\newline
Langagervej 2\newline
DK – 9220 Aalborg Ø\newline
Phone: +45 99407140\newline
aauf@forlag.aau.dk\newline
forlag.aau.dk
\strut\vfill
\noindent \copyright{} Copyright by Morten Kolbæk, except where otherwise stated \newline
\strut\vfill
\noindent Printed in Denmark by Rosendahls, 2018
\strut\vfill\vfill\vfill
\clearpage}

\chapter*{About the Author\markboth{About the Author}{About the Author}}\label{ch:cv}
\pagestyle{fancy}
\addcontentsline{toc}{chapter}{About the Author}
\begin{tabularx}{\textwidth}{@{}Xr}
    \Large Morten Kolbæk & \raisebox{-\totalheight/2}{\includegraphics[width=2cm]{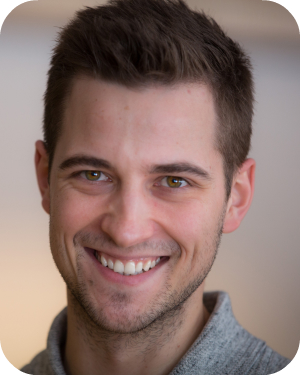}}\\
\end{tabularx}\par
\vspace{0.5cm}\noindent
Morten Kolbæk received the B.Eng. degree in electronic design at Aarhus University, Business and Social Sciences, AU Herning, Denmark, and the M.Sc. degree in signal processing and computing from Aalborg University, Denmark, in 2013 and 2015, respectively. 
He is currently pursuing his PhD degree at the section for Signal and Information Processing at the Department of Electronic Systems, Aalborg University, Denmark, under the supervision of Professor Jesper Jensen and Professor Zheng-Hua Tan.
His main research interests include single-microphone algorithms for speech enhancement and multi-talker speech separation, machine learning, deep learning in particular, and intelligibility improvement of noisy speech for hearing aids applications.

\chapter*{Abstract\markboth{Abstract}{Abstract}}\label{ch:Abstract}
\addcontentsline{toc}{chapter}{Abstract}

The cocktail party problem comprises the challenging task of listening to and understanding a speech signal in a complex acoustic environment, where multiple speakers and background noise signals simultaneously interfere with the speech signal of interest.
A signal processing algorithm that can effectively increase the speech intelligibility and quality of speech signals in such complicated acoustic situations is highly desirable. 
Especially for applications involving mobile communication devices and hearing assistive devices, increasing speech intelligibility and quality of noisy speech signals has been a goal for scientists and engineers for more than half a century.
Due to the re-emergence of machine learning techniques, today, known as deep learning, the challenges involved with such algorithms might be overcome.

In this PhD thesis, we study and develop deep learning-based techniques for two major sub-disciplines of the cocktail party problem: \emph{single-microphone speech enhancement} and \emph{single-microphone multi-talker speech separation}.

Specifically, we conduct in-depth empirical analysis of the generalizability capability of modern deep learning-based single-microphone speech enhancement algorithms. 
We show that performance of such algorithms is closely linked to the training data, and good generalizability can be achieved with carefully designed training data. 
Furthermore, we propose \acf{uPIT}, a deep learning-based algorithm for single-microphone speech separation and we report state-of-the-art results on a speaker-independent multi-talker speech separation task. 
Additionally, we show that uPIT works well for joint speech separation and enhancement without explicit prior knowledge about the noise type or number of speakers, which, at the time of writing, is a capability only shown by uPIT.     
Finally, we show that deep learning-based speech enhancement algorithms designed to minimize the classical short-time spectral amplitude mean squared error leads to enhanced speech signals which are essentially optimal in terms of \acf{STOI}, a state-of-the-art speech intelligibility estimator. 
This is important as it suggests that no additional improvements in STOI can be achieved by a deep learning-based speech enhancement algorithm, which is designed to maximize STOI.

\clearpage{\pagestyle{empty}\cleardoublepage}

\chapter*{Resumé\markboth{Resumé}{Resumé}}\label{ch:Resume}
\addcontentsline{toc}{chapter}{Resumé}
 
{\selectlanguage{danish}
Cocktailparty-problemet beskriver udfordringen ved at forstå et talesignal i et komplekst akustisk miljø, hvor stemmer fra adskillige personer, samtidig med baggrundsstøj, interferer med det ønskede talesignal.
En signalbehandlings algoritme, som effektivt kan øge taleforståeligheden eller talekvaliteten af støjfyldte talesignaler, er yderst eftertragtet. Specielt indenfor applikationer som vedrører mobil kommunikation eller høreapparater, har øgning af taleforståelighed eller talekvalitet af støjfyldte talesignaler været et mål for videnskabsfolk og ingeniører i mere end et halvt århundrede. 
Grundet en genopstået interesse for maskinlærings teknikker, som i dag er kendt som dyb læring, kan nogle af de udfordringer som er forbundet med sådanne algoritmer, måske blive løst.

I denne Ph.d.-afhandling studerer og udvikler vi dyb-læringsbaserede teknikker anvendeligt for to store underdiscipliner af cocktailparty-problemet: \emph{enkelt-mikrofon taleforbedring} og \emph{enkelt-mikrofon multi-taler taleseparation.}

Specifikt foretager vi dybdegående empiriske analyser af generaliserings-egenskaberne af moderne dyb-læringsbaserede enkelt-mikrofons taleforbed-ringsalgoritmer. Vi viser at ydeevnen af disse algoritmer er tæt forbundet med mængden og kvaliteten af træningsdata, og gode generaliseringsegenskaber kan opnås ved omhyggeligt designet træningsdata.
Derudover præsenterer vi utterance-level Permutation Invariant Training (uPIT), en dyb læringsbaseret algoritme til enkelt-mikrofon taleseparation og vi rapporterer state-of-the-art resultater for en taler-uafhængig multi-taler taleseparationsopgave. 
Ydermere viser vi, at uPIT fungerer godt til både taleseparation samt taleforbedring samtidigt, hvilket på tidspunktet for denne afhandling, er en egenskab, som kun uPIT har. 
Endelig viser vi, at dyb-læringsbaserede taleforbedrings algoritmer som er designet til at maksimere den klassiske short-time spectral amplitude mean squared error fører til forbedrede talesignaler, som essentielt er optimale med hensyn til} Short-Time Objective Intelligibility (STOI), {\selectlanguage{danish} en state-of-the-art taleforståelighedsprædiktor. Dette er vigtig, da det antyder at ingen yderligere forbedring af STOI kan opnås selv med dyb-læringsbaserede taleforbedrings algoritmer, som er designet til at maksimere STOI.
}


\cleardoublepage
\pdfbookmark[0]{Contents}{label:contents} 
\pagestyle{fancy} 
\tableofcontents

\chapter*{List of Abbreviations\markboth{List of Abbreviations}{List of Abbreviations}}\label{ch_listAcry} 
\pagestyle{fancy} 
\addcontentsline{toc}{chapter}{List of Abbreviations} 
\begin{acronym}[RASTA-PLPEEE]\setlength{\itemsep}{0.2ex} 
\acro{ADFD}{Akustiske Databaser for Dansk} 
\acro{AI}{Artificial Intelligence} 
\acro{AMS}{Amplitude Modulation Spectrogram} 
\acro{AM}{Amplitude Mask} 
\acro{ANN}{Artificial Neural Network} 
\acro{ASA}{Auditory Scene Analysis} 
\acro{ASR}{Automatic Speech Recognition} 
\acro{AUC}{Area Under ROC curve} 
\acro{BBRBM}{Bernoulli-Bernoulli RBM} 
\acro{BLSTM}{Bi-directional Long Short-Term Memory} 
\acro{BM}{Boltzmann Machine} 
\acro{BNN}{Biological Neural Network} 
\acro{BSS}{Blind-Source Separation} 
\acro{CASA}{Computational Auditory Scene Analysis} 
\acro{CC}{Closed-Condition} 
\acro{CDF}{Cumulative Distribution Function} 
\acro{CD}{Contrastive Divergence} 
\acro{CNN}{Convolutional Neural Network} 
\acro{CNTK}{Microsoft Cognitive Toolkit} 
\acro{CPU}{Central Processing Unit} 
\acro{DANet}{Deep Attractor Network} 
\acro{DBN}{Deep Belief Network} 
\acro{DBRNN}{Deep Bidirectional Recurrent Neural Network} 
\acro{DCT}{Discrete Cosine Transform} 
\acro{DFT}{Discrete Fourier Transform} 
\acro{DL}{Deep Learning} 
\acro{DNN}{Deep Neural Network} 
\acro{DPCL}{Deep Clustering} 
\acro{DRNN}{Deep Recurrent Neural Network} 
\acro{DTFT}{Discrete-Time Fourier Transform} 
\acro{EER}{Equal Error Rate} 
\acro{ELC}{Envelope Linear Correlation} 
\acro{EMSE}{Envelope Mean Squared Error} 
\acro{EM}{Expectation-Maximization} 
\acro{ERB}{Equivalent Rectangular Bandwidth} 
\acro{ESTOI}{Extended Short-Time Objective Intelligibility} 
\acro{EVD}{Eigen-Value Decomposition} 
\acro{FA}{False Alarm} 
\acro{FCT}{Fast Cosine Transform} 
\acro{FFT}{Fast Fourier Transform} 
\acro{FHMM}{Factorial Hidden Markov Model} 
\acro{FIR}{Finite Impulse Response} 
\acro{FLOPS}{Floating-point Operations Per Second} 
\acro{FNN}{Feed-forward Neural Network} 
\acro{GBRBM}{Gaussian-Bernoulli RBM} 
\acro{GFE}{Gammatone Filter bank Energies} 
\acro{GMM}{Gaussian Mixture Model} 
\acro{GPU}{Graphics Processor Unit} 
\acro{HINT}{Hearing In Noise Test} 
\acro{HIT-FA}{HIT minus False Alarm} 
\acro{HMM}{Hidden Markov Model} 
\acro{HPC}{High Performance Computing} 
\acro{IAM}{Ideal Amplitude Mask} 
\acro{IBM}{Ideal Binary Mask} 
\acro{IDCT}{Inverse Discrete Cosine Transform} 
\acro{IDFT}{Inverse Discrete Fourier Transform} 
\acro{IID}{Independent and Identically Distributed} 
\acro{IIR}{Infinite Impulse Response} 
\acro{INPSM}{Ideal Non-negative Phase Sensitive Mask} 
\acro{IPSF}{Ideal Phase Sensitive Filter} 
\acro{IPSM}{Ideal Phase Sensitive Mask} 
\acro{IRM}{Ideal Ratio Mask} 
\acro{ITS}{Inverse Transform Sampling} 
\acro{IUT}{Implementation Under Test} 
\acro{KLD}{Kullback-Leibler Divergence} 
\acro{KLT}{Karhunen-Loève Transform}
\acro{LC}{Local Criterion} 
\acro{LPC}{Linear Predictive Coding} 
\acro{LP}{Linear Prediction} 
\acro{LSTM}{Long Short-Term Memory} 
\acro{MAP}{Maximum a Posteriori} 
\acro{MCMC}{Markov Chain Monte Carlo} 
\acro{ME}{Miss Error} 
\acro{MFCC}{Mel-Frequency Cepstrum Coefficient} 
\acro{MIMD}{Multiple Instruction Multiple Data} 
\acro{MLP}{Multi-Layer Perceptron} 
\acro{ML}{Maximum Likelihood} 
\acro{MMELC}{Maximum Mean Envelope Linear Correlation} 
\acro{MMSE}{Minimum Mean Squared Error} 
\acro{MOS}{Mean Opinion Score} 
\acro{MRF}{Markov Random Field}  
\acro{MSE}{Mean Squared Error} 
\acro{NGDRNN}{Noise General Deep Recurrent Neural Network} 
\acro{NMF}{Non-negative Matrix Factorization} 
\acro{NN}{Neural Network} 
\acro{NSDRNN}{Noise Specific Deep Recurrent Neural Network} 
\acro{OC}{Open-Condition} 
\acro{OIM}{Objective Intelligibility Measure} 
\acro{PCD}{Persistent Contrastive Divergence} 
\acro{PDF}{Probability Density Function} 
\acro{PESQ}{Perceptual Evaluation of Speech Quality} 
\acro{PIT}{Permutation Invariant Training} 
\acro{PLP}{Perceptual Linear Prediction} 
\acro{PMF}{Probability Mass Function} 
\acro{PSA}{Phase Sensitive Approximation} 
\acro{PSD}{Power Spectral Density} 
\acro{PSF}{Phase Sensitive Filter} 
\acro{PSM}{Phase Sensitive Mask} 
\acro{PT}{Parallel Tempering} 
\acro{RASTA-PLP}{Relative Spectral Transform - Perceptual Linear Prediction} 
\acro{RASTA}{Relative Spectral} 
\acro{RBM}{Restricted Boltzmann Machine} 
\acro{RMS}{Root Mean Square} 
\acro{RNN}{Recurrent Neural Network} 
\acro{ROC}{Receiver Operating Characteristics} 
\acro{ReLU}{Rectified Linear Unit} 
\acro{ReLu}{Rectified Linear unit} 
\acro{SAR}{Source-to-Artifact Ratio} 
\acro{SDR}{Source-to-Distortion Ratio} 
\acro{SE}{Speech Enhancement} 
\acro{SGD}{Stochastic Gradient Descent} 
\acro{SIFT}{Scale-Invariant Feature Transform} 
\acro{SIMD}{Single Instruction Multiple Data} 
\acro{SIR}{Source-to-Interference Ratio} 
\acro{SISD}{Single Instruction Single Data} 
\acro{SI}{Speech Intelligibility} 
\acro{SNR}{Signal-to-Noise Ratio} 
\acro{SPP}{Speech Presence Probability} 
\acro{SQ}{Speech Quality} 
\acro{SR}{Speaker Recognition} 
\acro{SSN}{Speech Shaped Noise} 
\acro{STFT}{Short-Time Fourier Transform} 
\acro{STOI}{Short-Time Objective Intelligibility} 
\acro{STSA}{Short-Time Spectral Amplitude} 
\acro{SVD}{Singular-Value Decomposition} 
\acro{SVM}{Support Vector Machine} 
\acro{SV}{Speaker Verification} 
\acro{T-F}{Time-Frequency} 
\acro{TIMIT}{Texas Instruments Massachusetts Institute of Technology} 
\acro{UBM}{Universal Background Model} 
\acro{VAD}{Voice Activity Detection} 
\acro{WSJ0}{Wall Street Journal} 
\acro{WGN}{White Gaussian Noise} 
\acro{WSS}{Wide Sense Stationary} 
\acro{uPIT}{utterance-level Permutation Invariant Training} 
\end{acronym} 
\chapter*{List of Publications\markboth{List of Publications}{List of Publications}}\label{ch_listPub}
\pagestyle{fancy}
\addcontentsline{toc}{chapter}{List of Publications}
This main body (Part II) of this thesis consists of the following publications:
\begin{enumerate}
	\item[{[A]}] {\small M. Kolbæk, Z. H. Tan, and J. Jensen, “Speech Intelligibility Potential of General and Specialized Deep Neural Network Based Speech Enhancement Systems”, \textit{IEEE/ACM Transactions on Audio, Speech, and Language Processing}, vol. 25, no. 1, pp. 153–167, January 2017.}
	\item[{[B]}] {\small M. Kolbæk, Z.-H. Tan, and J. Jensen, “Speech Enhancement Using Long Short-Term Memory Based Recurrent Neural Networks for Noise Robust Speaker Verification”, \textit{IEEE Spoken Language Technology Workshop}, pp. 305-311, December 2016.}
	\item[{[C]}] {\small D. Yu, M. Kolbæk, Z.-H. Tan, and J. Jensen,  “Permutation Invariant Training of Deep Models for Speaker-Independent Multi-Talker Speech Separation”, \textit{IEEE International Conference on Acoustics, Speech, and Signal Processing}, pp 241-245, March 2017.}
	\item[{[D]}] {\small M. Kolbæk, D. Yu, Z.-H. Tan, and J. Jensen, “Multi-Talker Speech Separation With Utterance-Level Permutation Invariant Training of Deep Recurrent Neural Networks”, \textit{IEEE/ACM Transactions on Audio, Speech, and Language Processing}, vol. 25, no. 10, pp. 1901-1913, October 2017.}
	\item[{[E]}] {\small M. Kolbæk, D. Yu, Z.-H. Tan, and J. Jensen, “Joint Separation and Denoising of Noisy Multi-Talker Speech Using Recurrent Neural Networks and Permutation Invariant Training”, \textit{IEEE International Workshop on Machine Learning for Signal Processing}, pp. 1-6, September 2017.} 
	\item[{[F]}] {\small M. Kolbæk, Z.-H. Tan, and J. Jensen,  “Monaural Speech Enhancement Using Deep Neural Networks by Maximizing a Short-Time Objective Intelligibility Measure”, \textit{IEEE International Conference on Acoustics, Speech, and Signal Processing}, pp 5059-5063, April 2018.}
	\item[{[G]}]  {\small M. Kolbæk, Z.-H. Tan, and J. Jensen,  “On the Relationship Between Short-Time Objective Intelligibility and Short-Time Spectral-Amplitude Mean-Square Error for Speech Enhancement”, \textit{IEEE/ACM Transactions on Audio, Speech, and Language Processing}, vol. 27, no. 2, pp. 283-295, February 2019.}
\end{enumerate}

\chapter*{Preface\markboth{Preface}{Preface}}\label{ch:preface}
\pagestyle{fancy}
\addcontentsline{toc}{chapter}{Preface}
This thesis documents the scientific work carried out as part of the PhD project entitled \emph{Single-Microphone Speech Enhancement and Separation Using Deep Learning}. 
The thesis is submitted to the Technical Doctoral School of IT and Design at Aalborg University in partial fulfillment of the requirements for the degree of Doctor of Philosophy.
The project was funded by the Oticon Foundation\footnote{\url{http://www.oticonfoundation.com}}, and the work was carried out in the period from September 2015 to August 2018 within the Signal and Information Processing Section, in the Department of Electronic Systems, at Aalborg University.
Parts of the work was carried out during a four-month secondment at the Interactive Systems Design Lab at the University of Washington, Seattle USA, and at Microsoft Research, Redmond USA.

The thesis is structured in two parts: a general introduction and a collection of scientific papers.
The introduction review classical algorithms and deep learning-based algorithms for single-microphone speech enhancement and separation, and furthermore summarizes the scientific contributions of the PhD project. 
The introduction is followed by a collection of seven papers that are published in or submitted to peer-reviewed journals or conferences.

I would like to express my deepest gratitude to my two supervisors Jesper Jensen and Zheng-Hua Tan for their support and guidance throughout the project. 
In particular, I would like to thank Jesper Jensen for his sincere dedication to the project and for his abundant, and seemingly endless, supply of constructive criticism, which, although daunting at times, unarguably has improved all aspects of my work.   
Furthermore, I would like to give a special thanks to Dong Yu for a very giving and pleasant collaboration for which I am very grateful. 
Also, I would like to thank Les Atlas, Scott Wisdom, Tommy Powers and David Dolengewicz from the Interactive Systems Design Lab for their hospitality and helpfulness during my stay at University of Washington.   
Last, but not least, I wish to thank my family for their unconditional support. 
\vfill
\begin{flushright}
Morten Kolbæk\\
Bjerghuse, July 17, 2018
\end{flushright}

\cleardoublepage

\mainmatter
\part{Introduction}
\newcommand{\introtitle}{Introduction}
\chapter*{\introtitle}\label{sec:intro}
\addcontentsline{toc}{chapter}{\introtitle}
\acresetall

Most of us take it for granted and use it effortless on a daily basis; our ability to speak and hear. Nevertheless, the human speech production- and auditory systems are truly unique \cite{helmholtz_sensations_1954,fant_acoustic_1960,flanagan_speech_1972,schnupp_auditory_2011,moore_introduction_2013,fitch_evolution_2000,bitterman_ultra-fine_2008}.   

We are probably all familiar with the challenging situation at a dinner party when you attempt to converse with the person sitting across the table. Other persons, having their own conversations, are sitting around you, and you have to concentrate to hear the voice of the person you are trying to have a conversation with. 
Remarkably, the more you concentrate on the voice of your conversational partner, the more you understand and the less you feel distracted by the people talking loudly around you. 
This ability of selective auditory attention is one of the astonishing capabilities of the human auditory system. 
In fact, in 1953 it was proposed as an engineering discipline in the academic literature by Colin Cherry \cite{cherry_experiments_1953} when he asked: \\

\begin{adjustwidth}{10mm}{10mm}
{\centering \emph{How do we recognize what one person is saying when others are speaking at the same time (the "cocktail party problem")? On what logical basis could one design a machine ("filter") for carrying out such an operation?}}
\vspace{-3mm}
\begin{flushright} \emph{ -- Colin Cherry, 1953.} \end{flushright} 
\end{adjustwidth}

Ever since Colin Cherry coined the term \emph{cocktail party problem}, it has been, and still is, a very active topic of research within multiple scientific disciplines such as psychoacoustics, auditory neuroscience, electrical engineering, and computer science \cite{bregman_auditory_1990,bronkhorst_cocktail_2000,divenyi_speech_2005,wang_computational_2006,mcdermott_cocktail_2009,schnupp_auditory_2011,mesgarani_selective_2012,ziongolumbic_mechanisms_2013,bronkhorst_cocktail-party_2015,wang_supervised_2018}, and although Colin Cherry studied speech-interference signals in his seminal work in 1953, today, the cocktail party problem encompasses both speech and non-speech-interference signals \cite{haykin_cocktail_2005,wang_deep_2017}.   

In this PhD thesis, we study aspects of the cocktail party problem. 
Specifically, motivated by a re-emergence of a branch of machine learning, today, commonly known as \emph{deep learning} \cite{lecun_deep_2015}, we investigate how deep learning techniques can be used to address some of the challenges in two major sub-disciplines of the cocktail party problem: \emph{single-microphone speech enhancement} and \emph{single-microphone multi-talker speech separation}. 

\newcommand\barbelow[1]{\stackunder[1.2pt]{$#1$}{\rule{.8ex}{.075ex}}}

\section{Speech Enhancement and Separation}\label{sec:classical_se}
The common goal of single-microphone speech enhancement and single-microphone multi-talker speech separation algorithms is to improve some aspects, e.g. quality or intelligibility, of a single-microphone recording of one or more degraded speech signals \cite{deller_discrete-time_1993,divenyi_speech_2005,loizou_speech_2013,hendriks_dft-domain_2013}.
As the name implies, single-microphone algorithms process sound signals captured by a single microphone. Such algorithms are useful in applications where microphone arrays cannot be utilized, e.g. due to space, power, or hardware-cost restrictions, e.g. for in-the-ear hearing aids. 
Furthermore, since single-microphone algorithms do not rely on the spatial locations of target and interference signals, single-microphone algorithms compliment multi-microphone algorithms and can be used as a post-processing step for techniques such as beamforming, as those techniques are mainly effective, when target and interference signals are spatially separated \cite{brandstein_microphone_2001}.
Therefore, algorithms capable of enhancing or separating speech signals from single-microphone recordings are highly desirable.    

The main difference between speech enhancement and multi-talker speech separation algorithms is the number of target signals.  
If the target is only a single speech signal and all remaining sounds in the recording, both speech and non-speech sounds, are considered as noise, extracting that particular speech signal from the recording is considered as a speech enhancement task.
On the other hand, if the recording contains multiple speech signals, and possibly multiple non-speech sounds, and two or more of these speech signals are of interest, the task is a multi-talker speech separation task. 
In this sense, the speech enhancement problem may be seen as a special case of the multi-talker speech separation problem.

Applications for speech enhancement include mobile communication devices, e.g. mobile phones, or hearing assistive devices where usually only a single speech signal is the target. 
For these applications, successful algorithms have been developed, which e.g. rely on interference characteristics which are different than speech. Hence, these methods would not perform well for speech-like interference signals. 
Applications for multi-talker speech separation include automatic meeting transcription, multi-party human-machine interaction, e.g. for video games like Xbox or PlayStation, or automatic captioning for audio/video recordings, e.g. for YouTube or Facebook, all situations where overlapping speech is not uncommon. 
Since the interference signals for these applications are speech signals, single-microphone multi-talker speech separation possesses additional challenges compared to single-microphone speech enhancement. 
However, in theory, a perfect system for multi-talker speech separation would also be a perfect system for speech enhancement, but not the other way around.

\subsection{Classical Speech Enhancement Algorithms}\label{classe}
Let $x[n]$ be a sample of a clean time-domain speech signal and let a noisy observation $y[n]$ be defined as
\begin{equation}
y[n] = x[n] + v[n],
\label{eq1}
\end{equation}
where $v[n]$ is an additive noise sample representing any speech and non-speech, interference signal.
Then, the goal of single-microphone speech enhancement is to acquire an estimate $\hat{x}[n]$ of $x[n]$, which in some sense is "close to" $x[n]$ using $y[n]$ only. 

\begin{figure}
	\centering
	\begin{tikzpicture}[baseline=(current bounding box.north)]
	\tikzset{%
	do path picture/.style={%
		path picture={%
			\pgfpointdiff{\pgfpointanchor{path picture bounding box}{south west}}%
			{\pgfpointanchor{path picture bounding box}{north east}}%
			\pgfgetlastxy\x\y%
			\tikzset{x=\x/2,y=\y/2}%
			#1
		}
	},
	cross/.style={do path picture={    
			\draw [line cap=round] (-0.6,-0.6) -- (0.6,0.6) (-0.6,0.6) -- (0.6,-0.6);
	}},
}

\def\x{0.2}

\draw [thick, ->] (-2.2,2) -- (-1.5,2);
\node[above] at (-1.9,2) {\footnotesize $y[n]$};

\draw [thick, ->] (-0.3,2) -- (0.4,2);
\node[above] at (0.05,2) {\footnotesize $\barbelow{y}_m$};

\draw [thick, ->] (1.6,2) -- (2.3,2);
\node[above] at (3.45,3.2) {\footnotesize $r(k,m)$};

\draw [thick, ->]  (1.95,2) -- (1.95,3.2) -- (4.85,3.2) -- (4.85,2.16) ;


\draw [thick, ->] (3.5,2) -- (4.69,2);
\node[above] at (4.15,2.00) {\footnotesize $\hat{g}(k,m)$};

\node [draw,circle,cross,minimum width=0.2 ] at (4.85,2){}; 

\draw [thick, ->] (5.01,2) -- (6.2,2);
\node[above] at (5.61,2.00) {\footnotesize $\hat{a}(k,m)$};

\draw [thick, ->] (7.2,2) -- (8.1,2);
\node[above] at (7.75,2) {\footnotesize $\barbelow{\hat{x}}_m$};

\draw [thick, ->] (9.3,2) -- (10,2);
\node[above] at (9.7,2) {\footnotesize $\hat{x}[n]$};

\filldraw[fill=gray!20!white, draw=black] (-1.5 , 1.5) rectangle (-0.3 , 2.5) ;
\node at (-0.88,2) {\scriptsize {\begin{tabular}{c} Framing \end{tabular}}};

\filldraw[fill=gray!20!white, draw=black] (0.4 , 1.5) rectangle (1.6 , 2.5) ; 
\node at (1.0,2) {\scriptsize {\begin{tabular}{c} Analysis \end{tabular}}};

\filldraw[fill=gray!20!white, draw=black] (2.3 , 1.5) rectangle (3.5 , 2.5) ; 
\node at (2.9,2) {\scriptsize {\begin{tabular}{c} Gain \\ Estimator \end{tabular}}};

\filldraw[fill=gray!20!white, draw=black] (6.2 , 1.5) rectangle (7.4 , 2.5) ; 
\node at (6.8,2) {\scriptsize {\begin{tabular}{c} Synthesis \end{tabular}}};

\filldraw[fill=gray!20!white, draw=black] (8.1 , 1.5) rectangle (9.3 , 2.5) ; 
\node at (8.7,2) {\scriptsize {\begin{tabular}{c} Overlap- \\ add \end{tabular}}};

	\end{tikzpicture}
	\caption{Classical gain-based speech enhancement system. The noisy time-domain signal $y[n] = x[n] + v[n]$ is first segmented into overlapping frames $\barbelow{y}_m$. An analysis stage then applies a transform to arrive in a transform-domain $r(k,m)$ for time-frame $m$ and transform-coefficient $k$. A gain $\hat{g}(k,m)$ is then estimated and applied to $r(k,m)$ to arrive at an enhanced transform-coefficient $\hat{a}(k,m) = \hat{g}(k,m)r(k,m)$. Finally, a synthesis stage transforms the enhanced transform-coefficient into time domain and the final time-domain signal $\hat{x}[n]$ is obtained by overlap-add.}     
	\label{fig:se}
\end{figure}
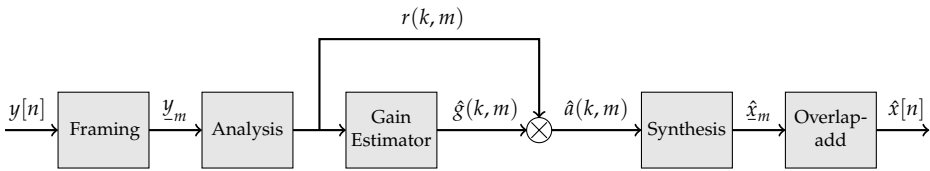

Throughout the years, a wide range of techniques have been proposed for estimating $x[n]$ and many of these techniques follow the gain-based approach shown in Fig.\;\ref{fig:se} , e.g. \cite{hendriks_dft-domain_2013,loizou_speech_2013}.   
%
%
First, the noisy time-domain signal $y[n]$ is segmented into overlapping frames $\barbelow{y}_m$ using a sliding window of length $N$. An analysis stage then applies a transform, e.g. the \ac{DFT}, to the frames to arrive in a transform-domain $r(k,m)$ for time-frame $m$ and transform-coefficient $k$. 
An estimator, to be further defined in the next sections, estimates a gain value $\hat{g}(k,m)$ that is applied to $r(k,m)$ to arrive at an enhanced transform-coefficient $\hat{a}(k,m) = \hat{g}(k,m)r(k,m)$. A synthesis stage then applies an inverse transform to the enhanced transform-coefficients to transform the coefficients back to time domain. Finally, the time-domain signal $\hat{x}[n]$ is obtained by overlap-adding the enhanced time-domain frames $\barbelow{\hat{x}}_m$ \cite{allen_short_1977}.

Although many speech enhancement algorithms follow the gain-based approach, their strategy for finding the gain value $\hat{g}(k,m)$, i.e. the design of the gain estimator, can be very different, and, in general, these techniques may be divided into four classes \cite{loizou_speech_2013}: 
1) Spectral subtractive-based algorithms (Sec.\,\ref{sec:se_sub1} ), 
2) Statistical model-based algorithms    (Sec.\,\ref{sec:se_sub2} ), 
3) Subspace based algorithms             (Sec.\,\ref{sec:se_sub3} ), and 
4) Machine learning-based algorithms     (Sec.\,\ref{sec:ml_cl_se}).    

\subsubsection{Spectral Subtraction Methods}\label{sec:se_sub1}
Speech enhancement algorithms based on spectral subtraction belong to the first class of algorithms proposed for speech enhancement and were developed in the late 1970s \cite{boll_suppression_1979,mcaulay_speech_1980,loizou_speech_2013}.
Specifically, let $y(k,m)$, $x(k,m)$, and $v(k,m)$ be the \ac{STFT} coefficients of the noisy signal $y[n]$, clean signal $x[n]$, and noise signal $v[n]$, from Eq.\;\eqref{eq1}, respectively. The spectral subtraction algorithm in its simplest form is then defined as    
\begin{equation}
\hat{x}(k,m) = \left[|y(k,m)| - |v(k,m)| \right] e^{j\phi_y{(k,m)}},
\label{eq2}
\end{equation}
where $|\cdot|$ denotes absolute value and $e^{j\phi_y{(k,m)}}$ is the phase of the noisy \ac{STFT} coefficients $y(k,m)$. 
From Eq.\;\eqref{eq2} it is clear why this algorithm is named "spectral subtraction" as the estimate $\hat{x}(k,m)$ is acquired simply by subtracting the noise magnitude $|v(k,m)|$ from the magnitude of the noisy signal $|y(k,m)|$ and appending the noisy phase $e^{j\phi_y{(k,m)}}$. Furthermore, by slightly rewriting Eq.\;\eqref{eq2}, we arrive at
\begin{equation}
\hat{x}(k,m) = g(k,m) |y(k,m)|e^{j\phi_y{(k,m)}},
\label{eq3}
\end{equation} 
where
\begin{equation}
g(k,m) = 1-\frac{|v(k,m)|}{|y(k,m)|}
\label{eq4}
\end{equation} 
is the gain function, which clearly shows that spectral subtraction as defined by Eq.\;\eqref{eq2} indeed belongs to the family of gain-based speech enhancement algorithms. 
Finally, although spectral subtraction as defined by Eq.\;\eqref{eq2} was primarily motivated heuristically \cite{boll_suppression_1979}, it was later shown \cite{mcaulay_speech_1980} that Eq.\;\eqref{eq2} is closely related to the maximum likelihood estimate of the clean speech \ac{PSD}, when speech and noise are modeled as independent stochastic processes \cite{mcaulay_speech_1980}. An assumption that is used heavily in later successful speech enhancement algorithms \cite{ephraim_speech_1984,hendriks_dft-domain_2013}. 

Although speech enhancement algorithms based on the spectral subtraction principle effectively reduce the noise in noisy speech signals, it has a few disadvantages. First, it requires an accurate estimate of the noise magnitude $|v(k,m)|$, which in general is not easily available and might be time varying. As a consequence, $|v(k,m)|$ was first estimated from non-speech periods prior to speech activity, e.g. using a \ac{VAD} algorithm \cite{loizou_speech_2013}. 
Furthermore, due to estimation errors of $|v(k,m)|$, $|\hat{x}(k,m)|$ might be negative, which by definition is an invalid magnitude spectrum. Several techniques have been proposed to alleviate this side-effect (e.g. \cite{boll_suppression_1979,hansen_use_1989,sondhi_improving_1981,haitian_spectral_2004,loizou_speech_2013}) and the simplest is to apply a half-wave rectifier and set all negative values to zero. Another technique is to set negative values to the value of adjacent non-negative frames, but regardless of the technique, spectral subtractive-based techniques are prone to signal distortions known as \emph{musical noise} due to estimation errors in the estimate of the noise magnitude $|v(k,m)|$.

\subsubsection{Statistically Optimal Methods}\label{sec:se_sub2}
Although spectral subtractive-based techniques are effective speech enhancement algorithms, they are primarily based on heuristics and not derived deliberately to be mathematically optimal. 
If, however, the speech enhancement problem is formulated as a statistical estimation problem with a well-defined optimality criterion and strictly defined statistical assumptions, a class of \emph{optimal} speech enhancement algorithms can be developed
\cite{hansen_iterative_1987,chen_new_2006,loizou_speech_2013,deller_discrete-time_1993,ephraim_statistical-model-based_1992,erkelens_minimum_2007,loizou_speech_2005,v_survey_2014,hendriks_dft-domain_2013,ephraim_speech_1984,ephraim_speech_1985,mcaulay_speech_1980}. 
One such class is the \ac{MMSE} estimators, for which two large sub-classes are the linear \ac{MMSE} estimators, commonly known as \emph{Wiener filters} after the mathematician Nobert Wiener \cite{wiener_extrapolation_1949}, and the non-linear \ac{STSA}-\ac{MMSE} estimators \cite{ephraim_speech_1984}.

\begin{figure}
	\centering
	\begin{tikzpicture}[baseline=(current bounding box.north)]
	\tikzset{%
	do path picture/.style={%
		path picture={%
			\pgfpointdiff{\pgfpointanchor{path picture bounding box}{south west}}%
			{\pgfpointanchor{path picture bounding box}{north east}}%
			\pgfgetlastxy\x\y%
			\tikzset{x=\x/2,y=\y/2}%
			#1
		}
	},
	plus/.style={do path picture={    
			\draw [line cap=round] (0,-0.6) -- (0,0.6) (-0.6,0) -- (0.6,-0);
	}},
	minus/.style={do path picture={    
		\draw [line cap=round] (-0.5,0) -- (0.5,-0);
}},
}

\draw [thick, ->] (-0.5,2) -- (0.8,2);
\node[above] at (-0.2,2) {\footnotesize $x[n]$};

\draw [thick, ->] (1,1) -- (1,1.8);
\node[above] at (0.65,0.9) {\footnotesize $v[n]$};

\draw [thick, ->]  (0.5,2) -- (0.5,3) -- (5,3) -- (5,2.2) ;

\node [draw,circle,plus,minimum width=0.2 ] at (1,2){}; 

\node[above] at (1.6,2) {\footnotesize $y[n]$};
\draw [thick, ->] (1.2,2) -- (2.2,2);

\filldraw[fill=gray!20!white, draw=black] (2.2 , 1.5) rectangle (3.7 , 2.5) ;
\node at (2.95,2) {\scriptsize {\begin{tabular}{c} Linear \\ Filter  \end{tabular}}};

\draw [thick, ->] (3.7,2) -- (4.8,2);
\node[above] at (4.3,2) {\footnotesize $\hat{x}[n]$};

\node [draw,circle,minus,minimum width=0.2 ] at (5,2){}; 

\draw [thick, ->] (5.2,2) -- (6.2,2);
\node[above] at (5.8,2) {\footnotesize $e[n]$};

	\end{tikzpicture}
	\caption{Linear estimation problem for which Wiener filters are optimal in a mean squared error sense.}     
	\label{fig:wien}
\end{figure}
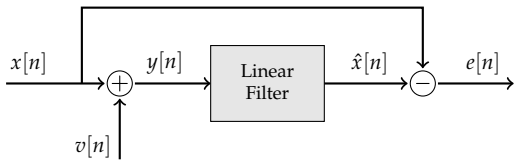

\paragraph{Basic Wiener Filters} \mbox{}\\
Wiener filters are minimum mean squared error optimal linear filters for the linear estimation problem shown in Fig.\,\ref{fig:wien}, where the observed signal $y[n]$ is given by $y[n] = x[n] + v[n]$, where $x[n]$ and $v[n]$ are assumed to be uncorrelated and stationary stochastic processes \cite{chen_new_2006,loizou_speech_2013,deller_discrete-time_1993}.
Wiener filters can have either a \ac{FIR} or an \ac{IIR} or be even non-causal. 
For the causal \ac{FIR} Wiener filter, the estimated signal $\hat{x}[n]$ is given by
\begin{equation}
\hat{x}[n] = \barbelow{h}^T_o \barbelow{y}(n),
\label{eq5}
\end{equation} 
where
\begin{equation}
\barbelow{h}_o = [h_1,\; h_2,\; \dots,\; h_L]^T
\label{eq6}
\end{equation} 
are the optimal filter coefficients and
\begin{equation}
\barbelow{y}(n) = [y[n],\; y[n-1],\; \dots,\; y[n-L+1]]^T
\label{eq7}
\end{equation} 
are the past $L$ samples of the observed signal. 
The optimal filter, $\barbelow{h}_o$, i.e. the Wiener filter, is then defined as
\begin{equation}
\barbelow{h}_o = \underset{\barbelow{h}}{\operatorname{arg\,min}} \;\; J_x(\barbelow{h}),
\label{eq8}
\end{equation} 
where $J_x(\barbelow{h})$ is the mean squared error given by
\begin{equation}
\begin{split}
J_x(\barbelow{h}) = \mathbb{E}\{e^2[n]\} =& \mathbb{E}\{(x[n]-\hat{x}[n])^2\},
\end{split}
\label{eq9}
\end{equation} 
and $\mathbb{E}\{\cdot\}$ denote mathematical expectation.
Finally, by differentiating Eq.\,\eqref{eq9} with respect to $\barbelow{h}$, equating to zero, and solving for $\barbelow{h}$, the optimal filter coefficients $\barbelow{h}_o$ are found to be
\begin{equation}
\begin{split}
\barbelow{h}_o & = 
(\barbelow{\barbelow{R}}_{\barbelow{x}\barbelow{x}} +\barbelow{\barbelow{R}}_{\barbelow{v}\barbelow{v}} )^{-1} \barbelow{r}_{x\barbelow{x}},
\end{split}
\label{eq10}
\end{equation} 
which is the well-known Wiener-Hopf solution%
\footnote{The Wiener-Hopf solution is usually on the form $\barbelow{\barbelow{R}}_{\barbelow{y}\barbelow{y}}^{-1} \barbelow{r}_{x\barbelow{y}}$ but since $x[n]$ and $v[n]$ are assumed uncorrelated, $\barbelow{\barbelow{R}}_{\barbelow{y}\barbelow{y}} = \barbelow{\barbelow{R}}_{\barbelow{x}\barbelow{x}} + \barbelow{\barbelow{R}}_{\barbelow{v}\barbelow{v}}$ and $\barbelow{r}_{x\barbelow{y}} = \barbelow{r}_{x\barbelow{x}}$.}
\cite{loizou_speech_2013,proakis_digital_2007}, where $\barbelow{\barbelow{R}}_{\barbelow{x}\barbelow{x}}$ and $\barbelow{\barbelow{R}}_{\barbelow{v}\barbelow{v}}$ denote the autocorrelation matrices of $\barbelow{x}$ and $\barbelow{v}$, respectively, and $\barbelow{r}_{x\barbelow{x}} = \mathbb{E}\{ x[n]\barbelow{x}\}$ denote the autocorrelation vector. 
From Eq.\,\eqref{eq10} it is seen that the optimal filter coefficients $\barbelow{h}_o$ are based on $\barbelow{\barbelow{R}}_{\barbelow{x}\barbelow{x}}$, $\barbelow{\barbelow{R}}_{\barbelow{v}\barbelow{v}}$, and $\barbelow{r}_{x\barbelow{x}}$, which are not directly available and must be estimated, for the filter to be used in practice. Since the noise process $v[n]$ is assumed to be stationary, accurate estimates of $\barbelow{\barbelow{R}}_{\barbelow{v}\barbelow{v}}$ might be acquired during non-speech periods and used during speech-active periods \cite{deller_discrete-time_1993,loizou_speech_2013}.

An alternative to the time-domain Wiener filter is the frequency-domain Wiener filter. If the filter $\barbelow{h}$ is allowed to be of infinite duration and non-causal, i.e.
$\barbelow{h}' = [ \dots , h'_{-1},\; h'_{0},\;h'_{1},\; \dots ]$,   
the Wiener filter can be defined in the frequency domain using a similar approach as just described. Let 
\begin{equation}
\hat{x}(\omega) = g(\omega) y(\omega), 
\label{eq11}
\end{equation} 
where $\hat{x}(\omega)$, $g(\omega)$, and $y(\omega)$ denote the \ac{DTFT} of the estimated speech signal $\hat{x}[n]$, the infinite duration time-domain filter $\barbelow{h}'$, and the noisy speech signal $y[n]$, respectively.  
The frequency domain Wiener filter is then given as \cite{loizou_speech_2013,deller_discrete-time_1993}
\begin{equation}
H(\omega) = \frac{P_{x}(\omega) }{P_{x}(\omega)  +P_{v}(\omega) },
\label{eq12}
\end{equation} 
where $P_{x}(\omega)$, and $P_{v}(\omega)$ are the \ac{PSD} of the clean speech signal $x[n]$, and noise signal $v[n]$, respectively.
Alternatively, the frequency domain Wiener filter can be formulated as 
\begin{equation}
H(\omega) = \frac{\xi_\omega  }{\xi_\omega + 1},
\label{eq13}
\end{equation} 
where
\begin{equation}
\xi_\omega = \frac{ P_{x}(\omega)  }{P_{v}(\omega) }
\label{eq14}
\end{equation} 
is known as the \emph{a priori} \ac{SNR} at frequency $\omega$.
From Eqs.\;\eqref{eq12} and \eqref{eq13} it is seen that the frequency-domain Wiener filter $g(\omega)$ is real, even, and non-negative and, consequently, does not modify the phase of $y(\omega)$, hence $\hat{x}(\omega)$ will have the same phase as $y(\omega)$, similarly to the spectral subtractive-based approaches \cite{gerkmann_phase_2012}. 
Furthermore, from Eq.\;\eqref{eq13} it can be deduced that the Wiener filter operates by suppressing signals with low \ac{SNR} relatively more than signals with higher \ac{SNR}.  
Finally, similarly to the time-domain Wiener filter, the frequency-domain Wiener filter, as formulated by Eqs.\;\eqref{eq12} and \eqref{eq13}, is not directly applicable in practice as speech may only be stationary during short time periods and information about the \emph{a priori} \ac{SNR} is not available in general. 
Consequently, $P_{x}(\omega)$ and $P_{v}(\omega)$ must be estimated using e.g. iterative techniques for short time periods where speech and noise are approximately stationary, e.g. \cite{loizou_speech_2013,deller_discrete-time_1993}.

\paragraph{Basic STSA-MMSE Estimators}  \mbox{}\\
Although the Wiener filter is considered the optimal complex spectral estimator, it is not the optimal spectral amplitude estimator, and based on the common belief at-the-time that phase was much less important than amplitude for speech enhancement (see e.g. \cite{wang_unimportance_1982,paliwal_importance_2011,gerkmann_phase_2012,gerkmann_phase_2015,mowlaee_phase_2014,mowlaee_phase-aware_2016} and references therein), it led to the development of optimal spectral amplitude estimators, commonly known as \ac{STSA}-\ac{MMSE} estimators \cite{ephraim_speech_1984}.  

Differently from the Wiener filters, \ac{STSA}-\ac{MMSE} estimators do not assume a linear relation between the observed data and the estimator. Instead, the \ac{STSA}-\ac{MMSE} estimators are derived using a Bayesian statistical framework, where explicit assumptions are made about the probability distributions of speech and noise \ac{DFT} coefficients. 

Specifically, let $A(k,m)$, and $R(k,m)$, $k = 1,\,2,\,\dots,\, K$,  $m=1,\,2,\,\dots,\, M $ denote random variables representing the $K$-point \ac{STFT} magnitude spectra for time frame $m$ of the clean speech signal $x[n]$, and noisy speech signal $y[n]$, respectively. Let $\hat{A}(k,m)$, and $V(k,m)$ be defined in a similar manner for the estimated speech signal $\hat{x}[n]$ and the noise signal $v[n]$, respectively. In the following the frame index $m$ will be omitted for convenience as all further steps apply for all time frames. 
Let
\begin{equation}
\barbelow{A} = \left[ A_1, A_2, \, \dots, \, A_K \right]^T,
\label{eq16}
\end{equation}
\begin{equation}
\barbelow{R} = \left[ R_1, R_2, \, \dots, \, R_K \right]^T,
\label{eq17}
\end{equation}
and
\begin{equation}
\barbelow{\hat{A}} = \left[ \hat{A}_1, \hat{A}_2, \, \dots, \, \hat{A}_K \right]^T,
\label{eq18}
\end{equation}
be the stack of these random variables into random vectors.
Also, let $p(\barbelow{A},\barbelow{R})$ denote the joint \ac{PDF} of clean and noisy spectral magnitudes and $p(\barbelow{A}|\barbelow{R})$, and $p(\barbelow{R})$ denote a conditional and marginal \ac{PDF}, respectively. 
Finally, let the Bayesian \ac{MSE} \cite{kay_fundamentals_2010,loizou_speech_2013} between the clean speech magnitude $\barbelow{A}$ and the estimated speech magnitude $\barbelow{\hat{A}}$, be defined as 
\begin{equation} 
\begin{split}
\mathcal{J}_{MSE}	& = \mathbb{E}_{\barbelow{A},\barbelow{R}}  \left\{ \left(\barbelow{A} - \barbelow{\hat{A}} \right)^2 \right\}.  \\
\end{split}
\label{eq19}
\end{equation}  
By minimizing the Bayesian \ac{MSE} with respect to $\barbelow{\hat{A}}$ it can be shown (see e.g. \cite{kay_fundamentals_2010,loizou_speech_2013}) that the optimal \ac{STSA}-\ac{MMSE} estimator is given as 
\begin{equation} 
\begin{split}
\barbelow{\hat{A}}	& = \mathbb{E}_{\barbelow{A}|\barbelow{R}}  \{ \barbelow{A}|\barbelow{R}  \}, \\
\end{split}
\label{eq20}
\end{equation}   
which is nothing more than the expected value of the clean speech magnitude $\barbelow{A}$ given the observed noisy speech magnitude $\barbelow{R}$.

From Eq.\;\eqref{eq20} a large number of estimators can be derived by considering different distributions of $p(\barbelow{A},\barbelow{R})$ \cite{hendriks_dft-domain_2013}. For example, in the seminal work of Ephraim and Malah in  \cite{ephraim_speech_1984}, the \ac{STFT} coefficients of the clean speech and noise were assumed to be statistically independent, zero-mean, Gaussian distributed random variables. 
This assumption is motivated by the fact that \ac{STFT} coefficients become uncorrelated, and under a Gaussian assumption therefore independent, with increasing frame length.
Based on these assumptions Eq.\,\eqref{eq20} simplifies \cite{ephraim_speech_1984,loizou_speech_2013} to    
\begin{equation} 
\begin{split}
\hat{A}(k)	& = G(\psi_k,\gamma_k) R(k), \\
\end{split}
\label{eq21}
\end{equation}  
where $G(\psi_k,\gamma_k)$ is a gain function that is applied to the noisy spectral magnitude $R(k)$, and 
\begin{equation} 
\begin{split}
\psi_k	& =  \frac{\mathbb{E}\{|A(k)|^2\}}{\mathbb{E}\{|V(k)|^2\}},\\
\end{split}
\label{eq24}
\end{equation}  
and
\begin{equation} 
\begin{split}
\gamma_k	& = \frac{R^2(k)}{\mathbb{E}\{|V(k)|^2\}}. \\
\end{split}
\label{eq25}
\end{equation}  
The term $\psi_k$ is referred to as \emph{a priori} \ac{SNR}, similarly to Eq.\,\eqref{eq14} since $\psi_k \approx \xi_\omega$%
\footnote{Equality only holds if \ac{DTFT} coefficients in Eq.\,\eqref{eq24} are computed for infinite sequences of stationary processes. Since they are \ac{DFT} coefficients computed based on finite sequences, it follows that $\psi_k \approx \xi_\omega$.}%
, and $\gamma_k$ is referred to as the \emph{a posteriori \ac{SNR}} as it reflects the \ac{SNR} of the observed, or noise corrupted, speech signal. 
As seen from Eq.\,\eqref{eq21} the \ac{STSA}-\ac{MMSE} gain is a function of \emph{a priori} and \emph{a posteriori} \ac{SNR}. However, although the Wiener gain in  Eq.\,\eqref{eq13} is also a function of \emph{a priori} \ac{SNR}, the \ac{STSA}-\ac{MMSE} gain in general introduces less artifacts at low SNR than the Wiener gain, partially due to the \emph{a posteriori} \ac{SNR} \cite{loizou_speech_2013,hu_subjective_2006}. 
In fact, at high \acp{SNR} (\ac{SNR} > 20 dB) the gains from the Wiener filter and STSA-MMSE estimator converges to the same value \cite{ephraim_speech_1984,loizou_speech_2013,chen_new_2006}. 

Since the first \ac{STSA}-\ac{MMSE} estimator was proposed using a Gaussian assumption, a large range of estimators have been proposed with different statistical assumptions, and cost functions, in an attempt to improve the performance by utilizing either more accurate statistical assumptions, which are more in line with the true probability distribution of speech and noise, or cost functions more in line with human perception \cite{loizou_speech_2005,martin_speech_2005,ephraim_speech_1985,ephraim_statistical-model-based_1992,hendriks_log-spectral_2009,hendriks_dft-domain_2013,erkelens_estimation_2008,erkelens_minimum_2007,cohen_relaxed_2005,cohen_speech_2005}.   
Finally, note that similarly to the Wiener filters, the \emph{a priori} \ac{SNR} has to be estimated, e.g. using noise \ac{PSD} tracking (see e.g. \cite{hendriks_dft-domain_2013} and references therein), in order to use the \ac{STSA}-\ac{MMSE} estimators in practice.

\subsubsection{Subspace Methods}\label{sec:se_sub3}
The third class of enhancement algorithms are known as subspace-based algorithms, as they are derived primarily using principles from linear algebra and not, to the same degree, on principles from signal processing and estimation theory, as the previously discussed algorithms were \cite{loizou_speech_2013}.  
The general underlying assumption behind these algorithms is that $K$-dimensional vectors of speech signals do not span the entire $K$-dimensional euclidean space, but instead are confined to a smaller $M$-dimensional subspace, i.e. $M < K$ \cite{ephraim_signal_1995,benesty_speech_2005}.   
Specifically, let a stationary stochastic process representing a clean speech signal $\barbelow{X} = \left[ X_1,\;X_2,\; \dots, \; X_K \right]^T$ be defined as
\begin{equation}
\barbelow{X} = \sum_{m=1}^{M} C_m \barbelow{p}_m = \barbelow{\barbelow{P}} \barbelow{C},  
\label{eq26}
\end{equation}
where $C_m$ are zero-mean, potentially complex, random variables and $\barbelow{p}_m$ are $K$-dimensional linearly independent, potentially complex, basis vectors, e.g. complex sinusoids \cite{ephraim_signal_1995}.    
Here,
\begin{equation}
\barbelow{C} = \left[ C_1,\, C_2, \, \dots, \, C_M \right]^T \in \mathbb{R}^{M},  
\label{eq26_1}
\end{equation}
and 
\begin{equation}
\barbelow{P} = \left[ \barbelow{p}_1,\, \barbelow{p}_2, \, \dots, \, \barbelow{p}_M \right] \in \mathbb{R}^{K \times M},  
\label{eq26_2}
\end{equation}
and if $M = K$, the transformation between $\barbelow{X}$ and $\barbelow{C}$ is always possible as it corresponds to a change of coordinate system \cite{ephraim_signal_1995}. 
However, for speech signals, such a transformation is often possible for $M < K$ \cite{ephraim_signal_1995}, which implies that $\barbelow{X}$ lies in a $M$-dimensional subspace spanned by the $M$ columns of $\barbelow{\barbelow{P}}$ in the $K$-dimensional Euclidean space. This subspace, is commonly referred to as the signal subspace.
Since the rank, denoted as $\mathcal{R}\{\cdot\}$, of $\barbelow{\barbelow{P}}$ is $\mathcal{R}\{\barbelow{\barbelow{P}}\} = M$, the covariance matrix of $\barbelow{X}$,
\begin{equation}
\barbelow{\barbelow{\Sigma}}_X = \mathbb{E}\{\barbelow{X}\barbelow{X}^T\} = \barbelow{\barbelow{P}} \barbelow{\barbelow{\Sigma}}_C \barbelow{\barbelow{P}}^T \in \mathbb{R}^{K \times K},
\label{eq27}
\end{equation}
where $\barbelow{\barbelow{\Sigma}}_C = \mathbb{E}\{\barbelow{C}\barbelow{C}^T\}$ is the covariance matrix of $\barbelow{C}$, will be rank deficient, $\mathcal{R}\{\barbelow{\barbelow{\Sigma}}_X\} = \mathcal{R}\{\barbelow{\barbelow{\Sigma}}_C\} = M < K$. 
Noting from the stationarity of $\barbelow{X}$ that $\barbelow{\barbelow{\Sigma}}_X \succeq 0$, it follows that $\barbelow{\barbelow{\Sigma}}_X$ only has non-negative eigenvalues. 
The fact that $\barbelow{\barbelow{\Sigma}}_X$ has some eigenvalues that are equal to zero is the key to subspace-based speech enhancement.

For convenience, let us rewrite our signal model from Eq.\,\eqref{eq1} in vector form, 
\begin{equation}
\barbelow{Y} = \barbelow{X} + \barbelow{V},
\label{eq28}
\end{equation}
where $\barbelow{Y}$, $\barbelow{X}$, and $\barbelow{V}$ are the $K$-dimensional stochastic vectors representing the time-domain noisy speech signal, the clean speech signal, and noise signal, respectively.  
Employing the standard assumption that speech $\barbelow{X}$ and noise signals $\barbelow{V}$ are stationary, uncorrelated, and zero-mean random processes \cite{ephraim_signal_1995,ephraim_speech_1984} it follows that 
\begin{equation}
\barbelow{\barbelow{\Sigma}}_Y = \barbelow{\barbelow{\Sigma}}_X + \barbelow{\barbelow{\Sigma}}_V,
\label{eq29}
\end{equation}
where $\barbelow{\barbelow{\Sigma}}_Y$, and $\barbelow{\barbelow{\Sigma}}_V$ are the covariance matrices of the noisy speech signal, and noise signal, respectively. 
Furthermore, with the additional assumption that the noise signal is white, with variance $\sigma_V^2$, Eq.\,\eqref{eq29} reduces to  
\begin{equation}
\barbelow{\barbelow{\Sigma}}_Y = \barbelow{\barbelow{\Sigma}}_X + \sigma_V^2 \barbelow{\barbelow{I}}_K,
\label{eq30}
\end{equation}
where $\barbelow{\barbelow{I}}_K$ is the $K$-dimensional identity matrix.
Now, consider the \ac{EVD} of Eq.\,\eqref{eq30} given as
\begin{equation}
\barbelow{\barbelow{U}} \barbelow{\barbelow{\Lambda}}_Y \barbelow{\barbelow{U}}^T = \barbelow{\barbelow{U}} \barbelow{\barbelow{\Lambda}}_X \barbelow{\barbelow{U}}^T + \barbelow{\barbelow{U}} \barbelow{\barbelow{\Lambda}}_V \barbelow{\barbelow{U}}^T,
\label{eq31}
\end{equation}
where $\barbelow{\barbelow{U}}$ is a matrix with the $K$ orthonormal eigenvectors of $\barbelow{\barbelow{\Sigma}}_Y$, and $\Lambda = \text{diag}(\lambda_{y,1}, \lambda_{y,2},\dots,\lambda_{y,K})$ is a diagonal matrix with the corresponding $K$ eigenvalues. 
Since it is assumed that $\mathcal{R}\{\barbelow{\barbelow{\Sigma}}_X\}$ is rank deficient (Eq.\,\eqref{eq26}) the eigenvalues of $\barbelow{\barbelow{\Sigma}}_Y$ can be partitioned in descending order based on their magnitude as
\begin{equation}
    \lambda_{y,k} = 
\begin{cases}
\lambda_{x,k} + \sigma_V^2 & \text{if } k=1,2,\dots, M\\
\sigma_V^2,              & \text{if } k=M+1, M+2, \dots, K.
\end{cases}
\label{eq32}
\end{equation}
Then, it follows \cite{ephraim_signal_1995,loizou_speech_2013} that the subspace spanned by the eigenvectors corresponding to the $M$ largest eigenvalues of  $\barbelow{\barbelow{\Sigma}}_Y$, i.e. the top line in Eq.\,\eqref{eq32}, corresponds to the subspace spanned by the eigenvectors of $\barbelow{\barbelow{\Sigma}}_X$, which is the same subspace spanned by the columns of $\barbelow{\barbelow{P}}$, i.e. the signal subspace.
Specifically, let, $\barbelow{\barbelow{U}}$ be partitioned as $\barbelow{\barbelow{U}} = [\barbelow{\barbelow{U}}_1 \;\barbelow{\barbelow{U}}_2]$ such that $\barbelow{\barbelow{U}}_1$ is a $K \times M$ matrix with the eigenvectors corresponding to the $M$ largest eigenvalues of $\barbelow{\barbelow{\Sigma}}_Y$, and $\barbelow{\barbelow{U}}_2$ is a $K \times (K-M)$ with the remaining $K-M$ eigenvectors, then $\barbelow{\barbelow{U}}_1\barbelow{\barbelow{U}}^T_1$ is a projection matrix that orthogonally projects its multiplicand onto the signal subspace. Similarly, $\barbelow{\barbelow{U}}_2\barbelow{\barbelow{U}}^T_2$ will be the projection matrix that projects its multiplicand onto the complementary orthogonal subspace, known as the noise subspace. Hence, it follows that a realization of the noisy signal can be decomposed as
\begin{equation}
\barbelow{y} = \barbelow{\barbelow{U}}_1\barbelow{\barbelow{U}}^T_1\barbelow{y} + \barbelow{\barbelow{U}}_2\barbelow{\barbelow{U}}^T_2\barbelow{y}.
\label{eq33}
\end{equation}   
Finally, since the noise subspace spanned by the columns of $\barbelow{\barbelow{U}}_2$ contains no components of the clean speech signal, the noise subspace can be nulled to arrive at an estimate of the clean speech signal given as
\begin{equation}
\hat{\barbelow{x}} = \barbelow{\barbelow{U}}_1\barbelow{\barbelow{U}}^T_1\barbelow{y}.
\label{eq34}
\end{equation}   
In fact, the solution in Eq.\,\eqref{eq34} can, similarly to the previously discussed methods, be viewed as a gain-based approach (see Fig.\,\ref{fig:se}) given by 
\begin{equation}
\hat{\barbelow{x}} = \barbelow{\barbelow{U}} \barbelow{\barbelow{G}}_M \barbelow{\barbelow{U}}^T_1\barbelow{y},
\label{eq36}
\end{equation}
where $\barbelow{\barbelow{G}}_M$ is simply the $M$-dimensional identity matrix.
In this form, a transformation $\barbelow{\barbelow{U}}^T_1\barbelow{y}$ is applied to the noisy time-domain speech signal $\barbelow{y}$, which in this case is the linear transformation matrix  $\barbelow{\barbelow{U}}^T_1$, known as the \ac{KLT}.
Then, a unit-gain $\barbelow{\barbelow{G}}_M$ is applied before an inverse \ac{KLT}, $\barbelow{\barbelow{U}}$, is used to reconstruct the enhanced signal to the time-domain. 

In fact, what differentiate most subspace-based speech enhancement methods is the choice of transform domain $\barbelow{\barbelow{U}}_1$ and the design of the gain matrix $\barbelow{\barbelow{G}}_M$. An alternative to the approach based on the \ac{EVD} of the covariance matrix, is the \ac{SVD} of time-domain signals ordered in either Toeplitz or Hankel matrices \cite{loizou_speech_2013}. 
Furthermore, the gain matrix can be designed with an explicitly defined trade-off between noise reduction and signal distortion and even to handle colored noise signals \cite{hermus_review_2006,benesty_speech_2005,loizou_speech_2013,lev-ari_extension_2003}.       

Finally, what most subspace-based speech enhancement algorithms have in common is the need for estimating the covariance matrix of the clean speech, or noise, signal and the, generally time-varying, dimension of the signal subspace $M$. 
Naturally, if $M$ is overestimated, some of the noise subspace is preserved, but if $M$ is underestimated some of the signal subspace is discarded. Consequently, the quality of these estimates highly influences the performance of subspace-based speech enhancement algorithms. 
Nevertheless, it has been shown that these algorithms are capable of improving speech intelligibility for hearing impaired listeners wearing cochlear implants \cite{loizou_subspace_2005}.

\subsubsection{Machine Learning Methods}\label{sec:ml_cl_se}
Common for all the previously discussed clean-speech estimators is that they are all, to some degree, derived using mathematical principles from probability theory, digital signal processing, or linear algebra. 
Consequently, they are based on various assumptions such as stationarity of the signals involved, uncorrelated clean-speech and noise signals, independence of speech and noise transform coefficients across time and frequency, etc. 
These assumptions are all trade-offs. On one hand, they must reflect the properties of real speech and noise signals, while, at the other hand, they must be simple enough that they allow mathematical tractable solutions.

Furthermore, they all require information about some, generally unknown, quantity such as the noise magnitude $|v(k,m)|$ for spectral subtractive-based techniques,  \emph{a priori} \ac{SNR} for the statistically optimal algorithms such as the Wiener filters or \ac{STSA}-\ac{MMSE} estimators, or the signal subspace dimension, or covariance matrices for the clean speech or noise signals, for the subspace-based techniques. 
These quantities need to be estimated, and their estimates are critical for the performance of the speech enhancement algorithm.
Finally, although these techniques are capable of improving the quality of a noisy speech signal, when the underlying assumptions are reasonably met  \cite{hu_subjective_2006}, they generally do not improve speech intelligibility for normal hearing listeners \cite{lim_enhancement_1979,kim_gain-induced_2011,hu_comparative_2007,hu_comparative_2007-1,luts_multicenter_2010,loizou_reasons_2011,jensen_spectral_2012,madhu_potential_2013,brons_effects_2014}.

A different approach to the speech enhancement task, a completely different paradigm in fact, is to consider the speech enhancement task as a supervised learning problem \cite{bishop_pattern_2006}. In this paradigm, it is believed that the speech enhancement task can be learned from observations of representative data, such as a large number of corresponding pairs of clean and noisy speech signals. 

Specifically, instead of designing a clean-speech estimator in closed-form using mathematical principles, statistical assumptions, and \emph{a priori} knowledge, the estimator is defined by a parameterized mathematical model, that represents a large function space, potentially with universal approximation properties such as \acp{GMM} \cite{mclachlan_mixture_1988}, \acp{ANN} \cite{hornik_multilayer_1989,cybenko_approximation_1989}, or \acp{SVM} \cite{cortes_support-vector_1995,hammer_note_2003}. The parameters of these machine learning models are then found as the solution to an optimization problem with respect to an objective function evaluated on a representative dataset.

This approach is fundamentally different from the previously described techniques since no restrictions, e.g. about linearity, or explicit assumptions, e.g. about stationarity or uncorrelated signals, are imposed on the model. 
Instead, signal features which are relevant for solving the task at hand, e.g. retrieving a speech signal from a noisy observation, are implicitly learned during the supervised learning process.
The potential big advantage of this approach is that less valid assumptions, made primarily for mathematical convenience, can be avoided and as we shall see in this section, and sections to come, such an approach might result in clean-speech estimators with a potential to exceed the performance of the non-machine learning based techniques proposed so far.

\pagebreak
\paragraph{Basic Principles} \mbox{}\\
The basic principle behind most machine learning based speech enhancement techniques can be formulated as  
\begin{equation}
\hat{\barbelow{o}} = \mathcal{F}(\barbelow{h}(\barbelow{y}),\barbelow{\theta}),
\label{eq37}
\end{equation}
where $\mathcal{F}(\cdot,\barbelow{\theta})$ denotes a parameterized model with parameters $\barbelow{\theta}$. The input signal $\barbelow{y}$ denotes the noisy speech signal and $\barbelow{h}(\cdot)$ is a vector-valued function that applies a feature transformation to the raw speech signal $\barbelow{y}$. The representation of the output $\hat{\barbelow{o}}$ depends on the application, but it could e.g. be the estimated clean-speech signal or the clean-speech \ac{STFT} magnitude.
The optimal parameters $\barbelow{\theta}^\ast$ are then found, without loss of generality, as the solution to the minimization problem given as 
\begin{equation}
\begin{aligned}
\barbelow{\theta^\ast} = \; & \underset{\barbelow{\theta}}{\text{argmin}}
& & \mathcal{J}(\mathcal{F}(\barbelow{h}(\barbelow{y}),\barbelow{\theta}),\barbelow{o}), \;\;\;  (\barbelow{y},\barbelow{o}) \in \mathcal{D}_{train}, \\
\end{aligned}
\label{eq38}
\end{equation} 
where $\mathcal{J}(\cdot,\cdot)$ is a non-negative objective function, and $(\barbelow{y},\barbelow{o})$ is an ordered pair, of noisy speech signals $\barbelow{y}$ and corresponding targets $\barbelow{o}$, e.g. clean-speech \ac{STFT} magnitudes, from a training dataset $\mathcal{D}_{train}$. 
In principle, the optimal parameters $\barbelow{\theta}$ are given such that $\mathcal{J}(\mathcal{F}(\barbelow{h}(\barbelow{y}),\barbelow{\theta}^\ast),\barbelow{o}) = 0$, i.e. $\hat{\barbelow{o}} = \barbelow{o} $. However, as datasets are incomplete, model capacity is finite, and learning algorithms non-optimal, achieving $\mathcal{J}(\mathcal{F}(\barbelow{h}(\barbelow{y}),\barbelow{\theta}^\ast),\barbelow{o}) = 0$, might not be possible.
In fact, it may not even be desirable as it may lead to a phenomena known as overfitting, where the model does not generalize, i.e. performs poorly, on data not experienced during training \cite{bishop_pattern_2006}.

Instead, what one typically wants in practice is to find a set of near-optimal parameters $\barbelow{\theta^\dagger}$ that achieve a low objective function value on the training set $\mathcal{D}_{train}$, but also on an unknown test dataset $\mathcal{D}_{test}$, where $\mathcal{D}_{test} \not\subset \mathcal{D}_{train}$, i.e. $\mathcal{D}_{test}$ is not a subset of $\mathcal{D}_{train}$, but still assumed to share the same underlying statistical distribution. Such a model is likely to generalize better, which ultimately enable the use of the model for practical applications, where the data is generally unknown. In fact, overfitting is the Achilles' heel of machine learning, and controlling the amount of overfitting and acquiring good generalization, is key to successfully applying machine learning based speech enhancement techniques in real-life applications.

\paragraph{Machine Learning for Enhancement}  \mbox{}\\
Machine learning has been applied to speech enhancement for several decades \cite{tamura_noise_1988,tamura_analysis_1989,tamura_improvements_1990,white_new_1990,kim_algorithm_2009,han_classification_2012,roux_ensemble_2013}, but until recently, not very successfully in terms of practical applicability.
In one of the first machine learning based speech enhancement techniques \cite{tamura_noise_1988} the authors proposed to use an \ac{ANN} (\acp{ANN} are described in detail in Sec.\,\ref{deeplearn}) to learn a mapping directly from a frame of the noisy speech signal $\barbelow{y}_m$ to the corresponding clean speech frame $\barbelow{x}_m$ as         
\begin{equation}
\hat{\barbelow{x}}_m = \mathcal{F}_{ANN}(\barbelow{y}_m,\barbelow{\theta}),
\label{eq39}
\end{equation}
where $\mathcal{F}_{ANN}(\cdot,\cdot)$ represents an \ac{ANN}. 
Although the technique proposed in \cite{tamura_noise_1988} was trained on only 216 words and with a small network, according to today's standard, their proposed technique slightly outperformed a spectral subtractive-based speech enhancement technique in terms of speech quality, but not speech intelligibility. 
Furthermore, the \ac{ANN} generalized poorly to speech and noise signals not part of the training set. Finally, it took three weeks to train the \ac{ANN} on a, at the time, modern super computer, which simply made it practically impossible to conduct experimental research using larger \acp{ANN} with larger datasets.
This might explain why little \ac{ANN} based speech enhancement literature exists from that time, compared to the previously discussed methods, such as Wiener filters or \ac{STSA}-\ac{MMSE} estimators, which, in general, require far less computational resources.

Almost two decades later, promising results were reported in \cite{kim_algorithm_2009}, where large improvements (more than 60\%) in speech intelligibility was achieved using a speech enhancement technique based on \acp{GMM}. Specifically, they followed a gain-based approach (see Fig.\,\ref{fig:se}), and estimated a \ac{T-F} gain $\hat{g}(k,m)$ for each frequency bin $k$ and time-frame $m$. The frequency decomposition of the time-domain speech signal was performed using a Gammatone filter bank with 25 channels \cite{patterson_complex_1992} and the gain was defined as
\begin{equation}
\hat{g}^{IBM}(k,m) = 
\begin{cases}
1 & \text{if } \mathcal{P}(\pi_1|r(k,m)) > \mathcal{P}(\pi_0|r(k,m))\\
0              & \text{otherwise},
\end{cases}
\label{eq40}
\end{equation}
where $\mathcal{P}(\pi_0|r(k,m))$ and $\mathcal{P}(\pi_1|r(k,m))$ denote the probabilities of the clean speech magnitude $|x(k,m)|$ belonging to one out of two classes. The two classes $\pi_0$, and $\pi_1$, denoted noise-dominated T-F units and speech-dominated T-F units, respectively, and were defined as
\begin{equation}
r(k,m) \in 
\begin{cases}
\pi_1 & \text{if } \frac{|x(k,m)|^2}{|v(k,m)|^2} > T_{SNR}(k)\\
\pi_0 & \text{otherwise},
\end{cases}
\label{eq41}
\end{equation}
where $\frac{|x(k,m)|^2}{|v(k,m)|^2}$ is the \ac{SNR} in frequency bin $k$ and time frame $m$ and $T_{SNR}(k)$ is an appropriately set frequency-dependent threshold. 
The probabilities $\mathcal{P}(\pi_0|r(k,m))$ and $\mathcal{P}(\pi_1|r(k,m))$ were estimated using two classifiers, one for each class, based on 256-mixture \acp{GMM}%
\footnote{Interestingly, in retrospect, they did attempt to use \acp{ANN}, but without good results.}
trained on 390 spoken utterances ($\approx 16$ min of speech) with a feature representation based on \ac{AMS} \cite{tchorz_snr_2003}.       
In fact, the binary gain defined by Eq.\,\eqref{eq40} is an estimate of the \ac{IBM}, which is simply defined by Eqs.\,\eqref{eq40} and \eqref{eq41} when oracle information about $|x(k,m)|^2$ and $|v(k,m)|^2$ is used.  
Furthermore, it has been shown that the \ac{IBM} can significantly improve intelligibility of noisy speech, even at very low \acp{SNR} \cite{wang_ideal_2005,brungart_isolating_2006,kjems_role_2009}, which makes the \ac{IBM} a highly desirable training target as speech intelligibility is likely to be increased if the mask is accurately estimated. 

This approach, first proposed in \cite{kim_algorithm_2009}, was reported to not only outperform classical methods such as the Wiener filter and \ac{STSA}-\ac{MMSE} estimator, it even achieved improvements in speech intelligibility at a scale not previously observed in the speech enhancement literature.    
Later, supporting results appeared in \cite{han_classification_2012,wang_exploring_2013} where even better performance was achieved using a binary classifier based on \acp{SVM}.

However, it was later discovered \cite{may_requirements_2014} that the great performance achieved by the systems proposed in \cite{kim_algorithm_2009,han_classification_2012} was primarily due to the reuse of the noise signal in both the training data and test data. This meant the systems in \cite{kim_algorithm_2009,han_classification_2012} were tested on realizations of the noise signal that were already used for training. 
In theory, it allowed the models to "memorize" the noise signal and simply subtract it from the noisy speech signal during test. 
This, obviously is not a possibility in real-life applications, where the exact noise signal-realization is generally not known in isolation. 

Regardless of the unrealistically good performance of the systems in \cite{kim_algorithm_2009,han_classification_2012} they, combined with the co-occurring Deep Learning revolution (described in detail in Sec.\;\ref{deeplearn}), reignited the interest in machine learning based speech enhancement.

\subsection{Classical Speech Separation Algorithms}\label{classep}
We now extend the formulation of the classical speech enhancement task (see Eq.\;\eqref{eq1}) to multi-talker speech separation.
Let $x_s[n]$ be a sample of a clean time-domain speech signal from speaker $s$, and let an observation of a mixture $y[n]$ be defined as
\begin{equation}
y[n] = \sum_{s = 1}^{S} x_s[n],
\label{eq42}
\end{equation}
where $S$ is the total number of speakers in the mixture.
Then, the goal of single-microphone multi-talker speech separation is to acquire estimates $\hat{x}_s[n]$ of $x_s[n]$, $s=1,\,2,\,\dots,\,S$, which in some sense are "close to" $x_s[n]$, $s=1,\,2,\,\dots,\,S$ using $y[n]$ only. 
In Sec.\,\ref{sec:classical_se} we have seen a large number of techniques proposed to solve the single-microphone speech enhancement task, and to some extent, they are fairly successful in doing so in practice. 
However, they all, except for the machine learning based techniques, rely heavily on specific statistical assumptions about the speech and noise signals. 
Specifically, in practice, the Wiener filters and \ac{STSA}-\ac{MMSE} estimators rely on accurate estimates of the noise \ac{PSD}.

Similarly, the subspace based techniques assume the noise signal is statistically white, or can be whitened, which in general requires additional information about the noise signal.     
Consequently, if the noise signal is a speech signal, which is non-stationary and colored, the methods described in Sec.\,\ref{sec:classical_se} perform poorly, as they rely on a signal model whose parameters are hard to estimate accurately.
This, in turn, might also explain why these techniques have not been successfully applied on the single-microphone multi-talker speech separation task given by Eq.\;\eqref{eq42}, as different techniques might be required to successfully handle speech-like interference signals, when the target signals themselves are speech.  
In this section we introduce some of the classical techniques, i.e. non-deep learning-based methods, that have been proposed for single-microphone multi-talker speech separation.

\subsubsection{Harmonic-Models}
Some of the early techniques for single-microphone speech separation were in fact more related to speech enhancement, than speech separation, as they aimed mainly at suppressing an interfering speaker, in a two-speaker mixture, than actually separating the speech signals (see e.g. \cite{parsons_separation_1976,hanson_speech_1983,hanson_harmonic_1984,quatieri_approach_1990}). 
Furthermore, compared to the speech enhancement techniques existing at the time, e.g. Wiener filters or \ac{STSA}-\ac{MMSE} estimators, the techniques proposed for speech separation were more involved compared to the simple gain-based approach used by the corresponding speech enhancement techniques. 
Finally, although more complicated, they were generally less successful as good performance could only be achieved by using \emph{a priori} knowledge generally not available in practice, such as information about the fundamental frequency of the interfering speaker \cite{hanson_speech_1983,hanson_harmonic_1984,quatieri_approach_1990}. 
For example, the techniques proposed in \cite{parsons_separation_1976,hanson_speech_1983,hanson_harmonic_1984} used the fact that voiced speech signals can be modeled by a sum of sinusoids and that two competing speakers generally have different fundamental frequency and consequently different harmonics. If knowledge about the fundamental frequency of the interfering speaker is known, one can, in theory, null that frequency and corresponding harmonics and suppress the interfering speaker. 
However, such an approach requires multi-pitch tracking as the individual pitch signals must be estimated from the noisy signal only, and such an approach only works if the fundamental frequencies of the speakers involved are sufficiently separated. 
Finally, as unvoiced speech does not possess any apparent harmonic structure \cite{hansen_speech_1991}, these techniques were only partly successful and not a general approach for single-microphone multi-talker speech separation \cite{parsons_separation_1976,hanson_speech_1983,hanson_harmonic_1984}.

\subsubsection{Computational Auditory Scene Analysis}
A different approach to the single-microphone multi-talker speech separation task is one based on principles from \ac{ASA} \cite{bregman_auditory_1990}. 
According to the \ac{ASA} paradigm, the auditory system works by decomposing acoustic signals into abstract objects known as auditory streams. 
For each acoustic source signal that impinge upon the eardrum, e.g. speech or environmental sounds, an auditory stream is produced, which enables the conscious or subconscious mind to focus on every one of them in isolation.
These auditory streams are thought to be produced in two steps \cite{bregman_auditory_1990}. First, a segmentation step decomposes the acoustic time-domain signal into individual units in a T-F domain representation, where it is assumed that each T-F unit primarily originates from a single source. Secondly, these T-F units are grouped into auditory streams based on grouping rules known as sequential grouping or simultaneous grouping.
Sequential grouping assumes T-F units that are similar across time belong to the same source and consequently are grouped into the same auditory stream, whereas simultaneous grouping merge T-F units that share similarities across frequency, e.g. harmonicity or common onsets and offsets.    
Designing algorithms to separate speech signals based on the principles of \ac{ASA}, is referred to as \ac{CASA}  \cite{brown_computational_1994,wang_computational_2006} and multiple techniques have been proposed to solve the single-microphone multi-talker speech separation task (see e.g. \cite{brown_computational_1994,hu_monaural_2004,bach_blind_2005,hu_auditory_2007,hu_tandem_2010,hu_unsupervised_2013,cooke_monaural_2010}).     

For example, in \cite{hu_monaural_2004} a \ac{CASA} system is proposed for co-channel separation of voiced speech using grouping rules based on cross-correlation analysis across cochlear channels (simultaneous grouping) and modulation analysis of cochlear-filter responses across time (temporal grouping). From these grouping rules, a binary mask is generated that is applied to the T-F representation of the mixture signal to separate the target speaker from the interference signal.   
In \cite{hu_unsupervised_2013} a system is proposed for co-channel speech separation of both voiced and unvoiced speech. 
Simultaneous grouping of the voiced parts of the co-channel signal is performed using an iterative pitch tracking algorithm that identifies the dominant fundamental frequency in the mixture. 
Sequential grouping is then formulated as a clustering problem that uses the information from the simultaneous grouping step and cluster T-F units across time that belong to the same speaker. 
Using information about the voiced segments of the co-channel signal the unvoiced parts of the mixture signal are identified using onset/offset analysis and a binary mask for the entire co-channel speech signal is computed and used to separate the speech signals.

Although \ac{CASA} based single-microphone multi-talker speech separation algorithms can be somewhat successful, they suffer from several drawbacks. 
For example, \ac{CASA} is to a large extent based on heuristics and manually-designed grouping and segmentation rules, which might not be optimal. 
Furthermore, these rules are primarily based on speech characteristics and, consequently, are not valid for non-speech signals. 
Finally, these systems typically lack the capability to learn from data.

\subsubsection{Non-Negative Matrix Factorization}
An alternative to the primarily heuristically based \ac{CASA} approaches are the more mathematically founded \ac{NMF} based algorithms \cite{lee_learning_1999}.  
The underlying assumption behind these algorithms is that a non-negative data matrix  $\barbelow{\barbelow{V}} \in \mathbb{R}^{K \times M}$ can be approximately factorized as
\begin{equation}
\barbelow{\barbelow{V}} \approx \barbelow{\barbelow{D}} \barbelow{\barbelow{H}},
\label{eq43}
\end{equation}
where $\barbelow{\barbelow{D}} \in \mathbb{R}^{K \times L}$, $\barbelow{\barbelow{H}} \in \mathbb{R}^{L \times M}$, and $L \ll M$ are non-negative matrices representing a dictionary matrix and an activation matrix, respectively. The dictionary matrix $\barbelow{\barbelow{D}}$ can be seen as a set of basis vectors for the data matrix $\barbelow{\barbelow{V}}$ and the columns of the activation matrix $\barbelow{\barbelow{H}}$ represent how much of each basis vector is needed to represent each column of $\barbelow{\barbelow{V}}$. The dimension $L$ is a tuning parameter that controls the accuracy of the approximation and is specified experimentally. Obviously, if $L=M$, the solution $\barbelow{\barbelow{V}} = \barbelow{\barbelow{D}} \barbelow{\barbelow{H}}$ can always be found, although such a decomposition is of no interest as no compact representation of $\barbelow{\barbelow{V}}$ is found.  
Factorizing $\barbelow{\barbelow{V}}$ into $\barbelow{\barbelow{D}}$ and $\barbelow{\barbelow{H}}$, where $L \ll M$, can be achieved in an iterative fashion \cite{lee_algorithms_2000} by updating 
\begin{equation}
	\barbelow{\barbelow{H}} =\barbelow{\barbelow{H}} \circ \frac{\barbelow{\barbelow{D}}^T\barbelow{\barbelow{V}}}{\barbelow{\barbelow{D}}^T\barbelow{\barbelow{D}}\barbelow{\barbelow{H}}},
	\label{eq44}
\end{equation}	
and
\begin{equation}
	\barbelow{\barbelow{D}} = \barbelow{\barbelow{D}} \circ \frac{\barbelow{\barbelow{V}}\barbelow{\barbelow{H}}^T}{\barbelow{\barbelow{D}}\barbelow{\barbelow{H}}\barbelow{\barbelow{H}}^T },
	\label{eq45}
\end{equation}	
where $\circ$ and $-$ denote element-wise multiplication and division, respectively. In fact, Eqs.\;\eqref{eq44} and \eqref{eq45} represent the solution to the least squares optimization problem given by
\begin{equation}
\begin{aligned}
& \underset{\barbelow{\barbelow{D}},\barbelow{\barbelow{H}}}{\text{minimize}} 
& &  \Vert \barbelow{\barbelow{V}} - \barbelow{\barbelow{D}}\barbelow{\barbelow{H}} \Vert ^2_F   \\
& \text{subject to}
& & \barbelow{\barbelow{D}},\barbelow{\barbelow{H}} \geq 0,
\end{aligned}
\label{eq46}
\end{equation}
where $\lVert \cdot \lVert^2_F$ is the squared Frobenius norm, and by iterating between Eqs.\;\eqref{eq44} and \eqref{eq45}, convergence to the optimal solution of Eq.\;\eqref{eq46} is guaranteed  \cite{lee_algorithms_2000}.

\ac{NMF} is a data-driven technique and signal-dependent dictionaries $\barbelow{\barbelow{D}}$ are generally required \cite{lee_learning_1999}. In the speaker separation case, a dictionary  $\barbelow{\barbelow{D}}_s$, $s=1,\,2,\,\dots,\,S$ is typically used for each speaker (see e.g. \cite{behnke_discovering_2003,schmidt_single-channel_2006,schmidt_nonnegative_2006,virtanen_monaural_2007,smaragdis_convolutive_2007}) and noise source if any is present \cite{thomsen_speaker-dependent_2016}. 
Let $r(k,m)$ and $a_s(k,m)$ denote \ac{STFT} magnitudes for time-frame $m$ and frequency bin $k$ for the noisy speech signal $y[n]$ and clean speech signal $x_s[n]$, from speaker $s$, respectively.    
Furthermore, let $\barbelow{\barbelow{R}} \in \mathbb{R}^{K \times M}$ and $\barbelow{\barbelow{A}}_s \in \mathbb{R}^{K \times M}$ denote spectrogram matrices populated with $r(k,m)$ and $a_s(k,m)$, for suitable ranges of variables $k$, $m$, respectively.  

From a matrix $\barbelow{\barbelow{A}}_s$, a speaker specific dictionary $\barbelow{\barbelow{D}}_s$, for speaker $s$, can be obtained using Eqs.\;\eqref{eq44} and \eqref{eq45} with $\barbelow{\barbelow{V}} = \barbelow{\barbelow{A}}_s$.
%
%
To acquire an estimate $\barbelow{\barbelow{\hat{A}}}_s$ from a mixture signal $\barbelow{\barbelow{R}}$ containing unknown realizations of multiple speakers, and noise sources, Eq.\;\eqref{eq44} is used with Eq.\,\eqref{eq45} being fixed, and with $\barbelow{\barbelow{V}} = \barbelow{\barbelow{R}}$ and $\barbelow{\barbelow{D}} = \barbelow{\barbelow{D}}_s$. When Eq.\;\eqref{eq44} has converged, the corresponding activation matrix $\barbelow{\barbelow{H}}_s$ is used to acquire and estimate of $\barbelow{\barbelow{A}}_s$ as          
\begin{equation}
\barbelow{\barbelow{\hat{A}}}_s = \barbelow{\barbelow{D}}_s \barbelow{\barbelow{H}}_s \;\;\ s=1,\,2,\,\dots,\,S.
\label{eq47}
\end{equation}	
Finally, using the phase of the mixture signal $\barbelow{\barbelow{R}}$, the overlap-add technique \cite{allen_short_1977}, and the \ac{IDFT}, the time-domain signals $\hat{x}_s[n]$, $s=1,\,2,\,\dots,\,S$, are obtained for each separated speaker.

\ac{NMF} is a simple but powerful technique for single-microphone multi-talker speech separation, or speech enhancement for that matter \cite{schmidt_wind_2007,roux_sparse_2015,grais_single_2011}. However, \ac{NMF} has multiple drawbacks. First, \ac{NMF} is a linear model and as such is limited in its model capacity. Second, \ac{NMF} generally requires speaker dependent dictionaries, making \ac{NMF} less suitable for speaker, or noise-type independent applications. Third, as signal-dependent activations $\barbelow{\barbelow{H}}_s$ are required in Eq.\;\eqref{eq47}, it is not straight forward to apply \ac{NMF} for real-time applications where low latency is critical. Finally, due to the computational complexity of the update equations (Eqs.\;\eqref{eq44} and \eqref{eq45}), \ac{NMF} does not scale well to large datasets.

\subsubsection{Generative Models}\label{genmodse}
It is well known that speech signals are highly structured with temporal dynamics on multiple levels \cite{fant_acoustic_1960,deller_discrete-time_1993,schnupp_auditory_2011}. 
For example, at the phone level, i.e. the physical speech sounds, structure exist due to e.g. prosody or physiological variations among humans (e.g. differences in fundamental, and formant frequencies), but also at the phoneme level, speech is structured due to e.g. grammar and language, but also due to phenomena like co-articulation. 

The speech separation algorithms discussed so far, such as \ac{CASA} or harmonic-model based techniques, consider this temporal structure only partially, while \ac{NMF} based algorithms do not take it into account at all.   
These drawbacks, of existing methods not fully utilizing the available information in speech signals, has motivated research in a different class of algorithms based on \acp{HMM}. 
Differently from the previously discussed methods, \acp{HMM} are generative stochastic models, and they have an internal discrete state representation that allow them to learn temporal dynamics of sequential data \cite{bishop_pattern_2006}.  
Specifically, a \ac{HMM} is a finite state machine that changes among a discrete number of states in a synchronous manner.
For each time step, the state of the \acp{HMM}, which is a latent variable, changes from one state to another based on a set of transition probabilities. 
Associated with each state is a set of observed stochastic variables, known as emission probabilities, which are typically represented as a \ac{GMM}. 
The parameters of a \ac{HMM}, i.e. transition and emission probabilities, are typically found such that they maximize a certain likelihood function with respect to a given dataset, 
and although a \ac{HMM} is a generative model, it is generally not used as such for speech processing tasks.
For example, for \ac{ASR}, phoneme-specific \acp{HMM} can be designed to model the distribution over speech signals containing certain phonemes. Then, to recognize an unknown phoneme from a speech signal, the conditional probability of the observed data, given a \ac{HMM}, is evaluated for each of the phoneme-specific \acp{HMM} and the phoneme associated with the \ac{HMM} of largest conditional probability, will be assigned as the phoneme in the speech signal \cite{deller_discrete-time_1993}.

If, however, more than one signal of interest is present in the speech signal, such as in the multi-talker speech separation task, the standard \ac{HMM} framework can be extended into what is known as factorial \acp{HMM} \cite{ghahramani_factorial_1997}. 
These factorial \ac{HMM} allow for more than one latent variable, hence allowing for generative models that can model speech mixtures containing multiple simultaneous sources, which ultimately led to the development of a large number of successful algorithms for single-microphone multi-talker speech separation \cite{roweis_one_2001,ozerov_factorial_2009,ephraim_statistical-model-based_1992,virtanen_speech_2006,hershey_audio-visual_2002,stark_source-filter-based_2011,mysore_non-negative_2010,kristjansson_super-human_2006,hershey_super-human_2010,cooke_monaural_2010,qian_past_2018}. 
In fact, the use of factorial \ac{HMM} led to a major milestone in speech separation research, as a single-microphone two-talker speech separation system was shown to be capable, in a narrow setting, to separate two-talker speech such that a machine, i.e. an \ac{ASR} system, could transcribe the speech signal better than humans \cite{kristjansson_super-human_2006,hershey_super-human_2010}.  

However, although factorial \acp{HMM} showed impressive results in \cite{kristjansson_super-human_2006,hershey_super-human_2010}, they do have some drawbacks. For example, computing the conditional probabilities required during training and test, is intractable \cite{mysore_non-negative_2010} and consequently these probabilities have to be estimated, which in general has a high computational complexity and scales poorly with the number of speakers \cite{qian_past_2018}. 
Also, as factorial \acp{HMM} for multi-talker speech separation require speaker-dependent \acp{HMM} for each speaker in the mixture, these techniques can only be applied in a speaker dependent context, where the identities of the speakers to be separated are known \emph{a priori}. 
This is a limitation that makes the factorial \ac{HMM} framework non-applicable in a range of real-world applications such as automatic meeting transcription or automatic captioning for audio/video recordings, where the identity and number of speakers are generally unknown. 
These limitations also explain why the techniques proposed in \cite{roweis_one_2001,ozerov_factorial_2009,ephraim_statistical-model-based_1992,virtanen_speech_2006,hershey_audio-visual_2002,stark_source-filter-based_2011,mysore_non-negative_2010,kristjansson_super-human_2006,hershey_super-human_2010,qian_past_2018} primarily considered two-talker speech separation of a limited number of \emph{known} speakers.  
Consequently, different techniques are required to enable multi-talker speech separation algorithms to work in such general applications, where only a limited amount of \emph{a priori} knowledge about the environment is available.
Potential candidates that might work in such environments are algorithms based on deep neural networks, which will be presented in detail in Sec.\,\ref{deepsep}. 

\subsection{Evaluation}
As mentioned in Sec.\,\ref{sec:classical_se}, the common goal of many speech enhancement, and multi-talker speech separation, algorithms is to improve either speech quality or intelligibility, of a degraded speech signal.
But how do you accurately evaluate if an algorithm-under-test really does improve one of these quantities?  

In general, the only way to truly evaluate if a speech processing algorithm in fact does improve speech quality or intelligibility, is by a listening test involving the end user, i.e. human test subjects. 
However, listening tests are involved as they require numerous human test subjects and the listening test itself, needs to be carefully planned based on whether the goal is to evaluate speech intelligibility or speech quality. 
Most people probably have an idea about what a good quality speech signal sounds like, and what would make the same signal a bad quality one, e.g. by introducing hiss or crackle sounds to the signal. Nevertheless, speech quality is highly subjective as it is primarily based on emotions and feelings. 
Speech intelligibility, on the other hand, is much more objective, if you will, as emotions and feelings in general do not influence your capability of understanding speech. Either you understand what is being said or you do not. Consequently, designing listening tests that truly evaluate speech quality or intelligibility, is no easy task \cite{deller_discrete-time_1993,loizou_speech_2013}.     

Therefore, to avoid these often tedious and time consuming listening tests, and to get a quick and somewhat accurate estimate of the listening-test result, a set of objective measures have been designed, which are based on mathematical functions that quantify the difference between clean and noisy/processed speech signals in a way that has a high correlation with listening-test results. 
In fact, in some cases, it is more desirable to use an objective measure, instead of a listening test involving human test subjects, as objective measures are fast, cheap, and consistently produce the same result for the same testing condition, whereas listening-test results might vary due to factors such as listener fatigue, or varying hearing ability among test subjects.      

In the following, three of the popular techniques for objective quality and intelligibility evaluation are briefly reviewed.

\subsubsection{Perceptual Evaluation of Speech Quality}
The \ac{PESQ} \cite{rix_perceptual_2001,noauthor_international_1990,noauthor_international_2003,noauthor_international_2005} measure is one of the most widely used objective measures for estimating speech quality \cite{loizou_speech_2013}. 
The \ac{PESQ} measure is designed to approximate the \ac{MOS}, which is a widely used listening test procedure for speech quality evaluation \cite{loizou_speech_2013,deller_discrete-time_1993,noauthor_international_1990,ieee_ieee_1969}.  
The \ac{MOS} is a very simple evaluation procedure, where the test subjects are asked to grade the speech signal they are hearing based on a scale with five discrete steps, with "$1$" representing a bad and very annoying sound quality, and "$5$" representing an excellent sound quality with imperceptible distortions. 
The final \ac{MOS} score, which is a single scalar between "$1$" and "$5$", is simply the average, or mean, of all the "opinion scores" for each test signal and for all test subjects, hence the name, mean opinion score.   
As mentioned, the \ac{PESQ} measure approximates \ac{MOS}, but the \ac{PESQ} algorithm is fairly complex as it consists of multiple steps involving pre-processing, time alignment, perceptual filtering, masking effects, etc. (see e.g. \cite[pp. 491-503]{loizou_speech_2013}). 
Nevertheless, \ac{PESQ} versions P.862.1/2 \cite{noauthor_international_2003,noauthor_international_2005} produce a number ranging from approximately $1$ to $4.5$, which allow comparisons between \ac{PESQ} and \ac{MOS}, and \ac{PESQ} has been found to be highly correlated with listening-test experiments based on \ac{MOS} \cite{rix_perceptual_2001,noauthor_international_2003}. 
In fact, although \ac{PESQ} was originally designed for evaluating speech coding algorithms, it was later shown that \ac{PESQ} correlated reasonably well with the quality of speech processed by commonly used speech enhancement algorithms \cite{hu_evaluation_2008}.
Also, \ac{PESQ} requires both the clean speech signal as well as the noisy/processed signal to estimate the perceived quality of the noisy/processed signal. This makes \ac{PESQ} an intrusive speech quality estimator, which limits its use to situations where the clean undistorted signal is available in isolation. For most applications of \ac{PESQ}, this is not a real limitation as \ac{PESQ} is usually used in laboratory conditions, where the clean signal is often available in isolation.

\subsubsection{Short-Time Objective Intelligibility} 
The \ac{STOI} \cite{taal_short-time_2010,taal_algorithm_2011} is, today, perhaps, the most widely used objective measure for estimating speech intelligibility.
Differently from \ac{PESQ}, \ac{STOI} is not designed to approximate any specific type of listening test, but merely designed to correlate well with listening tests evaluating speech intelligibility in general.  
Since intelligibility is binary in the sense that, either a given speech signal, say a word, is understood or it is not, listening-test results representing speech intelligibility can, most often, be quantified as a number between $0$ and $1$ that represents the percentage of words correctly understood \cite{loizou_speech_2013}.
To be comparable with such tests, \ac{STOI} is designed to produce a single scalar output in a similar range%
\footnote{In theory, \ac{STOI} can produce numbers in the interval $(-1,1)$, since \ac{STOI} is based on a correlation coefficient measure. However, in practice, negative numbers are rarely observed.}, with an output of $1$ indicating fully intelligible speech.  

Similarly to \ac{PESQ}, \ac{STOI} is an intrusive algorithm as it requires both the clean signal and the noise/processed signal in isolation.
Furthermore, \ac{STOI} is based on the assumption that modulation frequencies play an important role in speech intelligibility, and that all frequency bands in the cochlear filter are equally important. 
These are assumptions, which, to a certain degree, are justified empirically \cite{elliott_modulation_2009,schnupp_auditory_2011,drullman_effect_1994}. 
This also has the consequence that, compared to \ac{PESQ}, \ac{STOI} is a fairly simple algorithm.

Despite its simple formulation, \ac{STOI} has been found to be able to quite accurately predict the intelligibility of noisy/processed speech in a wide range of acoustic scenarios \cite{taal_algorithm_2011,jorgensen_speech_2015,falk_objective_2015,jensen_speech_2014,xia_evaluation_2012}.
Finally, an extension to \ac{STOI}, known as \ac{ESTOI}, has been proposed as a more accurate speech intelligibility predictor in the special cases where the noise sources are highly modulated \cite{jensen_algorithm_2016}.

\subsubsection{Blind Source Separation Evaluation}\label{secsdr}
When evaluating single-target signal speech processing algorithms, such as speech enhancement,  \ac{PESQ} and \ac{STOI} are useful, as these measures quantify how successful the algorithm-under-test process a degraded signal in a way that is perceptually desirable. 
If, however, multiple target signals exist, such as in a speech separation task, additional information about the processing artifacts might be desirable compared to what \ac{PESQ} and \ac{STOI} can provide \cite{vincent_first_2007,emiya_subjective_2011}. 
In other words, when a mixture signal that contains multiple speech and noise signals are processed by a speech separation algorithm, the enhanced or separated speakers might contain artifacts originating from multiple different sources. For example, these artifacts could originate from the noise signal itself, from processing artifacts, or due to "cross-talk", i.e. signal components from one target speaker appearing in the separated signal of the other.  

One of the most popular objective measures for evaluating speech separation algorithms that take these considerations into account, is the \ac{BSS} Eval toolkit \cite{vincent_performance_2006}. In the technique proposed in \cite{vincent_performance_2006}, the separated signals are decomposed into target-speaker components and three noise components known as interference, noise, and artifact. 
The interference component represents cross-talk from other target speakers. Noise and artifacts, represents environmental noise sources and processing artifacts, respectively. From this decomposition, energy-ratio measures are defined known as \ac{SDR}, \ac{SIR}, \ac{SAR}, and SNR, which each relate these decomposed elements of the separated signal in a way that provide useful information about the contribution of each of them. 
Finally, it has been found that these objective measures correlate well with listening test evaluating quality \cite{ward_bss_2018,cano_evaluation_2016}, and, obviously, the \ac{BSS} Eval toolkit only compliments other objective measures such as \ac{STOI} and \ac{PESQ}.

\section{Deep Learning}\label{deeplearn}
Section \ref{sec:classical_se} reviewed classical techniques that have been proposed to solve the single-microphone speech enhancement and single-microphone multi-talker speech separation tasks. 
Although, these techniques are very different, and try to solve different tasks, most of them rely heavily on one key component: \emph{domain knowledge}. 
For example, the Wiener filters and \ac{STSA}-\ac{MMSE} estimators reviewed in Sec.\,\ref{classe} are designed based on the assumption that speech and noise have different statistical characteristics that are governed by basic probability distributions. 
Similarly, techniques reviewed in Sec.\,\ref{classep} for multi-talker speech separation, such as the \ac{CASA}-based techniques, rely on detailed knowledge about the human auditory and speech production systems. 
Obviously, domain knowledge is extremely helpful when it is correct, and it has, and still is, used to solve many engineering problems. 
However, for complex tasks, utilizing domain knowledge might not be easy and it might even be destructive if the wrong assumptions are used. 
In other words, domain knowledge is useful for tasks that are well understood by the human engineers working on them. 
Instead, for complex tasks, it might lead to more successful solutions if the solution is learned, e.g. using a reinforcement strategy similarly to what is used in nature \cite{schultz_neural_1997}.
This philosophy, \emph{learning} the solution instead of designing it, is one of the defining principles in the deep learning paradigm \cite{bengio_learning_2009,deng_deep_2014,lecun_deep_2015} and combined with the other defining principle, \emph{depth}, it led to the deep learning revolution we know today, which is the topic of this section.    

\subsection{The Deep Learning Revolution}\label{deeplearnrev}
We are currently experiencing a deep learning revolution \cite{lecun_deep_2015,sejnowski_deep_2018}, and although \emph{deep learning} as an everyday term is less than a decade old, some of the fundamental principles used by deep learning algorithms today, dates back more than half a century \cite{lippmann_introduction_1987}.        
In fact, the first successful application of the prototypical learning model used in deep learning, the \acf{ANN} (to be introduced in Sec.\,\ref{sec:fnn}), was achieved by Frank Rosenblatt in the late 1950's with his \emph{perceptron} learning model \cite{rosenblatt_perceptron:_1958,rosenblatt_principles_1961}, and shortly after by Bernard Widrow with a model known as ADALINE \cite{widrow_adaptive_1960}. 
These models, heavily inspired by psychology and neuroscience \cite{mcculloch_logical_1943,hebb_organization_1949}, were designed in an attempt to model the neural networks comprising the human brain, hence the name \ac{ANN}. 

However, although these models were capable of solving simple pattern recognition tasks, a decade later it was proven by Marvin Minsky and Seymour Papert \cite{minsky_perceptrons:_1969,olazaran_sociological_1996} that perceptrons were, in fact, inherently limited due to their linear input-output mapping and could indeed solve only very simple tasks.   
Interestingly, it was already known at the time that a simple way to alleviate the limitations of the perceptron was to change the linear mapping to a more complex non-linear mapping by simply stacking multiple perceptrons into models known as \acp{MLP}. These more advanced, and deeper, \ac{ANN} models were much more capable in terms of representational capacity, and could potentially solve extremely complex problems \cite{minsky_perceptrons:_1969,lippmann_introduction_1987,olazaran_sociological_1996}. 
Unfortunately, at the time, training these \acp{MLP} was not feasible, partly due to lack of sufficient computational resources, but primarily due to lack of successful training algorithms.

It took almost two decades before an efficient algorithm for training \acp{MLP} were developed, which was popularized as \emph{back-propagation} by Rumelhart \emph{et al.} in 1986 \cite{rumelhart_learning_1986}%
\footnote{Although Rumelhart \emph{et al.} \cite{rumelhart_learning_1986} coined the term back-propagation, some argue (see e.g. \cite{schmidhuber_deep_2015-1,olazaran_sociological_1996}) that they did not invent the algorithm, they simply popularized it.}.
Not long after the invention of back-propagation for training \acp{MLP}, theoretical results were published, proving that \acp{MLP} can approximate practically any function, with any desired accuracy. These theoretical results are known as the \emph{universal approximation theorem} for MLPs \cite{hornik_multilayer_1989,cybenko_approximation_1989,hornik_approximation_1991}. 
The results were encouraging as they settled some of the speculations about the lack of potential of \acp{MLP} put forward by Minsky and Papert \cite{minsky_perceptrons:_1969,olazaran_sociological_1996} and proved, once and for all, that \acp{MLP} indeed did have the potential to solve complex tasks.

With the awareness of the back-propagation algorithm \cite{rumelhart_learning_1986} and the universal approximation theorem \cite{hornik_multilayer_1989,cybenko_approximation_1989,hornik_approximation_1991}, a natural question arises: why did it take two decades from the mid 1980s to the mid 2000s, before \acp{MLP}, or \acp{ANN} in general, became popular and practically applicable?
The short answer to this question is: due to lack of labeled data and computational resources \cite{halevy_unreasonable_2009,shazeer_outrageously_2017,goodfellow_deep_2016}.  
It was, however, a technique known as \emph{unsupervised pre-training} that ignited the deep learning revolution in 2006 with two seminal papers by Hinton \emph{et al.} \cite{hinton_reducing_2006,hinton_fast_2006}.

At the time, it was a general misconception that the optimization of \acp{MLP}, or \acp{DNN} as they are usually called today, got trapped in poor local minima, hence preventing \acp{DNN} with multiple non-linear layers to be efficiently trained \cite{bengio_greedy_2007}. 
In an attempt to alleviate this presumed challenge it was proposed to initialize the parameters of the \acp{DNN}, before back-propagation training, with the parameters of a generative model, known as a \ac{DBN} \cite{hinton_fast_2006,bengio_greedy_2007}.
The intuition behind this was that if you consider two random variables $X$ and $Y$, and wish to learn $P(Y|X)$ it might be useful to first learn $P(X)$ using a generative model.
\acp{DBN} are generative models constructed by stacking multiple \acp{RBM}, and then trained unsupervised using unlabeled data to model $P(X)$. RBMs are themselves generative models and belong to a broader class of undirected probabilistic graphical models, known as \acp{MRF} \cite{bishop_pattern_2006}. Inference in \acp{MRF}, however, is challenging as it requires the evaluation of a, generally intractable, partition function. 
However, Hinton \emph{et al.} showed in \cite{hinton_fast_2006} that \acp{RBM} can be combined into a \ac{DBN} and trained efficiently in a greedy layer-wise fashion using an approximate inference algorithm known as contrastive divergence \cite{hinton_training_2002}. The parameters of this \ac{DBN}, which is modeling $P(X)$ can then be used to initialize a \ac{DNN}, which is then "fine-tuned" using the traditional supervised back-propagation technique to model $P(Y|X)$.             

In the seminal paper \cite{hinton_fast_2006} Hinton \emph{et al.} showed that \acp{DNN} initialized with unsupervised pre-training and refined with supervised back-propagation \linebreak training, could achieve state-of-the-art results on a hand-written digits recognition task.
The results attracted a huge amount of attention from the academic community and ultimately sparked the renewed interest in \acp{DNN}.
It was, however, later recognized (see e.g. \cite{choromanska_loss_2014,kawaguchi_deep_2016,lecun_deep_2015,hinton_deep_2012,goodfellow_deep_2016,yu_roles_2010,nair_rectified_2010,ma_deep_2015})  that poor local minima in general was not a problem when training \acp{DNN} and similar or even better performance could be achieved without using unsupervised pre-training, especially for large labeled dataset. Consequently, today, unsupervised pre-training is a technique that is rarely used, but its influence and impact on the scientific field of \acp{DNN}, as a catalyst for \ac{DNN} research, cannot be overstated.   

Over the last decade, deep learning has truly revolutionized both academia and industry.
For example, deep learning technology has facilitated the development of algorithms that are close to, or even exceeding, human-level performance within multiple scientific disciplines such as automatic speech recognition \cite{amodei_deep_2015,xiong_achieving_2016,yu_recent_2017,saon_english_2017}, object recognition \cite{he_delving_2015}, face recognition \cite{taigman_deepface:_2014}, lip reading \cite{chung_lip_2017}, board and computer games \cite{silver_mastering_2016,silver_mastering_2017,moravcik_deepstack:_2017,mnih_human-level_2015}, and in healthcare applications  \cite{miotto_deep_2017,oakden-rayner_precision_2017} especially for cancer detection  \cite{esteva_dermatologist-level_2017,wu_breast_2017,liu_detecting_2017,bychkov_deep_2018,wang_searching_2017}. 

Furthermore, today, deep learning is the key technology of many companies, and although the deep learning revolution was initiated by Hinton \emph{et al.}, it was, in fact, the increase in low-cost computational resources made available by the general-purpose graphics processing unit \cite{noauthor_nvidia_nodate,krizhevsky_imagenet_2012,goodfellow_deep_2016} that really facilitated the success of deep learning and allowed \acp{DNN} to be applied on an industrial scale.     
For example, Facebook currently uses \acp{DNN} to predict and analyze user behavior 200 trillion, i.e. $200\times10^{12}$, times each day, something that was practically impossible just a decade ago \cite{noauthor_viva_2018,hazelwood_applied_2018}. 

Finally, in a recent study \cite{pricewaterhousecoopers_pwcs_2017} by PricewaterhouseCoopers, it is estimated that deep learning will contribute \$15.7 trillion to the global economy in 2030, which is more than the current output of China and India combined. These contributions will be within a wide range of sectors such as health care, automotive, financial services, transportation, logistics, retail, energy, and manufacturing, which also justifies why deep learning driven technology is believed to lead to \emph{the fourth industrial revolution} \cite{schwab_fourth_2015,noauthor_andrew_2017}. 

\subsection{Feed-Forward Neural Networks}\label{sec:fnn}
Sections \ref{deeplearn} and \ref{deeplearnrev} introduced deep learning without defining exactly what a \ac{DNN} is. In this section, and sections to come, we will introduce three of the most popular DNN models: \acp{FNN} (Sec.\,\ref{sec:fnn}), \acp{RNN} (Sec.\,\ref{sec:rnn}), and \acp{CNN} (Sec.\,\ref{sec:cnn}) \cite{goodfellow_deep_2016}. 

A \ac{FNN} is a machine learning model and is represented as a parameterized function given by    
\begin{equation}
\barbelow{\hat{o}} = f(\barbelow{y},\theta),
\label{eq201}
\end{equation}   
where $\barbelow{y}$ is an input vector, $\theta$ is a set of parameters and $\barbelow{\hat{o}}$ is the \ac{FNN} output, i.e. the map of $\barbelow{y}$ by $f(\cdot,\cdot)$. 
The most basic \ac{FNN} is a single-layer \ac{FNN} given by
\begin{equation}
\barbelow{\hat{o}} = f^{ \{1\} }(\barbelow{y},\theta) =   \phi ( \barbelow{\barbelow{W}} \barbelow{y} +  \barbelow{b}), \;\;\; \theta = \{\barbelow{\barbelow{W}}, \barbelow{b}\},
\label{eq202}
\end{equation}   
where $\barbelow{\barbelow{W}}$ and $\barbelow{b}$ are the parameters and $\phi (\cdot)$ is a, generally non-linear, function known as the activation function.
The vector $\barbelow{b}$ is known as the bias vector and allows the \ac{FNN} to apply an affine transformation to $\barbelow{y}$. %
The prefix \emph{feed-forward} in \acp{FNN} comes from the fact that information only flows in one direction in the model in Eq.\,\eqref{eq202}, i.e. there are no recurrent connections.  
Furthermore, if $\phi (\cdot)$ is a binary thresholding function, Eq.\,\eqref{eq202} resembles the perceptron \cite{rosenblatt_perceptron:_1958} or ADALINE models \cite{widrow_adaptive_1960} and as shown by Minsky and Papert \cite{minsky_perceptrons:_1969}, these models are inherently limited as the input to the binary thresholding function is a purely linear transformation of the input $\barbelow{y}$.
Instead, if these models are stacked as
\begin{equation}
\barbelow{\hat{o}} = f^{ \{L\} }( \dots f^{ \{2\} }(f^{ \{1\} }(\barbelow{y},\theta_1),\theta_2) \dots, \theta_L)  
\label{eq203}
\end{equation}   
they form \acp{MLP} or multi-layer \acp{FNN}, which according to the universal approximation theorems (see e.g. \cite{hornik_multilayer_1989,cybenko_approximation_1989,hornik_approximation_1991}) can model practically any function. 
In fact, $L=2$ is sufficient for the universal approximation theorem to apply, although it might require an exponentially wide network, i.e. number of rows in $\barbelow{\barbelow{W}}$, to approximate a certain function with a given accuracy. 
However, as the number of layers $L$ increases, the compositional structure allow multi-layer \acp{FNN} to construct exponentially more complex decision boundaries, with the same number of parameters, than \acp{FNN} with $L=2$ \cite{rojas_networks_2003,pascanu_number_2014,montufar_number_2014}.         
The fact that \acp{FNN} gets exponentially more efficient, with respect to the parameters, as the number of layers increase, is exactly what drives the deep learning research community for increasingly deeper networks, and although deep networks are not trivial to train, modern deep learning models have been trained successfully with more than 1000 layers \cite{huang_deep_2016,he_deep_2016}.

Similarly to the machine learning methods presented in Sec.\,\ref{sec:ml_cl_se}, the optimal parameters of a \ac{FNN} are typically given as the solution to an optimization problem on the form
\begin{equation}
\begin{aligned}
\theta^\ast = \; & \underset{\theta}{\text{argmin}}
& & \sum_{\mathcal{D}_{train}} \mathcal{J}(f^{ \{L\} }(\barbelow{y},\theta),\barbelow{o}), \;\;\;  (\barbelow{y},\barbelow{o}) \in \mathcal{D}_{train}, \\
\end{aligned}
\label{eq204}
\end{equation} 
where $\mathcal{J}(\cdot,\cdot) \in \mathbb{R}$ is a non-negative cost function (e.g. mean squared error), $f^{ \{L\} }(\barbelow{y},\theta)$ represents a FNN with $L$ layers, $\theta = \{ \theta_1, \theta_2, \dots, \theta_L \}$ is the set of parameters, and $(\barbelow{y},\barbelow{o})$ is an ordered pair, of input signals $\barbelow{y}$ and corresponding targets $\barbelow{o}$, from a training dataset $\mathcal{D}_{train}$. The objective is then to find parameters $\theta$ such that $\barbelow{\hat{o}} = f^{ \{L\} }(\barbelow{y},\theta) \approx \barbelow{o}$.  
However, until 1986 a solution to Eq.\,\eqref{eq204} was only known for $L=1$, but when Rumelhart \emph{et al.} \cite{rumelhart_learning_1986} proposed the back-propagation algorithm, Eq.\,\eqref{eq204} could be solved, in theory, for any $L$ using the optimization method gradient descent (e.g. \cite{ruder_overview_2016,smith_dont_2017,wilson_marginal_2017}). 

Specifically, Rumelhart \emph{et al.} proposed to update any weight $w$ in a multi-layer FNN using gradient descent defined as
\begin{equation}
w^{(n+1)} = w^{(n)} - \mu \frac{\partial \mathcal{J}}{ \partial w^{(n)} },
\label{eq207}
\end{equation} 
where $\mu$ is the learning rate, $w^{(n)}$ is a parameter in an arbitrary layer at iteration $n$, and $\frac{\partial \mathcal{J}}{ \partial w^{(n)} }$ is the partial derivative of the cost function $\mathcal{J}$ with respect to the weight. 
Rumelhart \emph{et al.} showed in \cite{rumelhart_learning_1986} that $\frac{\partial \mathcal{J}}{ \partial w^{(n)} }$ can be straightforwardly decomposed, using the chain rule of differentiation, into a chain of products of simple partial derivatives that, except for the activation function, only involved differentiation of linear functions.  
In fact, Rumelhart \emph{et al.} showed that if the activation function was given by the sigmoid function defined as
\begin{equation}
\phi(x) = \frac{1}{1+e^{-x}}, \;\;\; x \in \mathbb{R}, \;\;\; \phi(x) \in (0,1), 
\label{eq211}
\end{equation}
where 
\begin{equation}
\frac{\partial \phi(x)}{\partial x} = \phi(x)(1-\phi(x)),
\label{eq212}
\end{equation}
this chain of partial derivatives could easily and efficiently be evaluated for all parameters in a multi-layer FNN, consequently, allowing multi-layer FNNs to be trained successfully. 

However, although the back-propagation technique enabled multi-layer FNNs, with sigmoid activation functions, to be efficiently trained, the sigmoid function, at the same time, prohibited multi-layer FNNs with more than a few layers to be trained due to a phenomena known as the \emph{vanishing gradient problem}. 
The vanishing gradient problem occurs because $\frac{\partial \mathcal{J}}{ \partial w }$ is decomposed into a chain of products where $\frac{\partial \phi(x)}{\partial x}$ appears once for each layer and since $\frac{\partial \phi(x)}{\partial x} \leq 0.25$, the partial derivative $\frac{\partial \mathcal{J}}{ \partial w }$ progressively gets smaller and smaller as the number of layers increase, i.e., the gradient vanishes for FNNs with a large number of layers.

A way to alleviate this problem is to use an activation function whose first derivative is close to one. Such an activation function is the rectified linear unit\,(ReLU) \cite{nair_rectified_2010}, which is simply a half-wave rectifier given as $max(x,0)$. 
The ReLU has the very simple first sub-derivatives given as $ \frac{\partial }{ \partial w } max(x,0) = 0$ for $x<0$ and $\frac{\partial}{ \partial w } max(x,0) = 1$ for $x>0$, which effectively reduces the vanishing gradient problem. 
In fact, using the ReLU instead of the sigmoid activation function is one of the key differences that enable deep FNNs, with hundreds of layers and billions of parameters, to be trained successfully with back-propagation and without unsupervised pre-training \cite{glorot_deep_2011,zeiler_rectified_2013,lecun_deep_2015,he_delving_2015,he_deep_2016,shazeer_outrageously_2017}.

\subsection{Recurrent Neural Networks}\label{sec:rnn}
We now turn our attention to a \ac{DNN} architecture known as a \acf{RNN}. In Sec.\,\ref{sec:fnn} we saw that a \ac{FNN} is a universal function approximator and one can ask why another architecture is needed if a FNN can approximate any function. The answer is at least twofold. 
First, the universal approximation theorem of \acp{FNN} says only something about the representational capacity of \acp{FNN} with an unlimited number of parameters. It does not, however, say anything about the \ac{FNN} topology, i.e. the number of layers and the number of units per layer. 
Secondly, optimization of \acp{FNN}, and \acp{DNN} in general, is a non-convex problem and generally no guaranties exist in terms of convergence and optimality, when the parameters are found using gradient descent and back-propagation \cite{boyd_convex_2004,goodfellow_deep_2016}. 
Therefore, some \ac{DNN} architectures might be more efficient in terms of parameters, or superior in terms of performance, compared to \acp{FNN} trained using gradient descent and back-propagation, and indeed, such architectures exist. Two such popular architectures are the \ac{RNN}, to be described in this section, and the \ac{CNN}, which will be the topic of the next section.   

Specifically, the basic single-layer \ac{RNN} architecture is given as \cite{rumelhart_learning_1986-1}
\begin{equation}
	\begin{aligned}
		\barbelow{h}^{(n)} = \;& \phi (\barbelow{\barbelow{W}} \barbelow{x}^{(n)} + \barbelow{\barbelow{V}} \barbelow{h}^{(n-1)} +\barbelow{b} ), \\
	\end{aligned}
	\label{eq213}
\end{equation}
where $\phi(\cdot)$ is an activation function, $\barbelow{\barbelow{W}}$, and $\barbelow{\barbelow{V}}$ are parameter matrices, $\barbelow{b}$ is a bias vector, and $\barbelow{h}^{(n)}$ is the output of the RNN at time index $n$. From Eq.\,\eqref{eq213} it is seen that the RNN architecture operates with a time index $(n)$, and differently from the FNN architecture (Eq.\,\eqref{eq202}) the RNN has a recurrent connection that is shared between time-steps, hence the name RNN. 
This time index and weight sharing property, is exactly what differentiates RNNs from FNNs and what allows RNNs to be efficient models of sequential data with strong temporal structure, such as speech. 
Also, similarly to FNN architectures, RNNs can be stacked into deep RNNs, hence increasing the model capacity \cite{pascanu_how_2013}, and it can even be shown that RNNs can model any dynamical system with any required accuracy, which is known as the universal approximation theorem for RNNs  \cite{schafer_recurrent_2006}. 
Furthermore, as a consequence of the recurrent connection, for any finite $n$, i.e. $n=1,2,\dots,N$, a FNN architecture exists that has the exact same behavior as a RNN \cite{rumelhart_learning_1986-1}.   
In fact, this is exactly what is utilized when training RNNs using the back-propagation-through-time technique. First, the RNN is converted, or unrolled in time, into an $N$-layer FNN, and then, it is trained using the standard back-propagation technique as described in Sec.\,\ref{sec:fnn}. \cite{goodfellow_deep_2016}. 

However, similarly to the FNNs, if the network is deep, i.e. if $N$ is large, training usually fails due to the vanishing gradient problem, which, in practice, prohibits the use of RNNs for signals with long time dependencies \cite{pascanu_difficulty_2013}. 
Also, since the recurrent weights are shared across time, the vanishing gradient problem is even more severe for RNNs, as the weights are constant for all time steps. This is especially true in the case where $\lVert \barbelow{\barbelow{V}} \lVert < 1$, i.e. the matrix norm of $\barbelow{\barbelow{V}}$ is less than one.
A phenomena known as exploding gradients also exists, which can occur in the case when $\lVert \barbelow{\barbelow{V}} \lVert > 1$, although this is usually handled simply by clipping the gradients \cite{goodfellow_deep_2016}. 

Several techniques have been proposed to alleviate the vanishing gradient problem. 
For example, by constraining the recurrent weight matrix to be orthogonal the vanishing gradient problem can be reduced (see e.g. \cite{arjovsky_unitary_2015,wisdom_full-capacity_2016,henaff_orthogonal_2016,jing_tunable_2017,vorontsov_orthogonality_2017}). 
A different approach to avoid the vanishing gradient problem is to change the \ac{RNN} architecture into what is known as gated-\acp{RNN}  \cite{goodfellow_deep_2016,zhang_architectural_2016,zhou_minimal_2016-1,cho_properties_2014,cho_learning_2014,hochreiter_long_1997,greff_lstm:_2017}, and the most popular of these gated-\acp{RNN} is the \ac{LSTM}-RNN \cite{hochreiter_long_1997,greff_lstm:_2017} given by  
\begin{equation}
\begin{aligned}
\barbelow{i}^{(n)} = \;& \sigma \left( 	\barbelow{\barbelow{W}}_i \barbelow{x}^{(n)} + \barbelow{\barbelow{V}}_i \barbelow{h}^{(n-1)} + \barbelow{b}_i \right), \\
\barbelow{f}^{(n)} = \;& \sigma\left( 	\barbelow{\barbelow{W}}_f \barbelow{x}^{(n)} + \barbelow{\barbelow{V}}_f \barbelow{h}^{(n-1)} + \barbelow{b}_f \right), \\
\barbelow{o}^{(n)} = \;& \sigma\left( 	\barbelow{\barbelow{W}}_o \barbelow{x}^{(n)} + \barbelow{\barbelow{V}}_o \barbelow{h}^{(n-1)} + \barbelow{b}_o \right), \\
\barbelow{d}^{(n)} = \;& \tanh\left( 		\barbelow{\barbelow{W}}_c \barbelow{x}^{(n)} + \barbelow{\barbelow{V}}_c \barbelow{h}^{(n-1)} + \barbelow{b}_c \right), \\
\barbelow{c}^{(n)} = \;& \barbelow{f}^{(n)} \circ \barbelow{c}^{(n-1)}  + \barbelow{i}^{(n)} \circ \barbelow{d}^{(n)}, \\
\barbelow{h}^{(n)} = \;& \barbelow{o}^{(n)} \circ \tanh\left( \barbelow{c}^{(n)} \right),
\end{aligned}
\label{eq214}
\end{equation}
where $\barbelow{x}^{(n)}$ is the input, $\sigma(\cdot)$ and $\tanh(\cdot)$ denote the sigmoid and hyperbolic tangent functions, "$\circ$" denotes element-wise multiplication, and the subscripts "$\cdot_i$", "$\cdot_f$", "$\cdot_o$", and "$\cdot_c$", with an abuse of notation, denote the parameters associated with the input gate, forget gate, output gate, and the cell state, respectively  \cite{hochreiter_long_1997,greff_lstm:_2017}.  
The \ac{LSTM} architecture minimizes the vanishing gradient problem primarily due to the algorithm step $\barbelow{c}^{(n)} = \barbelow{f}^{(n)} \circ \barbelow{c}^{(n-1)}  + \barbelow{i}^{(n)} \circ \barbelow{d}^{(n)}$, which is known as the cell state. The cell state has a recurrent connection to itself with no activation function. Furthermore, the value of the cell state is controlled by "gates" that only act multiplicative on the cell state, hence, controlling the flow of information into the cell and out of the cell. 
Since the cell state has no activation function and no weight directly associated with it, its value will remain constant during gradient updates hence, avoiding the vanishing gradient problem. 
However, although the \ac{LSTM} given by Eq.\,\eqref{eq214}, is more computational complex compared to the basic \ac{RNN} given by Eq.\,\eqref{eq213}, it is far easier to train, and works better in practice when $N$ is large \cite{chung_empirical_2014,collins_capacity_2017}. Consequently, the \ac{LSTM}-\ac{RNN} architecture is currently the most popular \ac{RNN} architecture used for speech processing applications such as speech recognition, enhancement, and separation \cite{xiong_achieving_2016,hannun_deep_2014,amodei_deep_2015,wang_supervised_2017,zhang_advanced_2017,yu_automatic_2015,qian_past_2018,weninger_single-channel_2014,weninger_speech_2015}.

\subsection{Convolutional Neural Networks}\label{sec:cnn}
Similarly to the \ac{RNN}, the \acf{CNN} is an architecture that utilizes weight-sharing, but compared to the \ac{RNN}, in a fundamentally different way \cite{atlas_artificial_1987,lecun_backpropagation_1989,lecun_generalization_1989}.  
As seen from Eqs.\,\eqref{eq213} and \eqref{eq214}, for \ac{RNN} architectures the same weight matrix is shared for each time step.
This configuration is known as \emph{tied weights} as the connections between weights and inputs are constant.
Although this configuration leads to powerful models due to the universal approximation theorem, it also leads to computational demanding networks due to the dense matrix vector multiplications. 
If, however, a large number of weights are redundant, i.e. taking similar values, due to a certain general structure in the data, it is more efficient to reuse these parameters instead of having them stored in multiple different locations in a large matrix.
This is exactly the principle behind the \ac{CNN} architecture. \acp{CNN} use \emph{untied weights} that are shared for multiple inputs. Mathematically, this corresponds to the convolution between the input signal and the parameters, hence the name \ac{CNN}.   

For example, in the two-dimensional case, the convolution%
\footnote{Although this is technically speaking the cross-correlation, since we use positive increments and not negative, we follow the convention and refer to it as convolution \cite{goodfellow_deep_2016}. } 
between a signal matrix $\barbelow{\barbelow{X}} \in \mathbb{R}^{J \times I}$ and a parameter matrix $\barbelow{\barbelow{W}} \in \mathbb{R}^{M \times N}$, where $M < J$ and $N < I$, is simply given as
\begin{equation}
\begin{aligned}
s_{j,i} = \sum_{m=1}^{M} \sum_{n=1}^{N} x_{j+m,i+n} w_{m,n},
\end{aligned}
\label{eq215}
\end{equation}
where $x_{j,i}$ and $w_{m,n}$ denote entries $j,i$, and $m,n$ of $\barbelow{\barbelow{X}}$ and $\barbelow{\barbelow{W}}$, respectively.

Similarly to the \ac{FNN} and \ac{RNN} architectures, a \ac{CNN} also consists of multiple layers of non-linear mappings, where each layer is based on a non-linear activation function. 
Differently from the \acp{FNN} and \acp{RNN}, for \acp{CNN} the input to the activation function is the convolution between the input to the layer and the set of layer-specific parameters $\barbelow{\barbelow{W}}$, usually called filters or kernels. This is fundamentally different from the dense matrix vector product used by the \ac{FNN} and \ac{RNN} architectures.

Usually each layer consists of multiple kernels that are all convolved with the same input, hence producing a larger number of outputs than inputs, known as feature maps. To reduce the memory complexity, usually a stride larger than one is used, which means that $j,i$ are incremented with step sizes larger than one. Another step usually applied is known as pooling, where each feature map is down-sampled with a certain factor. This pooling step adds translational invariance to the \ac{CNN}, which usually is a desirable quality, while at the same time reducing the memory requirement. If the \ac{CNN} is used for classification, a \ac{FNN} is usually used as the output layer.    

Since the number of parameters is defined by $M$ and $N$ and not by $J$ and $I$, as with \acp{FNN} and \acp{RNN}, and since $M\ll J$ and $N\ll I$ for many practical systems, \acp{CNN} can potentially require far less parameters for the same performance, i.e. \acp{CNN} can be more parameter-efficient compared to \acp{FNN} and \acp{RNN} \cite{goodfellow_deep_2016}.   
This is especially true for applications involving natural images where features such as edges usually contain more information about the content of the image than solid-color regions do. In such cases the kernels  $\barbelow{\barbelow{W}}$ can easily extract such information with a low number of parameters using e.g. a $3 \times 3$ edge detector (e.g Sobel kernel) \cite{duda_pattern_2001}. A \ac{FNN} on the other hand, would potentially require several orders of magnitude more parameters to apply the same operation as the operation should be a matrix vector product.     
In fact, this is what makes \acp{CNN} a very powerful model for natural images. 

It has even been shown that \acp{CNN} trained on large datasets, containing natural images, learn very specific and intuitive kernels at each layer \cite{krizhevsky_imagenet_2012,zeiler_visualizing_2014}. At the first layer the kernels resemble simple edge detectors. At the next layer combinations of edge-detectors are combined into more abstract, although distinct, objects and at even higher layers these abstract objects become identifiable as the different target classes in the dataset, such as animals, persons, etc \cite{krizhevsky_imagenet_2012,zeiler_visualizing_2014}.       

Finally, since the convolution in Eq.\,\eqref{eq215} is differentiable, \acp{CNN} can similarly to the \ac{FNN} and \ac{RNN} architectures be trained efficiently using the back-propagation technique, and today, \acp{CNN} are by far the most successful \ac{DNN} architecture for image applications (see e.g. \cite{rawat_deep_2017,gu_recent_2018,caron_deep_2018}).

\section{Deep Learning for Enhancement and Separation}\label{deepsesep}
So far, classical non-deep learning-based methods for single-microphone \linebreak speech enhancement have not been able to improve speech intelligibility of noisy speech in realistic scenarios (see Sec.\;\ref{sec:classical_se}). 
Similarly, classical non-deep learning-based methods for single-microphone multi-talker speech separation have so far not been able to successfully separate an audio signal consisting of multiple speech signals into individual speech signals without prior knowledge about the speakers. 
With the emergence of deep learning some of the challenges faced by previous techniques can now be overcome.  
In this section, we will review some of these deep learning-based techniques for single-microphone speech enhancement and single-microphone multi-talker speech separation. 

\subsection{Deep Learning Based Speech Enhancement}\label{deepse}
After Hinton \emph{et al.} \cite{hinton_reducing_2006,hinton_fast_2006} showed that \acp{DNN} could be trained successfully on an image-recognition task, a renewed interest in deep learning-based single-microphone speech enhancement emerged. 
In general, these techniques can be divided into two types: mask approximation-based techniques and signal approximation-based techniques.

\subsubsection{Mask Approximation}
Let $\barbelow{x}_m$ and $\barbelow{y}_m$ denote time-frame $m$ of a time-domain clean-speech signal and noisy-speech signal, respectively. 
Furthermore, let $\barbelow{a}_m$ and $\barbelow{r}_m$ denote the \ac{STFT} spectral magnitude vectors of $\barbelow{x}_m$ and $\barbelow{y}_m$, respectively.
Also, let $h(\barbelow{y}_m)$ denote a feature transformation of $\barbelow{y}_m$.
Finally, let 
\begin{equation}
\hat{\barbelow{g}}_m = f_{DNN}(h(\barbelow{y}_m),\barbelow{\theta}),
\label{eq300}
\end{equation} 
denote a gain vector, estimated by a \ac{DNN}\footnote{Note, $\hat{\barbelow{g}}_m$ can also be a function of multiple input vectors, i.e. for multiple $m$, which usually leads to improved performance for feed-forward DNNs.}%
\,$f_{DNN}(\cdot,\cdot)$ with parameters $\barbelow{\theta}$, such that $\hat{\barbelow{a}}_m = \hat{\barbelow{g}}_m \circ \barbelow{r}_m$ is an estimate of the clean speech spectral magnitude $\barbelow{a}_m$, and $\circ$ is element-wise multiplication.
The enhanced time-domain speech signal $\hat{\barbelow{x}}_m$ is then acquired by \ac{IDFT} using the phase of the noisy signal.    

The goal of the mask approximation-based technique is then to find a set of \ac{DNN} parameters $\barbelow{\theta}^\ast$ such that 
\begin{equation}
\begin{aligned}
\theta^\ast = \; & \underset{\barbelow{\theta}}{\text{argmin}}
& & \sum_{\mathcal{D}_{train}} \mathcal{J}(\hat{\barbelow{g}}_m,\barbelow{g}_m), \;\;\;  (\barbelow{y}_m,\barbelow{g}_m) \in \mathcal{D}_{train}, \\
\end{aligned}
\label{eq301}
\end{equation} 
where $\mathcal{D}_{train}$ denotes a training dataset, $\mathcal{J}(\cdot,\cdot)$ denotes a cost function, $\barbelow{g}_m$ is a target gain vector, and the dependence on $\barbelow{y}_m$ and $\barbelow{\theta}$ is implicit via $\hat{\barbelow{g}}_m$.
That is, the mask approximation-based technique aims to minimize the difference as measured by $\mathcal{J}(\cdot,\cdot)$ between the target gain $\barbelow{g}_m$ and the estimated gain $\hat{\barbelow{g}}_m$ (see e.g. \cite{wang_ideal_2005,hummersone_ideal_2014,williamson_complex_2016,delfarah_feature_2016,delfarah_features_2017,chen_noise_2016,chen_long_2016,chen_long_2017}). In the following, we review the work related to two of the most popular target gains: the \acf{IBM} \cite{wang_ideal_2005} and the \ac{IRM} \cite{hummersone_ideal_2014}.  

\paragraph{Ideal Binary Mask} \mbox{}\\
In the \ac{STFT} domain the \ac{IBM} is defined as (see e.g. \cite{wang_training_2014}) 
\begin{equation}
\hat{g}^{IBM}(k,m) = 
\begin{cases}
1 & \text{if }  \frac{|x(k,m)|}{|v(k,m)|} > T_{SNR}(k)\\
0 & \text{otherwise} ,
\end{cases}
\label{eq303}
\end{equation}
where $|x(k,m)|$ and $|v(k,m)|$ denote \ac{STFT} spectral magnitudes for frequency bin $k$ and time-frame $m$ of the clean speech signal and the noise signal, respectively, and $T_{SNR}(k)$ denote a frequency-dependent tuning parameter (see Eqs.\,\eqref{eq40}, and \eqref{eq41}).

One of the first to use the \ac{IBM} target for \ac{DNN} based single-microphone speech enhancement was Wang \emph{et al.} \cite{wang_towards_2013}. 
Wang \emph{et al.} proposed to use \acp{FNN} to estimate the \ac{IBM} from a noisy speech signal. The \acp{FNN} were first trained using the unsupervised pre-training technique and then fine-tuned using back-propagation. However, instead of using the \ac{IBM} for speech enhancement, the output vector of the penultimate \ac{FNN} layer was used as a feature vector for training \acp{SVM}. 
Using these \ac{FNN}-generated feature vectors the \acp{SVM} were trained to estimate \acp{IBM} used for speech enhancement. This approach was very similar to the previous techniques from Sec.\,\ref{sec:classical_se}, where \acp{GMM} \cite{kim_algorithm_2009} and \acp{SVM} \cite{han_classification_2012} were used, but since \acp{FNN} were used as feature extractors, performance improved compared to previous techniques where hand-engineered features were used \cite{kim_algorithm_2009,han_classification_2012}.    
Motivated by Wang \emph{et al.} \cite{wang_towards_2013}, Healy \emph{et al.} \cite{healy_algorithm_2013} proposed to use a \ac{FNN}-estimated IBM for speech enhancement directly, i.e without using \acp{SVM}. 
This approach led to further improvements compared to previous systems \cite{kim_algorithm_2009,han_classification_2012,wang_towards_2013}, presumably due to an increased amount of training data. 
Healy \emph{et al.} \cite{healy_algorithm_2013} even reported large improvements in speech intelligibility for both normal hearing and hearing impaired listeners in a listening test. Similar conclusions were later reported in a subsequent study with a computationally more efficient system that did not use unsupervised pre-training \cite{healy_speech-cue_2014}.    
However, similarly to previous machine learning-based techniques \cite{kim_algorithm_2009,han_classification_2012}, Wang \emph{et al.} \cite{wang_towards_2013} and Healy \emph{et al.} \cite{healy_algorithm_2013} used prior knowledge generally not available in a real-life situation, as the same noise sequence was used during training and test. That is, although these systems achieved impressive performance in unrealistic conditions, they did not reveal any information about the performance to be expected in general real-life scenarios.

\paragraph{Ideal Ratio Mask} \mbox{}\\
Although \ac{IBM}-based speech enhancement systems can achieve good performance (see e.g. \cite{kim_algorithm_2009,han_classification_2012}) several studies suggested (see e.g. \cite{hummersone_ideal_2014,madhu_potential_2013,jensen_spectral_2012}) that a continuous mask might perform better as the binary T-F segmentation of the \ac{IBM}, as either speech or noise dominated, might be too coarse as speech and noise is likely to be present at the same time in the same T-F unit. Obviously, since the IBM is a special case of a general continuous mask, it is expected that a continuous mask can outperform a binary mask in terms of speech enhancement evaluation metrics \cite{jensen_spectral_2012}.   
One such continuous mask, which highly resembles the frequency domain Wiener filter (see Eq.\,\eqref{eq12}), is the \ac{IRM} defined as
\begin{equation}
g^{IRM}(k,m) = \frac{|x(k,m)|^\beta}{|x(k,m)|^\beta + |v(k,m)|^\beta},
\label{eq216}
\end{equation}
where $|x(k,m)|$ and $|v(k,m)|$ denote the clean-speech signal magnitude and noise signal magnitude in frequency bin $k$ and time frame $m$, respectively, and $\beta$ is a tuning parameter \cite{wang_training_2014}.  
Note, unlike the Wiener filter which is statistically optimal, the IRM is not optimal in any obvious way and was presumably motivated heuristically.

In \cite{narayanan_ideal_2013} the \ac{IRM} was proposed as a \ac{DNN} training target and it was reported that the \ac{IRM} outperformed the \ac{IBM}, when used in a speech enhancement front-end for an \ac{ASR} system. It was later shown that the \ac{IRM} also outperformed the \ac{IBM} in terms of objective evaluation metrics such as \ac{PESQ} and \ac{STOI}, when tested in various acoustic environments \cite{wang_training_2014}.    

Furthermore, using a similar technique, large improvements in speech intelligibility was reported in \cite{healy_algorithm_2015} for hearing impaired listeners and moderate improvements for normal hearing listeners. 
In fact, although the system in \cite{healy_algorithm_2015} was "narrow" in the sense that it was speaker and noise-type specific, i.e. trained and tested in matched speaker and noise type conditions, it was the first study to report significant improvements in speech intelligibility for hearing impaired and normal hearing listeners using a single-microphone speech enhancement algorithm.    
In a subsequent study \cite{wang_deep_2015}, generalizability with respect to unknown noise sources was investigated and it was reported in \cite{chen_large-scale_2016} that improvements in speech intelligibility could be achieved for noise types not seen during training, if a very large number of noise types were included in the training set. However, the improvement in speech intelligibility in \cite{chen_large-scale_2016} was significantly reduced compared to \cite{healy_algorithm_2015}, especially for normal hearing listeners where modest improvements were achieved for a babble noise type and practically no improvement for a cafeteria noise type. 
In addition to the promising results in \cite{healy_algorithm_2015,chen_large-scale_2016,bolner_speech_2016,monaghan_auditory_2017}, where the intelligibility improvements were reported for normal hearing and/or hearing impaired listeners using hearing-aids, promising results have also been reported for users of cochlear implants (see e.g. \cite{goehring_speech_2017,lai_deep_2017,lai_deep_2018}). 

Even though the studies in e.g. \cite{healy_algorithm_2015,chen_large-scale_2016,bolner_speech_2016,monaghan_auditory_2017,goehring_speech_2017,lai_deep_2017,lai_deep_2018} showed promising results, they generally only considered either a single noise type, a single speaker or a narrow range of SNRs. That is, these studies only revealed information about the performance to be expected by DNN based speech enhancement systems in non-general usage scenarios where either the noise type, speaker identity or SNR is known \emph{a priori}.

\subsubsection{Signal Approximation}
Differently from the mask approximation-based technique where the goal is to minimize the difference between an estimated gain and a target gain (see Eq.\,\eqref{eq301}), the goal of the signal approximation-based techniques is to minimize the difference between the clean speech, e.g. clean speech STFT magnitudes, and the estimated speech (see e.g. \cite{xu_experimental_2014,liu_experiments_2014,xu_regression_2015,wang_training_2014,weninger_discriminatively_2014,erdogan_phase-sensitive_2015,erdogan_deep_2017}).  

For example, in \cite{weninger_discriminatively_2014} it is proposed to use a cost function defined as 
\begin{equation}
\begin{aligned}
\theta^\ast = \; & \underset{\barbelow{\theta}}{\text{argmin}}
& & \sum_{\mathcal{D}_{train}} \mathcal{J}(\hat{\barbelow{a}}_m ,\barbelow{a}_m), \;\;\;  (\barbelow{r}_m,\barbelow{a}_m) \in \mathcal{D}_{train}, \\
\end{aligned}
\label{eq302}
\end{equation} 
where the goal is to find a gain vector $\hat{\barbelow{g}}_m$, which, when applied to the noisy magnitude $\barbelow{r}_m$ minimize the difference between the estimated speech signal magnitude $\hat{\barbelow{a}}_m = \hat{\barbelow{g}}_m \circ \barbelow{r}_m$ and the target signal magnitude $\barbelow{a}_m$. 
This is arguably more sensible than the mask approximation-based technique, as no target gain is explicitly defined, and the DNN is trained to estimate a gain that achieve the minimum cost with respect to the target $\barbelow{a}_m$, i.e. the clean speech magnitude.   
In fact, when training a DNN using Eq.\,\eqref{eq302} the gain that is indirectly estimated is given as  
\begin{equation}
g^{AM}(k,m) =   \frac{|x(k,m)|}{|y(k,m)|},
\label{eq304}
\end{equation}
where $|y(k,m)|$ denotes the noisy speech signal magnitude in frequency bin $k$ and time frame $m$, which ultimately allows for perfect reconstruction of the clean speech magnitude, i.e. $\barbelow{a}_m = \barbelow{g}^{AM}_m \circ \barbelow{r}_m$ \cite{weninger_discriminatively_2014}. 

However, since the phase of the noisy signal is typically used for reconstructing the enhanced speech signal in the time domain, perfect reconstruction of $\barbelow{a}_m$ only leads to perfect time domain signal reconstruction in the case when $|y(k,m)| = |x(k,m)| + |v(k,m)|$. This, unfortunately, is only true in the unlikely event when the clean speech and noise have identical phases, i.e. $\angle x(k,m) = \angle v(k,m)$. 
Since $|y(k,m)| \ne |x(k,m)| + |v(k,m)|$ in general, it was proposed in \cite{erdogan_phase-sensitive_2015,erdogan_deep_2017} to use the \ac{PSA} cost function defined as
\begin{equation}
\begin{aligned}
\theta^\ast = \; & \underset{\barbelow{\theta}}{\text{argmin}}
& & \sum_{\mathcal{D}_{train}} \mathcal{J}(\hat{\barbelow{a}}_m  \; , \; \barbelow{a}_m \circ \barbelow{\phi}_m), \;\;\;  (\barbelow{r}_m,\barbelow{a}_m) \in \mathcal{D}_{train}, \\
\end{aligned}
\label{eq305}
\end{equation} 
where $\barbelow{\phi}_m = [\phi(1,m), \phi(2,m), \dots, \phi(K,m)]^T$ and $\phi(k,m) = \cos(\angle x(k,m) - \angle y(k,m))$.   
In fact, the PSA gain that minimizes Eq.\,\eqref{eq305} is known as the \ac{PSF} and is given by
\begin{equation}
g^{PSF}(k,m) = Re \left[ \frac{x(k,m)}{y(k,m)}  \right] = \frac{|x(k,m)|}{|y(k,m)|} \cos(\angle x(k,m) - \angle y(k,m)),
\label{eq217}
\end{equation}
and is the optimal real-valued filter that minimizes $|g(k,m)y(k,m) - x(k,m)|$. Furthermore, except in the unlikely case when $\angle x(k,m) = \angle v(k,m)$, Eq.\,\eqref{eq217}, will lead to a higher \ac{SNR} compared to e.g. the \ac{IRM} \cite{erdogan_phase-sensitive_2015}.  
Finally, the \ac{PSA} cost function (Eq.\,\eqref{eq305}) is currently the training objective for DNN based speech enhancement that achieves the best performance in terms of speech enhancement evaluation metrics such as PESQ and STOI \cite{williamson_time-frequency_2017}.  
However, similarly to the studies evaluating the mask approximation technique for DNN based speech enhancement, the studies based on signal approximation (e.g. \cite{xu_experimental_2014,liu_experiments_2014,xu_regression_2015,wang_training_2014,weninger_discriminatively_2014,erdogan_phase-sensitive_2015,erdogan_deep_2017}) were in general trained and tested in narrow usage scenarios where either the noise type, speaker identity or SNR was known \emph{a priori}. That is, these studies also reveal limited information about how DNN based speech enhancement systems perform in general usage scenarios.

\subsection{Deep Learning Based Speech Separation}\label{deepsep}
Similarly to the renewed interest in single-microphone speech enhancement (Sec.\;\ref{deepse}), a renewed interest for DNN based single-microphone multi-talker speech separation emerged as well.
Some of the first to apply modern \acp{DNN} to single-microphone multi-talker speech separation were Du \emph{et al.} \cite{du_speech_2014} and Huang \emph{et al.} \cite{huang_deep_2014,huang_joint_2015}.  

In \cite{du_speech_2014} a multi-layer \ac{FNN} was trained, using unsupervised pre-training, to estimate the log-power spectrum for a target speaker from the log-power spectrum of a mixed signal consisting of two speakers. 
This approach is very similar to the enhancement techniques in \cite{xu_experimental_2014,liu_experiments_2014,xu_regression_2015} and the main difference is the interference signal, which is a speech signal in \cite{du_speech_2014} and not environmental noise signals as in \cite{xu_experimental_2014,liu_experiments_2014,xu_regression_2015}. Nevertheless, Du \emph{et al.} \cite{du_speech_2014} showed good separation performance of a known speaker in terms of \ac{STOI} and output-\ac{SNR} using a signal approximation-based approach. 

Differently from Du \emph{et al.} \cite{du_speech_2014}, Huang \emph{et al.} \cite{huang_deep_2014,huang_joint_2015} proposed to use multi-layer \acp{RNN} to separate a two-speaker mixture signal into the original two speech signals, i.e. Huang \emph{et al.} \cite{huang_deep_2014,huang_joint_2015} proposed to separate the two speech signals in the mixture signal, whereas Du \emph{et al.} \cite{du_speech_2014} simply extracted a known "target" speaker. 
Huang \emph{et al.} \cite{huang_deep_2014,huang_joint_2015} separated a mixture signal using a multi-layer \ac{RNN} with two output streams, i.e. the output vector was twice the original size. For each output stream the \ac{RNN} was trained using a signal approximation-based approach to minimize the \ac{MSE} between the true source signals and the separated signals. 
Although the system was speaker-dependent, i.e. the same speakers were used for training and testing, Huang \emph{et al.} \cite{huang_deep_2014,huang_joint_2015} showed that this approach works well for mixture signals containing both same-gender and opposite-gender speech signals.   

However, even though Du \emph{et al.} \cite{du_speech_2014} and Huang \emph{et al.} \cite{huang_deep_2014,huang_joint_2015} showed that \acp{DNN} could be used for single-microphone multi-talker speech separation and that they outperformed other existing methods based on \acp{GMM} and \ac{NMF}, their methods were still rather limited as they, similarly to the factorial \ac{HMM}-based techniques presented in Sec.\,\ref{genmodse}, were speaker-dependent as they required detailed \emph{a priori} knowledge about the speakers.

A few techniques \cite{weng_deep_2015,wang_unsupervised_2016} did manage to overcome this speaker dependence by introducing additional assumptions. 
For example, in \cite{weng_deep_2015} it was assumed that one speech signal always had an average energy level larger than the other speech signal, i.e. a mixture \ac{SNR} different from 0 dB. 
With this assumption, one multi-layer \ac{FNN} was trained to extract the speaker with the high average energy and another multi-layer FNN was trained to extract the speech signal with low average energy. With this approach a mixture signal containing two speakers of unknown identity could be separated somewhat successfully. 
In \cite{wang_unsupervised_2016} it was instead assumed that the mixture signal consisted of exactly one male and one female speaker. In this case, a multi-layer \ac{FNN} with two output streams was trained to separate the input mixture such that the female speech signal was assigned to e.g.  output stream one and the male speech signal to output stream two. In turn, this enabled the \ac{FNN} to separate unknown speakers of different gender. 
However, even though the techniques proposed in \cite{weng_deep_2015,wang_unsupervised_2016} could separate two-speaker speech signals without \emph{a priori} knowledge about the identity of the speakers, they were still rather limited as they were not easily scaled to more than two speakers of various gender.

\subsubsection{Label Permutation Problem}
The limited success of \ac{DNN} based techniques for speaker-independent multi-talker speech separation, and the reason why most techniques considered only known-two-speaker separation (see e.g. \cite{du_speech_2014,huang_deep_2014,huang_joint_2015,weng_deep_2015,wang_unsupervised_2016,zhang_deep_2016,healy_algorithm_2017,tu_speaker-dependent_2017,venkataramani_end--end_2017,delfarah_recurrent_2018,bramslow_improving_2018}), is partly due to a label permutation problem.          
When training a \ac{DNN} for speaker-independent multi-talker speech separation the permutation of the sources at the output of the \ac{DNN} is unknown. 
Specifically, for a two-speaker separation task, let $\barbelow{o}_1$ denote a target vector for speaker one, and let $\barbelow{o}_2$ denote a target vector for speaker two. 
Furthermore, let $\barbelow{o} = \left[ \barbelow{o}^T_1 \; \barbelow{o}^T_2 \right]^T$ denote a concatenated supervector.
Finally, let $\hat{\barbelow{o}}$, which is the output of a \ac{DNN}, denote the estimate of $\barbelow{o}$. Then, during training, the target vector can in principle be in one of two configurations: either as $\barbelow{o} = \left[ \barbelow{o}^T_1 \; \barbelow{o}^T_2 \right]^T$ or as $\barbelow{o} = \left[ \barbelow{o}^T_2 \; \barbelow{o}^T_1 \right]^T$.
Empirically, it has been observed that if $\barbelow{o}_1$ and $\barbelow{o}_2$ always represent the same speakers, or by speakers of the same gender, training with a predefined permutation, such as $\barbelow{o} = \left[ \barbelow{o}^T_1 \; \barbelow{o}^T_2 \right]^T$ or as $\barbelow{o} = \left[ \barbelow{o}^T_2 \; \barbelow{o}^T_1 \right]^T$, is possible, and is basically the technique used in \cite{delfarah_recurrent_2018,weng_deep_2015,wang_unsupervised_2016,huang_joint_2015,huang_deep_2014}. 
On the other hand, due to the label permutation problem, training simply fails, if the training set consist of many utterances spoken by many speakers of both genders.

\subsubsection{Deep Clustering}
The first successful technique, known as deep clustering, that solved the label permutation problem was proposed by Hershey \emph{et al.} \cite{hershey_deep_2016}. In deep clustering, the speech separation problem is cast as a clustering problem instead of a classification or regression problem as previous techniques. 
Hershey \emph{et al.} \cite{hershey_deep_2016} used \ac{LSTM}-\acp{RNN} to learn a mapping from each T-F unit in the mixture signal to a high dimensional embedding space, where embeddings of T-F units belonging to the same speaker are close (in some sense) and form speaker-specific clusters, which can then be used to separate the speech signals. 

More specifically, let $\barbelow{y} \in \mathbb{R}^{N}$ denote a feature vector of a mixture signal defined according to Eq.\,\eqref{eq42} containing $s = 1,2, \dots, S$ linearly mixed speakers. 
The feature representation can e.g. be given by a \ac{STFT} such that $N = K \times M$ denote the total number of T-F units, where $K$ is the total number of frequency bins and $M$ is the total number of time-frames. 
Furthermore, let $\barbelow{\barbelow{V}} \in \mathbb{R}^{N \times S}$ denote a target matrix with a row for each index in $\barbelow{y}$, i.e. for each T-F unit, and each row in $\barbelow{\barbelow{V}}$ is given by an $S$-dimensional one-hot encoded vector that indicates what speaker a given T-F unit belongs to. 
For example, for $S=3$, if the first T-F unit in $\barbelow{y}$ is dominated by, say, speaker one, the first row in $\barbelow{\barbelow{V}}$ will be given as $[1\; 0\; 0]$. The second row will be $[0\; 0\; 1]$, if the second entry in $\barbelow{y}$ is dominated by speaker three, and so on. The rows in $\barbelow{\barbelow{V}}$ can be viewed as a generalization of the \ac{IBM} to multiple speakers as the assignment of a T-F unit to a speaker is simply defined as the speaker with the most energy in the given T-F unit.   
The matrix $\barbelow{\barbelow{V}}$ can also be seen as an $S$-dimensional embedding of each entry in $\barbelow{y}$ and from  $\barbelow{\barbelow{V}}$ it is trivial to identify which T-F units in $\barbelow{y}$ belong to the same speaker, by simply applying a clustering algorithm, e.g. K-means \cite{macqueen_methods_1967}, to the rows of $\barbelow{\barbelow{V}}$.

Since $\barbelow{\barbelow{V}}$ is easily constructed in laboratory conditions, where speech mixtures can be synthetically mixed according to Eq.\,\eqref{eq42}, one can imagine to use $\barbelow{\barbelow{V}}$ as a training target for supervised learning and then estimate a matrix $\barbelow{\barbelow{\hat{V}}} \in \mathbb{R}^{N \times D}$ as
\begin{equation}
\barbelow{\barbelow{\hat{V}}} = f(\barbelow{y},\theta),
\label{eq218}
\end{equation}   
where $f(\barbelow{y},\theta)$ denote a parameterized learning model that maps each entry of $\barbelow{y}$ into a $D$-dimensional embedding space such that they are clustered similarly to the $S$-dimensional embeddings in $\barbelow{\barbelow{V}}$.

In fact, Hershey \emph{et al.} showed in \cite{hershey_deep_2016} that an estimate $\barbelow{\barbelow{\hat{V}}}$ can easily be acquired by a model $f(\barbelow{y},\theta)$ when it has been trained using a cost function given as  
\begin{equation}
\begin{aligned}
\mathcal{J}(\barbelow{\barbelow{\hat{V}}}, \barbelow{\barbelow{V}}) &= \left\lVert  \barbelow{\barbelow{\hat{V}}} \barbelow{\barbelow{\hat{V}}}^T - \barbelow{\barbelow{V}} \barbelow{\barbelow{V}}^T    \right\lVert_F^2, \\
\end{aligned}
\label{eq219}
\end{equation}   
where $\lVert \cdot \lVert_F^2$ is the squared Frobenius, and $\barbelow{\barbelow{\hat{V}}} \barbelow{\barbelow{\hat{V}}}^T \in \mathbb{R}^{N \times N}$ and $\barbelow{\barbelow{{V}}} \barbelow{\barbelow{{V}}}^T \in \mathbb{R}^{N \times N}$ denote affinity matrices that indicate if a pair of T-F units belong to the same speaker/cluster. 
If an estimate $\barbelow{\barbelow{\hat{V}}}$ is acquired from a well-trained model $f(\barbelow{y},\theta)$, the cluster assignments for all T-F units are easily found using e.g. K-means clustering, which can then be used to form a binary mask that can separate the mixture signal.  

Note, the matrices $\barbelow{\barbelow{\hat{V}}} \barbelow{\barbelow{\hat{V}}}^T$ and $\barbelow{\barbelow{{V}}} \barbelow{\barbelow{{V}}}^T$ are $N \times N$, which for long signals gets intractable to compute. For example, for a 10s audio signal with a 256-point \ac{STFT} using 10 ms frame hop, these matrices have more than 16 billion entries. 
However, as $S,D \ll N$, Hershey \emph{et al.} proposed to minimize the equivalent, but computationally tractable, cost function given by 
\begin{equation}
\begin{aligned}
\mathcal{J}(\barbelow{\barbelow{\hat{V}}}, \barbelow{\barbelow{V}}) &= \left\lVert  \barbelow{\barbelow{\hat{V}}}^T \barbelow{\barbelow{\hat{V}}}  \right\lVert_F^2  
- 2 \left\lVert  \barbelow{\barbelow{\hat{V}}}^T \barbelow{\barbelow{V}}        \right\lVert_F^2  +  
\left\lVert  \barbelow{\barbelow{V}}         \barbelow{\barbelow{V}}^T      \right\lVert_F^2  , 
\end{aligned}
\label{eq220}
\end{equation} 
which scales according to $\mathcal{O}(D^2)$, and not as $\mathcal{O}(N^2)$.  

As shown in Hershey \emph{et al.} \cite{hershey_deep_2016}, the label permutation problem is elegantly avoided when the speech separation problem is cast as a clustering problem. 
Furthermore, when $f(\barbelow{y},\theta)$ is modeled using \ac{LSTM}-\acp{RNN}, state-of-the-art results can be achieved. 

However, although Hershey \emph{et al.} \cite{hershey_deep_2016} reported unprecedented results on a speaker-independent single-microphone multi-talker speech separation task, the deep clustering approach had several drawbacks. 
For example, during inference, a clustering algorithm, e.g. K-means \cite{macqueen_methods_1967}, is required to separate the speakers and consequently the number of speakers $S$ needs to be known \emph{a priori}. 
Also, deep clustering as proposed by Hershey \emph{et al.} \cite{hershey_deep_2016} use a binary gain, which may not be optimal if a large number of speakers or noise sources are present in the mixture. 
Furthermore, in \cite{hershey_deep_2016} only clean speech is considered and it is not obvious how noise sources should be handled.  
Finally, as each T-F unit is represented by a D-dimensional embedding vector (in \cite{hershey_deep_2016} $D\approx40$), the output of a deep clustering model needs to be $D$-times larger than the input, which might be computationally demanding for long signals.   

As a final note, concurrently with the work presented in this thesis, the deep clustering technique has been improved in several aspects such as, soft-clustering \cite{isik_single-channel_2016}, regression based speech enhancement \cite{luo_deep_2017}, improved objective functions \cite{wang_alternative_2018}, and phase estimation \cite{wang_end--end_2018}, which have led to significant gains in performance when measured by \ac{SDR} (see Sec.\,\ref{secsdr}). Also, concurrently with our work, other competing techniques have been proposed such as the deep attractor network \cite{chen_deep_2017,luo_speaker-independent_2018} and source-contrastive estimation \cite{stephenson_monaural_2017-1}, which are both techniques inspired by deep clustering.    

\section{Scientific Contribution}\label{sec:contrib}
\begin{figure*}[ht] 
	\centering
	\resizebox {0.81\textwidth} {!} {
	\begin{tikzpicture}[baseline=(current bounding box.north)]
		

\node at (1.5,4.1) {\scriptsize {\begin{tabular}{c} Deep learning-based \\ single-microphone \\ speech enhancement \end{tabular}}};
\node at (5.5,4.1) {\scriptsize {\begin{tabular}{c} Deep learning-based \\ single-microphone \\ multi-talker speech separation \end{tabular}}};

\node[ rotate=90] at (-0.6,2.75) {\scriptsize {\begin{tabular}{c} Analyzing  \\ State-of-the-art \end{tabular}}};
\node[ rotate=90] at (-0.6,0.75) {\scriptsize {\begin{tabular}{c} Improving  \\ State-of-the-art \end{tabular}}};




\filldraw[fill=gray!10!white, draw=black] (0 , 0) rectangle (3 , 1.5) ;

\filldraw[fill=gray!10!white, draw=black] (4 , 0) rectangle (7 , 1.5) ;

\filldraw[fill=gray!10!white, draw=black] (0 , 2.0) rectangle (3 , 3.5) ;

\filldraw[fill=gray!10!white, draw=black] (4 , 2.0) rectangle (7 , 3.5) ;

\node at (1.5,2.75) {\scriptsize {\begin{tabular}{c} [A] \;\; [B] \;\; [G] \end{tabular}}};
\node at (1.5,0.75) {\scriptsize {\begin{tabular}{c} [F] \end{tabular}}};

\node at (5.5,2.75) {\scriptsize {\begin{tabular}{c} [E] \end{tabular}}};
\node at (5.5,0.75) {\scriptsize {\begin{tabular}{c} [C] \;\; [D] \end{tabular}}};

	\end{tikzpicture}}
	%
	%
	{\fontsize{6.2}{7.44}\selectfont
	\begin{multicols}{2}
	\begin{enumerate}
		\item[{[A]}] {“Speech Intelligibility Potential of General and Specialized Deep Neural Network Based Speech Enhancement Systems”,  \textit{IEEE/ACM Transactions on Audio, Speech, and Language Processing}, January 2017.}
		\item[{[B]}] {“Speech Enhancement Using Long Short-Term Memory Based Recurrent Neural Networks for Noise Robust Speaker Verification”,  \textit{IEEE Spoken Language Technology Workshop}, December 2016.}
		\item[{[C]}] {“Permutation Invariant Training of Deep Models for Speaker-Independent Multi-talker Speech Separation”, \textit{IEEE International Conference on Acoustics, Speech, and Signal Processing}, March 2017.}
		\item[{[D]}] {“Multi-talker Speech Separation With Utterance-Level Permutation Invariant Training of Deep Recurrent Neural Networks”,  \textit{IEEE/ACM Transactions on Audio, Speech, and Language Processing}, October 2017.}
	\end{enumerate}
		\columnbreak
	\begin{enumerate}
		\item[{[E]}] {“Joint Separation and Denoising of Noisy Multi-talker Speech Using Recurrent Neural Networks and Permutation Invariant Training”, \textit{IEEE International Workshop on Machine Learning for Signal Processing}, September 2017.} 
		\item[{[F]}] {“Monaural Speech Enhancement Using Deep Neural Networks by Maximizing a Short-Time Objective Intelligibility Measure”,  \textit{IEEE International Conference on Acoustics, Speech, and Signal Processing}, April 2018.}
		\item[{[G]}]  {“On the Relationship between Short-Time Objective Intelligibility and Short-Time Spectral-Amplitude Mean Squared Error for Speech Enhancement”, \textit{under major revision in IEEE/ACM Transactions on Audio, Speech, and Language Processing}, August 2018.}
		\vfill\null
	\end{enumerate}
	\end{multicols}
	}
\caption{Scientific contribution of the papers making up this thesis: 1) Papers [A], [B], and [G] analyze state-of-the-art speech enhancement and contribute with novel insights. 2) Paper [F] improves state-of-the-art speech enhancement with a novel algorithm. 3) Paper [E] analyzes state-of-the-art multi-talker speech separation and contributes with novel insights. 4) Papers [C] and [D] improve state-of-the-art multi-talker speech separation with novel algorithms.}
\label{fig:contrib}
\end{figure*}
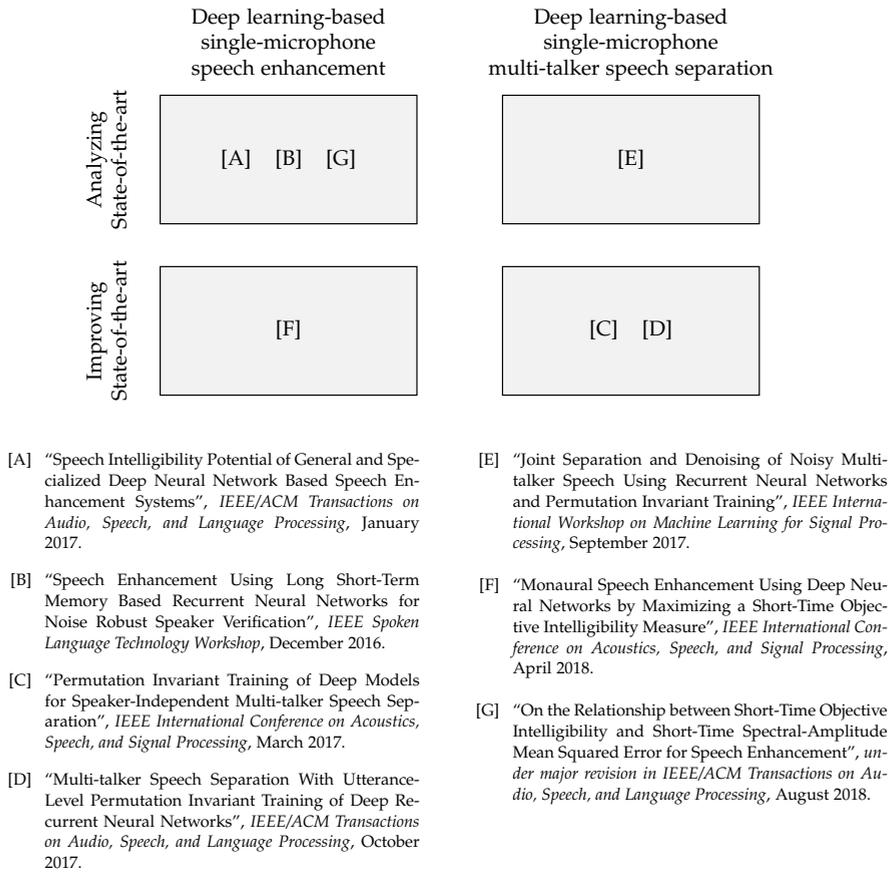

The main body of this thesis (Part II) consists of a collection of seven papers.  
These papers have contributed scientifically by analyzing state-of-the-art techniques, leading to novel insights, or improving state-of-the-art techniques, with novel algorithms, within two disciplines: Deep learning-based single-microphone speech enhancement and deep learning-based single-\linebreak microphone multi-talker speech separation.  
Figure \ref{fig:contrib} summarizes for each of the seven papers the type of scientific contribution and the discipline within which the contribution is made. 
%
%

\subsection{Specific Contributions}
In the following, we shortly summarize the main scientific contribution of each paper in Part II.

\subsubsection*{[A] Speech Intelligibility Potential of General and Specialized Deep Neural Network Based Speech Enhancement Systems}
In this paper, we study the generalizability capability of deep learning-based single-microphone speech enhancement algorithms.
Specifically, we investigate how speech enhancement systems based on FNNs perform in terms of PESQ and STOI when tested in acoustic scenarios that are either matched or unmatched, i.e. the noise type, speaker, or SNR used for testing are either similar or different from the noise type, speaker or SNR used for training. 
This is motivated by recent studies where large improvement in PESQ and STOI have been reported by DNN based speech enhancement systems that are narrowly trained.  

Not surprisingly, we find that one generally loses performance when a system that is trained in a narrow acoustic setting is tested in a more general and realistic acoustic scenario. 
We also find that matching the noise type is the most critical for acquiring good speech enhancement performance, whereas matching the SNR is less critical and good performance for unmatched speakers can be achieved if only a modest number of speakers are included in the training set.

\subsubsection*{[B] Speech Enhancement Using Long Short-Term Memory Based Recurrent Neural Networks for Noise Robust Speaker Verification}
In this paper, we study the generalizability capability of a deep learning-based speech enhancement algorithm with respect to noise robust speaker verification.
Specifically, we propose to use a LSTM-RNN based speech enhancement algorithm as a denoising front-end for a noise robust and male-speaker-independent speaker verification system. 

Compared to two baseline systems based on a STSA-MMSE estimator and NMF, we find that the denoising front-end based on the LSTM-RNN performed the best in terms of equal error rate on a speaker verification task, when tested using various noise types and SNRs.
Despite the fact that the LSTM-RNN was tested in unmatched male-speaker and noise type conditions, it outperformed the NMF based baseline even though this baseline utilized \emph{a priori} information about both the speaker and noise type.

\subsubsection*{[C] Permutation Invariant Training of Deep Models for Speaker-Indepen-\linebreak dent Multi-talker Speech Separation}
In this paper, we propose a deep learning-based technique for single-micro-phone speaker-independent multi-talker speech separation. 
Specifically, we propose the \ac{PIT} technique, which circumvent the label permutation problem, mentioned in Sec.\,\ref{deepsep}, and allow DNNs to be trained successfully for speaker-independent multi-talker speech separation.
We evaluate PIT using FNNs and CNNs for two-talker speech separation using both matched speakers, i.e. same speakers for training and test, and unmatched speakers, i.e. different speakers for training and test.  
 
We find that FNNs and CNNs trained with PIT and tested on a speaker-independent two-talker speech separation task achieved state-of-the-art results, and outperformed techniques based on CASA and NMF in terms of SDR. 
We also find that CNNs trained with PIT perform on par with the deep clustering technique proposed in \cite{hershey_deep_2016}, although the PIT models are computationally less complex. 
Finally, we find that models trained with PIT generalize well to a Danish dataset, although the models have only been trained on English speech.

\subsubsection*{[D] Multi-talker Speech Separation With Utterance-Level Permutation Invariant Training of Deep Recurrent Neural Networks}
In this paper, we propose the \ac{uPIT} technique, which is an extension of the PIT technique proposed in [C]. 
Although PIT allowed for deep learning-based speaker-independent multi-talker speech separation, PIT only works well in practice, when large signal contexts, i.e. a large number of frames, are used. This is a limitation for applications that require low latency.
The uPIT technique, on the other hand, does not have this drawback. 

In this paper we show that LSTM-RNNs trained using uPIT can achieve state-of-the-art results on a speaker-independent multi-talker speech separation task with both two-speaker and three-speaker mixed speech. 
Furthermore, we show that these models can achieve this performance using a small signal context of one frame, which is a dramatic reduction compared to the 50 or 100 frames required by PIT.
Also, similarly to PIT, uPIT perform on par with deep clustering even though uPIT is algorithmically much simpler than deep clustering. 
Finally, we show that a single LSTM-RNN successfully separates both two-speaker and three-speaker mixed speech without \emph{a priori} knowledge about the number of speakers.

\subsubsection*{[E] Joint Separation and Denoising of Noisy Multi-talker Speech Using Recurrent Neural Networks and Permutation Invariant Training}
In this paper, we study aspects of the uPIT technique. Specifically, we investigate how uPIT can be used for single-microphone speaker-independent multi-talker speech separation and enhancement, simultaneously. 
This is different from previously speech enhancement techniques (see Sec.\,\ref{deepse}) that only consider a single target speaker and existing speech separation techniques (see Sec.\,\ref{deepsep}), that consider only noise-free multi-talker mixtures. 

We show that LSTM-RNNs trained using uPIT in noisy environments can improve SDR as well as ESTOI, on a speaker-independent multi-talker speech separation and enhancement task, for various noise types and SNRs. 
We also show that a LSTM-RNN trained using uPIT generalize well to unmatched noise types and that a single model is capable of handling multiple noise types with only a slight decrease in performance. 
Finally, we show that a LSTM-RNN trained using uPIT can improve both SDR and ESTOI without \emph{a priori} knowledge about the exact number of speakers.

\subsubsection*{[F] Monaural Speech Enhancement Using Deep Neural Networks by Maximizing a Short-Time Objective Intelligibility Measure}
In this paper, we propose to use a STOI inspired cost function for training DNNs for single-microphone speech enhancement.
Since STOI has proven an accurate estimator of speech intelligibility, it is hypothesized that a DNN that is trained to estimate speech that maximizes STOI, might lead to speech with a large speech intelligibility. Specifically, compared to the standard \ac{STSA}-\ac{MSE} cost function, which does not have any obvious link to speech intelligibility, a cost function inspired by STOI might be advantageous.

We show that FNNs, trained with an approximate-STOI cost function, improve STOI when tested using matched and unmatched noise types, at multiple SNRs.   
More surprisingly, we observe that approximate-STOI optimal FNNs perform on par with FNNs based on the standard STSA-MSE cost function.  
Consequently, our results suggest that DNN based speech enhancement algorithms, based on the STSA-MSE cost function, might be essentially optimal in terms of estimated speech intelligibility as measured by STOI.

\subsubsection*{[G] On the Relationship between Short-Time Objective Intelligibility and Short-Time Spectral-Amplitude Mean Squared Error for Speech Enhancement}
In this paper, we analyze the surprising result of paper [F], where no apparent gain in STOI can be achieved by a FNN based speech enhancement system trained to maximize an approximate-STOI cost function, as compared to a system trained to minimize the classical STSA-MSE cost function. 

We show theoretically that the optimal Bayesian estimator that maximizes approximate-STOI, under certain general conditions, is asymptotically equivalent to the well-known STSA minimum mean square error estimator. 
Furthermore, through simulation experiments, we show that equality holds for practical FNN based speech enhancement systems. 
The theoretical and empirical results presented in this paper support the surprising result in paper [F] and optimizing for STSA-MSE leads to enhanced speech signals which are essentially optimal in terms of STOI.

\subsection{Summary of Contributions}
The main scientific outcomes of this thesis may be summarized as follows: 

1) In papers [A] and [B], in-depth empirical analysis of the generalizability capability of modern deep learning-based single-microphone speech enhancement algorithms have been conducted. 
In paper [A] it is found, not surprisingly, that performance generally decreases as the acoustic variability of the test data increases. However, it is found that good generalizability with respect to unmatched speakers can be achieved if a modest amount of speakers are included in the training set.  
Furthermore, in paper [B] it is shown that a DNN based speech enhancement system can generalize to unmatched speakers and achieve state-of-the-art speaker verification performance without \emph{a priori} knowledge about the speakers.
The findings in papers [A] and [B], besides from contributing with novel insights, can serve as guidelines in speech-enhancement-algorithm selection or in the design process of future deep learning-based speech enhancement systems.

2) In papers [C], [D], and [E], state-of-the-art techniques for deep learning-based single-microphone speaker-independent multi-talker speech separation were proposed. Specifically, the uPIT technique was proposed, which is algorithmically simpler, yet perform on par with state-of-the-art techniques, such as deep clustering and the deep attractor network. 
Furthermore, uPIT easily extends to multiple speakers and works well for joint speech separation and enhancement without explicit \emph{a priori} knowledge about the noise type or number of speakers, which, at the time of writing, is a capability only shown by uPIT.

3) In papers [F] and [G], it was hypothesized that DNN based speech enhancement systems, trained with an approximate-STOI cost function, would lead to estimated speech signals with an improved STOI score, compared to speech signals estimated by systems trained with the standard STSA-MSE cost function.  
However, supported by experimental and theoretical results, we conclude that this is not the case.
In fact, STSA-MSE leads to enhanced speech signals which are essentially optimal in terms of STOI, and additional improvements in STOI cannot be achieved by a STOI inspired cost function.

\section{Directions of Future Research}
During the last decade deep learning has evolved from a somewhat exotic academic discipline to a fairly mature technology that is widely accessible and is used at an industrial scale.  
Consequently, deep learning has received a tremendous amount of attention from both academia and industry and new deep learning theory, and applications of deep learning, are constantly being proposed.   
This also applies within the areas of speech enhancement and speech separation where an almost countless number of papers have been published over the last couple of years.  
Promising research directions in the area of deep learning based speech enhancement and separation include:

\subsection*{Scale Up While Scaling Down}
As apparent from Secs.\,\ref{deeplearn} and \ref{deepsesep}, DNNs must be fairly big, have multiple layers, a large amount of units, and be trained on a large amount of data before they can perform well. Obviously, one way to achieve better performance is simply to scale up and train with even more data and use even larger models \cite{halevy_unreasonable_2009,shazeer_outrageously_2017}. 
Indeed, this is a valid approach, and is one of the main innovations, if you will, behind many state-of-the-art deep learning based techniques. 
However, training such models is computationally demanding, but more importantly, the memory and computational requirements of DNNs might prohibit their use in applications where computational resources are limited, such as in small embedded devices like mobile phones or hearing aids. 
Therefore, a direction of future research, which is already very active, is on scaling down DNNs without compromising performance, e.g. by reducing the number of parameters in an informed way, increasing the number of layers, while decreasing the number of units, to make the model more parameter-efficient, or reducing the numerical precision of the weights (see e.g. \cite{bianchini_complexity_2014,gupta_deep_2015,tu_reducing_2016,garland_low_2016,kim_bitwise_2017,molchanov_pruning_2017,cintra_low-complexity_2018}).

\subsection*{Beyond Single-Microphone Algorithms}
In this thesis, we have focused purely on single-microphone algorithms. However, utilizing information from multiple microphones can be beneficial if the signal of interest is spatially separated from the interference signals. In such situations, improved performance might be achieved if this information is included.
Hence, a direction of future research is to study how signals from multiple microphones can be efficiently utilized in a deep learning framework.   
For example, one might extend the uPIT technique to work with multi-microphone signals. 
Several promising techniques have already been proposed, where DNNs are used in combination with multi-microphone techniques such as beamforming (see e.g. \cite{hoshen_speech_2015,erdogan_improved_2016,heymann_neural_2016,zhang_deep_2017,drude_tight_2017,boeddeker_exploring_2018,heymann_generic_2017,heymann_beamnet:_2017,boeddeker_optimizing_2017,wang_multi-channel_2018,heymann_performance_2018}).

\subsection*{Beyond Single-Modality Algorithms}
It is well-known that human auditory perception is strongly influenced by visual perception \cite{mcgurk_hearing_1976} and that speech intelligibility in noisy acoustic conditions increase if the listener can observe the face of the person who speaks \cite{sumby_visual_1954}.
Indeed, speech enhancement and separation algorithms can also benefit from such information (see e.g. \cite{hershey_audio-visual_2002}), and although fusing signals from multiple modalities is a challenging task, the emergence of deep learning has alleviated some of these challenges (see e.g. \cite{owens_audio-visual_2018,gabbay_seeing_2018,ephrat_looking_2018}).    
Therefore, studying how deep learning based techniques for speech enhancement and separation can benefit from e.g. visual data is an interesting direction for future research.

\subsection*{Beyond the Mean Squared Error Cost Function}
We have already shown in papers [F] and [G] (see Fig.\,\ref{fig:contrib}) that an approximate-STOI cost function is equivalent to the STSA-MSE cost function and no gain in terms of STOI can be achieved by maximizing an approximate-STOI cost function. 
This conclusion is supported by other very recent work \cite{zhao_perceptually_2018,zhang_training_2018,fu_end--end_2018,naithani_deep_2018}.    
However, it might be that improvements in speech intelligibility or quality can be achieved by optimizing other perceptually-inspired cost functions such as PESQ or Binaural STOI \cite{andersen_predicting_2016}, e.g. using deep reinforcement learning techniques (see e.g. \cite{koizumi_dnn-based_2017,fakoor_reinforcement_2017}). 
Therefore, an interesting direction of future research is to consider alternatives to the commonly used STSA-MSE cost function, which might lead to improved performance of DNN based speech enhancement systems. 

\subsection*{Towards Time-Domain End-to-End Systems}
One of the main advantages of deep learning-based techniques is that they do not require highly specialized, and hand-engineered, features as previous machine learning-based technique did. 
Today, the most used feature, and target, for speech processing applications is the STSA or log-STSA and most speech enhancement and separation algorithms apply the phase of the noisy signal for time-domain reconstruction. Obviously, this is sub-optimal since the noisy phase can only lead to distortions. 
Therefore, an interesting direction of future research is to study how deep learning models can operate directly on the time-domain signal%
\footnote{In fact, this is exactly what Tamura \emph{et al.} \cite{tamura_noise_1988} attempted in 1988 using DNNs, although, today, we might have the model architecture, data, and computational resources to actually succeed.}%
, which potentially can lead to improved performance over the current methods which rely on the noisy phase. 
In fact, a potential deep learning model for time-domain processing is the CNN, which has already shown promising results with respect to time-domain speech enhancement (see e.g. \cite{hoshen_speech_2015,park_fully_2017,fu_raw_2017,venkataramani_end--end_2017,bai_convolutional_2018,zeghidour_end--end_2018,fu_end--end_2018,lee_raw_2017}).

\pagebreak
{\small\bibliographystyle{bib/IEEEtran}\bibliography{bib/mybib}}

\appendix
\part{Papers}
\titleformat{
  \chapter
}[
display%
]{
  \normalfont\huge
}{
  \begin{center}\color{aaublue}\chaptertitlename\ \thechapter\end{center}
}{
  1cm
}{
  \thispagestyle{empty}\begin{center}\Large
}[
  \end{center}
]


  \cleardoublepage
  \setcounter{enumiii}{0}
  \setcounter{enumii}{0}
  \setcounter{enumiv}{0}
  \setcounter{enumi}{0}
  \setcounter{equation}{0}
  \setcounter{figure}{0}
  \setcounter{footnote}{0}
  \setcounter{mpfootnote}{0}
  \setcounter{paragraph}{0}
  \setcounter{parentequation}{0}
  \setcounter{part}{0}
  \setcounter{section}{0}
  \setcounter{subparagraph}{0}
  \setcounter{subsection}{0}
  \setcounter{subsubsection}{0}
  \setcounter{table}{0}
  \papertitlepage{%
  Speech Intelligibility Potential of General and Specialized Deep Neural Network Based Speech Enhancement Systems}{paper:paperA}{%
  Morten Kolbæk, Zheng-Hua Tan, and Jesper Jensen
}{%
  The paper has been published in\\
  \textit{IEEE/ACM Transactions on Audio, Speech, and Language Processing}, \\vol. 25, no. 1, pp. 153–167, January 2017.
}{%
  \noindent\copyright\ 2016 IEEE
}

\acresetall
\newcommand{\RNum}[1]{\lowercase\expandafter{\romannumeral #1\relax}}

\begin{abstract}
In this paper we study aspects of single microphone Speech Enhancement\,(SE) based on Deep Neural Networks\,(DNNs). Specifically, we explore the generalizability capabilities of state-of-the-art DNN based SE systems with respect to the background noise type, the gender of the target speaker, and the Signal-to-Noise Ratio\,(SNR). 
Furthermore, we investigate how specialized DNN based SE systems, which have been trained to be either noise type specific, speaker specific or SNR specific, perform relative to DNN based SE systems that have been trained to be noise type general, speaker general and SNR general.  
Finally, we compare how a DNN based SE system trained to be noise type general, speaker general and SNR general performs relative to a state-of-the-art Short-Time Spectral Amplitude Minimum Mean Square Error\,(STSA-MMSE) based SE algorithm.

We show that DNN based SE systems, when trained specifically to handle certain speakers, noise types and SNRs, are capable of achieving large improvements in estimated Speech Quality\,(SQ) and Speech Intelligibility\,(SI), when tested in matched conditions.   
Furthermore, we show that improvements in estimated SQ and SI can be achieved by a DNN based SE system when exposed to unseen speakers, genders and noise types, given a large number of speakers and noise types have been used in the training of the system.  
In addition, we show that a DNN based SE system that has been trained using a large number of speakers and a wide range of noise types outperforms a state-of-the-art STSA-MMSE based SE method, when tested using a range of unseen speakers and noise types. 
Finally, a listening test using several DNN based SE systems tested in unseen speaker conditions show that these systems can improve SI for some SNR and noise type configurations but degrade SI for others.    	
\end{abstract}

\section{Introduction} \label{sec:intro1}
\IEEEPARstart{I}{mproving} quality and intelligibility of noisy speech signals is of great interest in a vast amount of applications such as mobile communications, speech recognition systems, and hearing aids. 
In a single-microphone setting, improving \ac{SQ} and especially \ac{SI} is a challenging task and is an active topic of research \cite{hendriks_dft-domain_2013,c._loizou_speech_2013,wang_supervised_2015}. 
Traditionally, single microphone Speech Enhancement\,(SE) has been addressed by statistical model based methods such as the Wiener filter \cite{c._loizou_speech_2013} and Short-Time Spectral Amplitude Minimum Mean Square Error\,(STSA-MMSE) estimators, e.g., \cite{ephraim_speech_1985,martin_speech_2005,erkelens_minimum_2007}. 
However, recent advances in Computational Auditory Scene Analysis\,(CASA) and machine learning have introduced new methods, e.g.  Deep Neural Network\,(DNN), Gaussian Mixture Model\,(GMM), and Support Vector Machine\,(SVM) based methods, which address single-microphone SE and speech segregation in terms of advanced statistical estimators. These estimators aim at estimating either a clean speech \ac{T-F} representation directly or a T-F mask that is applied to the T-F representation of the noisy speech to arrive at an estimate of the clean speech signal \cite{wang_training_2014,healy_algorithm_2013,healy_algorithm_2015,kim_algorithm_2009,han_classification_2012,wang_towards_2013,chen_large-scale_2016,chen_noise_2016,han_towards_2013}.  
For some potential future applications, e.g. DNN based SE algorithms for hearing aids or mobile communications, the range of possible acoustic situations which can realistically occur is virtually endless. It is therefore important to understand how such methods perform in different acoustic situations, and how they perform, when they are exposed to "unseen" situations, i.e. acoustic scenarios not encountered during training. Despite the obvious importance of this generalizability question, it is currently not well understood.  

In this study we focus on situations where a single target speaker is present in additive noise and the aim of the SE algorithm is to enhance the speech signal and attenuate the noise using a single-microphone recording. 
Generally, when evaluating generalizability of machine learning based SE algorithms, there are at least three dimensions in which the input signal can vary: 
\RNum{1}) the noise type dimension, \RNum{2}) the speaker dimension and \RNum{3}) the Signal to Noise Ratio\,(SNR) dimension. Therefore, evaluation of DNN based SE methods should cover each of these dimensions in a way similar to what is expected to be experienced in a real life scenario. 
For mobile communication devices and hearing aid systems, evaluation should hence encompass a wide range of noise types, a wide range of speakers and a wide range of SNRs, in order to give a realistic estimate of the expected performance of the algorithm in real life scenarios.
On the other hand, for applications where the typical usage situation is much more well-defined, e.g. voice-controlled devices to be used in a car cabin situation, training and testing might involve only car cabin noises at a narrow SNR range for a single particular speaker.   

The exploration of these three dimensions is motivated by the fact that no matter how many noise types, SNRs, and speakers a SE system is exposed to during training, in a real life scenario, sooner or later the system will be exposed to an unseen noise type, an unseen speaker or an unseen SNR.   
However, if the system is well trained, one might expect that the system has captured some general acoustic characteristics from these dimensions and hence generalizes well to unseen conditions. 
Furthermore, if any \emph{a priori} knowledge about the noise type, speaker characteristics or SNR is available, it is important to know what performance gain can be achieved by utilizing this \emph{a priori} knowledge.

Several studies have investigated aspects of generalizability of SE algorithms based on DNNs, SVMs, and GMMs, e.g.
 \cite{chen_large-scale_2016,xu_regression_2015,lee_single_2015,huang_joint_2015,liu_experiments_2014,wang_training_2014,healy_algorithm_2015,wang_deep_2015,wang_towards_2013,chen_noise_2016,healy_algorithm_2013,kim_algorithm_2009,han_classification_2012,han_towards_2013,gonzalez_mask-based_2014,chen_long_2016,delfarah_feature_2016,kumar_speech_2016}. However, these models are fundamentally different in both training schemes and architectures \cite{bishop_pattern_2006} and since DNNs are currently state-of-the-art in a large number of applications \cite{lecun_deep_2015} and have outperformed SVMs and GMMs in SE tasks \cite{healy_algorithm_2013,healy_algorithm_2015,wang_towards_2013,kim_algorithm_2009}, SVMs and GMMs are less suited for the current study and are therefore not considered. 

\pagebreak
Common for all the studies, based on DNNs \cite{chen_large-scale_2016,xu_regression_2015,lee_single_2015,huang_joint_2015,liu_experiments_2014,wang_training_2014,healy_algorithm_2015,wang_deep_2015,wang_towards_2013,chen_noise_2016}, is that during training or testing one or more of the generalizability dimensions defined above are held fixed, while others are varied. To the authors knowledge no study exists which explores the simultaneous variation of all the three dimensions - a situation which is realistic for many real life applications.   
Furthermore, interpretation of existing studies is sometimes complicated by the fact that the training and test signals, for the dimensions which \emph{are} varied, are not described in all details. 
For example, the distribution of males and females is often not reported \cite{xu_regression_2015,huang_joint_2015,liu_experiments_2014,wang_training_2014} and it is hence not clear if the system is mostly a gender specific or gender general system. Several studies \cite{xu_regression_2015,huang_joint_2015,wang_training_2014} use the TIMIT corpus \cite{garofolo_darpa_1993}, which is approximately 70\% male and 30\% female. 
Furthermore, the duration of the different training noise types is typically not considered when the training data is constructed, hence the exact distribution of the noise types is unknown. For example, in \cite{healy_algorithm_2013,xu_regression_2015,huang_joint_2015,liu_experiments_2014} noise sequences with highly varying duration are used, which makes it unclear to which extent these systems are noise specific or noise general.  
Another issue related to the noise dimension is concerning the construction of training and test data. In several studies 
\cite{healy_algorithm_2013,huang_joint_2015,han_classification_2012,liu_experiments_2014,kim_algorithm_2009,wang_towards_2013}
, the exact same noise realizations were used for training and testing. In \cite{may_requirements_2014} this training/testing paradigm was analyzed, and it was found to erroneously give remarkably better performance compared to the realistic scenario, where the actual noise sequence is unknown. 
Furthermore, the systems presented in  \cite{chen_large-scale_2016,xu_regression_2015,lee_single_2015,huang_joint_2015,liu_experiments_2014,wang_training_2014,healy_algorithm_2015,wang_deep_2015,wang_towards_2013,chen_noise_2016} are based on various network architectures, training methods, testing methods, speech corpora, noise databases, feature representations, target representations etc. 
As a consequence of these differences, their results cannot be directly related and it is therefore unclear how a state-of-the-art DNN based SE algorithm perform when the generalizability dimensions mentioned above are considered simultaneously. 
Finally, it is unclear to what extent state-of-the-art DNN based SE algorithms provide improvements over existing non-DNN based SE methods. 
In \cite{xu_regression_2015,wang_deep_2015-1,wang_training_2014} a DNN based SE method similar to the one studied here outperforms several different non-DNN based methods such as statistical MMSE based methods \cite{cohen_noise_2003,cohen_speech_2001,erkelens_minimum_2007,erkelens_estimation_2008,hendriks_comparison_2008} and non-Negative Matrix Factorization\,(NMF) methods \cite{wang_training_2014,lee_algorithms_2000}.
However, since the DNNs used in \cite{xu_regression_2015,wang_deep_2015-1,wang_training_2014} have not been trained across all three generalizability dimensions, the comparison may not yield a true picture of the actual performance difference. This is particularly true with the statistical MMSE based methods \cite{c._loizou_speech_2013,hendriks_dft-domain_2013}, which have not been trained to handle any specific noise types or speakers but merely rely on basic statistical assumptions with respect to Short-Time Fourier Transform\,(STFT) coefficients and might perform worse than a system trained on a given speaker or noise type.    

The goal of this paper is therefore to conduct a systematic evaluation of the generalizability capabilities of a state-of-the-art DNN based SE algorithm in terms of estimated SQ and SI. Specifically, we investigate how a state-of-the-art DNN based SE method performs when it is trained to be noise type specific vs. noise type general, speaker specific vs. speaker general, and SNR specific vs. SNR general. 
Furthermore, we study the performance drop, if any, for systems which are specialized in one or more of the three generalizability dimensions, compared to a completely general DNN based SE system, which relies on essentially no prior knowledge with respect to speaker characteristics, noise type, and SNR. 
Additionally, it is investigated how this general system performs compared to a state-of-the-art non-machine learning based method namely the STSA-MMSE estimator employing generalized gamma priors as proposed in \cite{erkelens_minimum_2007,erkelens_estimation_2008,hendriks_comparison_2008}\footnote{http://insy.ewi.tudelft.nl/content/software-and-data}. This is of interest since the STSA-MMSE method relies on very little prior knowledge compared to conventional DNN based SE methods \cite{healy_algorithm_2013,healy_algorithm_2015,chen_large-scale_2016}.  
Furthermore, given that the computational and memory complexity associated with DNN type of systems is typically orders of magnitude larger than that associated with simple STSA-MMSE based systems it is of obvious interest to understand the performance gain of this technology.
Finally, a listening test is conducted, using both specialized and general DNN based SE systems, to investigate if such systems improve SI, when tested in different matched and unmatched conditions.

It is important to note that this paper emphasizes on the generalizability properties of DNN based SE algorithms in terms of estimated and measured SI, since these properties has not yet been rigorously investigated in the current literature \cite{chen_large-scale_2016,xu_regression_2015,lee_single_2015,huang_joint_2015,liu_experiments_2014,wang_training_2014,healy_algorithm_2015,wang_deep_2015,wang_towards_2013,chen_noise_2016}. 
To do so, we rely on a specific implementation of a feed-forward DNN, whose architecture and training procedure resemble those of a large range of existing DNN based SE methods \cite{healy_algorithm_2015,wang_training_2014,xu_regression_2015,liu_experiments_2014,xu_experimental_2014}. This allows us to expect that our findings are representative not only for our particular implementation but are generally valid for DNN based SE methods. The fact that the DNN based SE method under study is a representative member of a larger class of algorithms also implies that this particular implementation does not necessarily outperform all existing methods with respect to estimated SQ and SI.

Furthermore, obviously, the three chosen generalizability dimensions are not the only dimensions for which mixing scenarios can vary. Other such dimensions include reverberation conditions, e.g. in terms of varying room impulse responses, or digital signal processing conditions, e.g. in terms of signal sampling rate, number of bits with which each sample is represented, microphone characteristics, compression/coding schemes, etc. Furthermore, for DNN based SE algorithms the DNN architecture can also be varied and considered as a dimension. 
We have chosen the speaker dimension, the noise type dimension and the SNR dimension for this particular work since these are dimensions most often encountered in the SE literature  \cite{chen_large-scale_2016,xu_regression_2015,lee_single_2015,huang_joint_2015,liu_experiments_2014,wang_training_2014,healy_algorithm_2015,wang_deep_2015,wang_towards_2013,chen_noise_2016,healy_algorithm_2013,kim_algorithm_2009,han_classification_2012,han_towards_2013,gonzalez_mask-based_2014,chen_long_2016,delfarah_feature_2016,kumar_speech_2016}. 
Furthermore, in most papers related to DNN based SE algorithms only a single speaker is considered, so it is of interest to study how well these algorithms generalize to unknown speakers. Finally, the performance of non-machine learning based SE algorithms such as STSA-MMSE and Wiener filtering based approaches are known to be highly dependent on the noise type, and SNR, but not the speaker. Hence, it is of interest to study how a DNN based SE algorithm performs in a large range of noise types, speakers and SNRs.  

The paper is organized as follows: Section \ref{sec:se} describes the DNN architecture, training procedure and speech material used for conducting the desired experiments. Section \ref{sec:result} describes and discusses the experimental setups and results. Finally, in Section \ref{sec:conA} the findings are concluded.

\section{Speech Enhancement Using Neural Networks}
\label{sec:se}
\subsection{Speech Corpus and Noisy Mixtures}
The phonetically balanced Danish speech corpus \emph{\ac{ADFD}}\footnote{https://www.nb.no/sbfil/dok/nst\_taledat\_no.pdf} is used as target speech material for training and testing all DNN based SE systems considered in this paper. This corpus consists of two sets: One set (set 1) consisting of 56 speakers with 986 spoken utterances for each speaker and another set (set 2) with 560 speakers and 311 spoken utterances, and males and females are approximately equally distributed among the two sets. 
The majority of the text material is based on conversational speech, but also single words, numbers and sequences of numbers are included, and each utterance has an average duration of approximately 5 seconds.

The training, validation, and test sets, were constructed such that no sentence appears more than once in the combined training, validation, and test set. The sampling frequency was 16 kHz and all files were normalized to have unit \ac{RMS} power.

The noisy mixtures for all experiments were constructed by adding a noise signal to a clean speech signal at a certain SNR. The noise signal was scaled to achieve the desired SNR based on the speech active region of the speech signal, i.e. the silence parts in the beginning and in the end of the speech signal were omitted in SNR computation. 
Omitting the silence parts for the SNR computation is crucial since the inclusion of these parts will effectively decrease the energy estimate of the clean speech, hence a lower noise power is required to achieve the same SNR, than if the silence regions were omitted. 
The difference in SNR between these two approaches of constructing noisy mixtures can be more than one dB and is typically not considered in the literature \cite{healy_algorithm_2015,huang_joint_2015,xu_regression_2015}, even though it is of importance if results from different studies are to be related. 
Alternatively, a Voice Activity Detection\,(VAD) algorithm could have been used to exclude all silent regions, which would  be highly beneficial for practical applications. However, for simplicity and to be in-line with existing literature \cite{chen_large-scale_2016,xu_regression_2015,lee_single_2015,huang_joint_2015,liu_experiments_2014,wang_training_2014,healy_algorithm_2015,wang_deep_2015,wang_towards_2013,chen_noise_2016,healy_algorithm_2013,kim_algorithm_2009,han_classification_2012,han_towards_2013,gonzalez_mask-based_2014,chen_long_2016,delfarah_feature_2016,kumar_speech_2016}, we excluded the VAD for all experiments. As before, the global SNR based approach were chosen from a practical perspective and to be in-line with existing literature \cite{chen_large-scale_2016,xu_regression_2015,lee_single_2015,huang_joint_2015,liu_experiments_2014,wang_training_2014,healy_algorithm_2015,wang_deep_2015,wang_towards_2013,chen_noise_2016,healy_algorithm_2013,kim_algorithm_2009,han_classification_2012,han_towards_2013,gonzalez_mask-based_2014,chen_long_2016,delfarah_feature_2016,kumar_speech_2016}, where global SNR is by far the most common.
\subsection{Features and Labels}
The choice of training targets for supervised speech enhancement have been widely studied \cite{wang_time-frequency_2008,jensen_spectral_2012,hummersone_ideal_2014,wang_training_2014,c._loizou_speech_2013,wang_ideal_2005,madhu_potential_2013,williamson_complex_2015}. 
Recent studies \cite{healy_algorithm_2015,madhu_potential_2013,wang_training_2014,hummersone_ideal_2014} suggest that continuous targets such as the Ideal Ratio Mask\,(IRM) are preferable over binary targets such as the Ideal Binary Mask\,(IBM) \cite{wang_ideal_2005,madhu_potential_2013}. Therefore, the DNN studied in this paper is trained in a supervised fashion to estimate the IRM from a feature representation of a noisy speech signal.  

The T-F representation used to construct the IRM is based on a gammatone filter bank with 64 filters linearly spaced on a MEL frequency scale from 50 Hz to 8 kHz and with a bandwidth equal to one \ac{ERB} \cite{wang_computational_2006}\footnote{http://web.cse.ohio-state.edu/pnl/shareware/cochleagram}.  
The output of the filter bank is divided into 20 ms frames with 10 ms overlap and with a sampling frequency of 16 kHz, each T-F unit represents a vector of 320 samples. 

Let  $ \mathbf{x}(n, \omega) $ denote the Time-Frequency\,(T-F) unit of the clean speech signal at frame $n$ and frequency channel $\omega$, and let $ \mathbf{d}(n, \omega) $ denote the corresponding T-F unit of the noise signal. Then the IRM is computed as \cite{wang_training_2014} 
\begin{equation}
\nonumber
\text{IRM}(n, \omega)= \left(\frac{||\mathbf{x}(n, \omega)||^2}{||\mathbf{x}(n, \omega)||^2 + ||\mathbf{d}(n, \omega)||^2} \right)^\beta,  
\end{equation}
where $ ||\mathbf{x}(n, \omega)||^2 $ is the squared 2-norm, i.e. the clean speech energy, of T-F unit $n$ in frequency channel $\omega$. Likewise, $||\mathbf{d}(n, \omega)||^2$ is the noise energy of a T-F unit $n$ in frequency channel $\omega$. The variable $\beta$ is a tunable parameter and has for all experiments in this paper been set to $\beta = 0.5$, which in \cite{wang_training_2014} was found empirically to provide good results.

To have discriminative and noise robust features, each frame is transformed into a 1845-dimensional feature vector inspired by \cite{healy_speech-cue_2014,healy_algorithm_2013,healy_algorithm_2015,wang_towards_2013,chen_feature_2014,wang_supervised_2015,wang_exploring_2013}. Although, very recent works use only magnitude spectra \cite{chen_large-scale_2016,amodei_deep_2015,wang_deep_2015} a large context of several hundred milliseconds is used, which is undesirable for real time applications. The chosen feature vector was found to outperform features of magnitude spectra when these were based on only a small context.
The features used are 31 \ac{MFCC}, 15 Amplitude Modulation Spectrogram\,(AMS), 13 \ac{RASTA-PLP} and 64 \ac{GFE}. Furthermore delta and double delta features are computed and a context of 2 past and 2 future frames are utilized, hence arriving at a 1845-dimensional feature vector. All feature vectors are normalized to zero mean and unit variance.

\subsection{Network Architecture and Training}
The DNNs used in this paper follow a feed-forward architecture with a 1845-dimensional input
layer and three hidden layers, each with 1024 hidden units, and 64 output units \cite{healy_algorithm_2015,wang_training_2014}. The activation functions for the hidden units are \acp{ReLU} \cite{nair_rectified_2010} and for the output units the sigmoid function is applied. The hidden layers are initialized using the "GlorotUniform" approach \cite{glorot_understanding_2010}. 
Furthermore, the DNN has approximately 4 million tunable parameters in terms of weights and biases. The values of the parameters are found using \ac{SGD} following the AdaGrad  approach \cite{duchi_adaptive_2011}. The gradients are computed using backpropagation based on the Mean Square Error\,(MSE) error function using a batch size of 1024 \cite{bishop_pattern_2006}. Furthermore, 20\% dropout has been applied to all hidden layers during training to reduce overfitting \cite{hinton_improving_2012}.  
In order to further reduce overfitting, an early-stopping training scheme is applied, which stops the training, when the MSE of the validation set has not decreased with more than 1\% for more than 20 epochs.  
Although used in \cite{healy_algorithm_2013,xu_regression_2015}, unsupervised DNN pre-training using Deep Belief Networks \cite{lecun_deep_2015,mohamed_acoustic_2012,bengio_greedy_2007} was found not to significantly improve performance and has therefore not been applied in the reported results. 

Finally, it is well known that increasing the network size or changing the network architecture can improve performance of DNN based algorithms \cite{xu_regression_2015,wang_deep_2015,halevy_unreasonable_2009,hannun_deep_2014,amodei_deep_2015,chen_large-scale_2016}. However, it is not practically feasible to include network architecture as a dimension in our experiments. Furthermore, although absolute performance might be better with a different architecture, the conclusions drawn from using a fixed-sized feed-forward DNN are expected to be valid for a broader range of DNN architectures, since the underlying assumptions are practically the same.

\subsection{Signal Enhancement}
After DNN training, the IRM is estimated for a given test signal by forward propagating its feature representation, for all frames, through the DNN. The output of the DNN represents the estimated IRM, $\widehat{\text{IRM}}(n, \omega)$ for the given frame.  
The estimated IRM  can then be applied to the T-F representation of the noisy speech signal by multiplying the given entry of the mask to all 320 noisy signal samples of a T-F unit. 
All T-F units in each frequency channel are then concatenated and all overlapping parts are summed. Afterwards, the 64 frequency channels can  be synthesized into a time domain signal by first compensating for the different group delays in the different channels and then adding the frequency channels. 
The group delay compensation is performed by time reversing the signals, passing them through the gammatone filter bank and then time reversing the signals once again \cite{wang_computational_2006}.    

\subsection{Evaluation of Enhancement Performance}
Speech signals enhanced with the DNN based SE algorithm studied in this paper were evaluated using the Short-Time Objective Intelligibility\,(STOI) \cite{taal_algorithm_2011} measure and the wideband extension of the Perceptual Evaluation of Speech Quality\,(PESQ) measure \cite{rix_perceptual_2001,_p.862.2_2005}. The STOI measure estimates SI and PESQ estimates SQ and both have been found to be highly correlated with human listening tests \cite{taal_algorithm_2011,c._loizou_speech_2013}. STOI is defined in the range $[-1,1]$ and PESQ is defined in the range $[1,4.5]$ and for both measures higher is better. We used the implementations of STOI and PESQ available from \cite{taal_algorithm_2011} and \cite{c._loizou_speech_2013}, respectively.  

Although other performance measures exists such as Signal to Distortion Ratio\,(SDR), Signal to Interferences Ratio\,(SIR), and Signal to Artifact Ratio\,(SAR)\cite{c._loizou_speech_2013,fevotte_bss_2011} we report only PESQ and STOI to limit the number of tables. Furthermore, PESQ and STOI are by far the most used speech quality and speech intelligibility estimators in the literature \cite{chen_large-scale_2016,xu_regression_2015,lee_single_2015,huang_joint_2015,liu_experiments_2014,wang_training_2014,healy_algorithm_2015,wang_deep_2015,wang_towards_2013,chen_noise_2016,healy_algorithm_2013,kim_algorithm_2009,han_classification_2012,han_towards_2013,gonzalez_mask-based_2014,chen_long_2016,delfarah_feature_2016,kumar_speech_2016}.

\section{Experimental Results and Discussion}
\label{sec:result} 
To investigate the generalizability capability of DNN based SE systems with relation to: \RNum{1}) the noise type dimension, \RNum{2}) the speaker dimension and \RNum{3}) the SNR dimension, three experimental setups, one for each dimension, have been designed.  
When a dimension is explored the remaining two dimensions are held fixed. For example, when exploring the SNR dimension, the SNR dimension is varied but only a single speaker and a single noise type is used for both training and testing. 
Furthermore, a fourth setup has been constructed where a general system has been designed. This system was trained using a wide range of speakers, noise types and SNRs, hence the system relies on a minimum of \emph{a priori} knowledge. This "general" system is compared against the three experiments previously described, as well as a state-of-the-art non-machine learning based SE algorithm.

\subsection{SNR Dimension}\label{sec:snrdim}
The purpose of the SNR experiments is to investigate the impact on the performance of DNN based SE systems, when training is performed based on a single SNR vs. a wide range of SNRs, i.e. constructing a SNR specific or a SNR general system.   
The SNR dimension is explored using speech material based on 986 spoken utterances from a single female speaker from the ADFD set 1. These 986 utterances were divided such that 686 were used for training, 100 for validation and 200 for testing.   
Two noise types have been investigated, a stationary \ac{SSN} and a non-stationary Babble\,(BBL) noise. 
The SSN sequence is constructed by filtering a 50 min. Gaussian white noise sequence through a $12$th-order all-pole filter with coefficients found from Linear Predictive Coding\,(LPC) analysis of 100 randomly chosen TIMIT sentences \cite{garofolo_darpa_1993}.  
The BBL noise is also based on TIMIT. The corpus, which consists of a total of 6300 spoken sentences, is randomly divided  into 6 groups of 1050 concatenated utterances. Each group is then truncated to equal length followed by addition of the six groups. This results in a BBL noise sequence with a duration of over 50 min. 
The SSN and BBL sequences were both divided such that 40 min. were used for training, 5 min. were used for validation and 5 min. for testing, hence there is no overlapping samples in the noise segments used for training, validation and test. 
To investigate how the performance of DNN based SE systems depends on the SNR dimension, eight systems were trained with eight different SNR settings for both SSN and BBL noise. All 16 systems were tested using eight SNRs ranging from -15 dB to 20 dB with steps of 5 dB. 
For each noise source, the first system was trained using -5 dB since this is a commonly encountered SNR in the literature \cite{healy_algorithm_2015,healy_algorithm_2013} where SI is typically degraded and DNN based SE algorithms have been successfully applied \cite{healy_algorithm_2015,healy_algorithm_2013}. 
The next system was trained using SNRs from -5 dB to 0 dB with steps of 1 dB. In a similar fashion wider and wider SNR ranges were used for training the remaining systems with the widest range being from -15 dB to 20 dB. The precise intervals are given in Tables \ref{table:snr_ssn_stoi}, \ref{table:snr_ssn_pesq}, \ref{table:snr_bbl_stoi} and \ref{table:snr_bbl_pesq}.    
For all systems, each training utterance was mixed with different noise realizations 35 times in order to increase the amount of training data. For each noisy mixture, the SNR was drawn from a discrete uniform distribution defined within the given SNR range. Due to the large number of realizations, it is assumed that the distribution of drawn SNRs is approximately uniform. The noise signal used for each noisy mixture was extracted from the whole training noise sequence by using a starting index drawn from a discrete uniform distribution defined over the entire length of the noise sequence. If the starting index is such that there is no room for the whole utterance, the remaining samples are extracted from the beginning of the noise sequence.  
Following the same procedure, each validation utterance is mixed with different noise realizations 10 times. Using this form of training data augmentation, the total amount of training utterances, used for training each system, is increased to $686 \times 35=24010$, which is approximately equal to 33 hours of speech material and is approximately $65\%$ more data than used by \cite{healy_algorithm_2015}.  

The results of the SNR dimension experiments are presented in Tables \ref{table:snr_ssn_stoi} and \ref{table:snr_ssn_pesq} for SSN and in Tables \ref{table:snr_bbl_stoi} and \ref{table:snr_bbl_pesq} for BBL noise. 
From Tables \ref{table:snr_ssn_stoi} and \ref{table:snr_bbl_stoi} it is seen that the SNR specific system of SNR of -5 dB achieves relatively large STOI improvements, for test signal SNRs in the range -10 dB to 5 dB.
In general, it can be observed that inclusion of SNRs in the range from -15 dB to 5 dB has a larger positive impact on the performance than inclusion of SNRs above 5 dB.  
This might be explained partly by the fact that intelligibility is almost 100\% (STOI $\approx 1$) for test signal SNRs above 5 dB, and partly by the limited noise energy, which makes it more difficult for the DNN to actually learn important noise characteristics.   
Tables \ref{table:snr_ssn_pesq} and \ref{table:snr_bbl_pesq} show a somewhat similar picture. The inclusion of training signals with SNRs around 0 dB in general improves performance, but extending the training SNR range from 5 dB to 20 dB does not further improve performance. Furthermore, it is also seen that the system in general cannot improve PESQ for test signals with SNRs below -5 dB.  

Based on these experiments it can be concluded that there is generally a good correspondence between SNR ranges used in training and STOI improvement seen during testing. For example, the systems trained in the SNR range from -5 dB to 0 dB perform better at 0 dB than the systems trained using only -5 dB.  
Furthermore, even at -15 dB, where the noise energy is approximately 40 times larger than the speech energy, STOI is still improved with 0.074 and 0.093 for SSN and BBL noise, respectively, when this particular SNR is included in the training set, and the improvement is almost constant for SSN, and even slightly increasing for BBL noise, when a wider range of positive SNRs are included in the training set.  
Also, the system trained using the widest SNR range from -15 dB to 20 dB achieves almost similar performance as the -5 dB SNR specialized system, when tested at an SNR of -5 dB, and generally performs better at other SNRs.   
This observation is in line with related studies \cite{zhang_deep_2016} and is of large practical importance, as it suggests that DNN based SE systems should simply be trained using as large a training signal SNR range as practically possible.
%
%
\begin{table*}[h!]                             
	\centering 
	\caption{STOI improvement for the SNR dimension. Eight DNN based SE systems trained on different SNR ranges as indicated in the first row. 
		The noise type dimension is held constant using SSN only and the speaker dimension is held constant using a single female speaker.  
		The systems are evaluated using STOI for test signals with 8 different SNRs ranging from -15dB to 20dB. The second column presents the STOI score for the unprocessed noisy mixtures. Columns 3-10 present STOI improvements.} 
	\label{table:snr_ssn_stoi}
    \resizebox{\textwidth}{!}{%
	\begin{tabular}{lccccccccc}     
		\midrule \midrule       
		\textbf{} & {Noisy} & {--5dB} & {--5dB -- 0dB} & {--5dB -- 5dB} & {--10dB -- 5dB} & {--15dB -- 5dB} & {--15dB -- 10dB} & {--15dB -- 15dB} & {--15dB -- 20dB} \\ 
		\midrule
        \textbf{-15dB} & 0.354 & 0.016 & 0.019 & 0.028 & 0.063 & 0.074 & 0.075 & 0.075 & 0.072 \\ 
		\textbf{-10dB} & 0.417 & 0.170 & 0.166 & 0.165 & 0.186 & 0.186 & 0.185 & 0.183 & 0.179 \\ 
		\textbf{-5dB}  & 0.519 & \textbf{0.219} & \textbf{0.218} & \textbf{0.218} & \textbf{0.219} & \textbf{0.216} & \textbf{0.215} & \textbf{0.213} & \textbf{0.210} \\  
		\textbf{0dB}   & 0.642 & 0.180 & 0.186 & 0.187 & 0.185 & 0.183 & 0.183 & 0.182 & 0.181 \\  
		\textbf{5dB}   & 0.756 & 0.115 & 0.125 & 0.130 & 0.128 & 0.126 & 0.128 & 0.127 & 0.127 \\  
		\textbf{10dB}  & 0.844 & 0.058 & 0.070 & 0.078 & 0.077 & 0.076 & 0.079 & 0.079 & 0.078 \\  
		\textbf{15dB}  & 0.905 & 0.016 & 0.030 & 0.040 & 0.039 & 0.039 & 0.044 & 0.045 & 0.044 \\ 
		\textbf{20dB}  & 0.944 & -0.010 & 0.005 & 0.015 & 0.014 & 0.014 & 0.020 & 0.023 & 0.023 \\ 
        \midrule \midrule
	\end{tabular}  } 
\end{table*}    
%
%
\begin{table*}[h!]                                        
	\centering 
	\caption{As Table \ref{table:snr_ssn_stoi} but for PESQ.} 
	\label{table:snr_ssn_pesq}
    \resizebox{\textwidth}{!}{%
	\begin{tabular}{lccccccccc}     
		\midrule \midrule                 
		\textbf{} & {Noisy} & {--5dB} & {--5dB -- 0dB} & {--5dB -- 5dB} & {--10dB -- 5dB} & {--15dB -- 5dB} & {--15dB -- 10dB} & {--15dB -- 15dB} & {--15dB -- 20dB} \\ 
		\midrule
		\textbf{-15dB}  & 1.133 & -0.044 & -0.041 & -0.044 & -0.036 & -0.027 & -0.029 & -0.035 & -0.032 \\ 
		\textbf{-10dB}  & 1.115 & 0.025 & 0.024 & 0.025 & 0.038 & 0.044 & 0.042 & 0.041 & 0.041 \\         
		\textbf{-5dB}   & 1.115 & 0.198 & 0.190 & 0.192 & 0.202 & 0.196 & 0.196 & 0.191 & 0.187 \\          
		\textbf{0dB}    & 1.144 & 0.457 & 0.425 & 0.421 & 0.410 & 0.400 & 0.410 & 0.408 & 0.405 \\           
		\textbf{5dB}    & 1.234 & 0.700 & 0.691 & 0.655 & 0.643 & 0.638 & 0.630 & 0.636 & 0.642 \\           
		\textbf{10dB}   & 1.438 & \textbf{0.769} & \textbf{0.875} & 0.879 & 0.863 & 0.859 & 0.831 & 0.803 & 0.811 \\          
		\textbf{15dB}   & 1.811 & 0.583 & 0.830 & \textbf{0.942} & \textbf{0.925} & \textbf{0.911} & \textbf{0.948} & \textbf{0.902} & \textbf{0.878} \\          
		\textbf{20dB}   & 2.346 & 0.130 & 0.518 & 0.764 & 0.745 & 0.733 & 0.860 & 0.877 & 0.848 \\  
        \midrule \midrule
	\end{tabular}  }
\end{table*}    
%
%
%
%
%
%
%
\begin{table*}[h!]                                            
	\centering  
	\caption{As Table \ref{table:snr_ssn_stoi} but for BBL.} 
	\label{table:snr_bbl_stoi}
    \resizebox{\textwidth}{!}{%
	\begin{tabular}{lccccccccc}     
		\midrule \midrule                 
		\textbf{} & {Noisy} & {--5dB} & {--5dB -- 0dB} & {--5dB -- 5dB} & {--10dB -- 5dB} & {--15dB -- 5dB} & {--15dB -- 10dB} & {--15dB -- 15dB} & {--15dB -- 20dB} \\ 
		\midrule               
		\textbf{-15dB} & 0.292 & 0.048 & 0.034 & 0.033 & 0.070 & 0.093 & 0.095 & 0.094 & 0.096 \\ 
		\textbf{-10dB} & 0.369 & 0.161 & 0.150 & 0.146 & 0.170 & 0.173 & 0.173 & 0.174 & 0.174 \\ 
		\textbf{-5dB}  & 0.480 & \textbf{0.214} & \textbf{0.218} & \textbf{0.216} & \textbf{0.214} & \textbf{0.205} & \textbf{0.205} & \textbf{0.206} & \textbf{0.206} \\  
		\textbf{0dB}   & 0.608 & 0.187 & 0.200 & 0.202 & 0.194 & 0.188 & 0.189 & 0.191 & 0.191 \\   
		\textbf{5dB}   & 0.728 & 0.128 & 0.147 & 0.152 & 0.146 & 0.141 & 0.144 & 0.147 & 0.146 \\   
		\textbf{10dB}  & 0.823 & 0.070 & 0.091 & 0.098 & 0.095 & 0.091 & 0.096 & 0.098 & 0.097 \\  
		\textbf{15dB}  & 0.890 & 0.024 & 0.045 & 0.056 & 0.053 & 0.050 & 0.057 & 0.059 & 0.059 \\  
		\textbf{20dB}  & 0.934 & -0.008 & 0.013 & 0.026 & 0.023 & 0.021 & 0.029 & 0.032 & 0.033 \\ 
        \midrule \midrule
	\end{tabular}  }  
\end{table*}    
%
%
%
\begin{table*}[h!]                                    
	\centering 
	\caption{As Table \ref{table:snr_ssn_stoi} but for PESQ and BBL.} 
	\label{table:snr_bbl_pesq}
    \resizebox{\textwidth}{!}{%
	\begin{tabular}{lccccccccc}     
		\midrule \midrule                 
		\textbf{} & {Noisy} & {--5dB} & {--5dB -- 0dB} & {--5dB -- 5dB} & {--10dB -- 5dB} & {--15dB -- 5dB} & {--15dB -- 10dB} & {--15dB -- 15dB} & {--15dB -- 20dB} \\ 
		\midrule
		\textbf{-15dB} & 1.201 & -0.063 & -0.066 & -0.058 & -0.070 & -0.080 & -0.066 & -0.079 & -0.075 \\ 
		\textbf{-10dB} & 1.180 & -0.047 & -0.060 & -0.058 & -0.056 & -0.052 & -0.055 & -0.055 & -0.054 \\ 
		\textbf{-5dB}  & 1.143 & 0.086 & 0.090 & 0.089 & 0.081 & 0.074 & 0.072 & 0.079 & 0.075 \\          
		\textbf{0dB}   & 1.162 & 0.289 & 0.312 & 0.319 & 0.294 & 0.280 & 0.279 & 0.284 & 0.293 \\           
		\textbf{5dB}   & 1.270 & 0.493 & 0.571 & 0.580 & 0.543 & 0.511 & 0.516 & 0.527 & 0.538 \\           
		\textbf{10dB}  & 1.478 & \textbf{0.636} & 0.772 & 0.805 & 0.770 & 0.732 & 0.740 & 0.745 & 0.741 \\          
		\textbf{15dB}  & 1.829 & 0.621 & \textbf{0.826} & \textbf{0.914} & \textbf{0.884} & \textbf{0.844} & \textbf{0.872} & \textbf{0.872} & \textbf{0.854} \\          
		\textbf{20dB}  & 2.312 & 0.426 & 0.691 & 0.863 & 0.835 & 0.794 & 0.863 & 0.871 & 0.851 \\  
        \midrule \midrule
	\end{tabular}  }  
\end{table*}

\subsection{Noise Dimension}\label{sec:noiseDim}
The purpose of the noise dimension experiments is to investigate the performance impact when DNN based SE systems are trained on a single noise type vs. a wide range of noise types. In other words, this allows us to compare a noise specific vs. a noise general system.    
The noise dimension has been explored using the same 986 spoken utterances from the same single female speaker as used in the SNR experiments. Likewise, the partition of the speech material into training, validation and test set is also identical to the SNR experiments.  
To explore the noise dimension, six distinct noise types were used: SSN\,(N1) and BBL\,(N2) from the SNR experiments and four additional noises: street\,(N3), pedestrian\,(N4), cafe\,(N5) and bus\,(N6), from the CHiME3 dataset \cite{barker_third_2015}.  
Furthermore, 1260 randomly selected sound effect noises from soundbible.com\footnote{http://soundbible.com/free-sound-effects-1.html} were used to construct a seventh noise type referred to as the \emph{mix}\,(N7) noise type. These 1260 noises were first truncated to have a maximum duration of 3 seconds each and then concatenated into one large noise sequence. The sound effects include sounds from animals, singing humans, explosions, airplanes, slamming doors etc.  
All seven (N1 -- N7) noise types used for the noise experiments were first truncated to have a total duration of 50 min. and then divided into a 40 min. training set, a 5 min. validation set and a 5 min. test set, hence there is no overlapping samples in the noise segments used for training, validation and test.

To investigate how the performance of the DNN based SE system depends on the noise dimension, eight systems were trained with eight different noise combinations all at an SNR of -5 dB. Two systems were trained with only one noise type, namely the stationary SSN\,(N1) and non-stationary BBL\,(N2). The remaining six systems were trained with an increasing number of noise types starting with N1 -- N2 and ending with N1 -- N7 as indicated in the second row in Tables \ref{table:ntstoi} and \ref{table:ntpesq}.  
When noise types were combined, the 40 min. noise sequences were concatenated and similar to the SNR experiment, a noise sequence was extracted based on a randomly chosen starting index within this concatenated noise sequence. Similarly to the SNR experiment, each utterance in the training set was mixed with a randomly chosen noise sequence 35 times, hence a total of $686 \times 35=24010$ noise mixtures were constructed. The large number of mixtures and the identical duration of the noise sequences ensures that the noise distribution within the training data is approximately uniform, hence a noise-general system is constructed.  
All eight systems have been tested with speech signals contaminated by all seven noise types, which ensures that all but the system trained with all seven noises will be tested with at least one unseen and at the most 6 unseen noise types.     

The results are presented in Tables \ref{table:ntstoi} and \ref{table:ntpesq} where the first column represents the noise types used for testing and the second row represents the noises used for training.   
Table \ref{table:ntstoi} shows that when a system is trained using SSN only (N1) it achieves a relatively large STOI improvement of 0.22, when tested on that particular noise type, but generalizes poorly on the majority of the unseen test noises. Similarly, when a system is trained on BBL (N2), the performance is good in the matched noise case, but the system generalizes poorly to other noise types.  
Furthermore, when both SSN and BBL noise types are included equally in the training set (N1-N2), the system performs almost as good as the individual noise specific systems. 
However, the system does not generalize as well to the unseen noises as N1 did alone, except for the \emph{mix} noise type, that similarly to BBL is highly non-stationary.  
It is also interesting to notice that the SSN and BBL specific systems achieve very similar performance for test signals contaminated by SSN and BBL, respectively. This is in contrast to STFT-based methods for which non-stationary noise is much more challenging \cite{hendriks_dft-domain_2013}.  
A different picture is seen when a third noise (N3) is added in the training set (N1-N3). This system performs similarly well in the matched noise type setting, but also for the unseen noises the performance has increased considerably. Similar behavior is seen when the remaining noise types are included in the training set. 
Furthermore, even though new noise types are included in the training set, the performance of the system is almost constant in the matched noise type setting. 
One can argue that \emph{str}, \emph{ped}, \emph{caf} and \emph{bus} are quite similar noise types, but it is seen that the system trained with signals contaminated by all but the mix noise type (N1-N6) generalizes relatively well to the mix noise type, which is a noise type radically different from the others.  
From Table \ref{table:ntpesq} a similar behavior is observed where relatively large PESQ scores are achieved for all testing noises, already after noise type N1 -- N3 have been included in the training set.  
Similar for both Tables \ref{table:ntstoi} and \ref{table:ntpesq} is that there is generally a good correspondence between noise types used for training and STOI and PESQ improvements seen during testing. For example, the systems performing best on SSN and BBL noise are the systems that have been trained on only these noise types. However, a system trained on both noise types show only a slightly decrease in performance.       
Furthermore, the noise general system (N1 -- N7), where all seven noise types are used for training, achieves on average the best performance across all seven noise types, while still being comparable in performance to the more specialized systems where only a single or a few noise types have been used for training. This is similar to the SNR experiments where no particular degradation in performance was observed by extending the SNR range used for training.  
%
%
%
\begin{table*}[h!]                                              
	\centering 
	\caption{STOI improvement for the noise type dimension. Eight DNN based SE systems have been trained with different combinations of seven different noise types (N1-N7) as given by the first row. 
		The SNR dimension is held constant at -5dB and the speaker dimension is held constant using a single female speaker. 
		The systems have been evaluated using STOI and test signals corrupted by all seven noise types. The second column presents the STOI score for the noisy unprocessed mixtures. Columns 3-10 present STOI improvements.} 
	\label{table:ntstoi}
    \resizebox{0.85\textwidth}{!}{%
	\begin{tabular}{lccccccccc}     
		\midrule \midrule           
		\textbf{} & {Noisy} & {N1} & {N2} & {N1--N2} & {N1-N3} & {N1--N4} & {N1--N5} & {N1--N6} & {N1--N7}\\   
        \midrule        
		\textbf{N1: ssn} & 0.519 & \textbf{0.220} & 0.083 & 0.207 & 0.209 & \textbf{0.208} & \textbf{0.206} & \textbf{0.206} & \textbf{0.203} \\   
		\textbf{N2: bbl} & 0.482 & 0.029 & \textbf{0.217} & \textbf{0.210} & \textbf{0.211} & 0.204 & 0.202 & 0.203 & 0.199 \\     
		\textbf{N3: str} & 0.590 & 0.122 & -0.079 & 0.080 & 0.174 & 0.172 & 0.172 & 0.173 & 0.171 \\ 
		\textbf{N4: ped} & 0.504 & 0.095 & -0.008 & 0.078 & 0.139 & 0.157 & 0.161 & 0.160 & 0.158 \\ 
		\textbf{N5: caf} & 0.572 & 0.072 & -0.007 & 0.065 & 0.143 & 0.155 & 0.165 & 0.167 & 0.165 \\  
		\textbf{N6: bus} & 0.703 & 0.071 & -0.058 & 0.003 & 0.112 & 0.114 & 0.118 & 0.130 & 0.128 \\  
		\textbf{N7: mix} & 0.685 & 0.015 & 0.028 & 0.038 & 0.072 & 0.078 & 0.092 & 0.093 & 0.119 \\             
        \midrule \midrule
	\end{tabular}  }      
\end{table*}
%
%
%
\begin{table*}[h!]                                       
	\centering  
	\caption{As Table \ref{table:ntstoi} but for PESQ.} 
	\label{table:ntpesq}
    \resizebox{0.85\textwidth}{!}{%
	\begin{tabular}{lccccccccc}     
		\midrule \midrule                  
		\textbf{} & {Noisy} & {N1} & {N2} & {N1--N2} & {N1-N3} & {N1--N4} & {N1--N5} & {N1--N6} & {N1--N7}\\   
        \midrule
		\textbf{N1: ssn} & 1.112 & \textbf{0.197} & -0.012 & \textbf{0.175} & 0.186 & 0.174 & 0.175 & 0.178 & 0.173 \\   
		\textbf{N2: bbl} & 1.174 & -0.072 & 0.060 & 0.048 & 0.054 & 0.047 & 0.039 & 0.046 & 0.032 \\   
		\textbf{N3: str} & 1.069 & 0.114 & -0.002 & 0.071 & 0.302 & 0.294 & 0.294 & 0.294 & 0.298 \\    
		\textbf{N4: ped} & 1.099 & 0.033 & -0.025 & 0.005 & 0.095 & 0.118 & 0.125 & 0.125 & 0.120 \\         
		\textbf{N5: caf} & 1.081 & 0.030 & -0.003 & 0.025 & 0.191 & 0.224 & 0.237 & 0.247 & 0.242 \\            
		\textbf{N6: bus} & 1.083 & 0.125 & 0.010 & 0.036 & \textbf{0.329} & \textbf{0.336} & \textbf{0.351} & \textbf{0.421} & \textbf{0.415} \\    
		\textbf{N7: mix} & 1.143 & 0.002 & \textbf{0.067} & 0.059 & 0.126 & 0.144 & 0.161 & 0.180 & 0.293 \\               
        \midrule \midrule
	\end{tabular}  }      
\end{table*}
\subsection{Speaker Dimension}\label{sec:sprkTest}
The purpose of the speaker dimension experiments is to study the impact of using a single speaker vs. a wide range of speakers in the training material, i.e. constructing a speaker specific or speaker general system.  

The speaker dimension is explored using speech material based on 311 spoken utterances from 41 male and 41 female speakers from the ADFD set 2. The utterances from the 41 females are referenced as F-ID1 -- F-ID41 and similarly, the utterances from the 41 males are referenced as M-ID1 -- M-ID41.   
For each of the speakers of interest, (F/M-ID1 -- F/M-ID21) 231 utterances were used for training, 30 for validation and 50 for testing. Furthermore, 50 utterances from each of the 40 remaining speakers (F/M-ID22 -- F/M-ID41) were used as testing material for unseen speaker testing. The text material used for the 50 test utterances from each speaker was the same for all 82 speakers used for these experiments.    

A total of 10 systems were trained. Five systems using speech material corrupted with SSN at an SNR of $-5$ dB and five systems with speech material corrupted with BBL noise at an SNR of $-5$ dB.  
For each noise type, speakers F-ID1 and MID-1 were used to train two individual speaker specific systems. Furthermore, speakers F-ID2 -- 21 and M-ID2 -- 21 were used to train two individual gender specific systems and finally the speakers F-ID2 -- 21 and M-ID2 -- 21 were combined (F/M-ID2 -- 21) and used to train a single speaker general system.   

All systems were evaluated in both a seen speaker and an unseen speaker scenario using the test material from speaker F/M-ID1 -- F/M-ID21 and F/M-ID22 -- F/M-ID41, respectively. However, the systems trained using only one speaker is tested using 20 speakers, instead of only one speaker, to give more realistic unseen-speaker results. 
Since the number of distinct utterances used for training vary between the different systems, due to the varying number of speakers, a fixed total number of 18480 training utterances were used for training all systems. This is done to ensure that all systems are presented to the same amount of noise material. Using the same argument a total number of 1200 utterances were used for validation during the training of all systems.
To achieve 18480 training mixtures, and 1200 validation mixtures, for each system, each distinct utterance was mixed with unique noise realizations multiple times as given by Table \ref{tab:dataAugment}.  
\begin{table}[h!] 
	\centering
	\caption{Training and validation data augmentation scheme used for results reported in subsection \ref{sec:sprkTest} to ensure all systems use the same amount of data. The format is the following: $\#speakers \times \#utterances \times \#repetitions = \#mixtures$}
	\label{tab:dataAugment}
    \resizebox{0.7\textwidth}{!}{%
	\begin{tabular}{lcc}     
		\midrule \midrule  
		\multicolumn{1}{c}{{System}} & {\#Training Utterances} & {\#Validation Utterances} \\ 
        \midrule
		Speaker Specific  & $1 \times 231 \times 80 = 18480$   & $ 1 \times 30 \times 40  = 1200$  \\ 
		Gender Specific   & $20 \times 231 \times 4 = 18480$       & $20 \times 30 \times 2  = 1200$  \\ 
		Speaker General & $40 \times 231 \times 2 = 18480$      & $40 \times 30 \times 1 = 1200$   \\ 
        \midrule \midrule
	\end{tabular}  } 
\end{table}

The results with SSN are presented in Tables \ref{table:sprk_ssn_stoi} and \ref{table:sprk_ssn_pesq}, and the results with BBL noise are presented in Tables \ref{table:sprk_bbl_stoi} and \ref{table:sprk_bbl_pesq}. The first column presents the speaker IDs used for testing and the second row represents speaker IDs used for training.  

From Table \ref{table:sprk_ssn_stoi} it is seen that speaker specific systems trained on a single speaker achieves a STOI improvement of $0.168$ and $0.204$ for same-gender-same-speaker testing, for the female (F-ID1) and male (M-ID1) specific systems, respectively. However, if these systems are tested with new speakers of same gender, i.e. same-gender-new-speaker testing, the STOI improvements are reduced to $0.127$ and $0.114$ for the female (F-ID1) and male (M-ID1) specific systems, respectively. 
Furthermore, if the systems are tested on opposite gender the STOI improvement decreases to $0.067$ and $0.062$ for the female (F-ID1) and male (M-ID1) specific systems, respectively. 
Similar behavior, but with larger variations, is seen from Table \ref{table:sprk_bbl_stoi} where the systems have been trained using utterances corrupted with BBL noise instead of SSN. Table \ref{table:sprk_bbl_stoi} shows that systems trained using F-ID1 and M-ID2 improve STOI with $0.131$ and $0.184$ for same-gender-same-speaker testing, for the female (F-ID1) and male (M-ID1) systems, respectively. However, these improvements reduce to $0.046$ and $0.039$ for same-gender-new-speakers testing, and to $-0.093$ and $-0.107$ for new-gender testing, for the female (F-ID1) and male (M-ID1) systems, respectively.              

From these results it can be concluded that systems which are trained using only a single speaker generalizes very well to unseen utterances from the same speaker but not as good to unseen utterances from new speakers of same gender and even worse to opposite gender. Especially for BBL noise, the systems even degrade the signals when evaluated using opposite gender. 

If gender specific systems are trained with 20 speakers instead of only a single speaker it is seen from Table \ref{table:sprk_ssn_stoi} that the STOI improvements in the same-gender-same-speakers testing case are $0.175$ and $0.174$ for the female (F-ID2 -- F-ID21) and male (M-ID2 -- M-ID21) systems, respectively.  
Furthermore, the STOI improvements in the same-gender-new-speakers testing case are $0.170$ and $0.160$ for the female (F-ID2 -- F-ID21) and male (M-ID2 -- M-ID21) systems, respectively. Compared to the systems trained using a single speaker, the systems trained using 20 speakers of same gender generalize considerably better to the same-gender-new-speaker testing case. Also in the new-gender testing case Table \ref{table:sprk_ssn_stoi} shows STOI improvements of $0.124$ and $0.119$ for the female (F-ID2 -- F-ID21) and male (M-ID2 -- M-ID21) systems, respectively. However, Table \ref{table:sprk_bbl_stoi} shows that STOI is degraded when the female (F-ID2 -- F-ID21) and male (M-ID2 -- M-ID21) systems are tested in the new-gender testing case. 
Finally, if a speaker general system is trained using both males and females (F/M-ID2 -- 21) and is tested in an unseen speaker setting based on both genders (F/M-ID22 -- 41) using respectively SSN and BBL noise, the STOI improvements are $0.164$ and $0.111$, respectively.  
This shows that the speaker general system in terms of STOI generalizes considerably better than the speaker specific and gender specific systems to unseen speakers of both genders, for both a stationary and non-stationary noise type. Importantly, it is seen that the loss from a gender specific system to a gender general system is almost zero.

One interesting observation is the decrease in performance, when compared to the experiments exploring the noise type dimension in subsection \ref{sec:noiseDim}. For example, Table \ref{table:ntstoi} shows that a system specialized to a single female speaker using BBL noise at an SNR of -5 dB achieves a STOI improvement of $0.217$. 
Table \ref{table:sprk_bbl_stoi} shows that a similar system (F-ID1) trained with a different female speaker using BBL noise at an SNR of -5 dB achieves a STOI improvement of $0.131$, which is a considerable difference.
There is one major difference between these two systems. For the experiments used to produce Table \ref{table:ntstoi}, the speaker was represented by 686 distinct spoken utterances, whereas for the experiments used to produce Table \ref{table:sprk_bbl_stoi} only 231 distinct spoken utterances were used. 
This indicates that not only the number of speakers but also the variability in speech material from each speaker is crucial to achieve good generalizability. 

In general, it can be observed that a DNN based SE system trained using a single speaker becomes speaker specific and performs well, in terms of estimated SI, when evaluated using the same speaker. If a large number of speakers, of the same gender, are used for training, the system becomes gender specific and generalizes well to unseen speakers of same gender. Furthermore, if a large number of male and female speakers are used for training, the system becomes speaker general and generalizes well to unseen speakers of both genders. This applies for systems trained using training signals corrupted with either SSN or BBL noise. In terms of estimated SQ a similar behavior can only be observed for systems trained with training signals corrupted with SSN whereas for the systems trained using training signals corrupted with BBL noise no, or only minor, improvements were found as shown by Tables \ref{table:sprk_ssn_pesq} and \ref{table:sprk_bbl_pesq}.           
\begin{table*}[h!]                                              
	\centering 
	\caption{STOI improvement for the speaker dimension. Five DNN based SE systems have been trained on a varying number of speakers of both genders as given by the first row. 
		The systems have been tested in both speaker matched and unmatched conditions. 
		The noise type dimension is held constant using SSN for training and testing and the SNR dimension is held constant using an SNR of -5dB for training and testing. 
		The systems have been evaluated using STOI. The second column presents the STOI score for the noisy unprocessed mixtures. Columns 3-7 present STOI improvements}
	\vspace{-2 mm}
	\label{table:sprk_ssn_stoi}
    \resizebox{0.8\textwidth}{!}{%
	\begin{tabular}{lcccccccc}     
		\midrule \midrule                              
		{Test $\backslash$ Train}        &{Noisy} &{F-ID1} & {M-ID1} & {F-ID2 -- 21} & {M-ID2 -- 21} & {F/M-ID2 -- 21} \\
		\textbf{F-ID1} 				& 0.564 & \textbf{0.168} & --    & --    & --    & --    \\                       
		\textbf{M-ID1} 				& 0.460 & --    & \textbf{0.204} & --    & --    & -- \\                           
		\textbf{F-ID2-21} 			& 0.532 & --	& --    & \textbf{0.175} & --    & \textbf{0.170} \\            
		\textbf{F-ID22-41} 			& 0.530 & 0.127 & 0.062 & 0.170 & 0.119 & 0.166 \\          
		\textbf{M-ID2-21} 			& 0.543 & --    & --    & --    & \textbf{0.174} & 0.167 \\              
		\textbf{M-ID22-41} 			& 0.538 & 0.067 & 0.114 & 0.124 & 0.160 & 0.163 \\            
		\textbf{F/M-ID2--21} 	    & 0.538 & --    & --    & --    & --    & 0.167 \\   
		\textbf{F/M-ID22--41} 	    & 0.535 & 0.097 & 0.089 & 0.147 & 0.140 & 0.164 \\   
        \midrule \midrule
	\end{tabular}  }   
\end{table*}    
\vspace{4 mm}
\begin{table*}[h!] 
	\centering 
	\caption{As Table \ref{table:sprk_ssn_stoi} but for PESQ}
	\vspace{-2 mm}
	\label{table:sprk_ssn_pesq}
    \resizebox{0.8\textwidth}{!}{%
	\begin{tabular}{lcccccccc}     
		\midrule \midrule                 
		\textbf{Test $\backslash$ Train}        &{Noisy} &{F-ID1} & {M-ID1} & {F-ID2 -- 21} & {M-ID2 -- 21} & {F/M-ID2 -- 21} \\
		\midrule
        \textbf{F-ID1} 				& 1.062 & \textbf{0.160} & --    & --    & --    & --    \\                     
		\textbf{M-ID1} 				& 1.078 & --    & \textbf{0.185} & --    & --    & --    \\                          
		\textbf{F-ID2-21} 			& 1.068 & --    & --    & \textbf{0.168} & --    & 0.158 \\          
		\textbf{F-ID22-41} 			& 1.065 & 0.108 & 0.043 & 0.160 & 0.058 & 0.150 \\          
		\textbf{M-ID2-21} 			& 1.110 & --    & --    & --    & \textbf{0.219} & 0.208 \\              
		\textbf{M-ID22-41} 			& 1.118 & 0.070 & 0.149 & 0.141 & 0.199 & \textbf{0.213} \\            
		\textbf{F/M-ID2--21} 	    & 1.096 & --    & --    & --    & --    & 0.175 \\   
		\textbf{F/M-ID22--41} 	    & 1.093 & 0.087 & 0.095 & 0.149 & 0.126 & 0.180 \\ 
        \midrule \midrule
	\end{tabular}  }     
\end{table*}
\vspace{4 mm}
\begin{table*}[h!]          
	\centering 
	\caption{As Table \ref{table:sprk_ssn_stoi} but for BBL}
	\vspace{-2 mm}
	\label{table:sprk_bbl_stoi}
    \resizebox{0.8\textwidth}{!}{%
	\begin{tabular}{lcccccccc}     
		\midrule \midrule                                                             
		{Test $\backslash$ Train}        &{Noisy} &{F-ID1} & {M-ID1} & {F-ID2 -- 21} & {M-ID2 -- 21} & {F/M-ID2 -- 21} \\
		\midrule
        \textbf{F-ID1} 				& 0.535 & \textbf{0.131} & --    & --    & --    & --   \\                       
		\textbf{M-ID1} 				& 0.433 & --    & \textbf{0.184} & --    & --    & --    \\                         
		\textbf{F-ID2-21} 			& 0.498 & --    & --    & \textbf{0.121} & --    & 0.110 \\            
		\textbf{F-ID22-41} 			& 0.496 & 0.046 & -0.107 & 0.117 & -0.059 & 0.108 \\          
		\textbf{M-ID2-21} 			& 0.511 & --    & --    & --    & \textbf{0.140} & 0.112 \\              
		\textbf{M-ID22-41} 			& 0.507 & -0.093 & 0.039 & -0.007 & 0.125 & \textbf{0.115} \\            
		\textbf{F/M-ID2--21} 	    & 0.505 & --    & --    & --    & --    & 0.110 \\   
		\textbf{F/M-ID22--41} 	    & 0.501 & -0.025 & -0.034 & 0.054 & 0.032 & 0.111 \\ 
        \midrule \midrule
	\end{tabular}  }     
\end{table*}
%
%
\begin{table*}[h!]              
	\centering 
	\caption{As Table \ref{table:sprk_ssn_stoi} but for PESQ and BBL}
	\vspace{-2 mm}
	\label{table:sprk_bbl_pesq}
    \resizebox{0.8\textwidth}{!}{%
	\begin{tabular}{lcccccccc}     
		\midrule \midrule                                                   
		{Test $\backslash$ Train}        &{Noisy} &{F-ID1} & {M-ID1} & {F-ID2 -- 21} & {M-ID2 -- 21} & {F/M-ID2 -- 21} \\
		\midrule
        \textbf{F-ID1} 				& 1.094 & \textbf{0.065}  & -- 	& --  & --  & --  \\                       
		\textbf{M-ID1} 				& 1.159 & -- & \textbf{0.029} & -- & -- & -- \\                        
		\textbf{F-ID2-21} 			& 1.129 & --    & --    & \textbf{0.001} & --    & -0.007 \\             
		\textbf{F-ID22-41} 			& 1.130 & -0.029 & -0.049 & -0.007 & -0.064 & -0.017 \\          
		\textbf{M-ID2-21} 			& 1.168 & --    & --    & --    & \textbf{0.024} & -0.005 \\               
		\textbf{M-ID22-41} 			& 1.181 & -0.074 & -0.040 & -0.077 & 0.002 & -0.007 \\             
		\textbf{F/M-ID2--21} 	    & 1.141 & --    & --    & --    & --    & \textbf{0.001} \\    
		\textbf{F/M-ID22--41} 	    & 1.164 & -0.059 & -0.051 & -0.050 & -0.037 & -0.019 \\  
        \midrule \midrule
	\end{tabular}  }      
\end{table*}
\subsection{Combined Dimensions}\label{comDim}
The purpose of the combined dimension experiments is twofold.  
First, we wish to determine the performance decrease, if any, of a general DNN based SE system vs. the specialized systems considered in the three previous subsections, where only one dimension was varied at a time. 
Such experiments can be used to relate results previously reported in the literature, where at least one dimension has been fixed, to the more general case where all three dimensions are varied.        
Secondly, we wish to investigate how such a general DNN based SE method performs relative to a state-of-the-art non-machine learning based method, namely the STSA-MMSE method proposed in \cite{erkelens_minimum_2007}. This is done in an attempt to give a realistic picture of the performance difference between these two classes of algorithms, which utilize different kinds of prior knowledge.  

Alternatively, we could have compared the performance with a NMF based SE approach, which is another popular SE algorithm. However, several studies 
\cite{wang_supervised_2015,williamson_deep_2015,kolbaek_speech_2016,williamson_estimating_2015} 
show that DNN based SE algorithms outperform NMF based approaches on several tasks. Furthermore, the NMF based approach can be viewed as a single hidden layer DNN. Hence, comparing the performance of the DNN based SE algorithm investigated in this paper to a NMF based SE algorithm is less interesting than a comparison with the STSA-MMSE based SE approach, which is from a completely different class of algorithms. 

STSA-MMSE type of methods such as \cite{ephraim_speech_1984,erkelens_minimum_2007} are very general and make only few assumptions about the target and noise signals and are therefore often used in practice \cite{hendriks_dft-domain_2013}. Furthermore, the performance of these simple non-machine learning based algorithms in terms of speech intelligibility improvements are well studied in the literature, e.g. \cite{madhu_potential_2013,hu_comparative_2007,luts_multicenter_2010,jensen_spectral_2012}. 
Although deep neural network based speech enhancement algorithms have shown impressive performance, they are often trained and tested in narrow settings using either a few noise types \cite{healy_algorithm_2015,healy_speech-cue_2014} or a single speaker \cite{chen_large-scale_2016}. It is therefore of interest to identify if/when a deep neural network based speech enhancement algorithm can outperform a non-machine learning based method, when approximately the same type of general a priori information is utilized: given that the computational and memory complexity associated with deep neural network type of systems is typically orders of magnitude larger than that associated with simple STSA-MMSE based systems, it is of obvious interest to understand the performance gain one can expect from the increased memory and computational complexity.

The comparison is based on a "General" DNN based SE system trained using all the noise types from the noise dimension experiments at all the SNRs from the SNR dimension experiments, and using all the speakers from the speaker dimension experiments. This means that the system is trained using 7 different and equally distributed noise types mixed with 20 female and 20 male speakers at SNRs from -15 dB to 20 dB. To encompass the increased variability of this dataset compared to the previous datasets the training set size is increased to $40 \times 231 \times 12 = 110880$ utterances.  
To make a fair comparison to the STSA-MMSE method, which does not strongly rely on prior speaker, SNR, or noise type knowledge, 10 unseen noises, 20 unseen females and 20 unseen males are used for evaluating the performance at SNRs from -10 dB to 20 dB. The 10 noises are taken from the DEMAND noise database\footnote{http://parole.loria.fr/DEMAND} and represent a wide range of both stationary and non-stationary noise types.   

The STSA-MMSE method relies on the assumption that noise free Discrete Fourier Transform\,(DFT) coefficients are distributed according to a generalized gamma distribution with parameters $\gamma = 2$ and $\nu = 0.15$ \cite{erkelens_minimum_2007,hendriks_dft-domain_2013}. The \emph{a priori} SNR estimator used by the STSA-MMSE method is the Decision-Directed approach \cite{ephraim_speech_1984} using a smoothing factor of $0.98$ and a noise Power Spectral Density\,(PSD) estimate based on the noise PSD tracker reported in \cite{hendriks_mmse_2010}\footnote{http://insy.ewi.tudelft.nl/content/software-and-data}. 
For each utterance, the noise tracker was initialized using a noise PSD estimate based on a noise only region prior to speech activity.      

The results of the experiments are presented in Tables \ref{table:noise_mmse_stoi} and \ref{table:noise_mmse_pesq} for the STSA-MMSE method and in Tables \ref{table:noise_widesys_stoi} and \ref{table:noise_widesys_pesq} for the general  DNN based SE system. The performance scores for the noisy unprocessed mixtures are given in parenthesis and the average across all 10 noises at each SNR is given in the last row. 
From Tables \ref{table:noise_mmse_stoi} and \ref{table:noise_widesys_stoi} it is seen that for all SNRs, the DNN based SE system outperforms the STSA-MMSE method in terms of STOI. Similar behavior is seen from Tables \ref{table:noise_mmse_pesq} and \ref{table:noise_widesys_pesq}, where the systems are evaluated using PESQ. 
However, at high SNRs the STSA-MMSE method achieves comparable results with the DNN based SE method and for some noise types such as \emph{DM station} and \emph{DM traffic}, the STSA-MMSE even achieves slightly better PESQ scores at SNRs above 5 dB. 
This might be explained by the fact that the STSA-MMSE algorithm uses prior knowledge in terms of an ideal noise PSD estimate based on a noise only signal region prior to speech activity. This prior knowledge could be particularly beneficial for stationary noise types, where the initial noise PSD estimate remains correct throughout the utterance. The DNN based SE method explored in this paper does not utilize such prior knowledge. 
However, in \cite{seltzer_investigation_2013,xu_regression_2015} noise PSD estimates obtained prior to speech activity were used in combination with traditional features to train a DNN based SE system and it was shown that performance was improved, when such prior knowledge was utilized. 
It is also seen that the STSA-MMSE method on average does not improve STOI, whereas the general DNN based SE method does. For some conditions such as \emph{DM station} at an SNR of -5 dB the improvement is as high as 0.096.  
In general, it can be observed that a DNN based SE system trained across all three generalizability dimensions using a large number of noise types, speakers and SNRs, outperforms a state-of-the-art non-machine learning based method, even though this method utilizes prior knowledge in terms of ideal initial noise PSD estimates. However, the performance of the general DNN based SE system is on average considerably reduced compared to the specialized systems where only one generalizability dimension was varied at a time.  
From this, it can be concluded that if the usage situation of a SE algorithm is well-defined e.g., in terms of speaker characteristics, noise type, or SNR range, considerably performance improvements can be achieved using a DNN based SE algorithm that has been specifically trained to fit the application. 
On the other hand, for more general applications where the acoustic usage situation cannot be narrowed down in one or more of these dimensions, the advantage of DNN based SE methods is much smaller, while they may still offer improvements over current state-of-the-art non-machine learning based methods.
\begin{table*}[h!]                                  
	\centering     
	\caption{Average STOI performance improvement scores using a state-of-the-art STSA-MMSE estimator. The score in the parenthesis is for the noisy unprocessed signals. The test material is based on 2000 utterances evenly distributed among 20 males and 20 females mixed with 10 different noise types from the DEMAND noise corpus at seven SNRs in the range from -10 dB to 20 dB.}                              
	\label{table:noise_mmse_stoi}                                        
    \resizebox{\textwidth}{!}{%
	\begin{tabular}{lccccccc}     
		\midrule \midrule                         
		\textbf{} & {-10dB}& {-5dB} & {0dB} & {5dB} & {10dB} & {15dB} & {20dB} \\  
        \midrule
		\textbf{DM bus} &       -0.003 (0.819) & -0.003 (0.884) & -0.003 (0.927) & -0.003 (0.955) & -0.003 (0.972) & -0.003 (0.983) & -0.003 (0.991) \\    
		\textbf{DM cafe} &      -0.026 (0.521) & -0.011 (0.643) & -0.003 (0.756) & -0.001 (0.843) & -0.002 (0.902) & -0.003 (0.939) & -0.003 (0.962) \\   
		\textbf{DM cafeter} &   -0.043 (0.459) & -0.022 (0.58) & -0.006 (0.706) & -0.001 (0.811) & -0.002 (0.884) & -0.003 (0.928) & -0.003 (0.955) \\ 
		\textbf{DM car} &       0.008 (0.913) & 0.005 (0.945) & 0.002 (0.966) & 0.000 (0.979) & -0.001 (0.987) & -0.002 (0.992) & -0.002 (0.996) \\        
		\textbf{DM metro} &     0.002 (0.62) & 0.007 (0.73) & 0.006 (0.821) & 0.002 (0.886) & 0.000 (0.929) & -0.001 (0.955) & -0.002 (0.972) \\        
		\textbf{DM resto} &     -0.054 (0.395) & -0.031 (0.496) & -0.012 (0.623) & -0.004 (0.746) & -0.003 (0.84) & -0.004 (0.902) & -0.004 (0.939) \\   
		\textbf{DM river} &     0.011 (0.55) & 0.020 (0.655) & 0.020 (0.755) & 0.013 (0.838) & 0.006 (0.897) & \textbf{0.002} (0.936) &  \textbf{0.000} (0.961) \\         
		\textbf{DM square} &    -0.008 (0.651) & -0.001 (0.761) & -0.000 (0.846) & -0.002 (0.904) & -0.003 (0.94) & -0.003 (0.962) & -0.003 (0.977) \\  
		\textbf{DM station} &   0.008 (0.496) & \textbf{0.022} (0.614) & \textbf{0.023} (0.733) & \textbf{0.016} (0.829) & \textbf{0.008} (0.894) & \textbf{0.002} (0.934) &  \textbf{0.000} (0.958) \\      
		\textbf{DM traffic} &   \textbf{0.019} (0.611) & 0.021 (0.724) & 0.016 (0.819) & 0.009 (0.887) & 0.003 (0.93) & 0.001 (0.957) & -0.001 (0.974) \\       
		\textbf{Average} &      -0.009 (0.604) & 0.001 (0.703) & 0.004 (0.795) & 0.003 (0.868) & 0.000 (0.917) & -0.001 (0.949) & -0.002 (0.968) \\       
        \midrule \midrule
	\end{tabular}  }                                                                   
\end{table*} 
\vspace{3 mm}
\begin{table*}[h!]                              
	\centering           
	\caption{As Table \ref{table:noise_mmse_stoi}  but for a state-of-the-art DNN based SE algorithm.}                          
	\label{table:noise_widesys_stoi}                                        
    \resizebox{\textwidth}{!}{%
	\begin{tabular}{lccccccc}     
		\midrule \midrule                      
		\textbf{} & {-10dB}& {-5dB} & {0dB} & {5dB} & {10dB} & {15dB} & {20dB} \\      
        \midrule        
		\textbf{DM bus} &       0.033 (0.819) & 0.021 (0.884) & 0.011 (0.927) & 0.004 (0.955) & -0.001 (0.972) & -0.004 (0.983) & -0.006 (0.991) \\  
		\textbf{DM cafe} &      0.034 (0.521) & 0.058 (0.643) & 0.054 (0.756) & 0.038 (0.843) & 0.022 (0.902) & 0.010 (0.939) & 0.002 (0.962) \\    
		\textbf{DM cafeter} &   -0.011 (0.459) & 0.038 (0.58) & 0.056 (0.706) & 0.045 (0.811) & 0.027 (0.884) & 0.014 (0.928) & 0.005 (0.955) \\ 
		\textbf{DM car} &       0.018 (0.913) & 0.008 (0.945) & 0.002 (0.966) & -0.002 (0.979) & -0.004 (0.987) & -0.006 (0.992) & -0.007 (0.996) \\ 
		\textbf{DM metro} &     0.040 (0.62) & 0.046 (0.73) & 0.038 (0.821) & 0.024 (0.886) & 0.012 (0.929) & 0.004 (0.955) & -0.002 (0.972) \\    
		\textbf{DM resto} &     -0.017 (0.395) & 0.046 (0.496) & 0.078 (0.623) & \textbf{0.069} (0.746) & \textbf{0.043} (0.84) & \textbf{0.022} (0.902) & \textbf{0.009} (0.939) \\   
		\textbf{DM river} &     0.077 (0.55) & 0.089 (0.655) & 0.074 (0.755) & 0.048 (0.838) & 0.026 (0.897) & 0.011 (0.936) & 0.001 (0.961) \\    
		\textbf{DM square} &    0.064 (0.651) & 0.054 (0.761) & 0.036 (0.846) & 0.021 (0.904) & 0.010 (0.94) & 0.003 (0.962) & -0.003 (0.977) \\  
		\textbf{DM station} &   0.076 (0.496) & \textbf{0.096} (0.614) & \textbf{0.080} (0.733) & 0.051 (0.829) & 0.027 (0.894) & 0.013 (0.934) & 0.004 (0.958) \\ 
		\textbf{DM traffic} &   \textbf{0.085} (0.611) & 0.072 (0.724) & 0.048 (0.819) & 0.027 (0.887) & 0.013 (0.93) & 0.004 (0.957) & -0.002 (0.974) \\ 
		\textbf{Average} &      0.040 (0.604) & 0.053 (0.703) & 0.048 (0.795) & 0.032 (0.868) & 0.018 (0.917) & 0.007 (0.949) & 0.000 (0.968) \\    
        \midrule \midrule
	\end{tabular}  }                                                                  
\end{table*}  
\begin{table*}[h]                                  
	\centering     
	\caption{As Table \ref{table:noise_mmse_stoi}  but for PESQ.}                         
	\label{table:noise_mmse_pesq}                                        
    \resizebox{\textwidth}{!}{%
	\begin{tabular}{lccccccc}     
		\midrule \midrule                       
		\textbf{} & {-10dB}& {-5dB} & {0dB} & {5dB} & {10dB} & {15dB} & {20dB} \\   
        \midrule
		\textbf{DM bus} &       0.172 (1.26) & 0.278 (1.51) & 0.325 (1.93) & 0.336 (2.46) & 0.264 (3.05) & 0.129 (3.61) & -0.001 (4.04) \\      
		\textbf{DM cafe} &      -0.035 (1.13) & 0.013 (1.11) & 0.067 (1.20) & 0.142 (1.42) & 0.206 (1.83) & 0.222 (2.37) & 0.174 (2.97) \\      
		\textbf{DM cafeter} &   -0.061 (1.17) & -0.001 (1.11) & 0.056 (1.16) & 0.129 (1.32) & 0.206 (1.65) & 0.228 (2.14) & 0.177 (2.73) \\ 
		\textbf{DM car} &       \textbf{0.363} (1.21) & \textbf{0.500} (1.45) & \textbf{0.682} (1.81) & \textbf{0.742} (2.35) & 0.677 (2.94) & 0.459 (3.53) & 0.192 (4.01) \\       
		\textbf{DM metro} &     0.019 (1.13) & 0.134 (1.17) & 0.264 (1.33) & 0.356 (1.64) & 0.370 (2.12) & 0.297 (2.70) & 0.180 (3.28) \\      
		\textbf{DM resto} &     -0.130 (1.25) & -0.046 (1.13) & 0.029 (1.11) & 0.113 (1.20) & 0.227 (1.43) & 0.305 (1.84) & 0.295 (2.40) \\     
		\textbf{DM river} &     0.009 (1.07) & 0.061 (1.09) & 0.183 (1.17) & 0.410 (1.36) & 0.605 (1.75) & 0.632 (2.30) & 0.500 (2.92) \\      
		\textbf{DM square} &    0.039 (1.08) & 0.102 (1.14) & 0.203 (1.29) & 0.303 (1.61) & 0.344 (2.10) & 0.330 (2.68) & 0.247 (3.29) \\     
		\textbf{DM station} &   -0.008 (1.09) & 0.093 (1.07) & 0.271 (1.13) & 0.511 (1.30) & 0.688 (1.63) & \textbf{0.705} (2.14) & \textbf{0.565} (2.75) \\   
		\textbf{DM traffic} &   0.054 (1.07) & 0.188 (1.09) & 0.406 (1.19) & 0.615 (1.43) & \textbf{0.706} (1.86) & 0.700 (2.43) & 0.558 (3.05) \\   
		\textbf{Average} &      0.042 (1.14) & 0.132 (1.19) & 0.249 (1.33) & 0.366 (1.61) & 0.429 (2.04) & 0.401 (2.57) & 0.289 (3.14) \\      
        \midrule \midrule
	\end{tabular}  }                                                                   
\end{table*} 
\vspace{3 mm}
\begin{table*}[h!]                          
	\centering    
	\caption{As Table \ref{table:noise_mmse_stoi} but for PESQ with a state-of-the-art DNN based SE algorithm.}                            
	\label{table:noise_widesys_pesq}                                        
    \resizebox{\textwidth}{!}{%
	\begin{tabular}{lccccccc}     
		\midrule \midrule                       
		\textbf{} & {-10dB}& {-5dB} & {0dB} & {5dB} & {10dB} & {15dB} & {20dB} \\  
        \midrule
		\textbf{DM bus} &       0.414 (1.26) & 0.550 (1.51) & 0.596 (1.93) & 0.543 (2.46) & 0.401 (3.05) & 0.225 (3.61) & 0.080 (4.04) \\     
		\textbf{DM cafe} &      0.007 (1.13) & 0.149 (1.11) & 0.300 (1.20) & 0.425 (1.42) & 0.479 (1.83) & 0.467 (2.37) & 0.372 (2.97) \\      
		\textbf{DM cafeter} &   -0.047 (1.17) & 0.066 (1.11) & 0.196 (1.16) & 0.350 (1.32) & 0.471 (1.65) & 0.499 (2.14) & 0.423 (2.73) \\ 
		\textbf{DM car} &       \textbf{0.811} (1.21) & \textbf{1.021} (1.45) & \textbf{1.119} (1.81) & \textbf{1.005} (2.35) & \textbf{0.763} (2.94) & 0.447 (3.53) & 0.167 (4.01) \\      
		\textbf{DM metro} &     0.095 (1.13) & 0.242 (1.17) & 0.373 (1.33) & 0.463 (1.65) & 0.483 (2.12) & 0.408 (2.70) & 0.274 (3.28) \\     
		\textbf{DM resto} &     -0.140 (1.25) & -0.009 (1.13) & 0.157 (1.11) & 0.338 (1.20) & 0.499 (1.43) & 0.571 (1.84) & \textbf{0.523} (2.4) \\    
		\textbf{DM river} &     0.111 (1.07) & 0.268 (1.09) & 0.464 (1.17) & 0.639 (1.36) & 0.682 (1.75) & \textbf{0.615} (2.30) & 0.445 (2.92) \\     
		\textbf{DM square} &    0.170 (1.08) & 0.327 (1.14) & 0.484 (1.29) & 0.572 (1.61) & 0.564 (2.10) & 0.489 (2.69) & 0.343 (3.29) \\    
		\textbf{DM station} &   0.059 (1.08) & 0.199 (1.07) & 0.356 (1.13) & 0.513 (1.30) & 0.610 (1.63) & 0.603 (2.14) & 0.478 (2.75) \\   
		\textbf{DM traffic} &   0.172 (1.07) & 0.347 (1.09) & 0.513 (1.19) & 0.620 (1.43) & 0.636 (1.86) & 0.575 (2.43) & 0.427 (3.05) \\  
		\textbf{Average} &      0.165 (1.14) & 0.316 (1.19) & 0.456 (1.33) & 0.547 (1.61) & 0.559 (2.04) & 0.490 (2.57) & 0.353 (3.14) \\     
        \midrule \midrule
	\end{tabular}  }                                                                    
\end{table*} 
\subsection{Listening Test}
To investigate how the DNN based SE system performs in practice, an intelligibility test, using 10 normal-hearing Danish graduate students, has been conducted. The gender distribution among the 10 students was 3 females and 7 males with ages from 20 to 28 years and a mean age of 24.    
Five systems have been designed for the SI test and their training specifications are given by Table \ref{tab_sitest}. 
\begin{table}[]
	\centering
	\caption{DNN based SE systems used for the intelligibility test presented in Figs. \ref{fig_sitest1} and \ref{fig_sitest2}. The first colum shows the system ID and the remaining columns show the training criteria.}
	\label{tab_sitest}
    \resizebox{0.8\textwidth}{!}{%
	\begin{tabular}{lccc}     
		\midrule \midrule   
		\multicolumn{1}{c}{{System ID}} & {Noise Dim.} & {SNR Dim.} & {Speaker Dim. }  \\
        \midrule        
		DNN-1  & SSN  & -5 dB             & 20 Female \\ 
		DNN-2  & SSN  & -15 dB -- 20 dB   & 20 Female \\ 
		DNN-3  & BBL  & -5 dB             & 20 Female \\ 
		DNN-4  & BBL  & -15 dB -- 20 dB   & 20 Female \\ 
		DNN-5  & N3--N7, WGN, BBL-ADFD & -15 dB -- 20 dB   & 20 Female, 20 Male \\ 
        \midrule \midrule
	\end{tabular}  }  
\end{table}
The systems are designed to investigate if a female specific system, in different noise and SNR conditions (DNN-1 -- DNN-4), can improve SI, when exposed to an unseen female speaker. This is an extension of the experiments in \cite{healy_algorithm_2015} where the system was tested in matched speaker and matched SNR conditions only. 
 
Furthermore, DNN-5, which is a "general" system that has been trained on a wide range of speakers, noise types and SNRs, is included in the experiments to investigate if such a general system can improve SI, when exposed to both an unseen speaker and noise type. 

The noise types used for training DNN-5 include \ac{WGN}, babble noise (BBL-ADFD) and N3 - N7 from the noise dimension tests described in subsection \ref{sec:noiseDim}. The BBL-ADFD noise is constructed using the procedure for BBL, as described in subsection \ref{sec:snrdim}, but with three males and three females from the unused part of the ADFD corpus.   
Each test subject was exposed to five repetitions of 32 test conditions (2 noise types $\times$ 4 SNRs $\times$ 4 processing conditions), hence each test subject was exposed to a total of 160 sentences. The two noise types are SSN (N1) and BBL (N2) noise and the four SNRs are -13 dB, -9 dB, -5 dB and -1 dB. This SNR range was chosen to cover SNRs where SI is close to $0\%$ (-13 dB) and close to $100\%$ (-1 dB). 
The four processing conditions for each noise type were unprocessed corrupted speech, and corrupted speech processed by DNN-1, DNN-2, and DNN-5, for SSN and DNN-3, DNN-4, and DNN-5, for BBL noise. 
Immediately prior to the listening test, each test subject performed a familiarization test using 24 noisy utterances from a left out test set.     
The speech material used for the SI test was based on the Danish Dantale-II speech corpus \cite{wagener_design_2003}. 
Each utterance, which is spoken by a female, consists of five words from five different word classes appearing in the following order: name, verb, numeral, adjective and a noun and the test subject was asked to identify the spoken words via a computer interface. There are a total of 10 different words within each word class, hence the Dantale-II corpus is based on a total of 50 different words. All sentences are constructed such that they are syntactically correct but semantically unlikely, which makes it difficult to predict one word based on another, hence the corpus is suitable for intelligibility tests.
The SI test was performed in an audiometric booth using a set of beyerdynamic DT 770 headphones and a Focusrite Scarlett 2i2 sound card 

The results are presented in Figs. \ref{fig_sitest1} and \ref{fig_sitest2} for SSN and BBL noise, respectively. 
Figs. \ref{fig_sitest1} and \ref{fig_sitest2} show that DNN-5, which is the speaker, noise type, and SNR general system, is unable to improve SI at any of the four SNRs of BBL noise as well as the SNRs at -13 dB, -9 dB, and -1 dB of SSN. 
A paired-sample t-test shows that this SI degradation is statistical significant, i.e. $p < 0.05$, for all these results. It is also seen that DNN-5 improves SI with a small amount for SSN at an SNR of -5 dB. However, this improvement is not statistically significant $(p = 0.44)$. 
For DNN-2 and DNN-4, which are the female and noise type specific, but SNR general systems, a somewhat different picture is observed. In general both DNN-2 and DNN-4 perform better than DNN-5. For SSN, DNN-2 manages to improve SI over the unprocessed signals at SNR -9 dB, while DNN-4 improves SI at SNRs of -5 dB and -1 dB. However, none of these improvements are statistical significant ($p = 0.10, p = 0.10, p = 0.25$, respectively)

Finally, for DNN-1 and DNN-2, which are the female, noise type, and SNR specific systems, DNN-1 improves over DNN-2, whereas DNN-3 in general performs worse than DNN-4. Especially at an SNR of -5 dB DNN-3 performs significantly ($p<0.001$) worse than DNN-4 $(p = 0.10)$ relative to the unprocessed signals. 
This is surprising since DNN-3 is trained at only -5 dB SNR, while DNN-4 had been trained using the SNR range from -15 dB to 20 dB.
Furthermore, the observed SI improvement, especially for DNN-4 and DNN-5 using BBL, is lower than one would expect based on the STOI scores for related models in Sec. \ref{sec:result}. This discrepancy between STOI scores and observed SI, especially for highly modulated noise signals, has previously been observed \cite{healy_algorithm_2015,chen_large-scale_2016,jorgensen_effects_2015,taal_algorithm_2011}.
For DNN-1 a statistically significant improvement of $10.4$ percentage points $(p=0.011)$ in SI is observed at an SNR of -5 dB, which also corresponds to the SNR at which DNN-1 is trained. 
To the authors knowledge, SI improvements achieved by a female specific DNN based SE system tested on an unseen female speaker has not yet been reported. Furthermore, the system outperforms a wide range of previously reported SI test results by non-machine learning based methods reported in \cite{luts_multicenter_2010,hu_comparative_2007} and is comparable with the SI results reported in \cite{jensen_spectral_2012} where a single continuous-gain MMSE method was used.
\begin{figure}
    \centering
    \begin{minipage}{0.48\textwidth}
        \centering
        \includegraphics[width=1\textwidth]{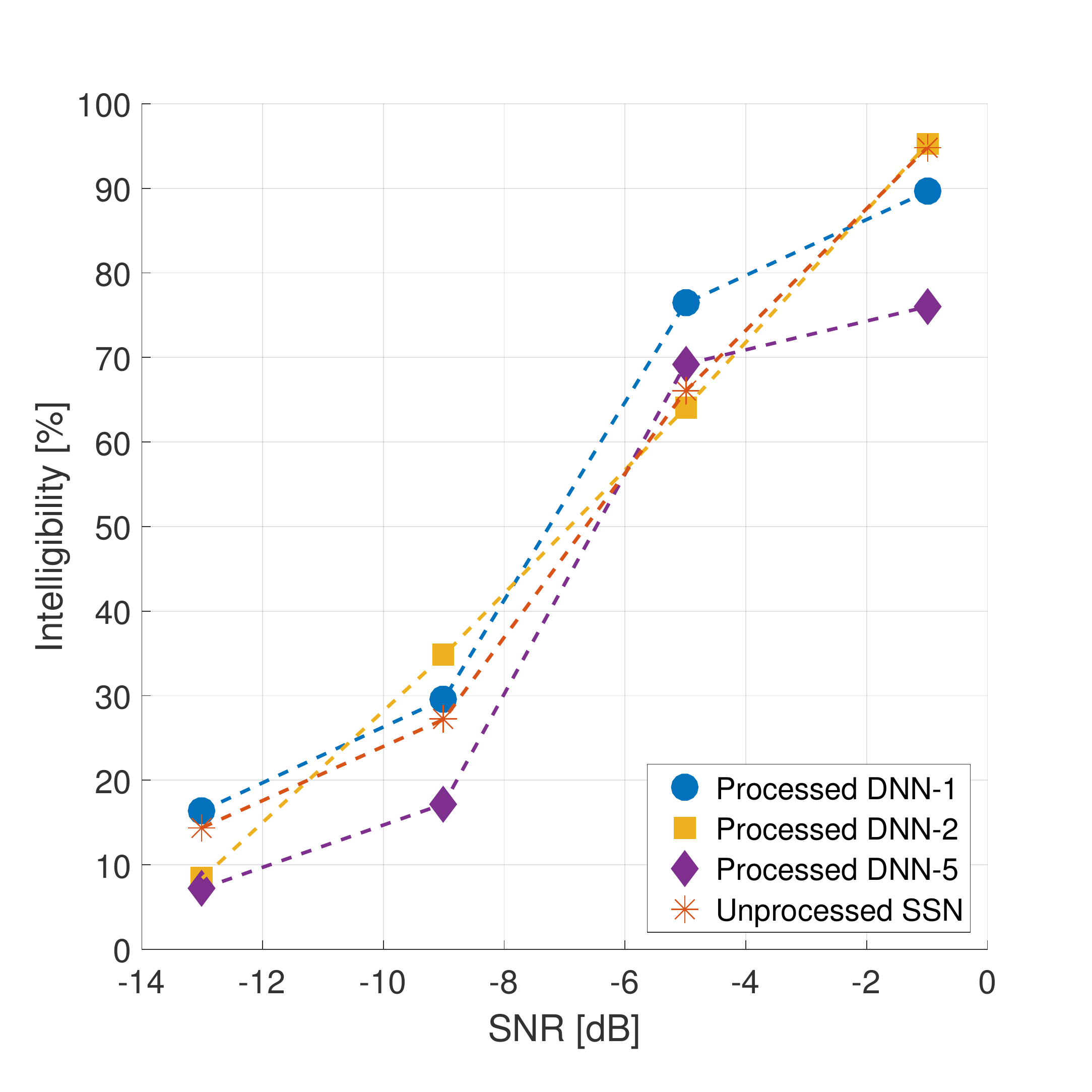}
        \caption{SI test results for 3 different DNN based SE systems processing SSN corrupted speech signals based on 10 Danish test subjects.}
        \label{fig_sitest1}
    \end{minipage}\hfill
    \begin{minipage}{0.48\textwidth}
        \centering
        \includegraphics[width=1\textwidth]{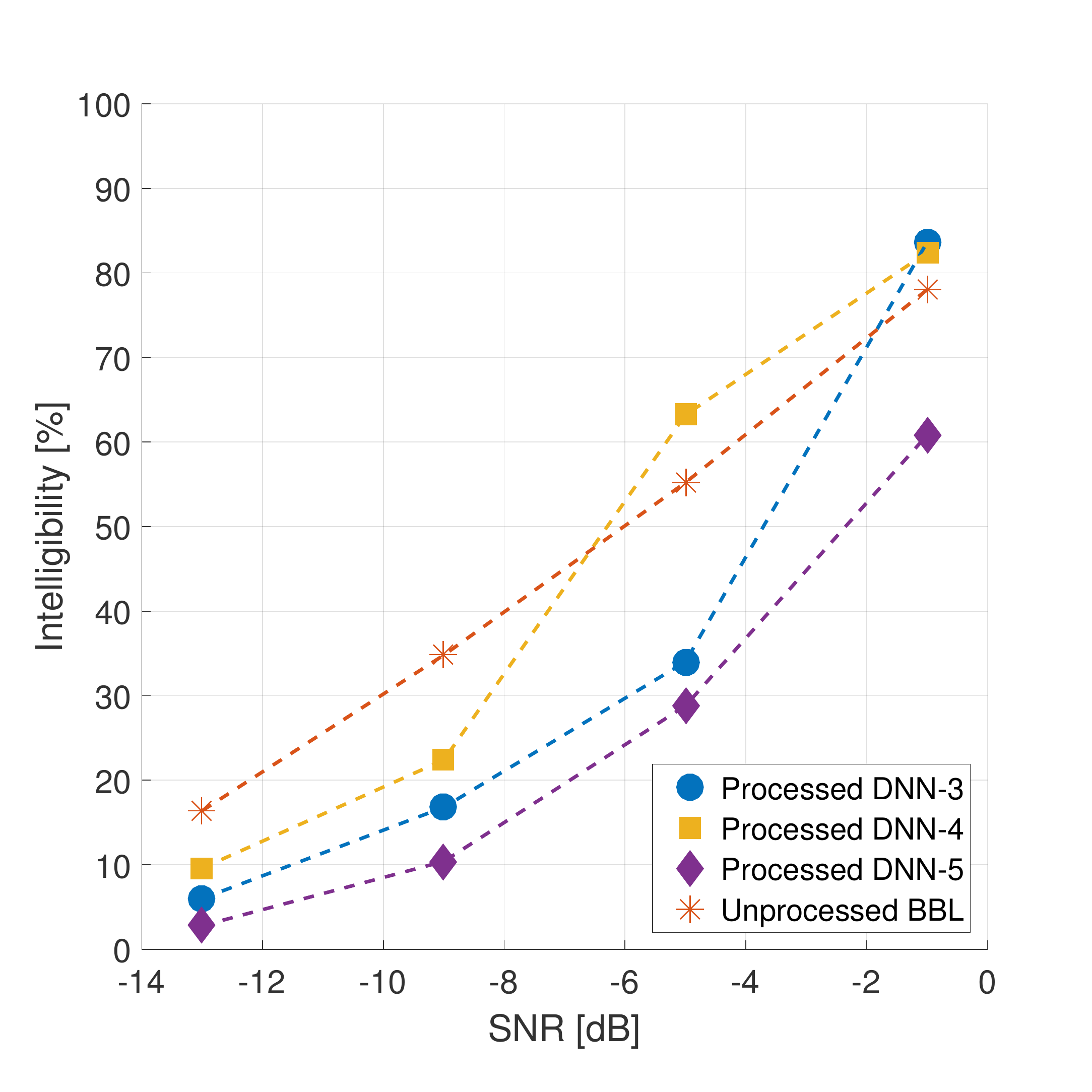}
        \caption{SI test results for 3 different DNN based SE systems processing BBL corrupted speech signals based on 10 Danish test subjects.}
        \label{fig_sitest2}
    \end{minipage}
\end{figure}

\section{Conclusion}
\label{sec:conA}
In this paper the generalizability of a state-of-the-art Deep Neural Network (DNN) based Speech Enhancement\,(SE) method has been investigated. Specifically, it has been investigated how noise specific, speaker specific and Signal-to-Noise Ratio\,(SNR) specific systems perform in relation to noise general, speaker general and SNR general systems, respectively. Furthermore, it has been investigated how such systems perform in relation to a single DNN based SE system which has been designed to be speaker, noise type and SNR general. Also, a comparison between this general DNN based SE system and a state-of-the-art Short-Time Spectral Amplitude Minimum Mean Square Error\,(STSA-MMSE) based SE method has been conducted.   
In general, a positive correspondence between training data variability and generalization was observed. 
Specifically, it was found that DNN based SE systems generalize well to both unseen speakers and unseen noise types given a large number of speakers and noise types were included in the training set.  
Furthermore, it was found that specialized DNN based SE systems trained on only one noise type, one speaker or one SNR, outperformed DNN based SE systems trained on a wide range of noise types, speakers, and SNRs in terms of both estimated Speech Quality\,(SQ) and estimated Speech Intelligibility\,(SI). 
In addition, a general DNN based SE algorithm trained using a large number of speakers, a large number of noise types at a large range of SNRs, outperformed a state-of-the-art STSA-MMSE SE algorithm in terms of estimated SQ and SI. However, the performance of this general DNN based SE system, was considerably reduced compared to the specialized systems, that have been optimized to only a single noise type, a single speaker or a single SNR.           
Finally, it was found that a DNN based SE system trained to be female, noise type and SNR specific, was able to improve SI when tested with an unseen female speaker for particular SNR and noise type configurations, although degrading SI for others.  

In general, it can be concluded that DNN based SE systems do have potential to improve SI in a broader range of usage situations than investigated in \cite{healy_algorithm_2015,chen_large-scale_2016}.
Furthermore, the experiments conducted in this paper, indicate that matching the noise type is critical in acquiring good performance for DNN based SE algorithms, whereas matching the SNR dimension is the least critical followed by the speaker dimension for which good generalization can be achieved with a modest amount of training speakers. 
Also, it can be concluded that considerable improvement in performance can be achieved if the usage situation is limited such that the DNN based SE method can be optimized towards a specific application.

Even though the results reported in this paper are considered general, there is some experimental evidence \cite{xu_regression_2015,wang_deep_2015,halevy_unreasonable_2009,hannun_deep_2014,amodei_deep_2015,chen_large-scale_2016} showing that generalizability performance of DNN based SE algorithms, and DNNs in general, improves when more data and larger networks are being applied, hence SQ and SI performance of DNN based SE systems are expected to improve in the future, when more data and computational resources become available.

%
%
%
%
%
%
\section*{Acknowledgment}
The authors would like to thank Asger Heidemann Andersen for providing software used to conduct the SI tests, and NVIDIA Corporation for the donation of a Titan X GPU.  

{\small\bibliographystyle{bib/IEEEtran}\bibliography{bib/mybibA}}


  \cleardoublepage
  \setcounter{enumiii}{0}
  \setcounter{enumii}{0}
  \setcounter{enumiv}{0}
  \setcounter{enumi}{0}
  \setcounter{equation}{0}
  \setcounter{figure}{0}
  \setcounter{footnote}{0}
  \setcounter{mpfootnote}{0}
  \setcounter{paragraph}{0}
  \setcounter{parentequation}{0}
  \setcounter{part}{0}
  \setcounter{section}{0}
  \setcounter{subparagraph}{0}
  \setcounter{subsection}{0}
  \setcounter{subsubsection}{0}
  \setcounter{table}{0}
  \papertitlepage{%
  Speech Enhancement Using Long Short-Term Memory Based Recurrent Neural Networks for Noise Robust Speaker Verification}{paper:paperB}{%
  Morten Kolbæk, Zheng-Hua Tan, and Jesper Jensen
}{%
  The paper has been published in\\
  \textit{Proceedings IEEE Spoken Language Technology Workshop}, \\pp. 305-311, December 2016.
}{%
  \noindent\copyright\ 2016 IEEE
}

\acresetall
\begin{abstract}
In this paper we propose to use a state-of-the-art \ac{DRNN} based Speech Enhancement\,(SE) algorithm for noise robust \ac{SV}.
Specifically, we study the performance of an i-vector based SV system, when tested in noisy conditions using a DRNN based SE front-end utilizing a Long Short-Term Memory\,(LSTM) architecture. We make comparisons to systems using a Non-negative Matrix Factorization\,(NMF) based front-end, and a Short-Time Spectral Amplitude Minimum Mean Square Error\,(STSA-MMSE) based front-end, respectively.

We show in simulation experiments that a male-speaker and text-independent DRNN based SE front-end, without specific \emph{a priori} knowledge about the noise type outperforms a text, noise type and speaker dependent NMF based front-end as well as a STSA-MMSE based front-end in terms of equal error rates for a large range of noise types and signal to noise ratios on the RSR2015 speech corpus.       
\end{abstract}
\section{Introduction}
\label{sec:intro3}

Biometric technologies, such as speaker verification\,(SV), are a secure, fast and convenient alternative to traditional authentication methods such as\linebreak typed passwords. In fact, the global market for biometric technologies is rapidly growing and is expected to reach \$41.5 billion in 2020 with annual growth rates of more than 20\% \cite{cumming_adoption_2016}.
However, before biometric technologies can be completely adopted and applied in practice, they must, among other things, be robust against external interferences.
This implies that the SV systems are reliable in a broad range of acoustic settings including different noisy environments, competing talkers, recording devices, etc.

In recent years, the branch of machine learning known as \ac{DL} has gained a tremendous amount of attention in both academia and industry. 
DL is a term covering a wide range of machine learning techniques such as Deep Neural Networks\,(DNN), Recurrent Neural Networks or Convolutional Neural Networks\,(CNN) \cite{goodfellow_deep_2016}. Techniques, which have revolutionized a wide range of applications.  \cite{he_delving_2015,bojarski_end_2016,dahl_context-dependent_2012,amodei_deep_2015,taigman_deepface:_2014}.

Especially Automatic Speech Recognition\,(ASR) has been improved using DL \cite{yu_automatic_2015,hannun_deep_2014} and, although DL has revolutionized ASR, DL based noise robust SV has not attained much attention \cite{zhao_robust_2014,medina_wavelet_2003,nugraha_single-channel_2014,du_dnn_2015}. 

In general, SV is the task of verifying the identity of a person based on the voice of the speaker. Specifically, a SV system records via a single or multiple microphones an utterance from a speaker, and the task of the SV system is to verify or reject the claimed identity based on this spoken utterance. If the spoken text is known \textit{a priori}, it is referred as text-dependent SV, while if the spoken text is unknown it is referred as text-independent SV.  

In this paper, noise robustness of a text-dependent SV system is investigated using a state-of-the-art Long Short-Term Memory\,(LSTM) based Deep Recurrent Neural Network\,(DRNN) applied as a denoising front-end using single microphone recordings. 
It should be mentioned that \ac{SR} and SV is very related and differ in principle only in the way the system is evaluated or applied, i.e. for either identity verification or recognition. 
Although, we focus on SV in this paper the proposed front-end denoising techniques could just as well be applied for SR. For the same reasons the referenced literature focuses on both SR and SV. 

Typically, noise robust SV systems can be achieved by modifying either the back-end
or the front-end of the SV system 
\cite{zhao_robust_2014,wang_robust_2007,omid_sadjadi_assessment_2010,medina_wavelet_2003,nugraha_single-channel_2014,du_dnn_2015,moreno-daniel_acoustic_2007,medina_robust_2003,nolazco-flores_enhancing_2008,ming_speaker_2006,baek_thomsen_speaker-dependent_2016,garcia-romero_multicondition_2012,villalba_handling_2013,ming_robust_2007,dat_robust_2015,saeidi_interspeech_2013,shepstone_total_2016,hansen_speaker_2015,hurmalainen_noise_2015,hurmalainen_exemplar-based_2012,hurmalainen_similarity_2015}. 
The back-end constitutes the \ac{UBM}, i-vector extractor and scoring, whereas the front-end constitutes preprocessing steps in terms of denoising of the microphone signal, and feature extraction, prior to back-end processing.     
Even though noise robust SV has been intensively studied in the literature 
\cite{garcia-romero_multicondition_2012,villalba_handling_2013,ming_robust_2007,dat_robust_2015,saeidi_interspeech_2013,zhao_robust_2014,wang_robust_2007,omid_sadjadi_assessment_2010,medina_wavelet_2003,nugraha_single-channel_2014,du_dnn_2015,moreno-daniel_acoustic_2007,medina_robust_2003,nolazco-flores_enhancing_2008,ming_speaker_2006,baek_thomsen_speaker-dependent_2016,hansen_speaker_2015,hurmalainen_noise_2015,hurmalainen_exemplar-based_2012,hurmalainen_similarity_2015}, 
only a few studies \cite{zhao_robust_2014,medina_wavelet_2003,nugraha_single-channel_2014,du_dnn_2015} have applied DNNs in a denoising context. Furthermore, none of these studies apply a DRNN as a SE front-end and compare these methods with existing SE approaches.

In a recent study \cite{baek_thomsen_speaker-dependent_2016} it was shown that if \emph{a priori} knowledge about the noise type is available, a Non-negative Matrix Factorization\,(NMF) based SE front-end outperforms a Wiener filtering based SE front-end \cite{c._loizou_speech_2013} as well as a Short-Time Spectral Amplitude Minimum Mean Square Error\,(STSA-MMSE) based SE front-end \cite{erkelens_minimum_2007}.
However, in the SE literature, several studies \cite{wang_supervised_2015,williamson_deep_2015,williamson_estimating_2015} show that DNN based SE algorithms outperform NMF based SE methods in terms of estimated speech quality and estimated speech intelligibility. Hence, a natural question to ask is whether DNN based SE algorithms also outperform NMF based SE in a SV context. This is the question addressed in this paper. 

The paper is organized as follows. In Sec. \ref{sec:snm} the speech corpora and noise data used for training and testing the NMF dictionaries, the DRNN models as well as the SV system are described. In Sec. \ref{sec:dnn} the proposed DRNN based SE front-ends are presented and in Sec. \ref{sec:base} the baseline systems are presented, which are the NMF based SE front-ends, the STSA-MMSE based SE front-end as well as the SV baseline system. In Sec. \ref{sec:res} the experimental design and results are discussed and finally the paper is concluded in Sec. \ref{sec:con1}.

\section{Speech and Noise Data}
\label{sec:snm}
The denoising task performed by the SE front-ends investigated in this paper can be described by the linear model given by  
\begin{equation}
{y}(n) =  x(n) + d(n),
\label{eq:model}
\end{equation}
where $y(n)$, $x(n)$, and $d(n)$ are the noisy speech signal, the clean speech signal and the additive noise signal, respectively. The task of the denoising front-ends, further described in the following sections, is to estimate $x(n)$ based on observations of $y(n)$. 

\subsection{Speech Corpora}  
In all simulation experiments (reported in Sec. \ref{sec:res}) the clean speech signal $x(n)$, is based on the male part of the RSR2015 corpus \cite{larcher_text-dependent_2014} and the data is allocated among the SV speaker models, the NMF, and DRNN front-ends according to Table. \ref{tab:datalloc}. 

For training SV speaker models, text ID 1 and sessions 1, 4, and 7 from male speakers from m002 to m050 and m052 are selected, and for testing, sessions 2, 3, 5, 6, 8, and 9 are used. Hence, the SV system is text-dependent and is based on 50 male speakers and each speaker is enrolled in the system using 3 utterances and tested with 6 utterances. Furthermore, sessions 1, 4, and 7 have been recorded using a Samsung Nexus smartphone, whereas sessions 2, 5, and 8 and 3, 6, and 9 have been recorded using a Samsung Galaxy S and a HTC Desire smartphone, respectively. That is, the SV system is tested in an unmatched microphone/recording device setting. 

For training the speaker dependent dictionaries used by the NMF based SE front-ends, text ID 1 and sessions 1, 4, and 7 are used. Hence, the NMF font-ends are similarly tested in an unmatched microphone/recording setting.        

The DRNN based front-ends are trained using text IDs 2 -- 30 and sessions 1 -- 9 from male speakers from m053 -- m142, and validated in terms of an early stopping scheme using the same utterances and session IDs, but using speakers from m143 -- m157. 
Although the DRNN front-ends are tested in a matched microphone setting, since they are trained on all nine sessions, they are tested in an unmatched text and an unmatched speaker setting, which is considered a considerably more challenging task.     
\begin{table}[h!]
	\centering
	\caption{Allocation of RSR2015 male-speaker speech data used for training and testing the SV system, as well as the NMF and DRNN front-ends.  }
	\label{tab:datalloc}
    \resizebox{0.6\textwidth}{!}{%
	\begin{tabular}{lcccc}     
	\midrule \midrule  
	\textbf{System} & \textbf{Cond.} & \textbf{Text ID.} & \textbf{Sess. ID} & \textbf{Sprk. ID}    \\  
    \midrule
	SV              & Train              & 1                 & 1, 4, 7           & 2 -- 50 \& 52     \\
	SV              & Test               & 1                 & 2,3,5,6,8,9       & 2 -- 50 \& 52     \\
	NMF             & Train              & 1                 & 1, 4, 7           & 2 -- 50 \& 52     \\
	NMF             & Test               & 1                 & 2,3,5,6,8,9       & 2 -- 50 \& 52     \\
	DRNN            & Train              & 2--30              & 1--9               & 53 -- 142         \\
	DRNN            & Val                & 2--30              & 1--9               & 143 -- 157        \\
	DRNN            & Test               & 1                 & 2,3,5,6,8,9       & 2 -- 50 \& 52   \\ 
        \midrule \midrule
	\end{tabular}  }  
\end{table}

\subsection{Noise Data}
The noise signal $d(n)$ (as given by Eq.\;\eqref{eq:model}) is used to simulate real-life noisy environments, such that the noise robustness of the SV system can be evaluated. For this evaluation the following 6 noise types are used: bus\,(BUS), cafeteria\,(CAF), street\,(STR) and pedestrian\,(PED) from the CHiME3 dataset \cite{barker_third_2015}, as well as a babble\,(BBL) noise, and a Speech Shaped Noise\,(SSN) created by the authors.

The SSN sequence is constructed by filtering a 50 min. Gaussian white noise sequence through a $12$th-order all-pole filter with coefficients found from \ac{LPC} analysis of 100 randomly chosen sentences from a Danish speech corpus known as \emph{Akustiske Databaser for Dansk} (ADFD)\footnote{https://www.nb.no/sbfil/dok/nst\_taledat\_no.pdf}.  

The BBL noise is similarly based on the ADFD corpus. From the ADFD test set, four male and four female speakers are randomly selected. Each speaker is represented by 986 utterances which are normalized to unit Root Mean Square\,(RMS) following the removal of any silent segments using a Voice Activity Detection\,(VAD) algorithm. Then, all 986 utterances from each speaker is concatenated into 8 signals, following truncation to equal length and addition of the eight signals into a single eight speaker babble noise signal.

All six (BUS, CAF, STR, PED, BBL, SSN) noise types were first truncated to have a total duration of 50 min. and then divided into a 40 min. training set, 5 min. validation set and a 5 min. test set. Hence, there are no overlapping noise segments between the training, validation, and test noise.

The noisy mixtures at different Signal to Noise Ratios\;(SNRs) were constructed using the model in Eq.\;\eqref{eq:model} by scaling the noise signal $d(n)$ accordingly. The noise signal was scaled to achieve the desired SNR based on the duration of the entire speech signal $x(n)$. 

Furthermore, a sampling frequency of 16 kHz is used throughout the paper and all audio files are normalized to unit RMS.

\section{Speech Enhancement Using Deep Recurrent Neural Networks}
\label{sec:dnn}
Speech enhancement algorithms based on DNNs have in recent years gained a large amount of attention and showed impressive performance in terms of improving speech quality and speech intelligibility \cite{wang_supervised_2015,williamson_deep_2015,williamson_estimating_2015,chen_large-scale_2016,huang_joint_2015}. Common for these algorithms is that they use a DNN as a regression model to estimate a ratio mask that is applied to the Time-Frequency\,(T-F) representation of the noisy speech signal to acquire an estimate of the clean speech signal. 

A related approach will be adopted in this paper. Specifically, a DNN is employed, but it is improved using LSTM layers \cite{hochreiter_long_1997} and a training criterion that indirectly constructs a ratio mask by minimizing the Mean Square Error\,(MSE) between the desired clean speech signal and the noise. 
In this way, the model learns to separate speech from noise, which is the real desired goal, rather than minimizing the MSE between an ideal mask and an estimated mask, as is typically done \cite{chen_large-scale_2016,healy_algorithm_2015,wang_training_2014}. 

The T-F representation used for the DRNN based SE front-end is a\linebreak $N_{STFT}=512$ point Short-Time Fourier Transform\,(STFT) using a frame width of 32 ms and a frame spacing of 16 ms \cite{c._loizou_speech_2013}. In this way, a frequency dimension of $N = N_{STFT}/2+1 = 257$, covering positive frequencies, is achieved. When the estimated ratio mask has been applied to the noisy speech signal, the time domain representation is achieved by applying an Inverse Short-Time Fourier Transform\,(ISTFT) using the phase from the noisy signal.

\subsection{DRNN Architecture and Training}
All DRNN based front-ends are based on an architecture constituting two LSTM layers and a single fully connected feed-forward output layer with sigmoid activation functions.
The input to the DRNN is the magnitude of the STFT coefficients of the noisy mixture $y(n)$, including a context of 15 past frames and 15 future frames, hence arriving at a final input dimension of $N \times 31= 257 \times 31=7967$.
The output constitutes a ratio mask for a single frame, and the dimension is therefore related to the size of the STFT, i.e. 257 (STFT order is 512).

The training criterion used for training the DRNNs is defined as follows:
Let $|x(n,\omega)|$, $|d(n,\omega)|$ and $|y(n,\omega)|$ denote the magnitude of the STFT of the clean speech signal, the noise signal and the noisy mixture, respectively. Furthermore let $\hat{x}(n,\omega)$ and $\hat{d}(n,\omega)$ denote the estimate of the magnitude of the clean speech signal and noise signal, respectively. 
Finally, let $o(n,\omega)$ denote the output of the DRNN, and let $m_x(n,\omega)$ and $m_d(n,\omega)$ denote the ratio mask representing the speech signal and noise signal, respectively. Since the DRNN has one sigmoid output layer, the speech ratio mask $m_x(n,\omega)$ for a single T-F unit is simply defined as    
\begin{equation}
m_x(n,\omega) = o(n,\omega),
\end{equation}
and $m_d(n,\omega)$ as
\begin{equation}
m_d(n,\omega) = 1 - o(n,\omega).
\label{eq:eq1}
\end{equation}
Furthermore, $\hat{x}(n,\omega)$ is defined as
\begin{equation}
\hat{x}(n,\omega) =  m_x(n,\omega) \times  |y(n,\omega)|,
\label{eq:eq2}
\end{equation}
and $\hat{d}(n,\omega)$ as
\begin{equation}
\hat{d}(n,\omega) =  m_d(n,\omega) \times  |y(n,\omega)|.
\label{eq:eq3}
\end{equation}
Finally, the DRNN MSE training criteria for a single training example $(d(n,\omega)$, $x(n,\omega))$ is defined as: 
\begin{equation}
\begin{split}
MSE(n) &= \frac{1}{N} \sum_{\omega = 1}^{N} (\hat{d}(n,\omega) - |d(n,\omega)|)^2 \\ & + \frac{1}{N} \sum_{\omega = 1}^{N} (\hat{x}(n,\omega) - |x(n,\omega)|)^2.
\label{eq:eq4}
\end{split}
\end{equation}

By using the training criteria given by Eq.\;\eqref{eq:eq4}, it is ensured that the MSE between $\hat{d}(n,\omega)$ and $|d(n,\omega)|$, as well as $\hat{x}(n,\omega)$ and $|x(n,\omega)|$ is minimized, while still ensuring that $m_d(n,\omega)$ and $m_x(n,\omega)$ represents a valid ratio mask, i.e. $m_x(n,\omega) + m_d(n,\omega) = 1$.     

Although $m_x(n,\omega)$ and $m_d(n,\omega)$ are not explicitly used, in this particular work, since the current DRNN directly estimates $|x(n,\omega)|$, the formulation in Eq.\;\eqref{eq:eq4} allows the output layer to be straightforwardly extended to multiple outputs by extending the dimension of the output layer and applying a softmax to ensure all outputs are correctly normalized, hence separating e.g. multiple speakers \cite{yu_permutation_2016}.

The DRNNs used for all experiments in this paper are implemented in CNTK\footnote{https://www.cntk.ai} \cite{agarwal_introduction_2014} and are trained using stochastic gradient descent with truncated backpropagation through time, using 10 time steps and a momentum term of $0.9$ for all epochs. The learning rate is initially set to $0.1$, but is reduced with a factor of $2$, when the validation error has not decreased for one epoch. During training, $20\%$ dropout \cite{hinton_improving_2012} is used for the LSTM layers and the training is aborted, when the learning rate becomes less than $1^{-10}$. When the learning rate is decreased, the training continues from the previous best model.  

\subsection{DRNN Based SE Front-Ends}
A total of seven DRNN based SE front-ends are investigated: Six Noise Specific DRNN (NSDRNN) front-ends, one for each noise type, and one Noise General DRNN (NGDRNN) front-end trained on all six noise types. 

The NSDRNN front-ends are each trained on a particular noise type, hence, at test time, \emph{a priori} knowledge about the noise type is required. This is similar to the NMF front-ends, which also rely on this prior knowledge. 

For the NGDRNN front-end, only a single model is trained using a combination of all six noise types. This front-end therefore utilizes only a minimum amount of \emph{a priori} information, since it is unaware of the actual noise type. The NGDRNN front-end is included to investigate the performance that can be achieved, if less \emph{a priori} knowledge about the noise type is available.

For all DRNN front-ends, $10^5$ noisy mixtures are used for training. The mixtures are generated by drawing a SNR at random from a discrete uniform distribution defined within the SNR range from -5 dB -- 20 dB. Due to the large number of realizations, it is assumed that the distribution of drawn SNRs is approximately uniform. The noise signal used for each noisy mixture was extracted from the whole training noise sequence by using a starting index drawn from a discrete uniform distribution defined over the entire length of the noise sequence. If the starting index is such that there is no room for the whole utterance, the remaining samples are extracted from the beginning of the noise sequence. 
For the NGDRNN front-end, the training noise sequence is constructed by concatenating the six individual noise type sequences, hence the $10^5$ noisy mixtures contain all six noise types evenly distributed, whereas for the NSDRNN front-ends the $10^5$ noisy mixtures contain only a single noise type. A similar approach is used for generating the mixtures used for validation and test.

\section{Baseline Systems}
\label{sec:base}
This section describes the SV baseline as well as the SE baseline front-ends, namely the NMF based SE front-ends and the STSA-MMSE based SE front-end.

\subsection{NMF Baseline}
The basic observation behind NMF is that a non-negative matrix $\mathbf{V} \in \mathbb{R}^{m \times n}$ can be approximately factorized into a product of two non-negative matrices $\mathbf{D} \in \mathbb{R}^{m \times k}$ and $\mathbf{H} \in \mathbb{R}^{k \times n}$ \cite{lee_algorithms_2000} as given by
\begin{equation}
\mathbf{V} \approx \mathbf{D} \mathbf{H}, 
\label{eq:nmf11}
\end{equation}
where $\mathbf{D}$ is known as the dictionary and $\mathbf{H}$ is the activation matrix. The activation matrix $\mathbf{H}$ is used to identify what parts of the dictionary are required to accurately approximate $\mathbf{V}$. The number of columns $k$ in the dictionary $\mathbf{D}$ is a tuning parameter used to adjust the representational power of the factorization.   

The dictionary $\mathbf{D}$ and the activations $\mathbf{H}$ can be found by solving the constrained and regularized least squares minimization problem given by
\begin{equation}
\begin{aligned}
	& \underset{\mathbf{D},\mathbf{H}}{\text{minimize}} 
	& & \frac{1}{2} \Vert \mathbf{V} - \mathbf{D} \mathbf{H} \Vert ^2_F + \alpha \Vert \mathbf{H} \Vert _1  \\
	& \text{subject to}
	& & \mathbf{D},\mathbf{H} \geq 0,
\end{aligned}
\label{eq:nmf}
\end{equation}
where $\Vert \cdot \Vert ^2_F $ is the squared Frobenius norm, $\Vert \cdot \Vert_1 $ is the $\ell1$-norm, and $\alpha > 0$ is a sparsity parameter \cite{schmidt_wind_2007}. 
Equation \eqref{eq:nmf} can be solved in an iterative fashion using  
\begin{subequations}
	\begin{equation}
		\mathbf{H} = \mathbf{H} \circ \frac{\mathbf{D}^T\mathbf{V}}{\mathbf{D}^T\mathbf{D}\mathbf{H} + \alpha},
		\label{eq:nmfupdateH}
	\end{equation}	
	\begin{equation}
		\mathbf{D} = \mathbf{D} \circ \frac{\mathbf{V}\mathbf{H}^T}{\mathbf{D}\mathbf{H}\mathbf{H}^T },
		\label{eq:nmfupdateD}
	\end{equation}	
\label{eq:nmfupdate}
\end{subequations}
where $\circ$ is the Hadamard product, i.e. element-wise multiplication. The solution to Eq.\;\eqref{eq:nmf} is found by alternating between update rule \eqref{eq:nmfupdateH} and \eqref{eq:nmfupdateD} until the value of the cost function given by Eq.\;\eqref{eq:nmf} is below a predefined threshold \cite{schmidt_wind_2007,lee_algorithms_2000}.

When NMF is applied for SE using a model given by Eq.\;\eqref{eq:model}, $\mathbf{V}$ is the STFT magnitudes of a noisy speech signal $\mathbf{V_y}$, and is on the following form:
\begin{equation}
\begin{split}
\mathbf{V_y} & \approx \mathbf{D} \mathbf{H} =
\begin{bmatrix}
\mathbf{D_x} \;
\mathbf{D_d}
\end{bmatrix}
\begin{bmatrix}
\mathbf{H_x} \\
\mathbf{H_d} \\
\end{bmatrix},
\end{split}
\label{eq:nmf13}
\end{equation}
where $\mathbf{D_x}$ and $\mathbf{D_d}$ are speech and noise dictionaries, respectively and $\mathbf{H_x}$ and $\mathbf{H_d}$ are their corresponding activations. The dictionaries $\mathbf{D_x}$ and $\mathbf{D_d}$ are found using the approach given by Eq.\;\eqref{eq:nmfupdate}, in an offline training procedure, prior to test time. 

At test time, using the already trained $\mathbf{D_x}$ and $\mathbf{D_d}$ and a test sample $\mathbf{V_y}$, the corresponding activations $\mathbf{H_x}$ and $\mathbf{H_d}$ are found jointly using Eqs.\;\eqref{eq:nmf13} and \eqref{eq:nmfupdateH}, and the estimate of the clean speech STFT magnitudes $\hat{\mathbf{X}}$ are acquired by \cite{baek_thomsen_speaker-dependent_2016} 
\begin{equation}
\hat{\mathbf{X}} = \mathbf{Y} \circ \frac{\frac{\mathbf{D_x}\mathbf{H_x}}{\mathbf{D_d}\mathbf{H_d}}}{1+\frac{\mathbf{D_x}\mathbf{H_x}}{\mathbf{D_d}\mathbf{H_d}}}.
\label{eq:nmfres}
\end{equation}
The time domain signal is finally achieved by ISTFT of $\hat{\mathbf{X}}$ using the phase of the noisy mixture. 

For the experiments conducted in this paper, one NMF dictionary $\mathbf{D_x}$, is trained for each speaker and each noise type, hence the NMF front-ends are speaker, text, and noise type dependent. This is similar to the study in \cite{baek_thomsen_speaker-dependent_2016}, which enables the use of a NMF based denoising front-end with only a small amount of training data. However, it requires \emph{a priori} knowledge about the noise type at test time.

Furthermore, similarly to \cite{baek_thomsen_speaker-dependent_2016}, the speaker dictionaries $\mathbf{D_x}$ have a fixed size of $64$ columns, i.e. $k = 64$ and are trained using speech-only regions by removing all frames with a sample variance less than $3 \times 10^{-5}$. 
Finally, the NMF training is terminated when the value of the cost function given in Eq.\;\eqref{eq:nmf} is less than  $10^{-4}$ or the number of iterations exceed $500$.

\subsection{STSA-MMSE Baseline}
The STSA-MMSE front-end is a statistical based SE method, which relies on the assumption that noise free Discrete Fourier Transform\,(DFT) coefficients are distributed according to a generalized gamma distribution with parameters $\gamma = 2$ and $\nu = 0.15$ \cite{erkelens_minimum_2007,hendriks_dft-domain_2013}. The \emph{a priori} SNR estimator used by the STSA-MMSE front-end is the Decision-Directed approach \cite{ephraim_speech_1984} using a smoothing factor of $0.98$ and a noise Power Spectral Density\,(PSD) estimate based on the noise PSD tracker reported in \cite{hendriks_mmse_2010}\footnote{http://insy.ewi.tudelft.nl/content/software-and-data}. 
For each utterance, the noise tracker was initialized using a noise PSD estimate based on the first 1000 samples i.e.\;62.5 ms, which is assumed to be a noise-only region. Since the STSA-MMSE front-end only relies on simple statistical assumptions, it is basically text, speaker, and noise type independent and is therefore the method that relies on the least amount of \emph{a priori} knowledge compared to the NMF, NSDRNN, and NGDRNN front-ends.

\subsection{Speaker Verification Baseline}
\label{sec:sv}

The SV baseline is a Gaussian Mixture Model\,(GMM)-UBM i-vector based system \cite{dehak_front-end_2011} and is similar to the system investigated in \cite{baek_thomsen_speaker-dependent_2016}. The baseline is implemented using the Kaldi Speech Recognition Toolkit \cite{povey_kaldi_2011}.

The 4380 male speaker utterances from the TIMIT corpus \cite{garofolo_darpa_1993} are used for obtaining the GMM-UBM as well as the total variability matrix used for i-vector extraction. The used features are 13 Mel Frequency Cepstrum Coefficients\,(MFCC) based on voice only regions of frames with a duration of 25 ms and a frequency range from 0 -- 8 kHz and with cepstral liftering disabled. The energy threshold and the energy mean scale of the Kaldi VAD function are set to their default values of $5.5$ and $0.5$, respectively. 

During enrollment, an i-vector of dimension $400$ is generated for each of the three enrollment utterances, for each speaker. The final speaker model is constructed as the average of these three i-vectors. 

During test time, the cosine distance between each speaker model i-vector and all test utterance i-vectors is computed, and since 50 speakers are enrolled in the SV system and each speaker is represented by 6 test utterances, a total number of $50 \times 6 \times 50=15000$ trials are conducted for each evaluation. When a model is chosen as the target speaker, the remaining $49$ models are used as imposters, hence the last multiplication with $50$. 

From the $15000$ cosine scores the \ac{ROC} curve is constructed, and the \ac{EER} is identified and used as the final evaluation score. The EER is the location on the ROC curve where the false positive rate is equal to the false rejection rate, i.e. one minus the true positive rate. For EERs, lower is better, and a flawless SV system will achieve an EER of zero. 

\section{Experimental Results and Discussion}
\label{sec:res}

The performance of the SV system with different denoising front-ends is presented in Tables \ref{tab:resbbl} -- \ref{tab:resstr}. The SV system is evaluated using noisy mixtures contaminated with the six noise types described in Sec. \ref{sec:snm} at SNRs in the range from -5 dB -- 20 dB. The system is also evaluated using the clean speech signals without any noise, in order to investigate how the denoising front-ends operate in noise-free conditions.

For each noise type and SNR, the SV system is evaluated using the following five front-ends: No front-end processing (No Proc.), STSA-MMSE based front-end processing, NMF based front-end processing, NSDRNN based front-end processing, and finally NGDRNN based front-end processing. 

It should be mentioned that both the NSDRNN and NGDRNN front-ends are tested in unmatched text and speaker conditions, while the NMF method is tested in matched text and speaker conditions.

Furthermore, since the STSA-MMSE, NSDRNN and NGDRNN front-ends are speaker independent the same front-end can be used for both target speakers and imposters. However, since the NMF based front-ends are\linebreak speaker dependent the NMF front-end with speaker ID similar to the target speaker is used to process all trials for that particular speaker, i.e. for both target speaker and imposters. This is done to account for the situation where the claimed speaker ID is false, i.e. an imposter. In these situations it should be ensured that the NMF processing cannot induce a false positive by using a front-end not matched to the speaker ID.  
\begin{table}[h!]
	\caption{EER for SV system using BBL noise corrupted speech.}
	\vspace{-2mm}
	\label{tab:resbbl}
	\centering
	\resizebox{0.6\textwidth}{!}{%
		\begin{tabular}{lccccc}     
			\midrule \midrule  
			SNR & \begin{tabular}[c]{@{}c@{}}No Proc.\end{tabular} & \begin{tabular}[c]{@{}c@{}}STSA-\\MMSE\end{tabular} & \begin{tabular}[c]{@{}c@{}}NMF\end{tabular} & \begin{tabular}[c]{@{}c@{}}NS\\DRNN\end{tabular} & \begin{tabular}[c]{@{}c@{}}NG\\DRNN\end{tabular} \\ 
			\midrule
			-5 dB    & 46.0     & 44.8 	& 40.1    & \textbf{28.9}  & 33.6      \\ 
			0 dB     & 37.9    	& 36.6 	& 32.2    & \textbf{19.6}  & 21.0   	  \\ 
			5 dB     & 26.6    	& 27.4 	& 23.8    & \textbf{14.6}  & 14.8   	  \\ 
			10 dB    & 17.6   	& 18.3 	& 17.0    & \textbf{12.0}  & 13.0      \\ 
			15 dB    & 11.6     & 12.1 	& 14.0    & 10.7  & \textbf{10.5}   	  \\ 
			20 dB    & \textbf{9.26}     & 10.3 	& 11.6    & 9.39  & 9.67   	  \\ 
			Clean    & \textbf{6.67}     & 11.7 	& 14.5    & 10.7  & 11.7   	  \\ \midrule
			Average  & 22.2     & 23.0 	& 21.9    & \textbf{15.1}  & 16.3   	  \\ 
			\midrule \midrule
	\end{tabular}  }  
\end{table}
\begin{table}[h!]
	\caption{EER for SV system using BUS noise corrupted speech.}
	\vspace{-2mm}
	\label{tab:resbus}
	\centering
	\resizebox{0.6\textwidth}{!}{%
		\begin{tabular}{lccccc}     
			\midrule \midrule 
			SNR & \begin{tabular}[c]{@{}c@{}}No Proc.\end{tabular} & \begin{tabular}[c]{@{}c@{}}STSA-\\MMSE\end{tabular} & \begin{tabular}[c]{@{}c@{}}NMF\end{tabular} & \begin{tabular}[c]{@{}c@{}}NS\\DRNN\end{tabular} & \begin{tabular}[c]{@{}c@{}}NG\\DRNN\end{tabular} \\ 
			\midrule
			-5 dB    & 32.0     & 28.0 	& 26.1    & 17.3  & \textbf{16.9}      \\ 
			0 dB     & 27.3    	& 23.5 	& 17.9    & 14.2  & \textbf{12.9}   	  \\ 
			5 dB     & 21.7    	& 21.4 	& 13.4    & \textbf{11.1}  & 11.6   	  \\ 
			10 dB    & 16.3   	& 16.7 	& 10.3    & \textbf{8.67}  & 11.3      \\ 
			15 dB    & 11.3     & 12.6  & 9.07    & \textbf{8.75}  & 10.0   	  \\ 
			20 dB    & 8.49     & 10.9 	& \textbf{7.54}    & 7.75  & 9.87   	  \\ 
			Clean    & \textbf{6.67}     & 11.7 	& 8.07    & 8.94  & 11.7   	  \\ \midrule
			Average  & 17.7     & 17.8 	& 13.2    & \textbf{11.0}  & 12.0   	  \\ 
			\midrule \midrule
	\end{tabular}  }  
\end{table}
\begin{table}[h!]
	\caption{EER for SV system using CAF noise corrupted speech.}
	\vspace{-2mm}
	\label{tab:rescaf}
	\centering
	\resizebox{0.6\textwidth}{!}{%
		\begin{tabular}{lccccc}     
			\midrule \midrule 
			SNR & \begin{tabular}[c]{@{}c@{}}No Proc.\end{tabular} & \begin{tabular}[c]{@{}c@{}}STSA-\\MMSE\end{tabular} & \begin{tabular}[c]{@{}c@{}}NMF\end{tabular} & \begin{tabular}[c]{@{}c@{}}NS\\DRNN\end{tabular} & \begin{tabular}[c]{@{}c@{}}NG\\DRNN\end{tabular} \\ 
			\midrule
			-5 dB    & 39.9     & 40.0 	& 36.8    & \textbf{24.7}  & 25.6      \\ 
			0 dB     & 34.0    	& 33.0 	& 29.9    & \textbf{17.5}  & 19.2   	  \\ 
			5 dB     & 26.7    	& 26.6 	& 22.8    & \textbf{14.0}  & 15.1   	  \\ 
			10 dB    & 18.8   	& 19.2 	& 18.0    & \textbf{11.7}  & 12.1      \\ 
			15 dB    & 12.8     & 13.8 	& 14.3    & \textbf{9.95}  & 11.2   	  \\ 
			20 dB    & \textbf{8.90}     & 11.1 	& 13.0    & 9.35  & 10.6   	  \\ 
			Clean    & \textbf{6.67}     & 11.7 	& 11.7    & 11.2  & 11.7   	  \\ \midrule
			Average  & 21.1     & 22.2 	& 20.9    & \textbf{14.1}  & 15.1   	  \\ 
			\midrule \midrule
	\end{tabular}  }  
\end{table}
\begin{table}[h!]
	\caption{EER for SV system using PED noise corrupted speech.}
	\vspace{-2mm}
	\label{tab:resped}
	\centering
	\resizebox{0.6\textwidth}{!}{%
		\begin{tabular}{lccccc}     
			\midrule \midrule 
			SNR & \begin{tabular}[c]{@{}c@{}}No Proc.\end{tabular} & \begin{tabular}[c]{@{}c@{}}STSA-\\MMSE\end{tabular} & \begin{tabular}[c]{@{}c@{}}NMF\end{tabular} & \begin{tabular}[c]{@{}c@{}}NS\\DRNN\end{tabular} & \begin{tabular}[c]{@{}c@{}}NG\\DRNN\end{tabular} \\ 
			\midrule
			-5 dB    & 43.1     & 38.6 	& 40.7    & \textbf{29.6}  & 30.3      \\ 
			0 dB     & 35.6    	& 31.1 	& 32.9    & 22.0  & \textbf{20.2}   	  \\ 
			5 dB     & 26.3    	& 22.0 	& 24.0    & 15.4  & \textbf{13.7}   	  \\ 
			10 dB    & 18.3   	& 15.5 	& 17.4    & 12.6  & \textbf{10.8}      \\ 
			15 dB    & 11.9     & 12.1 	& 12.2    & \textbf{8.55}  & 9.56   	  \\ 
			20 dB    & 8.57     & 10.3 	& 10.3    & \textbf{8.49}  & 10.9   	  \\ 
			Clean    & \textbf{6.67}     & 11.7 	& 11.3    & 12.3  & 11.7   	  \\ \midrule
			Average  & 21.5     & 20.2 	& 21.3    & 15.6  & \textbf{15.3}   	  \\ 
			\midrule \midrule
	\end{tabular}  }  
\end{table}
\begin{table}[h!]
	\caption{EER for SV system using SSN noise corrupted speech.}
	\vspace{-2mm}
	\label{tab:resssn}
	\centering
	\resizebox{0.6\textwidth}{!}{%
		\begin{tabular}{lccccc}     
			\midrule \midrule 
			SNR & \begin{tabular}[c]{@{}c@{}}No Proc.\end{tabular} & \begin{tabular}[c]{@{}c@{}}STSA-\\MMSE\end{tabular} & \begin{tabular}[c]{@{}c@{}}NMF\end{tabular} & \begin{tabular}[c]{@{}c@{}}NS\\DRNN\end{tabular} & \begin{tabular}[c]{@{}c@{}}NG\\DRNN\end{tabular} \\ 
			\midrule
			-5 dB    & 44.4     & 35.7 	& 37.8    & \textbf{20.5}  & 21.6      \\ 
			0 dB     & 34.9    	& 25.9 	& 26.9    & \textbf{14.5}  & 16.0   	  \\ 
			5 dB     & 25.5    	& 18.0 	& 17.7    & \textbf{11.9}  & 13.2   	  \\ 
			10 dB    & 16.1   	& 11.6 	& 12.1    & \textbf{10.9}  & 11.4      \\ 
			15 dB    & 10.3     & 9.70 	& \textbf{9.51}    & 10.0  & \textbf{9.51}   	  \\ 
			20 dB    & \textbf{7.40}     & 10.3 	& 8.17    & 9.52  & 9.48   	  \\ 
			Clean    & \textbf{6.67}     & 11.7 	& 10.5    & 10.3  & 11.7   	  \\ \midrule
			Average  & 20.8     & 17.6 	& 17.5    & \textbf{12.5}  & 13.3   	  \\ 
			\midrule \midrule
	\end{tabular}  }  
\end{table}
\begin{table}[h!]
	\caption{EER for SV system using STR noise corrupted speech.}
	\vspace{-2mm}
	\label{tab:resstr}
	\centering
	\resizebox{0.6\textwidth}{!}{%
		\begin{tabular}{lccccc}     
			\midrule \midrule 
			SNR & \begin{tabular}[c]{@{}c@{}}No Proc.\end{tabular} & \begin{tabular}[c]{@{}c@{}}STSA-\\MMSE\end{tabular} & \begin{tabular}[c]{@{}c@{}}NMF\end{tabular} & \begin{tabular}[c]{@{}c@{}}NS\\DRNN\end{tabular} & \begin{tabular}[c]{@{}c@{}}NG\\DRNN\end{tabular} \\ 
			\midrule
			-5 dB    & 41.1     & 34.6 	& 35.4    & \textbf{22.3}  & 24.3      \\ 
			0 dB     & 33.7    	& 26.5 	& 27.3    & 17.4  & \textbf{16.8}   	  \\ 
			5 dB     & 26.2    	& 21.5 	& 19.0    & 14.6  & \textbf{13.8}   	  \\ 
			10 dB    & 18.3   	& 15.7 	& 14.1    & 12.4  & \textbf{10.9}      \\ 
			15 dB    & 12.0     & 11.9 	& 12.2    & 9.61  & \textbf{9.40}   	  \\ 
			20 dB    & 9.01     & 11.1 	& 10.0    & 9.10  & \textbf{8.61}   	  \\ 
			Clean    & \textbf{6.67}     & 11.7 	& 10.5    & 10.7  & 11.7   	  \\ \midrule
			Average  & 21.0     & 19.0 	& 18.4    & 13.7  & \textbf{13.6}   	  \\ 
			\midrule \midrule
	\end{tabular}  }  
\end{table}

It is seen from Tables \ref{tab:resbbl} -- \ref{tab:resstr} that the NSDRNN and NGDRNN front-ends achieve the lowest EER for the majority of the test conditions and outperforms the NMF and STSA-MMSE front-ends with a large margin, especially at SNRs below 10 dB. However, no front-end achieves the EER of $6.67$ for the clean condition, hence it seems that all methods introduce some distortion at high SNRs. For practical applications it might be beneficial to incorporate an SNR estimator, such that the SE front-ends only are used when needed, i.e. for low SNRs.    

A somewhat surprising observation is that the NGDRNN front-end in general performs well and not only outperforms the NMF and STSA-MMSE front-ends for the majority of noise types and SNRs, but also the NSDRNN front-ends for several SNRs and noise types. 

This is an observation of practical importance, since it shows that using a single DRNN based front-end, which is both text, SNR, male-speaker and noise type independent eliminates the need for noise type classification and speaker dependent front-ends as would be required by the NMF front-ends.   

The advantage of NMF based front-ends is that they can efficiently utilize small amounts of data. In \cite{baek_thomsen_speaker-dependent_2016} it is shown that using only three utterances from a speaker, a NMF based SE front-end can be designed which outperforms a STSA-MMSE based SE front-end, a Wiener filtering based SE front-end and a spectral subtraction based SE front-end. 

Since DNNs typically require a large amount of data, constructing speaker specific front-ends is not practically feasible, since it would require that each SV user should record large amount of enrollment speech. 
The results presented in Tables \ref{tab:resbbl} -- \ref{tab:resstr}  show that conventional speech corpora, such as RSR2015, can be used to design a male-speaker and text-independent SE front-end that achieves state-of-the-art performance for a number of noise types and SNRs, hence the NGDRNN front-end can be used for noise robust text-dependent and text-independent speaker verification. 
\section{Conclusion}
\label{sec:con1}
In this paper a Deep Recurrent Neural Network\,(DRNN) based Speech Enhancement\,(SE) algorithm has been studied in the context of noise-robust text-dependent Speaker Verification\,(SV). Specifically, a state-of-the-art Long-Short Term Memory\,(LSTM) based DRNN, trained to be either noise type specific or noise type general as well as text and male-speaker independent is used as denoising front-ends for an i-vector based SV system. Finally, the SV performance of the DRNN based SE front-ends are compared against speaker, text, and noise type dependent Non-negative Matrix Factorization\,(NMF) based SE front-ends as well as a Short-Time Spectral Amplitude Minimum Mean Square Error\,(STSA-MMSE) based SE front-end, which is speaker, text and noise type independent.     

We show that the noise type specific DRNN based SE front-ends outperform both the NMF based front-ends as well as the STSA-MMSE based front-end for an SNR range from -5 dB -- 10 dB, for six different noise types. Furthermore, we show that a text, male-speaker and noise type independent DRNN based SE front-end similarly outperforms both the NMF based SE front-ends and the STSA-MMSE based SE front-end at SNRs below 15 dB. This is a result of great practical importance, since it shows that a single DRNN based SE front-end can achieve state-of-the-art SV performance in a variety of noisy environments, hence eliminating the need for noise type classification and speaker dependent front-ends.

\section{Acknowledgment}
\label{sec:ack}
The authors would like to thank Nicolai Bæk Thomsen for assistance and software used for the speaker verification and non-negative matrix factorization baseline systems. Also, we would like to thank Dong Yu for useful discussions regarding CNTK and DNN training. Finally, we gratefully acknowledge the support of NVIDIA Corporation with the donation of the Titan X GPU used for this research. 

The paper reflects some results from the OCTAVE Project (\#647850),\linebreak funded by the Research European Agency (REA) of the European Commission, in its framework program Horizon 2020. The views expressed in this paper are those of the authors and do not engage any official position of the European Commission.
 
\pagebreak

{\small\bibliographystyle{bib/IEEEtran}\bibliography{bib/mybibB}}


  \cleardoublepage
  \setcounter{enumiii}{0}
  \setcounter{enumii}{0}
  \setcounter{enumiv}{0}
  \setcounter{enumi}{0}
  \setcounter{equation}{0}
  \setcounter{figure}{0}
  \setcounter{footnote}{0}
  \setcounter{mpfootnote}{0}
  \setcounter{paragraph}{0}
  \setcounter{parentequation}{0}
  \setcounter{part}{0}
  \setcounter{section}{0}
  \setcounter{subparagraph}{0}
  \setcounter{subsection}{0}
  \setcounter{subsubsection}{0}
  \setcounter{table}{0}

\iftrue
\chapter[Permutation Invariant Training of Deep Models for Speaker-\\Independent Multi-Talker Speech Separation]{Permutation Invariant Training of Deep Models for Speaker-Independent Multi-Talker Speech Separation}\label{paper:paperC}
\chaptermark{}
\vspace{2cm}
\begin{center}
\large Dong Yu, Morten Kolbæk, Zheng-Hua Tan, and Jesper Jensen
\end{center}
\vspace{3cm}
\begin{center}
\normalsize The paper has been published in\\
\textit{Proceedings IEEE International Conference on Acoustics, Speech, and Signal Processing}, pp. 241-245, 2017.
\end{center}
\vspace*{\fill}
\newpage\thispagestyle{empty}
\vspace*{\fill}
\noindent\copyright\ 2017 IEEE\par
\noindent{\em The layout has been revised.}
\vspace*{\fill}
\cleardoublepage
\fi

\acresetall
\begin{abstract}
We propose a novel deep learning training criterion, named \ac{PIT}, for speaker independent multi-talker speech separation, commonly known as the cocktail-party problem. Different from the multi-class regression technique and the \ac{DPCL} technique, our novel approach minimizes the separation error directly. This strategy effectively solves the long-lasting label permutation problem, that has prevented progress on deep learning based techniques for speech separation. We evaluated PIT on the WSJ0 and Danish mixed-speech separation tasks and found that it compares favorably to \ac{NMF}, \ac{CASA}, and DPCL and generalizes well over unseen speakers and languages. Since PIT is simple to implement and can be easily integrated and combined with other advanced techniques, we believe improvements built upon PIT can eventually solve the cocktail-party problem.
\end{abstract}
\section{Introduction}

Despite the significant progress made in dictating single-speaker speech in the recent years \cite{PretrainVSFineTune-Yu2010,CD-DNN-HMM-dahl2012,CD-DNN-HMM-SWB-seide2011,DNN4ASR-hinton2012}, the progress made in multi-talker mixed speech separation and recognition, often referred to as the cocktail-party problem \cite{Cherry53,AuditorySceneAnalysis-bregman1994}, has been less impressive. Although human listeners can easily perceive separate sources in an acoustic mixture, the same task seems to be extremely difficult for automatic computing systems, especially when only a single microphone recording of the mixed-speech is available \cite{MonauralSpeechSepChallenge-Cooke2010,SingleChannelSep-Weng2015}.

Nevertheless, solving the cocktail-party problem is critical to enable scenarios such as automatic meeting transcription, automatic captioning for audio/video recordings (e.g., YouTube), and multi-party human-machine interactions (e.g., in the world of Internet of things (IoT)), where speech overlapping is commonly observed.

Over the decades, many attempts have been made to attack this problem. Before the deep learning era, the most popular technique was \acf{CASA} \cite{CASA-cooke2005,CASA-ellis1996}. In this approach, certain segmentation rules based on perceptual grouping cues \cite{PerceptualCuesInCASA-wertheimer1938} are (often semi-manually) designed to operate on low-level features to estimate a time-frequency mask that isolates the signal components belonging to different speakers. This mask is then used to reconstruct the signal. \acf{NMF} \cite{sparseNMF-schmidt2006,NMF-SpeechSep-smaragdis2007,SparseNMF-le2015} is another popular technique which aims to learn a set of non-negative bases that can be used to estimate mixing factors during evaluation. Both CASA and NMF led to very limited success in separating sources in multi-talker mixed speech \cite{MonauralSpeechSepChallenge-Cooke2010}. The most successful technique before the deep learning era is the model based approach \cite{IBM-SuperHuman-kristjansson2006,SpeechSepWithFactorialHMM-virtanen2006,SpeechSepWithAdaptedModel-weiss2007}, such as factorial GMM-HMM \cite{FactorialHMM-ghahramani1997}, that models the interaction between the target and competing speech signals and their temporal dynamics. Unfortunately this model assumes and only works under closed-set speaker condition. 

Motivated by the success of deep learning techniques in single-talker \ac{ASR} \cite{PretrainVSFineTune-Yu2010,CD-DNN-HMM-dahl2012,CD-DNN-HMM-SWB-seide2011,DNN4ASR-hinton2012}, researchers have developed many deep learning techniques for speech separation in recent years. Typically, networks are trained based on parallel sets of mixtures and their constituent target sources \cite{SpeechSepTrainingTargets-wang2014,SpeechEnhanceWithDNN-xu2014,SpeechSepWithLSTM-weninger2015,JointMaskDNN-Huang2015}. The networks are optimized to predict the source belonging to the target class, usually for each time-frequency bin. Unfortunately, these works often focus on, and only work for, separating speech from (often challenging) background noise (or music) because speech has very different characteristics than noise/music. Note that there are indeed works that are aiming at separating multi-talker mixed speech (e.g. \cite{JointMaskDNN-Huang2015}). However, these works rely on speaker-dependent models by assuming that the (often few) target speakers are known during training.

The difficulty in speaker-independent multi-talker speech separation\linebreak comes from the label ambiguity or permutation problem (which will be described in Section~\ref{sec:problem1}). Only two deep learning based works \cite{SingleChannelSep-Weng2015,DeepClustering-hershey2015,DeepClustering2-isik2016} have tried to address and solve this harder problem. In Weng et al. \cite{SingleChannelSep-Weng2015}, which achieved the best result on the dataset used in 2006 monaural speech separation and recognition challenge \cite{MonauralSpeechSepChallenge-Cooke2010}, the instantaneous energy was used to solve the label ambiguity problem and a two-speaker joint-decoder with speaker switching penalty was used to separate and trace speakers. This approach tightly couples with the decoder and is difficult to scale up to more than two speakers due to the way labels are determined. Hershey et al. \cite{DeepClustering-hershey2015,DeepClustering2-isik2016} made significant progress with their \acf{DPCL} technique. In their work, they trained an embedding for each time-frequency bin to optimize a segmentation (clustering) criterion. During evaluation, each time-frequency bin was first mapped into the embedding space upon which a clustering algorithm was used to generate a partition of the time-frequency bins. Impressively, their systems trained on two-talker mixed-speech perform well on three-talker mixed-speech. However, in their approach it is assumed that each time-frequency bin belongs to only one speaker (i.e., a partition) due to the clustering step. Although this is often a good approximation, it is known to be sub-optimal. Furthermore, their approach is hard to combine with other techniques such as complex-domain separation.

In this paper, we propose a novel training criterion, named \acf{PIT}, for speaker independent multi-talker speech separation. Most prior arts treat speech separation as either a multi-class regression problem or a segmentation (or clustering) problem. PIT, however, considers it a \emph{separation} problem (as it should be) by minimizing the separation error. More specifically, PIT first determines the best output-target assignment and then minimizes the error given the assignment. This strategy, which is directly implemented inside the network structure, elegantly solves the long-lasting label permutation problem that has prevented progress on deep learning based techniques for speech separation.

We evaluated PIT on the WSJ0 and Danish mixed-speech separation tasks. Experimental results indicate that PIT compares favorably to NMF, CASA, and DPCL and generalizes well over unseen speakers and languages. In other words, through the training process PIT learns acoustic cues for source separation, which are both speaker and language independent, similar to humans. Since PIT is simple to implement and can be easily integrated and combined with other advanced techniques we believe improvements built upon PIT can eventually solve the cocktail-party problem.

\section{Monaural Speech Separation}\label{sec:problem1}

The goal of monaural speech separation is to estimate the individual source signals in a linearly mixed, single-microphone signal, in which the source signals overlap in the time-frequency domain. Let us denote the \(S\) source signal sequences in the time domain as \(\mathbf{x}_s(t), s=1,\cdots,S\) and the mixed signal sequence as \(\mathbf{y}(t)=\sum_{s=1}^{S} \mathbf{x}_s(t)\). The corresponding \acf{STFT} of these signals are \(\mathbf{X}_s(t,f)\) and  \(\mathbf{Y}(t,f)=\sum_{s=1}^{S} \mathbf{X}_s(t,f)\), respectively, for each time \(t\) and frequency \(f\). Given \(\mathbf{Y}(t,f)\), the goal of monaural speech separation is to recover each source \(\mathbf{X}_s(t,f)\). 

In a typical setup, it is assumed that only STFT magnitude spectra is available. The phase information is ignored during the separation process and is used only when recovering the time domain waveforms of the sources.

Obviously, given only the magnitude of the mixed spectrum \(|\mathbf{Y}(t,f)|\), the problem of recovering \(|\mathbf{X}_s(t,f)|\) is ill-posed, as there are an infinite number of possible \(|\mathbf{X}_s(t,f)|\) combinations that lead to the same \(|\mathbf{Y}(t,f)|\). To overcome this core problem, the system has to learn from some training set \(\mathbb{S}\) that contains pairs of \(|\mathbf{Y}(t,f)|\) and \(|\mathbf{X}_s(t,f)|\) to look for regularities. More specifically, we train a deep learning model \(g(\cdot)\) such that \(g\left(f(|\mathbf{Y}|\right);\theta)={|\tilde{\mathbf{X}}_s|, s=1,\cdots,S}\), where \(\theta\) is a model parameter vector, and \(f(|\mathbf{Y}|)\) is some feature representation of \(|\mathbf{Y}|\). For simplicity and clarity we have omitted, and will continue to omit, time-frequency indexes when there is no ambiguity.

It is well-known (e.g., \cite{SpeechSepTrainingTargets-wang2014}) that better results can be achieved if, instead of estimating \(|\mathbf{X}_s|\) directly, we first estimate a set of masks \(\mathbf{M}_s(t,f)\) using a deep learning model \(h\left(f(|\mathbf{Y}|);\theta \right)=\tilde{\mathbf{M}}_s(t,f)\) with the constraint that \(\tilde{\mathbf{M}}_s(t,f) \geq 0\) and \(\sum_{s=1}^S \tilde{\mathbf{M}}_s(t,f) = 1\)  for all time-frequency bins \((t,f)\). This constraint can be easily satisfied with the softmax operation. We then estimate \(|\mathbf{X}_s|\) as \(|\tilde{\mathbf{X}}_s| = \tilde{\mathbf{M}}_s \circ |\mathbf{Y}| \), where \( \circ\) is the element-wise product of two operands. This strategy is adopted in this study.

\pagebreak
Note that since we first estimate masks, the model parameters can be optimized to minimize the \acf{MSE} between the estimated mask \(\tilde{\mathbf{M}}_s\) and the \acf{IRM} \(\mathbf{M}_s=\frac{|\mathbf{X}_s|}{|\mathbf{Y}|}\),
\[J_m=\frac{1}{T \times F \times S}\sum_{s=1}^S \|\tilde{\mathbf{M}}_s - \mathbf{M}_s\|^2,\] 
where $T$ and $F$ denote the number of time frames and frequency bins, respectively. This approach comes with two problems. First, in silence segments, \(|\mathbf{X}_s|=0\) and \(|\mathbf{Y}|=0\), so that \(\mathbf{M}_s\) is not well defined. Second, what we really care about is the error between the estimated magnitude and the true magnitude of each source, while a smaller error on masks may not lead to a smaller error on magnitude. 

To overcome these limitations, recent works \cite{SpeechSepTrainingTargets-wang2014} directly minimize the \ac{MSE} 
\[J_x=\frac{1}{T \times F \times S}\sum_{s=1}^S \|\tilde{|\mathbf{X}}_s| - |\mathbf{X}_s|\|^2\] 
between the estimated magnitude and the true magnitude. Note that in silence segments \(|\mathbf{X}_s|=0\) and \(|\mathbf{Y}|=0\), and so the accuracy of mask estimation does not affect the training criterion for those segments. In this study, we estimate masks $\tilde{\mathbf{M}}_s$ which minimize \(J_x\).

\section{Permutation Invariant Training}\label{sec:train}

\begin{figure}[ht]
  \centering
   \includegraphics[width=1.0\linewidth]{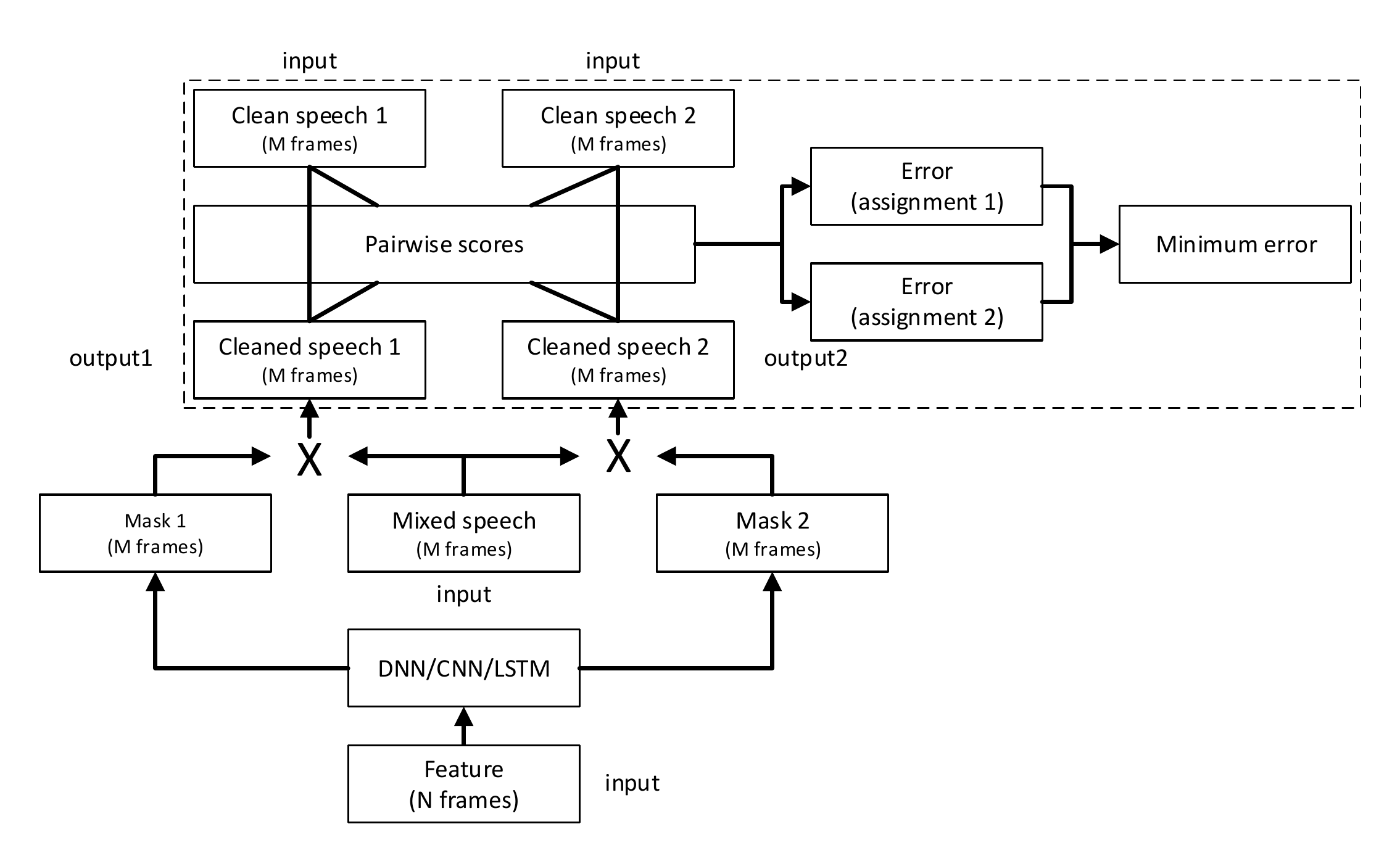}
  \caption{The two-talker speech separation model with permutation invariant training.}\label{fig:modelpit}
\end{figure}

Except DPCL \cite{DeepClustering-hershey2015,DeepClustering2-isik2016}, all other recent speech separation works treat the separation problem as a multi-class regression problem. In their architecture, $N$ frames of feature vectors of the mixed signal \(|\mathbf{Y}|\) are used as the input to deep learning models, such as \acfp{DNN}, \acfp{CNN}, and Long Short-Term Memory\,(LSTM) \acfp{RNN}, to generate one (often the center) frame of masks for each talker. These masks are then used to construct one frame of single-source speech \(|\tilde {\mathbf{X}}_1|\) and \(|\tilde {\mathbf{X}}_2|\), for source 1 and 2, respectively. 

During training we need to provide the correct reference (or target) magnitude \(|\mathbf{X}_1|\) and \(|\mathbf{X}_2|\) to the corresponding output layers for supervision. Since the model has multiple output layers, one for each mixing source, and they depend on the same input mixture, reference assigning can be tricky especially if the training set contains many utterances spoken by many speakers. This problem is referred to as the label ambiguity (or permutation) problem in \cite{SingleChannelSep-Weng2015,DeepClustering-hershey2015}. Due to this problem, prior arts perform poorly on speaker-independent multi-talker speech separation. It was believed that speaker-independent multi-talker speech separation is not feasible \cite{SpeechSepTutorial-wang2016}.

The solution proposed in this work is illustrated in Figure~\ref{fig:modelpit}. There are two key inventions in this novel model: permutation invariant training and segment-based decision making. 

In our new model the reference source streams are given as a set instead of an ordered list. In other words, the same training result is obtained, no matter in which order these sources are listed. This behavior is achieved with PIT highlighted inside the dashed rectangular in Figure~\ref{fig:modelpit}.  
In order to associate references to the output layers, we first determine the (total number of \(S!\)) possible assignments between the references and the estimated sources. %
We then compute the total MSE for each assignment, which is defined as the combined pairwise MSE between each reference \(|\mathbf{X}_s|\) and the estimated source \(|\tilde {\mathbf{X}}_s|\). 
The assignment with the least total MSE is chosen and the model is optimized to reduce this particular MSE. 
In other words we simultaneously conduct label assignment and error evaluation. Similar to the prior arts, PIT uses as input \(N\) successive frames (i.e., an input {\em meta-frame}) of features to exploit the contextual information. Different from the prior arts, the output of the PIT is also a window of frames. With PIT, we directly minimize the separation error at the meta-frame level. 
Although the number of speaker assignments is factorial in the number of speakers, the pairwise MSE computation is only quadratic, and more importantly the MSE computation can be completely ignored during evaluation.

During inference,  the only information available is the mixed speech. Speech separation can be directly carried out for each input meta-frame, for which an output meta-frame with \(M\) frames of speech is estimated for each stream. The input meta-frame is then shifted by one or more frames. Due to the PIT training criterion, output-to-speaker assignment may change across frames. In the simplest setup, we can just assume they do not change when reconstructing sources. Better performance may be achieved if a speaker-tracing algorithm is applied on top of the output of the network.

Once the relationship between the outputs and source streams are determined for each output meta-frame, the separated speech can be estimated, taking into account all meta-frames by, for example, averaging the same frame across meta-frames. 

\section{Experimental Results}
\label{sec:exp1}

\subsection{Datasets}
\label{subsec:datasets1}

We evaluated PIT on the WSJ0-2mix and Danish-2mix datasets. The WSJ0-2mix dataset was introduced in \cite{DeepClustering-hershey2015} and was derived from WSJ0 corpus \cite{wsj0}. The 30h training set and the 10h validation set contains two-speaker mixtures generated by randomly selecting speakers and utterances from the WSJ0 training set si\_tr\_s, and mixing them at various signal-to-noise ratios (SNRs) uniformly chosen between 0 dB and 5 dB. The 5h test set was similarly generated using utterances from 16 speakers from the WSJ0 validation set si\_dt\_05 and evaluation set si\_et\_05.

The Danish-2mix dataset was constructed from the Danish corpus \cite{dkCorpus}, which consists of approximately 560 speakers each speaking 312 utterances with average utterance duration of approximately 5 sec. The dataset was constructed by randomly selecting a set of 45 male and 45 female speakers from the corpus, and then allocating 232, 40, and 40 utterances from each speaker to generate mixed speech in the training, validation and \acf{CC} (seen speaker) test set, respectively. 40 utterances from each of another 45 male and 45 female speakers were randomly selected to construct the \acf{OC} (unseen speaker) test set. Speech mixtures were constructed in the way similar to the WSJ0-2mix dataset, but all mixed with 0 dB - the hardest condition. We constructed 10k and 1k mixtures in total in the training and validation set, respectively, and 1k mixtures for each of the CC and OC test sets. The Danish-3mix (three-talker mixed speech) dataset was constructed similarly.

In this study we focus on the WSJ0-2mix dataset so that we can directly compare PIT with published state-of-the-art results obtained using other techniques.

\subsection{Models}

Our models were implemented using the \acf{CNTK} \cite{CNTK2014}. The feed-forward DNN (denoted as DNN) has three hidden layers each with 1024 ReLU units. In (inChannel, outChannel)-(strideW, strideH) format, the CNN model has one $(1,64)-(2,2)$, four $(64,64)-(1,1)$, one $(64,128)-(2,2)$, two $(128,128)-(1,1)$, one $(128,256)-(2,2)$, and two $(256,256)-(1,1)$ convolution layers with $3 \times 3$ kernels, a pooling layer and a 1024-unit ReLU layer. The input to the models is the stack (over multiple frames) of the 257-dim STFT spectral magnitude of the speech mixture, computed using STFT with a frame size of 32ms and 16ms shift. There are $S$ output streams for $S$-talker mixed speech. Each output stream has a dimension of $257 \times M$, where $M$ is the number of frames in the output meta-frame. In our study, the validation set is only used to control the learning rate.

\subsection{Training Behavior}

\begin{figure}[ht] 
  \centering
   \centerline{\includegraphics[width=1\linewidth]{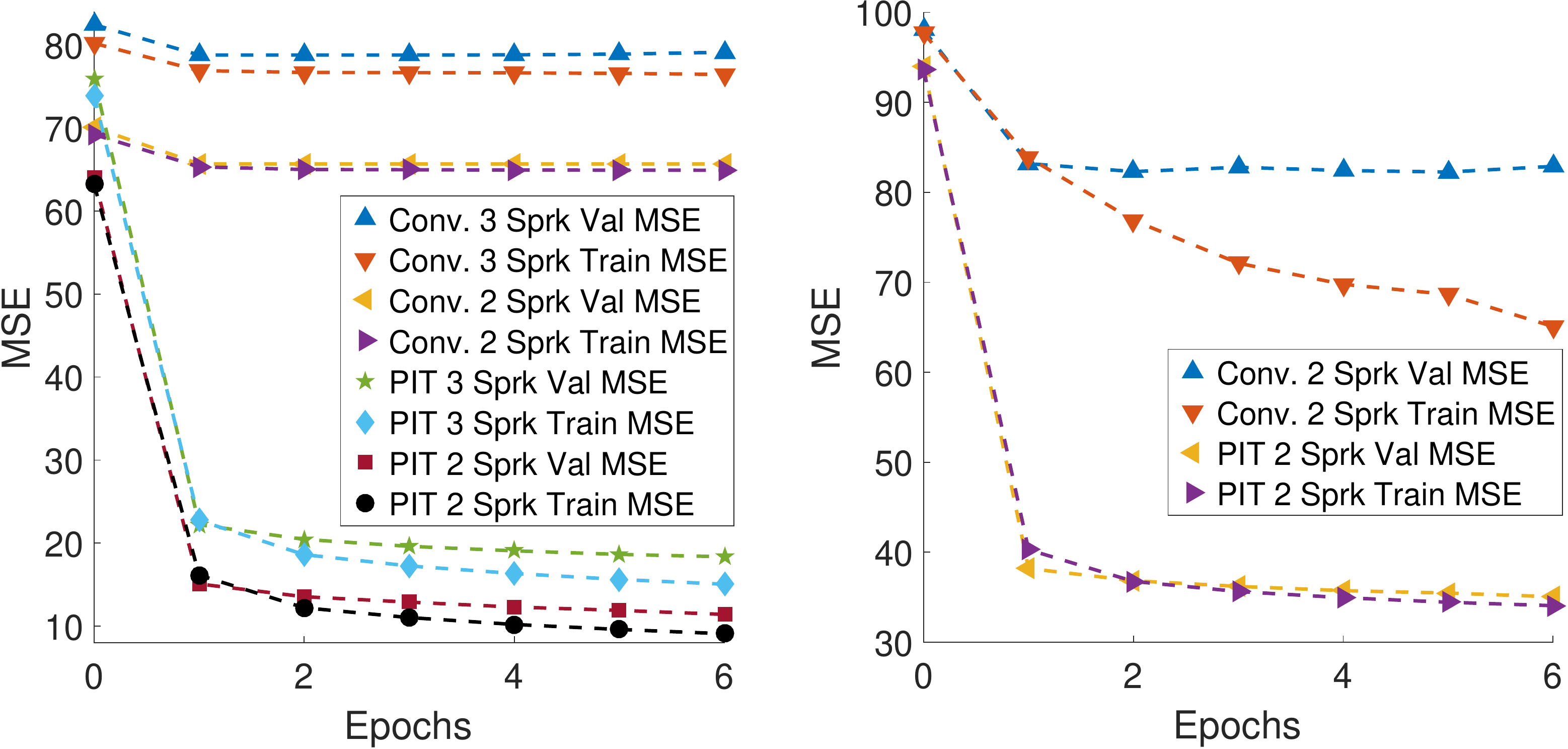}}
  \caption{MSE over epochs on the  Danish (left) and WSJ0 (right) training and validation sets with conventional training and PIT.}\label{fig:mse1}
\end{figure}

In Figure~\ref{fig:mse1} we plotted the DNN training progress as measured by the MSE on the training and validation set with conventional training and PIT on the mixed speech datasets described in subsection \ref{subsec:datasets1}. From the figure we can see clearly that the validation MSE hardly decreases with the conventional approach due to the label permutation problem discussed in \cite{SingleChannelSep-Weng2015,DeepClustering-hershey2015}. In contrast, training converges quickly to a much better MSE for both two- and three-talker mixed speech when PIT is used.

\subsection{Signal-to-Distortion Ratio Improvement}

We evaluated PIT on its potential to improve the \acf{SDR} \cite{vincent_performance_2006}, a metric widely used to evaluate speech enhancement performance.
\begin{table}[t]
\caption{SDR improvements (dB) for different separation methods on the WSJ0-2mix dataset.}
\label{tab:WSJ0-2mix}
\centering
\resizebox{0.65\textwidth}{!}{%
\begin{tabular}{l|c|cc|cc}    
\midrule \midrule  
Method & Input\textbackslash Output & \multicolumn{2}{c|} {Opt. Assign} & \multicolumn{2}{c} {Def. Assign}\\ 
       &  window                    		& CC & OC		&  CC 	& OC	\\
\midrule
Oracle NMF \cite{DeepClustering-hershey2015}	& -	& -	& -		& 5.1  & -     \\ 
CASA \cite{DeepClustering-hershey2015}   	& -  & - & - 	& 2.9  	& 3.1   \\
DPCL \cite{DeepClustering-hershey2015} & 100\textbackslash100 & 6.5 & 6.5& 5.9  	& 5.8   \\
DPCL+ \cite{DeepClustering2-isik2016}   & 100\textbackslash100 & -	& -	&  -  	& 10.3     \\ 
\midrule
PIT-DNN    & 101\textbackslash101    		& 6.2 	&     6.0  	&     5.3 	&     5.2     \\ 
PIT-DNN    & 51\textbackslash51    			& 7.3 	&     7.2  	& \bf{5.7} 	& \bf{5.6}     \\ 
PIT-DNN    & 41\textbackslash7    			& 10.1 	&     10.0  &    -0.3 	&    -0.6     \\ 
PIT-DNN    & 41\textbackslash5    		& \bf{10.5} & \bf{10.4}  &   -0.6 	&    -0.8     \\ 
\midrule
PIT-CNN    & 101\textbackslash101   		& 8.4 	&     8.6  	& \bf{7.7} 	& \bf{7.8}     \\ 
PIT-CNN    & 51\textbackslash51      		& 9.6 	&     9.7  	&     7.5 	&     7.7     \\ 
PIT-CNN    & 41\textbackslash7     			& 10.7 	&     10.7  &    -0.6 	&    -0.7     \\ 
PIT-CNN    & 41\textbackslash5      	& \bf{10.9} & \bf{10.9}  &   -0.8 	&    -0.9     \\ 
\midrule
IRM  		& -     						& 12.3 	& 12.5 	& 12.3 	& 12.5     \\ 
    \midrule \midrule
\end{tabular}  }  
\end{table}

In Table~\ref{tab:WSJ0-2mix} we summarized the SDR improvement in dB from different separation configurations for two-talker mixed speech in CC and OC. In these experiments each frame was reconstructed by averaging over all output meta-frames that contain the same frame. In the default assignment setup it is assumed that there is no output-speaker switch across frames (which is not true). This is the improvement achievable using PIT without any speaker tracing. In the optimal assignment setup, the output-speaker assignment for each output meta-frame is  determined based on mixing streams. This reflects the separation performance within each segment (meta-frame) and is the improvement achievable when the speakers are correctly traced. 
The gap between these two values indicates the contribution from speaker tracing. 
As a reference, we also provided the IRM result which is the oracle and upper bound achievable on this task. 

From the table we can make several observations. First, without speaker tracing (def. assign) PIT can achieve similar and better performance than the original DPCL \cite{DeepClustering-hershey2015}, respectively, with DNN and CNN, but under-performs the more complicated DPCL+ \cite{DeepClustering2-isik2016}. Note that, PIT is much simpler than even the original (simpler) DPCL and we did not fine-tune architectures and learning procedures as done in \cite{DeepClustering2-isik2016}. Second, as we reduce the output window size we can improve the separation performance within each window and achieve better SDR improvement if speakers are correctly traced (opt. assign). However, when output window size is reduced, the output-speaker assignment changes more frequently as indicated by the poor default assignment performance. Speaker tracing thus becomes more important given the larger gap between the opt. assign and def. assign. Fourth, PIT generalizes well on unseen speakers since the performances on the open and closed conditions are very close. Fifth, powerful models such as CNN consistently outperforms DNNs but the gain diminishes when the output window size is small.

\begin{table}[t]
\caption{SDR improvements (dB) based on optimal assignment for DNNs trained with Danish-2mix.}
\label{tab:twotalker}
\centering
\resizebox{0.5\textwidth}{!}{%
\begin{tabular}{ccccccccc}    
\midrule \midrule  
Method & \begin{tabular}[c]{@{}c@{}}Input\textbackslash Output\\ window\end{tabular} & CC &  OC & \begin{tabular}[c]{@{}c@{}}WSJ0\\ OC\end{tabular} \\ 
 \midrule
IRM	   &  -						    & 17.2      	& 17.3  & 13.2  \\ 
\midrule
PIT-DNN    & 101\textbackslash101       & 9.00   	  & 8.61 	& 4.29  \\ 
PIT-DNN    & 61\textbackslash61         & 9.87   	  & 9.44  & 5.17  \\ 
PIT-DNN    & 31\textbackslash31         & 11.1   	  & 10.7  & 6.18  \\ 
PIT-DNN    & 31\textbackslash7          & 14.0   	  & 13.8  & 9.03  \\
PIT-DNN    & 31\textbackslash5          & 14.1   	  & 13.9	& 9.29  \\ 
    \midrule \midrule
\end{tabular}  } 
\end{table}

In Table~\ref{tab:twotalker} we summarized the SDR improvement in dB with optimal assignment from different  configurations for DNNs trained on Danish-2mix. We also report SDR improvement using a dataset constructed identical to Danish-2mix but based on the si\_tr\_s data from WSJ0. Besides the findings obtained in Table~\ref{tab:WSJ0-2mix}, an interesting observation is that although the system has never seen English speech, it performs remarkably well on this WSJ0 dataset when compared to the IRM (oracle) values. These results indicate that the separation ability learned with PIT generalizes well not only across speakers but also across languages.

\section{Conclusion and Discussion}\label{sec:conclusion1}

In this paper, we have described a novel permutation invariant training technique for speaker-independent multi-talker speech separation. To the best of our knowledge this is the first successful work that employs the separation view (and criterion) of the task\footnote{Hershey et al. \cite{DeepClustering-hershey2015} tried PIT (called permutation free training in their paper) but failed to make it work. They retried after reading the preprint of this work and now got positive results as well.}, instead of the multi-class regression or segmentation view that are used in prior arts. This is a big step towards solving the important cocktail-party problem in a real-world setup, where the set of speakers are unknown during the training time.

Our experiments on two-talker mixed speech separation tasks demonstrate that PIT trained models generalize well to unseen speakers and languages. Although our results are mainly on two-talker separation tasks, PIT can be easily and effectively extended to the three-talker case as shown in figure \ref{fig:mse1}.

In this paper we focused on PIT - the key technique that enables training for the separation of multi-talker mixed speech. PIT is much simpler yet performs better than the original DPCL \cite{DeepClustering-hershey2015} that contains separate embedding and clustering stages. 

Since PIT, as a training technique, can be easily integrated and combined with other advanced techniques, it has great potential for further improvement. We believe improvements can come from work in the following areas:

First, due to the change of output-speaker assignment across frames, there is a big performance gap between the optimal output-speaker assignment and the default assignment, especially in the same-gender case and when the output window size is small. This gap can be reduced with separate speaker tracing algorithms that exploit the overlapping frames and speaker characteristics (e.g., similarity) in output meta-frames. It is also possible to train an end-to-end system in which speaker tracing is directly built into the model, e.g., by applying PIT at utterance level. We will report these results in other papers.

Second, we only explored simple DNN/CNN structures in this work. More powerful models such as bi-directional LSTMs, CNNs with deconvolution layers, or even just larger models may further improve the performance. Hyper-parameter tuning will also help and sometimes lead to significant performance gain.

Third, in this work we reconstructed source streams from spectral magnitude only. Unlike DPCL, PIT can be easily combined with reconstruction techniques that exploit complex-valued spectrum to further boost performance.

Fourth, the acoustic cues learned by the model are largely speaker and language independent. It is thus possible to train a universal speech separation model using speech in various speakers, languages, and noise conditions. 

Finally, although we focused on monaural speech separation in this work, the same technique can be deployed in the multi-channel setup and combined with techniques such as beamforming due to its flexibility. In fact, since beamforming and PIT separate speech using different information, they complement with each other. For example, speaker tracing may be much easier when beamforming is available.

\section{Acknowledgment}
We thank Dr. John Hershey at MERL and Zhuo Chen at Columbia University for sharing the WSJ0-2mix data list and for valuable discussions.

{\small\bibliographystyle{bib/IEEEtran}\bibliography{bib/mybibC}}


  \cleardoublepage
  \setcounter{enumiii}{0}
  \setcounter{enumii}{0}
  \setcounter{enumiv}{0}
  \setcounter{enumi}{0}
  \setcounter{equation}{0}
  \setcounter{figure}{0}
  \setcounter{footnote}{0}
  \setcounter{mpfootnote}{0}
  \setcounter{paragraph}{0}
  \setcounter{parentequation}{0}
  \setcounter{part}{0}
  \setcounter{section}{0}
  \setcounter{subparagraph}{0}
  \setcounter{subsection}{0}
  \setcounter{subsubsection}{0}
  \setcounter{table}{0}
  \papertitlepage{%
  Multi-Talker Speech Separation With Utterance-Level Permutation Invariant Training of Deep Recurrent Neural Networks
}{paper:paperD}{%
  Morten Kolbæk, Dong Yu,  Zheng-Hua Tan, and Jesper Jensen
}{%
  The paper has been published in\\
\textit{IEEE/ACM Transactions on Audio, Speech, and Language Processing}, \\vol. 25, no. 10, pp.~1901-1913, March 2017.
}{%
  \noindent\copyright\ 2017 IEEE
}

\acresetall

\begin{abstract}
	In this paper we propose the utterance-level Permutation Invariant Training\;(uPIT) technique. 
	uPIT is a practically applicable, end-to-end, deep learning based solution for speaker independent multi-talker speech separation. 
		Specifically, uPIT extends the recently proposed Permutation Invariant Training\;(PIT) technique with an utterance-level cost function, hence eliminating the need for solving an additional permutation problem during inference, which is otherwise required by frame-level PIT. 
		We achieve this using Recurrent Neural Networks\;(RNNs) that, during training, minimize the utterance-level separation error, hence forcing separated frames belonging to the same speaker to be aligned to the same output stream.    
		In practice, this allows RNNs, trained with uPIT, to separate multi-talker mixed speech without any prior knowledge of signal duration, number of speakers, speaker identity or gender.

		We evaluated uPIT on the WSJ0 and Danish two- and three-talker mixed-speech separation tasks and found that uPIT outperforms techniques based on Non-negative Matrix Factorization\;(NMF) and Computational Auditory Scene Analysis\;(CASA), and compares favorably with Deep Clustering\;(DPCL) and the \ac{DANet}. Furthermore, we found that models trained with uPIT generalize well to unseen speakers and languages.
		Finally, we found that a single model, trained with uPIT, can handle both two-speaker, and three-speaker speech mixtures.	
\end{abstract}

\section{Introduction}\label{sec:intro4}
Having a conversation in a complex acoustic environment, with multiple noise sources and competing background speakers, is a task humans are remarkably good at \cite{haykin_cocktail_2005,bronkhorst_cocktail_2000}. 
The problem that humans solve when they focus their auditory attention towards one audio signal in a complex mixture of signals is commonly known as the cocktail party problem \cite{haykin_cocktail_2005,bronkhorst_cocktail_2000}.
Despite intense research for more than half a century, a general machine based solution to the cocktail party problem is yet to be discovered \cite{cherry_experiments_1953,bronkhorst_cocktail_2000,haykin_cocktail_2005,cooke_monaural_2010}. 
A machine solution to the cocktail party problem is highly desirable for a vast range of applications. These include automatic meeting transcription, automatic captioning for audio/video recordings (e.g. YouTube), multi-party human-machine interaction (e.g. in the world of Internet of things (IoT)), and advanced hearing aids, where overlapping speech is commonly encountered.

Since the cocktail party problem was initially formalized \cite{cherry_experiments_1953}, a large number of potential solutions have been proposed \cite{divenyi_speech_2005}, and the most popular techniques originate from the field of Computational Auditory Scene Analysis\;(CASA) \cite{ellis_prediction-driven_1996,cooke_modelling_2005,wang_computational_2006,shao_model-based_2006,hu_unsupervised_2013}. 
In CASA, different segmentation and grouping rules are used to group Time-Frequency\;(T-F) units that are believed to  belong to the same speaker. The rules are typically hand-engineered and based on heuristics such as pitch trajectory, common onset/offset, periodicity, etc. The grouped T-F units are then used to extract a particular speaker from the mixture signal.
Another popular technique for multi-talker speech separation is Non-negative Matrix Factorization\;(NMF) \cite{schmidt_single-channel_2006,smaragdis_convolutive_2007,roux_sparse_2015,lee_algorithms_2000}. The NMF technique uses non-negative dictionaries to decompose the spectrogram of the mixture signal into speaker specific activations, and from these activations an isolated target signal can be approximated using the dictionaries.
For multi-talker speech separation, both CASA and NMF have led to limited success \cite{cooke_monaural_2010,divenyi_speech_2005} and the most successful techniques, before the deep learning era, are based on probabilistic models \cite{kristjansson_super-human_2006,virtanen_speech_2006,stark_source-filter-based_2011}, such as factorial GMM-HMM \cite{ghahramani_factorial_1997}, that model the temporal dynamics and the complex interactions of the target and competing speech signals. Unfortunately, these models assume and only work under closed-set speaker conditions, i.e. the identity of the speakers must be known \emph{a priori}. 

More recently, a large number of techniques based on deep learning \cite{goodfellow_deep_2016} have been proposed, especially for Automatic Speech Recognition\;(ASR) \cite{yu_roles_2010,dahl_context-dependent_2012,seide_conversational_2011,hinton_deep_2012,xiong_achieving_2016,saon_english_2017}, and speech enhancement  \cite{wang_towards_2013,wang_training_2014,xu_experimental_2014,weninger_speech_2015,huang_joint_2015,chen_large-scale_2016,kolbaek_speech_2017,du_speech_2014,goehring_speech_2017}. Deep learning has also been applied in the context of multi-talker speech separation (e.g.  \cite{huang_joint_2015}), although successful work has, similarly to NMF and CASA, mainly been reported for closed-set speaker conditions.      

The limited success in deep learning based speaker independent multi-talker speech separation is partly due to the label permutation problem\linebreak (which will be described in detail in Sec.~\ref{sec:PIT}).
To the authors knowledge only four deep learning based works \cite{weng_deep_2015,hershey_deep_2016,chen_deep_2017,yu_permutation_2017} exist, that have tried to address and solve the harder speaker independent multi-talker speech separation task. 

In Weng \emph{et al.} \cite{weng_deep_2015}, which proposed the best performing system in the 2006 monaural speech separation and recognition challenge \cite{cooke_monaural_2010}, the instantaneous energy was used to determine the training label assignment, which alleviated the label permutation problem and allowed separation of unknown speakers. Although this approach works well for two-speaker mixtures, it is hard to scale up to mixtures of three or more speakers.   

Hershey \emph{et al.} \cite{hershey_deep_2016} have made significant progress with their Deep Clustering (DPCL) technique. In their work, a deep Recurrent Neural Network (RNN) is used to project the speech mixture into an embedding space, where T-F units belonging to the same speaker form a cluster. In this embedding space a clustering algorithm (e.g. K-means) is used to identify the clusters. Finally, T-F units belonging to the same clusters are grouped together and a binary mask is constructed and used to separate the speakers from the mixture signal. 
To further improve the model \cite{isik_single-channel_2016}, another RNN is stacked on top of the first DPCL RNN to estimate continuous masks for each target speaker. 
Although DPCL show good performance, the technique is potentially limited because the objective function is based on the affinity between the sources in the embedding space, instead of the separated signals themselves. 
That is, low proximity in the embedding space does not necessarily imply perfect separation of the sources in the signal space.    

Chen \emph{et al.} \cite{chen_deep_2017,chen_single_2017} proposed a related technique called Deep Attractor Network\;(DANet). Following DPCL, the DANet approach also learns a high-dimensional embedding of the mixture signals. Different from DPCL, however, it creates attractor points (cluster centers) in the embedding space, which attract the T-F units corresponding to each target speaker. The training is conducted in a way similar to the Expectation Maximization\;(EM) principle. The main disadvantage of DANet over DPCL is the added complexity associated with estimating attractor points during inference.

Recently, we proposed the Permutation Invariant Training\;(PIT) technique%
\footnote{In \cite{hershey_deep_2016}, a related permutation free technique, which is similar to PIT for exactly two-speakers, was evaluated with negative results and conclusion.}
\cite{yu_permutation_2017} for attacking the speaker independent multi-talker speech separation problem and showed that PIT effectively solves the label permutation problem.
However, although PIT solves the label permutation problem at training time, PIT does not effectively solve the permutation problem during inference, where the permutation of the separated signals at the frame-level is unknown. We denote the challenge of identifying this frame-level permutation, as the \emph{speaker tracing problem}. 

In this paper, we extend PIT and propose an utterance-level Permutation Invariant Training\;(uPIT) technique, which is a practically applicable, end-to-end, deep learning based solution for speaker independent multi-talker speech separation.
Specifically, uPIT extends the frame-level PIT technique \cite{yu_permutation_2017} with an utterance-level training criterion that effectively eliminates the need for additional speaker tracing or very large input/output contexts, which is otherwise required by the original PIT \cite{yu_permutation_2017}. 
We achieve this using deep Long Short-Term Memory\;(LSTM) RNNs \cite{hochreiter_long_1997} that, during training, minimize the utterance-level separation error, hence forcing separated frames belonging to the same speaker to be aligned to the same output stream. 
This is unlike other techniques, such as DPCL and DANet, that require a distinct clustering step to separate speakers during inference. Furthermore, the computational cost associated with the uPIT training criterion is negligible compared to the computations required by the RNN during training and is zero during inference. 
We evaluated uPIT on the WSJ0 and Danish two- and three-talker mixed-speech separation tasks and found that uPIT outperforms techniques based on NMF and CASA, and compares favorably with DPCL and DANet. 
Furthermore, we show that models trained with uPIT generalize well to unseen speakers and languages, and finally, we found that a single model trained with uPIT can separate both two-speaker, and three-speaker speech mixtures.
	
\pagebreak	
The rest of the paper is organized as follows. In Sec.~\ref{sec:problem} we describe the monaural speech separation problem. In Sec.~\ref{sec:mask} we extend popular optimization criteria used in separating single-talker speech from noises, to multi-talker speech separation tasks. 
In Sec.~\ref{sec:PIT} we discuss the label permutation problem and present the PIT framework. 
In Sec.~\ref{sec:integrated} we introduce uPIT and show how an utterance-level permutation criterion can be combined with PIT. We report series of experimental results in Sec.~\ref{sec:exp} and conclude the paper in Sec.~\ref{sec:conclusion}.

	\section{Monaural Speech Separation}\label{sec:problem}
	
	The goal of monaural speech separation is to estimate the individual source signals ${x}_s[n], \; s=1,2,\cdots,S$ in a linearly mixed single-microphone signal  
	\begin{equation}
		y[n]=\sum_{s=1}^{S} {x}_s[n],
	\end{equation}	
	based on the observed signal ${y}[n]$ only. 
	In real situations, the received signals may be reverberated, i.e., the underlying clean signals are filtered before being observed in the mixture. In this condition, we aim at recovering the reverberated source signals ${x}_s[n]$, i.e., we are not targeting the dereverberated signals.
	
	The separation is usually carried out in the T-F domain, in which the task can be cast as recovering the Short-Time discrete Fourier Transformation\;(STFT) of the source signals  ${X}_s(t,f)$ for each time frame $t$ and frequency bin $f$, given the mixed speech 
	\begin{equation}
	\begin{split}
	{Y}(t,f) =\sum_{n=0}^{N-1} {y}[n+tL]w[n]\exp(-j2{\pi}nf/N),
	\end{split}
	\end{equation}
	where $w[n]$ is the analysis window of length $N$, the signal is shifted by an amount of $L$ samples for each time frame $t=0,1,\cdots,T-1$, and each frequency bin $f=0,1,\cdots,N-1$ is corresponding to a frequency of $(f/N)f_s$ [Hz] when the sampling rate is $f_s$ [Hz]. 

	From the estimated STFT ${\hat{X}}_s(t,f)$ of each source signal, an inverse Discrete Fourier Transform\;(DFT) 
	\begin{equation}
		{\hat{x}}_{s,t}[n] =\frac{1} {N} \sum_{f=0}^{N-1} \hat{X}_s(t,f)\exp(j2{\pi}nf/N)    
	\end{equation}
	can be used to construct estimated time-domain frames, and the overlap-add operation
	\begin{equation}
		{\hat{x}_s}[n] =\sum_{t=0}^{T-1} v[n-tL] \hat{x}_{s,t}[n-tL]              
	\end{equation}
	can be used to reconstruct the estimate $\hat{x}_s[n]$ of the original signal, where $v[n]$ is the synthesis window.

	In a typical setup, however, only the STFT magnitude spectrum $A_s(t,f) \triangleq |{X}_s(t,f)|$ is estimated from the mixture during the separation process, and the phase of the mixed speech is used directly, when recovering the time domain waveforms of the separated sources. This is because phase estimation is still an open problem in the speech separation setup \cite{williamson_complex_2016,erdogan_deep_2017}.
	Obviously, given only the magnitude of the mixed spectrum, $R(t,f)\triangleq |{Y}(t,f)|$, the problem of recovering $A_s(t,f)$ is under-determined, as there are an infinite number of possible $A_s(t,f)$, $s=1, \dots ,S$ combinations that lead to the same $R(t,f)$. To overcome this problem, a supervised learning system has to learn from some training set $\mathbb{S}$ that contains corresponding observations of $R(t,f)$ and $A_s(t,f)$, $s=1, \dots ,S$.

    Let $\mathbf{a}_{s,i} = \left[ {{A_s}(i,1)} ,\; {{A_s}(i,2)} ,\; \dots\; {{A_s}(i,\frac{N}{2}+1)} \right]^T \in \mathbb{R}^{\frac{N}{2}+1}$ denote the single-sided magnitude spectrum for source $s$ at frame $i$.
    Furthermore, let $\mathbf{A}_s \in \mathbb{R}^{\left(\frac{N}{2}+1\right) \times T}$ be the single-sided magnitude spectrogram for source $s$ and all frames $i=1, \dots ,T$, defined as $\mathbf{A}_s = \left[\mathbf{a}_{s,1} ,\; \mathbf{a}_{s,2} ,\; \dots\;, \mathbf{a}_{s,T}\right]$.  
    Similarly, let $\mathbf{r}_{i} = \left[ {{R}(i,1)} ,\; {{R}(i,2)} ,\; \dots\; {{R}(i,\frac{N}{2}+1)} \right]^T$ be the single-sided magnitude spectrum of the observed signal at frame $i$ and let $\mathbf{R} = \left[\mathbf{r}_{1} ,\; \mathbf{r}_{2} ,\; \dots\;, \mathbf{r}_{T}\right] \in \mathbb{R}^{\left(\frac{N}{2}+1\right) \times T}$ be the single-sided magnitude spectrogram for all frames $i=1, \dots ,T$.

	Furthermore, let us denote a supervector $\mathbf{z}_{i} \hspace{-0.5mm}  =  \hspace{-1mm} \left[ \mathbf{a}_{1,i}^T , \mathbf{a}_{2,i}^T , \dots , \mathbf{a}_{S,i}^T \right]^T \hspace{-3mm} \in \mathbb{R}^{S\left(\frac{N}{2}+1\right)}$, consisting of the stacked source magnitude spectra for each source $s = 1, \dots, S$ at frame $i$ and let $\mathbf{Z} = \left[\mathbf{z}_{1} ,\; \mathbf{z}_{2} ,\; \dots\;, \mathbf{z}_{T}\right] \in \mathbb{R}^{S\left(\frac{N}{2}+1\right) \times T}$ denote the matrix of all $T$ supervectors.
	Finally, let $\mathbf{y}_{i} = \left[ {{Y}(i,1)} ,\; {{Y}(i,2)} ,\; \dots\; {{Y}(i,\frac{N}{2}+1)} \right]^T \in \mathbb{C}^{\frac{N}{2}+1}$ be the single-sided STFT of the observed mixture signal at frame $i$ and $\mathbf{Y} = \left[\mathbf{y}_{1} ,\; \mathbf{y}_{2} ,\; \dots\;, \mathbf{y}_{T}\right] \in \mathbb{C}^{\left(\frac{N}{2}+1\right) \times T} $ be the STFT of the mixture signal for all $T$ frames.  

	Our objective is then to train a deep learning model $g(\cdot)$, parameterized by a parameter set $\mathbf{\Phi}$, such that $g\left(d\left(\mathbf{Y}\right);\mathbf{\Phi}\right)=\mathbf{Z}$, where $d(\mathbf{Y})$ is some feature representation of the mixture signal: In a particularly simple situation, $d(\mathbf{Y}) = \mathbf{R}$, i.e., the feature representation is simply the magnitude spectrum of the observed mixture signal.    
	
	It is possible to directly estimate the magnitude spectra $\mathbf{Z}$ of all sources using a deep learning model. However, it is well-known (e.g. \cite{wang_training_2014, erdogan_deep_2017}), that better results can be achieved if, instead of estimating $\mathbf{Z}$ directly, we first estimate a set of masks ${M_s}(t,f)$, $s=1, \dots ,S$. 

	Let $\mathbf{m}_{s,i} = \left[ {{M_s}(i,1)} \;,\; {{M_s}(i,2)} \;,\; \dots\; {{M_s}(i,\frac{N}{2}+1)} \right]^T \in \mathbb{R}^{\frac{N}{2}+1}$ be the ideal mask (to be defined in detail in Sec.\;\ref{sec:mask}) for speaker $s$ at frame $i$, and let $\mathbf{M}_s = \left[\mathbf{m}_{s,1} ,\; \mathbf{m}_{s,2} ,\; \dots\;, \mathbf{m}_{s,T}\right] \in \mathbb{R}^{\left(\frac{N}{2}+1\right) \times T}$ be the ideal mask for all $T$ frames, 	
	such that $\mathbf{A}_s = \mathbf{M}_s \circ \mathbf{R} $, where $\circ$ is the Hadamard product, i.e. element-wise product of two operands.
	Furthermore, let us introduce the mask supervector $\mathbf{u}_{i} = \left[ \mathbf{m}_{1,i}^T \; \mathbf{m}_{2,i}^T \; \dots \; \mathbf{m}_{S,i}^T \right]^T \in \mathbb{R}^{S\left(\frac{N}{2}+1\right)}$ and the corresponding mask matrix $\mathbf{U} = \left[\mathbf{u}_{1} ,\; \mathbf{u}_{2} ,\; \dots\;, \mathbf{u}_{T}\right] \in \mathbb{R}^{S\left(\frac{N}{2}+1\right) \times T}$.
    Our goal is then to find an estimate $\hat{\mathbf{U}}$ of $\mathbf{U}$, using a deep learning model,
 	$h\left(\mathbf{R};\mathbf{\Phi} \right)=\hat{\mathbf{U}}$. 
 	Since,  $\hat{\mathbf{U}} = \left[\hat{\mathbf{u}}_{1} ,\; \hat{\mathbf{u}}_{2} ,\; \dots\;, \hat{\mathbf{u}}_{T}\right]$ and $\hat{\mathbf{u}}_{i} = \left[ \hat{\mathbf{m}}_{1,i}^T \; \hat{\mathbf{m}}_{2,i}^T \; \dots \; \hat{\mathbf{m}}_{S,i}^T \right]^T$, the model output is easily divided into output streams corresponding to the estimated masks for each speaker $\hat{\mathbf{m}}_{s,i}$, and their resulting magnitudes are estimated as $\hat{\mathbf{a}}_{s,i} = \hat{\mathbf{m}}_{s,i} \circ \mathbf{r}_{i}$.  
 	The estimated time-domain signal for speaker $s$ is then computed as the inverse DFT of $\hat{\mathbf{a}}_{s,i}$ using the phase of the mixture signal $\mathbf{y}_i$.

	\section{Masks and Training Criteria}\label{sec:mask}
	Since masks are to be estimated as an intermediate step towards estimating magnitude spectra of source signals, we extend in the following three popular masks defined for separating single-talker speech from noises to the multi-talker speech separation task at hand.

	\subsection{Ideal Ratio Mask}
	The Ideal Ratio Mask\;(IRM) \cite{wang_training_2014} for each source is defined as 
	\begin{equation}
		{M}_s^{irm}(t,f) =\frac {|{X}_s(t,f)|} { \sum_{s=1}^{S} |{X}_s(t,f)|}  .           
	\end{equation}
	When the phase of $\mathbf{Y}$ is used for reconstruction, the IRM achieves the highest Signal to Distortion Ratio\;(SDR) \cite{vincent_performance_2006}, when all sources have the same phase, (which is an invalid assumption in general). 
	IRMs are constrained to $0 \leq {{M}}_s^{irm}(t,f) \leq 1$ and $\sum_{s=1}^S {{M}}_s^{irm}(t,f) = 1$  for all T-F units. This constraint can easily be satisfied using the softmax activation function. 

	Since $\mathbf{Y}$ is the only observed signal in practice and $\sum_{s=1}^{S} |{X}_s(t,f)|$ is unknown during separation, the IRM is not a desirable target for the problem at hand. Nevertheless, we report IRM results as an upper performance bound since the IRM is a commonly used training target for deep learning based monaural speech separation \cite{chen_large-scale_2016,kolbaek_speech_2017}.

	\subsection{Ideal Amplitude Mask}
	Another applicable mask is the \ac{IAM} (known as FFT-mask in \cite{wang_training_2014}), or simply \ac{AM}, when estimated by a deep learning model. The IAM is defined as 
	\begin{equation}
		{M}_s^{iam}(t,f) =\frac {|{X}_s(t,f)|} {|{Y}(t,f)|}.             
	\end{equation}
	
	Through IAMs we can construct the exact $|{X}_s(t,f)|$ given the magnitude spectra of the mixed speech $|{Y}(t,f)|$. If the phase of each source equals the phase of the mixed speech, the IAM achieves the highest SDR. Unfortunately, as with the IRM, this assumption is not satisfied in most cases. IAMs satisfy the constraint that $0 \leq {{M}}_s^{iam}(t,f) \leq \infty$, although we found empirically that the majority of the T-F units are in the range of $0 \leq {{M}}_s^{iam}(t,f) \leq 1$. For this reason, softmax, sigmoid and ReLU are all possible output activation functions for estimating IAMs.

    \subsection{Ideal Phase Sensitive Mask}
	Both IRM and IAM do not consider phase differences between source signals and the mixture. This leads to sub-optimal results, when the phase of the mixture is used for reconstruction.  
	The \ac{IPSM} \cite{erdogan_phase-sensitive_2015,erdogan_deep_2017}
	\begin{equation}
		{M}_s^{ipsm}(t,f) =\frac {|{X}_s(t,f)|\cos(\theta_y(t,f)-\theta_s(t,f))} {|{Y}(t,f)|},             
	\end{equation}
	however, takes phase differences into consideration, where $\theta_y$ and $\theta_s$ are the phases of mixed speech ${Y}(t,f)$ and source ${X}_s(t,f)$, respectively. Due to the phase-correcting term, the IPSM sums to one, i.e. $\sum_{s=1}^S {{M}}_s^{ipsm}(t,f) = 1$. Note that since $|\cos(\cdot)| \leq 1$ the IPSM is smaller than the IAM, especially when the phase difference between the mixed speech and the source is large. 
	
	Even-though the IPSM in theory is unbounded, we found empirically that the majority of the IPSM is in the range of $0 \leq {M}_s^{ipsm}(t,f) \leq 1$. Actually, in our study we have found that approximately $20\%$ of IPSMs are negative. However, those negative IPSMs usually are very close to zero. To account for this observation, we propose the \ac{INPSM}, which is defined as 
	\begin{equation}
		{M}_s^{inpsm}(t,f) =max(0, {M}_s^{ipsm}(t,f)).             
	\end{equation}
	For estimating the IPSM and INPSM, Softmax, Sigmoid, tanh, and ReLU are all possible activation functions, and similarly to the IAM, when the IPSM is estimated by a deep learning model we refer to it as \ac{PSM}.

	\subsection{Training Criterion}
	Since we first estimate masks, through which the magnitude spectrum of each source can be estimated, the model parameters can be optimized to minimize the \acf{MSE} between the estimated mask $\hat{M}_s$ and one of the target masks defined above as
	\begin{equation}
		J_m=\frac{1}{B}\sum_{s=1}^S \|\hat{\mathbf{M}}_s - \mathbf{M}_s\|_F^2,
	\end{equation}
	where $B=T \times N \times S$ is the total number of T-F units over all sources and $\|\cdot\|_F$ is the Frobenius norm. This approach comes with two problems. First, in silence segments, $|{X}_s(t,f)|=0$ and $|{Y}(t,f)|=0$, so that the target masks ${M}_s(t,f)$ are not well defined. Second, what we really care about is the error between the reconstructed source signal and the true source signal. 
	
	To overcome these limitations, recent works \cite{wang_training_2014} directly minimize the MSE
	\begin{equation}
	\begin{split}
	J_a & =\frac{1}{B}\sum_{s=1}^S \| \hat{\mathbf{A}}_s - \mathbf{A}_s \|_F^2  \\
	& =\frac{1}{B}\sum_{s=1}^S \| \hat{\mathbf{M}}_s \circ \mathbf{R} - \mathbf{A}_s \|_F^2 
	\end{split}
	\label{eq:am}
	\end{equation}
	between the estimated magnitude, i.e. $\hat{\mathbf{A}}_s = \hat{\mathbf{M}}_s \circ \mathbf{R}$ and the true magnitude $\mathbf{A}_s$. Note that in silence segments $A_s(t,f)=0$ and $R(t,f)=0$,  so the accuracy of mask estimation does not affect the training criterion for those segments. 
	Furthermore, using Eq.\;\eqref{eq:am} the IAM is estimated as an intermediate step. 
	
	When the PSM is used, the cost function becomes 
	\begin{equation}
		\begin{split}
			J_{psm}  = \frac{1}{B}\sum_{s=1}^S \|\hat{\mathbf{M}}_s  \circ \mathbf{R} - \mathbf{A}_s  \circ \cos(\boldsymbol{\theta}_y - \boldsymbol{\theta}_s)\|_F^2. 
		\end{split}
		\label{eq:psm}
	\end{equation}
	
	In other words, using PSMs is as easy as replacing the original training targets with the phase discounted targets. 
	Furthermore, when Eq.\;\eqref{eq:psm} is used as a cost function, the IPSM is the upper bound achievable on the task \cite{erdogan_deep_2017}.

	\section{Permutation Invariant Training}\label{sec:PIT}

	\subsection{Conventional Multi-Talker Separation}
	A natural, and commonly used, approach for deep learning based speech separation is to cast the problem as a multi-class \cite{huang_joint_2015,tu_deep_2014,weng_deep_2015} regression problem as depicted in Fig.~\ref{fig:conventional}.  
	
	For this conventional two-talker separation model, $J$ frames of feature vectors of the mixed signal $\mathbf{Y}$ are used as the input to some deep learning model e.g. a feed-forward Deep Neural Network\,(DNN), Convolutional Neural Network\;(CNN), or LSTM RNN, to generate $M$ frames of masks for each talker.  
	Specifically, if $M=1$, the output of the model can be described by the vector $\hat{\mathbf{u}}_{i} = \left[ \hat{\mathbf{m}}_{1,i}^T \; \hat{\mathbf{m}}_{2,i}^T \right]^T$ and the  sources are separated as $\hat{\mathbf{a}}_{1,i} = \hat{\mathbf{m}}_{1,i} \circ \mathbf{r}_{i}$ and $\hat{\mathbf{a}}_{2,i} = \hat{\mathbf{m}}_{2,i} \circ \mathbf{r}_{i}$, for sources $s = 1,2$, respectively.

	\begin{figure}[ht]
		\centering
		\includegraphics[width=0.9\linewidth]{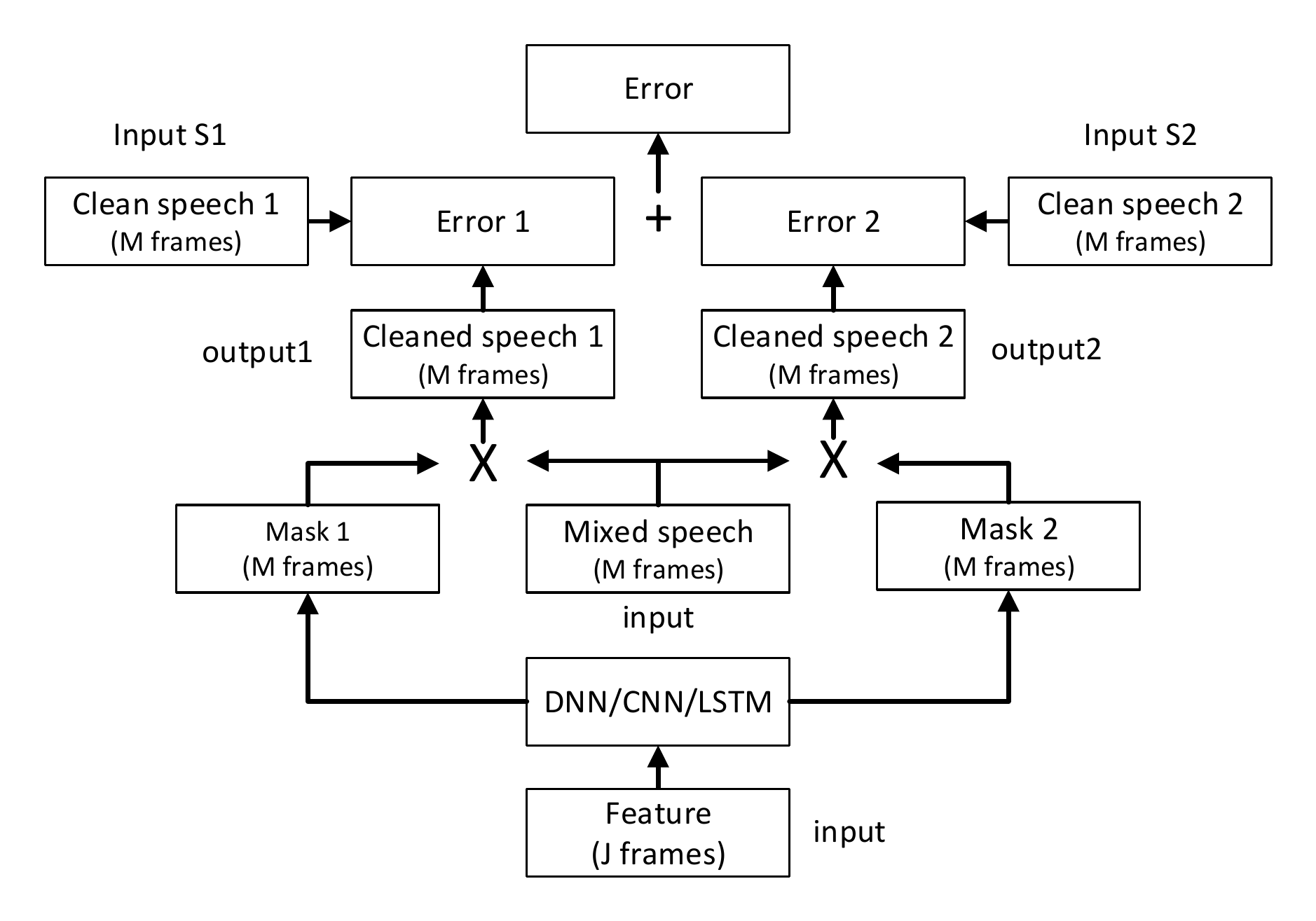}
		\caption{The conventional two-talker speech separation model.}\label{fig:conventional}
	\end{figure}

	\subsection{The Label Permutation Problem}
	During training, the error (e.g. using Eq.\;\eqref{eq:psm}) between the clean magnitude spectra $\mathbf{a}_{1,i}$ and $\mathbf{a}_{2,i}$ and their estimated counterparts $\hat{\mathbf{a}}_{1,i}$ and $\hat{\mathbf{a}}_{2,i}$ needs to be computed.
    However, since the model estimates the masks $\hat{\mathbf{m}}_{1,i}$ and $\hat{\mathbf{m}}_{2,i}$ simultaneously, and they depend on the same input mixture, it is unknown in advance whether the resulting output vector $\hat{\mathbf{u}}_{i}$ is ordered as $\hat{\mathbf{u}}_{i} = \left[ \hat{\mathbf{m}}_{1,i}^T \; \hat{\mathbf{m}}_{2,i}^T \right]^T$ or $\hat{\mathbf{u}}_{i} = \left[ \hat{\mathbf{m}}_{2,i}^T \; \hat{\mathbf{m}}_{1,i}^T \right]^T$. That is, the permutation of the output masks is unknown. 
    
    A na\"{i}ve approach to train a deep learning separation model, without exact knowledge about the permutation of the output masks, is to use a constant permutation as illustrated by Fig.~\ref{fig:conventional}. 
    Although such a training approach works for simple cases e.g. female speakers mixed with male speakers, in which case \emph{a priori} convention can be made that e.g. the first output stream contains the female speaker, while the second output stream is paired with the male speaker, the training fails if the training set consists of many utterances spoken by many speakers of both genders.

	This problem is referred to as the label permutation (or ambiguity) problem in \cite{weng_deep_2015,hershey_deep_2016}. Due to this problem, prior arts perform poorly on speaker independent multi-talker speech separation.

	\subsection{Permutation Invariant Training}
	Our solution to the label permutation problem is illustrated in Fig.~\ref{fig:model} and is referred to as Permutation Invariant Training\;(PIT) \cite{yu_permutation_2017}. 
	\begin{figure}[ht]
		\centering
		\includegraphics[width=0.9\linewidth]{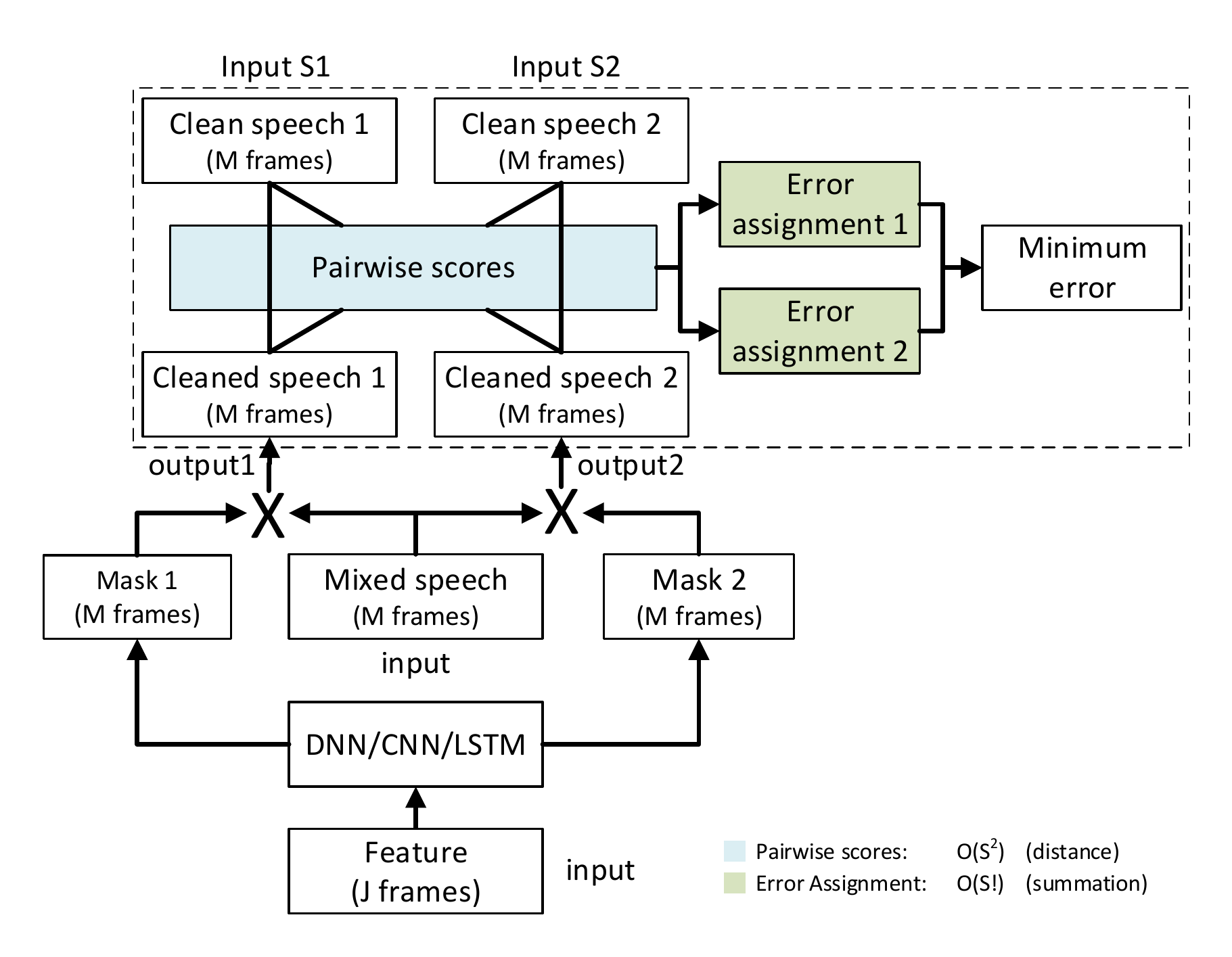}
		\caption{The two-talker speech separation model with permutation invariant training.}\label{fig:model}
	\end{figure}

	In the model depicted in Fig.~\ref{fig:model} (and unlike the conventional model in Fig.\;\ref{fig:conventional}) the reference signals are given as \emph{a set} instead of an ordered list. In other words, the same training result is obtained, no matter in which order these references are listed. This behavior is achieved with PIT highlighted inside the dashed rectangle in Fig.~\ref{fig:model}. 
	Specifically, following the notation from Sec.\;\ref{sec:problem}, we associate the reference signals for speaker one and two, i.e. $\mathbf{a}_{1,i}$ and $\mathbf{a}_{2,i}$, to the output masks $\hat{\mathbf{m}}_{1,i}$ and $\hat{\mathbf{m}}_{2,i}$, by computing the (total of $S^2$) pairwise MSEs between each reference signal $\mathbf{a}_{s,i}$ and each estimated source $\hat {\mathbf{a}}_{s,i}$. We then determine the (total of $S!$) possible permutations between the references and the estimated sources, and compute the \emph{per-permutation-loss} for each permutation. 
	That is, for the two-speaker case in Fig.~\ref{fig:model} we compute the \emph{per-permutation-loss} for the two candidate output vectors $\hat{\mathbf{u}}_{i} = \left[ \hat{\mathbf{m}}_{1,i}^T \; \hat{\mathbf{m}}_{2,i}^T \right]^T$ and $\hat{\mathbf{u}}_{i} = \left[ \hat{\mathbf{m}}_{2,i}^T \; \hat{\mathbf{m}}_{1,i}^T \right]^T$.   
	The permutation with the lowest MSE is chosen and the model is optimized to reduce this least MSE. In other words, we simultaneously conduct label assignment and error evaluation. Similarly to prior arts, we can use $J$, and $M$ successive input, and output frames, respectively, (i.e., a {\em meta-frame}) to exploit the contextual information. 
    Note that only $S^2$ pairwise MSEs are required (and not $S!$) to compute the \emph{per-permutation-loss} for all $S!$ possible permutations. Since $S!$ grows much faster than $S^2$, with respect to $S$, and the computational complexity of the pairwise MSE is much larger than the \emph{per-permutation-loss} (sum of pairwise MSEs), PIT can be used with a large number of speakers, i.e. $S\gg2$.
	
	During inference, the only information available is the mixed speech, but speech separation can be directly carried out for each input meta-frame, for which an output meta-frame with $M$ frames of speech is estimated. Due to the PIT training criterion, the permutation will stay the same for frames inside the same output meta-frame, but may change across output meta-frames. 
	In the simplest setup, we can just assume that permutations do not change across output meta-frames, when reconstructing the target speakers. However, this usually leads to unsatisfactory results as reported in \cite{yu_permutation_2017}. To achieve better performance, speaker tracing algorithms, that identify the permutations of output meta-frames with respect to the speakers, need to be developed and integrated into the PIT framework or applied on top of the output of the network.

	\section{Utterance-Level PIT}\label{sec:integrated}
	Several ways exist for identifying the permutation of the output meta-frames, i.e. solving the tracing problem. 
	For example, in CASA a related problem referred to as the Sequential Organization Problem has been addressed using a model-based sequential grouping algorithm \cite{shao_model-based_2006}. Although moderately successful for co-channel speech separation, where prior knowledge about the speakers is available, this method is not easily extended to the speaker independent case with multiple speakers. Furthermore, it is not easily integrated into a deep learning framework.      
	
	A more straight-forward approach might be to determine a change in permutation by comparing MSEs for different permutations of output masks measured on the overlapping frames of adjacent output meta-frames. However, this approach has two major problems. First, it requires a separate tracing step, which may complicate the model. Second, since the permutation of later frames depends on that of earlier frames, one incorrect assignment at an earlier frame would completely switch the permutation for all frames after it, even if the assignment decisions for the remaining frames are all correct.

	In this work we propose utterance-level Permutation Invariant Training (uPIT), a simpler yet more effective approach to solve the tracing problem and the label permutation problem than original PIT. Specifically, we extend the frame-level PIT technique with the following utterance-level cost function: 
	\begin{equation}
	J_{\phi^\ast} = \frac{1}{B}\sum_{s=1}^S \|\hat{\mathbf{M}}_s  \circ \mathbf{R} - \mathbf{A}_{\phi^\ast(s)}  \circ \cos( \boldsymbol{\theta}_y - \boldsymbol{\theta}_{\phi^\ast(s)})\|_F^2,	
	\label{eqPITutt1}
	\end{equation}	
	where $\phi^\ast$ is the permutation that minimizes the utterance-level separation error defined as
	\begin{equation}
	\phi^\ast = \underset{\phi\in \mathcal{P}}{\text{argmin}} \sum_{s=1}^S  \| \hat{\mathbf{M}}_s  \circ \mathbf{R} -  \mathbf{A}_{\phi(s)}\circ \cos(\boldsymbol{\theta}_y - \boldsymbol{\theta}_{\phi(s)}) \|_F^2,
	\label{eqPITutt}
	\end{equation}
	and $\mathcal{P}$ is the symmetric group of degree $S$, i.e. the set of all $S!$ permutations.

	In original PIT, the optimal permutation (in MSE sense) is computed and applied \emph{for each} output meta-frame. This implies that consecutive meta-frames might be associated with different permutations, and although PIT solves the label permutation problem, it does not solve the speaker tracing problem.  
	With uPIT, however, the permutation corresponding to the minimum utterance-level separation error is used \emph{for all} frames in the utterance. 
	In other words, the pair-wise scores in Fig.~\ref{fig:model} are computed for the whole utterance assuming all output frames follow the same permutation.
	Using the same permutation \emph{for all} frames in the utterance might imply that a non-MSE-optimal permutation is used for individual frames within the utterance. However, the intuition behind uPIT is that since the permutation resulting in the minimum utterance-level separation error is used, the number of non-optimal permutations is small and the model sees enough correctly permuted frames to learn an efficient separation model.
	For example, the output vector $\hat{\mathbf{u}}_{i}$ of a perfectly trained two-talker speech separation model, given an input utterance, should ideally be $\hat{\mathbf{u}}_{i} = \left[ \hat{\mathbf{m}}_{1,i}^T \; \hat{\mathbf{m}}_{2,i}^T \right]^T$, or $\hat{\mathbf{u}}_{i} = \left[ \hat{\mathbf{m}}_{2,i}^T \; \hat{\mathbf{m}}_{1,i}^T \right]^T \forall \; i=1, \dots , T$, i.e. the output masks should follow the same permutation for all $T$ frames in the utterance. 
	Fortunately, using Eq.\;\eqref{eqPITutt1} as a training criterion, for deep learning based speech separation models, this seems to be the case in practice (See Sec.\;\ref{sec:exp} for examples). 

	Since utterances have variable length, and effective separation presumably requires exploitation of long-range signal dependencies, models such as DNNs and CNNs are no longer good fits. Instead, we use deep LSTM RNNs and \ac{BLSTM} RNNs together with uPIT to learn the masks.
	Different from PIT, in which the input layer and each output layer has $N \times T$  and  $N \times M$ units, respectively, in uPIT, both input and output layers have $N$ units (adding contextual frames in the input does not help for LSTMs). With deep LSTMs, the utterance is evaluated frame-by-frame exploiting the whole past history information at each layer. When BLSTMs are used, the information from the past and future (i.e., across the whole utterance) is stacked at each layer and used as the input to the subsequent layer. 
	With uPIT, during inference we don't need to compute pairwise MSEs and errors of each possible permutation and no additional speaker tracing step is needed. We simply assume a constant permutation and treat the same output mask to be from the same speaker for all frames. This makes uPIT a simple and attractive solution.

	\section{Experimental Results}
	\label{sec:exp}
	
	We evaluated uPIT on various setups and all models were implemented using the Microsoft Cognitive Toolkit (CNTK) \cite{yu_computational_2015,agarwal_introduction_2014}\footnote{Available at: \url{https://www.cntk.ai/}}. The models were evaluated on their potential to improve the Signal-to-Distortion Ratio\;(SDR) \cite{vincent_performance_2006} and the Perceptual Evaluation of Speech Quality\;(PESQ) \cite{rix_perceptual_2001} score, both of which are metrics widely used to evaluate speech enhancement performance for multi-talker speech separation tasks.
	
	\subsection{Datasets}
	\label{subsec:datasets}
	
	We evaluated uPIT on the WSJ0-2mix, WSj0-3mix\footnote{Available at: \url{http://www.merl.com/demos/deep-clustering}} and Danish-2mix datasets using 129-dimensional STFT magnitude spectra computed with a sampling frequency of 8 kHz, a frame size of 32 ms and a 16 ms frame shift.
	
	The WSJ0-2mix dataset was introduced in \cite{hershey_deep_2016} and was derived from the WSJ0 corpus \cite{garofolo_csr-i_1993}. The 30h training set and the 10h validation set contain two-speaker mixtures generated by randomly selecting from 49 male and 51 female speakers and utterances from the WSJ0 training set si\_tr\_s, and mixing them at various Signal-to-Noise Ratios\;(SNRs) uniformly chosen between 0\,dB and 5\,dB. The 5h test set was similarly generated using utterances from 16 speakers from the WSJ0 validation set si\_dt\_05 and evaluation set si\_et\_05. The WSJ0-3mix dataset was generated using a similar approach but contains mixtures of speech from three talkers. 
	
	The Danish-2mix dataset is based on a corpus\footnote{Available at: {\url{http://www.nb.no/sbfil/dok/nst_taledat_dk.pdf}} } with  approximately 560 speakers each speaking 312 utterances with average utterance duration of approximately 5 sec. The dataset was constructed by randomly selecting a set of 45 male and 45 female speakers from the corpus, and then allocating 232 and 40 utterances from each speaker to generate mixed speech in the training, and validation set, respectively. A number of 40 utterances from each of another 45 male and 45 female speakers were randomly selected to construct the \ac{OC} (unseen speaker) test set. Speech mixtures were constructed similarly to the WSJ0-2mix with SNRs selected uniformly between 0\,dB and 5\,dB. Similarly to the WSJ0-2mix dataset we constructed 20k and 5k mixtures in total in the training and validation set, respectively, and 3k mixtures for the OC test set. 
	
	In our study, the validation set is used to find initial hyper-parameters and to evaluate \ac{CC} (seen speaker) performance, similarly to \cite{hershey_deep_2016,isik_single-channel_2016,yu_permutation_2017}.
	
	\subsection{Permutation Invariant Training}
	We first evaluated the original frame-level PIT on the two-talker separation dataset WSJ0-2mix, and differently from \cite{yu_permutation_2017}, we fixed the input dimension to 51 frames, to isolate the effect of a varying output dimension.
	In PIT, the input window and output window sizes are fixed. For this reason, we can use DNNs and CNNs. The DNN model has three hidden layers each with 1024 ReLU units. In (inChannel, outChannel)-(strideW, strideH) format, the CNN model has one $(1,64)-(2,2)$, four $(64,64)-(1,1)$, one $(64,128)-(2,2)$, two $(128,128)-(1,1)$, one $(128,256)-(2,2)$, and two $(256,256)-(1,1)$ convolution layers with $3 \times 3$ kernels, a $7\times17$ average pooling layer and a 1024-unit ReLU layer. The input to the models is the stack (over multiple frames) of the 129-dimensional STFT spectral magnitude of the speech mixture. 
	The output layer $\hat{\mathbf{u}}_{i}$ is divided into $S$ output masks/streams for $S$-talker mixed speech as $\hat{\mathbf{u}}_{i} = \left[ \hat{\mathbf{m}}_{1,i} \;;\; \hat{\mathbf{m}}_{2,i} \;;\; \dots \;;\; \hat{\mathbf{m}}_{S,i} \right]^T$. Each output mask vector $\hat{\mathbf{m}}_{s,i}$ has a dimension of $129 \times M$, where $M$ is the number of frames in the output meta-frame.

	\begin{figure}[ht] 
	\centering
	\centerline{\includegraphics[width=0.8\linewidth]{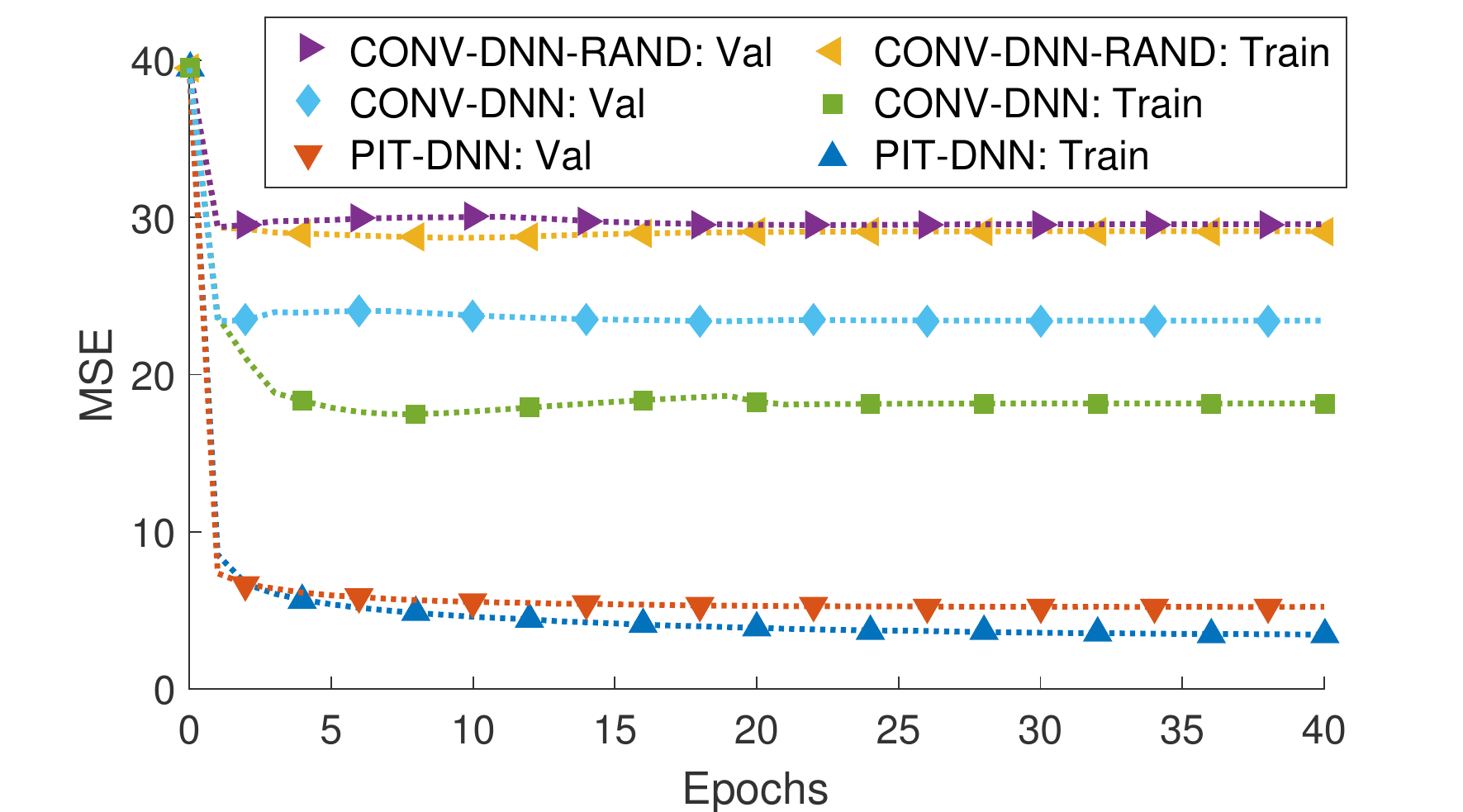}}
		\caption{MSE over epochs on the WSJ0-2mix training and validation sets with conventional training and PIT.}\label{fig:mse}
	\end{figure}

	In Fig.~\ref{fig:mse} we present the DNN training progress as measured by the MSE on the training and validation set with conventional training (CONV-DNN) and PIT on the WSJ0-2mix datasets described in subsection \ref{subsec:datasets}. We also included the training progress for another conventionally trained model but with a slightly modified version of the WSJ0-2mix dataset, where speaker labels have been randomized (CONV-DNN-RAND). 
	
	The WSJ0-2mix dataset, used in \cite{hershey_deep_2016}, was designed such that speaker one was always assigned the most energy, and consequently speaker two the lowest, when scaling to a given SNR. 
	Previous work \cite{weng_deep_2015} has shown that such speaker energy patterns are an effective discriminative feature, which is clearly seen in Fig.~\ref{fig:mse}, where the CONV-DNN model achieves considerably lower training and validation MSE than the CONV-DNN-RAND model, which hardly decreases in either training or validation MSE due to the label permutation problem \cite{weng_deep_2015,hershey_deep_2016}. 
	In contrast, training converges quickly to a very low MSE when PIT is used.
	\begin{table}[t]
		\caption{SDR improvements (dB) for different separation methods on the WSJ0-2mix dataset using PIT.}
		\label{tab:WSJ0-2mix-PIT}
		\centering
		\resizebox{0.60\textwidth}{!}{%
		\begin{tabular}{lccccc}
			\midrule \midrule
			Method &   \multirow{2}{*}{\makecell{ Input\textbackslash Output \\ window}} & \multicolumn{2}{c} {Opt. Assign.} & \multicolumn{2}{c} {Def. Assign. }\\ \cmidrule(l){3-4} \cmidrule(l){5-6}
			&                      		& CC & OC		&  CC 	& OC	\\
			\midrule
			PIT-DNN    & 51\textbackslash51    		& 6.8 & 6.7  & \bf{5.2} & \bf{5.2}     \\ 
			PIT-DNN    & 51\textbackslash5    		& \bf{10.3} & \bf{10.2}  & -0.8 & -0.8     \\ 
			PIT-CNN    & 51\textbackslash51     	& 9.6 & 9.6  & \bf{7.6} & \bf{7.5}     \\
			PIT-CNN    & 51\textbackslash5      	& \bf{10.9} & \bf{11.0}  & -1.0 & -0.9     \\
			IRM 	& - & 12.4 & 12.7 & 12.4 & 12.7 \\
			IPSM 	& - & 14.9 & 15.1 & 14.9 & 15.1 \\
			\midrule \midrule
		\end{tabular} }
	\end{table}
	
	In Table~\ref{tab:WSJ0-2mix-PIT} we summarize the SDR improvement in dB from different frame-level PIT separation configurations for two-talker mixed speech in closed condition\;(CC) and open condition\;(OC). In these experiments each frame was reconstructed by averaging over all output meta-frames that contain the same frame. In the default assignment (def. assign.) setup, a constant output mask permutation is assumed across frames (which is an invalid assumption in general). 
	This is the maximum achievable SDR improvement using PIT without the utterance-level training criterion and without an additional tracing step. 
	In the optimal assignment (opt. assign.) setup, the output-mask permutation for each output meta-frame is determined based on the true target, i.e. oracle information. This reflects the separation performance within each segment (meta-frame) and is the improvement achievable when the speakers are correctly separated. The gap between these two values indicates the possible contribution from speaker tracing. As a reference, we also provided the IRM and IPSM results.

	From the table we can make several observations. First, PIT can already achieve 7.5 dB SDR improvement (def. assign.), even though the model is very simple. Second, as we reduce the output window size, we can improve the separation performance within each window and achieve better SDR improvement, if speakers are correctly traced (opt. assign.). However, when output window size is reduced, the output mask permutation changes more frequently as indicated by the poor default assignment performance. Speaker tracing thus becomes more important given the larger gap between the optimal assignment and default assignment. Third, PIT generalizes well to unseen speakers, since the performances on the open and closed conditions are very close. Fourth, powerful models such as CNNs consistently outperform DNNs, but the gain diminishes when the output window size is small.  
	
	\subsection{Utterance-Level Permutation Invariant Training}
	As indicated by Table~\ref{tab:WSJ0-2mix-PIT}, an accurate output mask permutation is critical to further improve the separation quality. In this subsection we evaluate the uPIT technique as discussed in Sec.~\ref{sec:integrated} and the results are summarized in Table~\ref{tab:WSJ0-2mix-uPIT}.

	\begin{table}[t]
		\caption{SDR improvements (dB) for different separation methods on the WSJ0-2mix dataset using uPIT.}
		\label{tab:WSJ0-2mix-uPIT}
		\centering
		\resizebox{0.75\textwidth}{!}{%
		\begin{tabular}{lcccccc}
			\midrule \midrule
			Method & Mask Type &  \multirow{2}{*}{\makecell{ Activation \\ Function}}   & \multicolumn{2}{c} {Opt. Assign.} & \multicolumn{2}{c} {Def. Assign.}\\   \cmidrule(l){4-5} \cmidrule(l){6-7}
			&                      		&  & CC		&  OC 	& CC & OC	\\
			\midrule
			uPIT-BLSTM    & AM & softmax   		& \bf{10.4} &     \bf{10.3}	& 	\bf{9.0} 	& 	\bf{8.7}  \\  
			uPIT-BLSTM    & AM & sigmoid      	& 8.3 		&     8.3  	&   7.1 		&   7.2     \\ 
			uPIT-BLSTM    & AM & ReLU     		& 9.9 		&     9.9      &   8.7 		&   8.6     \\ 
			uPIT-BLSTM    & AM & Tanh      	    & 8.5  	    &     8.6      &   7.5 		&   7.5     \\ 
			uPIT-BLSTM    & PSM & softmax   		& 10.3 		&     10.2  	& 	9.1 		& 	9.0     \\
			uPIT-BLSTM    & PSM & sigmoid      	& 10.5 		&     10.4  	& 	9.2 		& 	9.1     \\
			uPIT-BLSTM    & PSM & ReLU     		& \bf{10.9}	&     \bf{10.8}	& 	\bf{9.4} 	& 	\bf{9.4}     \\
			uPIT-BLSTM    & PSM & Tanh      	    & 10.4 		&     10.3  	& 	9.0 		& 	8.9     \\ 
			uPIT-BLSTM    & NPSM & softmax   		& 8.7 		&     8.6  	& 	7.5 		& 	7.3     \\
			uPIT-BLSTM    & NPSM & sigmoid      	& \bf{10.6}	& \bf{10.6}	& 	\bf{9.4} 	& 	\bf{9.3}     \\
			uPIT-BLSTM    & NPSM & ReLU     		& 8.8 		&     8.8  	& 	7.6 		& 	7.6     \\
			uPIT-BLSTM    & NPSM & Tanh      	    & 10.1 		&     10.0  	& 	8.9 		& 	8.8     \\
			uPIT-LSTM    & PSM & ReLU   			& \bf{9.8} & \bf{9.8} & 7.0 & \bf{7.0} \\ 
			uPIT-LSTM    & PSM & sigmoid      		& \bf{9.8} & 9.6	& \bf{7.1} & 6.9 \\ 
			uPIT-LSTM    & NPSM & ReLU     		& \bf{9.8} & \bf{9.8} & \bf{7.1} & \bf{7.0} \\ 
			uPIT-LSTM    & NPSM & sigmoid      	& 9.2 		& 9.2 	&  6.8 		& 6.8 \\ 
			PIT-BLSTM    & PSM & ReLU   			& \bf{11.7} & \bf{11.7} & \bf{-1.7} & -1.9   \\ 
			PIT-BLSTM    & PSM & sigmoid      	& \bf{11.7} & \bf{11.7} & \bf{-1.7} & \bf{-1.7} \\ 
			PIT-BLSTM    & NPSM & ReLU     		& \bf{11.7} & \bf{11.7} & \bf{-1.7} & -1.8 \\ 
			PIT-BLSTM    & NPSM & sigmoid      	& 11.6 & 11.6 & -1.6 & \bf{-1.7} \\ 
			\midrule 
			IRM  & - & -  & 12.4 & 12.7 & 12.4 & 12.7 \\
			IPSM & - & -  & 14.9 & 15.1 & 14.9 & 15.1 \\
			\midrule \midrule
		\end{tabular} }
	\end{table}

	Due to the formulation of the uPIT cost function in Eq. \eqref{eqPITutt1} and Eq. \eqref{eqPITutt}, and to utilize long-range context, RNNs are the natural choice, and in this set of experiments, we used LSTM RNNs. All the uni-directional LSTMs (uPIT-LSTM) evaluated have 3 LSTM layers each with 1792 units and all the bi-directional LSTMs (uPIT-BLSTM) have 3 BLSTM layers each with 896 units, so that both models have similar number of parameters. 
	
	All models contain random dropouts when fed from a lower layer to a higher layer and were trained with a dropout rate of 0.5. Note that, since we used Nvidia's cuDNN implementation of LSTMs, to speed up training, we were unable to apply dropout across time steps, which was adopted by the best DPCL model \cite{isik_single-channel_2016} and is known to be more effective, both theoretically and empirically, than the simple dropout strategy used in this work \cite{gal_theoretically_2015}. 
	
	In all the experiments reported in Table~\ref{tab:WSJ0-2mix-uPIT} the maximum epoch is set to 200 although we noticed that further performance improvement is possible with additional training epochs. Note that the epoch size of 200 seems to be significantly larger than that in PIT as indicated in Fig.~\ref{fig:mse}. This is likely because in PIT each frame is used by $T$ ($T=51$) training samples (input meta-frames) while in uPIT each frame is used just once in each epoch. 
	
	The learning rates were set to $2 \times 10^{-5}$ per sample initially and scaled down by $0.7$ when the training objective function value increases on the training set. The training was terminated when the learning rate got below $10^{-10}$. Each minibatch contains 8 randomly selected utterances.

	As a related baseline, we also include PIT-BLSTM results in Table~\ref{tab:WSJ0-2mix-uPIT}. These models were also trained using LSTMs with whole utterances instead of meta-frames. The only difference between these models and uPIT models is that uPIT models use the utterance-level training criterion defined in Eqs.\,\eqref{eqPITutt1} and \eqref{eqPITutt}, instead of the meta-frame based criterion used by PIT.
	%
	\subsubsection{uPIT Training Progress}
	
	In Fig. \ref{fig:mse-uPIT} we present a representative example of the BLSTM training progress, as measured by the MSE of the two-talker mixed speech training and validation set, using Eq.\;\eqref{eqPITutt1}. We see that the training and validation MSEs are both steadily decreasing as function of epochs, hence uPIT, similarly to PIT, effectively solves the label permutation problem.
	
	\begin{figure}[ht] 
		\centering
		\centerline{\includegraphics[width=0.8\linewidth]{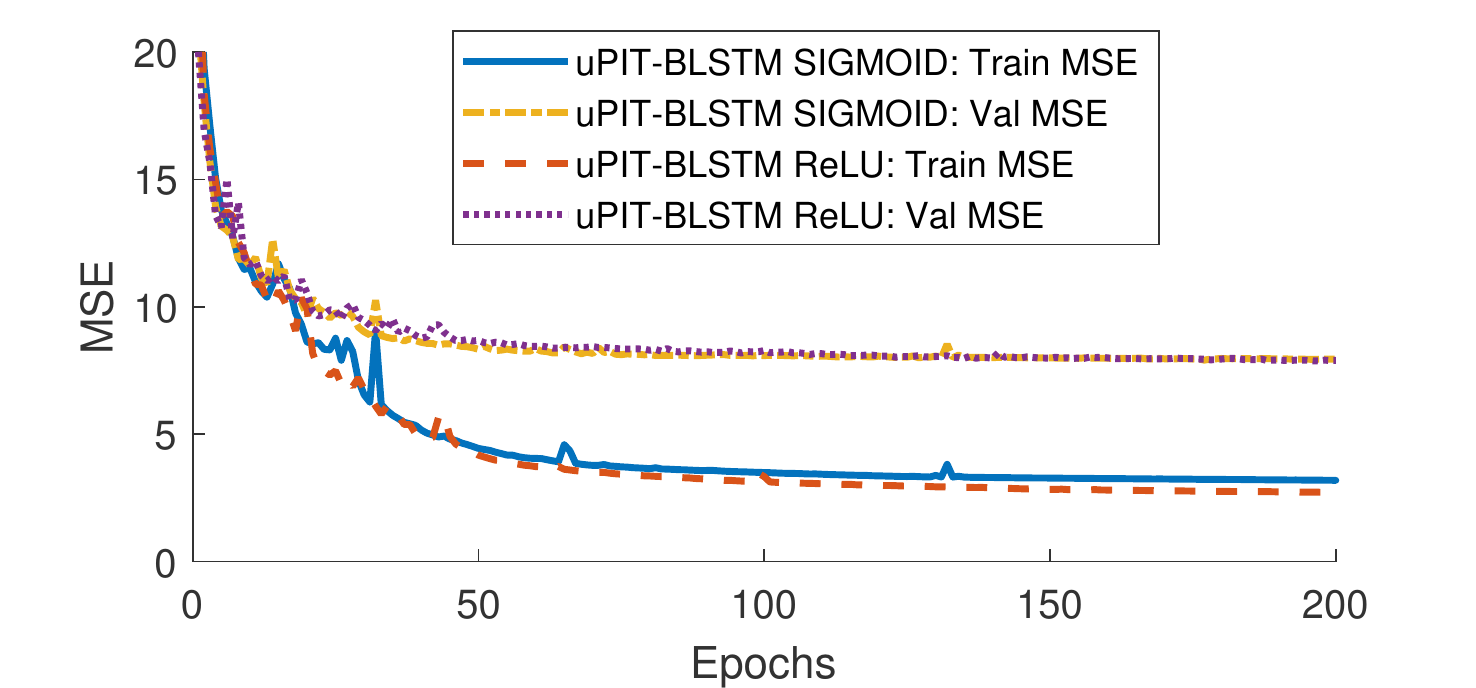}}
		\caption{MSE over epochs on the WSJ0-2mix PSM training and validation sets wit uPIT.}\label{fig:mse-uPIT}
	\end{figure}
	
	\subsubsection{uPIT Performance for Different Setups}
	From Table~\ref{tab:WSJ0-2mix-uPIT}, we can notice several things. First, with uPIT, we can significantly improve the SDR with default assignment over original PIT. In fact, a 9.4\,dB SDR improvement on both CC and OC sets can be achieved by simply assuming a constant output mask permutation (def. assign.), which compares favorably to 7.6\,dB (CC) and 7.5\,dB (OC) achieved with deep CNNs combined with PIT. 
	We want to emphasize that this is achieved through Eqs.\;\eqref{eqPITutt1} and \eqref{eqPITutt}, and not by using BLSTMs because the corresponding PIT-BLSTM default assignment results are so much worse, even though the optimal assignment results are the best among all models. 
	The latter may be explained from the PIT objective function that attempts to obtain a constant output mask permutation at the meta-frame-level, which for small meta-frames is assumed easier compared to the uPIT objective function, that attempts to obtain a constant output mask permutation throughout the whole utterance. 
	Second, we can achieve better SDR improvement over the AM using PSM and NPSM training criteria. 
	This indicates that including phase information does improve performance, even-though it was used implicitly via the cosine term in Eq.\;\eqref{eqPITutt1}. 
	Third, with uPIT the gap between optimal assignment and default assignment is always less than 1.5\,dB across different setups, hence additional improvements from speaker tracing algorithms is limited to 1.5\,dB. 	
	\subsubsection{Two-Stage Models and Reduced Dropout Rate}
	It is well known that cascading DNNs can improve performance for certain deep learning based applications \cite{zhang_deep_2016,nie_deep_2014,wang_recurrent_2017,isik_single-channel_2016}.
	In Table~\ref{tab:WSJ0-2mix-Further} we show that a similar principle of cascading two BLSTM models into a two-stage model\linebreak (-ST models in Table~\ref{tab:WSJ0-2mix-Further}) can lead to improved performance over the models presented in Table~\ref{tab:WSJ0-2mix-uPIT}. 
	In Table~\ref{tab:WSJ0-2mix-Further} we also show that improved performance, with respect to the same models, can be achieved with additional training epochs combined with a reduced dropout rate (-RD models in Table~\ref{tab:WSJ0-2mix-Further}).
	Specifically, if we continue the training of the two best performing models from Table~\ref{tab:WSJ0-2mix-uPIT} (i.e. uPIT-BLSTM-PSM-ReLU and uPIT-BLSTM-NPSM-Sigmoid) with 200 additional training epochs at a reduced dropout rate of 0.3, we see an improvement of $0.1$\;dB.
	Even larger improvements can be achieved with the two-stage approach, where an estimated mask is computed as the average mask from two BLSTM models as 
	\begin{equation}
	\begin{split}
	\hat{\mathbf{M}}_s &= \frac {\hat{\mathbf{M}}_s^{(1)} + \hat{\mathbf{M}}_s^{(2)}} {2}.
	\end{split}
	\end{equation}
	The mask $\hat{\mathbf{M}}_s^{(1)}$ is from an -RD model that serves as a first-stage model, and $\hat{\mathbf{M}}_s^{(2)}$ is the output mask from a second-stage model. The second-stage model is trained using the original input features as well as the mask $\hat{\mathbf{M}}_s^{(1)}$ from the first-stage model.
	The intuition behind this architecture is that the second-stage model will learn to correct the errors made by the first-stage model.   
	Table~\ref{tab:WSJ0-2mix-Further} shows that the two-stage models (-ST models) always outperform the single-stage models (-RD models) and overall, a 10\,dB SDR improvement can be achieved on this task using a two-stage approach.  
	\begin{table}[t]
		\caption{Further improvement on the WSJ0-2mix dataset with additional training epochs with reduced dropout (-RD) or stacked models (-ST)}
		\vspace{-1mm}
		\label{tab:WSJ0-2mix-Further}
		\centering
		\resizebox{0.7\textwidth}{!}{%
		\begin{tabular}{lcccccc}
			\midrule \midrule
			Method & \multirow{2}{*}{\makecell{ Mask \\ Type}} &  \multirow{2}{*}{\makecell{ Activation \\ Function}}   & \multicolumn{2}{c} {Opt. Assign.} & \multicolumn{2}{c} {Def. Assign.}\\   \cmidrule(l){4-5} \cmidrule(l){6-7}
			&                      		&  & CC		&  OC 	& CC & OC	\\
			\midrule
			uPIT-BLSTM-RD  & PSM 	& ReLU  	   	& 11.0 	&     11.0 	& 9.5 	& 9.5     \\ 
			uPIT-BLSTM-ST  & PSM 	& ReLU   	    & \bf{11.7} & \bf{11.7} & {10.0} & \bf{10.0}     \\ 
			uPIT-BLSTM-RD  & NPSM 	& Sigmoid  	   	& 10.7 	&     10.7  	& 9.5 	& 9.4     \\ 
			uPIT-BLSTM-ST  & NPSM 	& Sigmoid       & {11.5} 	& {11.5}  	& \bf{10.1} & \bf{10.0}     \\  
			\midrule
			IRM 		   & -  	& -  			& 12.4 & 12.7 & 12.4 & 12.7 \\
			IPSM 		   & - 		& -				& 14.9 & 15.1 & 14.9 & 15.1 \\
			\midrule \midrule
		\end{tabular} }
	\end{table}
	\subsubsection{Opposite Gender vs. Same Gender.}
	\begin{table}[t]
		\caption{SDR (dB) improvements on test sets of WSJ0-2mix divided into same and opposite gender mixtures}
		\vspace{-1mm}
		\label{tab:WSJ0-2mix-gender}
		\centering
		\resizebox{0.7\textwidth}{!}{%
		\begin{tabular}{lccccc}
			\midrule \midrule
			Method & Config & \multicolumn{2}{c} {CC} & \multicolumn{2}{c} {OC}\\ \cmidrule(l){3-4} \cmidrule(l){5-6}
			&                      		& Same & Opp.		&  Same 	& Opp.	\\
			\midrule
			uPIT-BLSTM-RD   & PSM-ReLU    		& 7.5 	& 11.5  	& 7.1 	& 11.6     \\ 
			uPIT-BLSTM-ST	 & PSM-ReLU   		& 7.8 	& \bf{12.1} & \bf{7.5} & \bf{12.2}    \\ 
			uPIT-BLSTM-RD   & NPSM-Sigmoid    	& 7.5 	& 11.5  	& 7.0 	& 11.5     \\ 
			uPIT-BLSTM-ST   & NPSM-Sigmoid   	& \bf{8.0} & \bf{12.1}  	& \bf{7.5} 	& 12.1  \\ 
			\midrule
			IRM  	& -							& 12.2  	& 12.7	& 12.4 	& 12.9   \\ 
			IPSM 	& - 						& 14.6 		& 15.1 	& 14.9 	& 15.3 \\
			\midrule \midrule
		\end{tabular} }
	\end{table}
	
	Table~\ref{tab:WSJ0-2mix-gender} reports SDR (dB) improvements on test sets of WSJ0-2mix divided into opposite-gender\;(Opp.) and same-gender\;(Same). From this table we can clearly see that our approach achieves much better SDR improvements on the opposite-gender mixed speech than the same-gender mixed speech, although the gender information is not explicitly used in our model and training procedure. In fact, for the opposite-gender condition, the SDR improvement is already very close to the IRM result. These results agree with breakdowns from other works \cite{hershey_deep_2016,isik_single-channel_2016} and generally indicate that same-gender mixed speech separation is a harder task.

	\subsubsection{Multi-Language Models}
	To further understand the properties of uPIT, we evaluated the uPIT-BLSTM-PSM-ReLU model trained on WSJ0-2mix (English) on the Danish-2mix test set. The results of this is reported in Table~\ref{tab:twolanguage}. 
	\begin{table}[t]
		\caption{SDR (dB) and PESQ improvements on WSJ0-2mix and Danish-2mix with uPIT-BLSTM-PSM-ReLU trained on WSJ0-2mix and a combination of two languages.}
		\label{tab:twolanguage}
		\centering
		\resizebox{0.5\textwidth}{!}{%
		\begin{tabular}{lcccc}
			\midrule \midrule
			Trained on & \multicolumn{2}{c} {WSJ0-2mix} & \multicolumn{2}{c} {Danish-2mix}\\  \cmidrule(l){2-3} \cmidrule(l){4-5}
			& SDR & PESQ		&  SDR & PESQ	\\
			\midrule
			WSJ0-2mix    	& 9.4 	&     0.62  	&     8.1 	&     0.40     \\ 
			+Danish-2mix    & 8.8 	&     0.58  	&     10.6	&     0.51     \\ 
			\midrule
			IRM	   		    & 12.7  & 2.11      	& 15.2  & 1.90  \\ 
			IPSM 			& 15.1 	& 2.10 		& 17.7 	& 1.90 \\
			\midrule \midrule
		\end{tabular} }
	\end{table}
	An interesting observation, is that although the system has never seen Danish speech, it performs remarkably well in terms of SDR, when compared to the IRM (oracle) values. These results indicate, that the separation ability learned with uPIT generalizes well, not only across speakers, but also across languages. 
	In terms of PESQ, we see a somewhat larger performance gap with respect to the IRM. This might be explained by the fact that SDR is a waveform matching criteria and does not necessarily reflect perceived quality as well as PESQ. Furthermore, we note that the PESQ improvements are similar to what have been reported for DNN based speech enhancement systems \cite{kolbaek_speech_2017}.
	
	We also trained a model with the combination of English and Danish datasets and evaluated the models on both languages. The results of these experiments are summarized in Table~\ref{tab:twolanguage}. Table~\ref{tab:twolanguage}, indicate that by including Danish data, we can achieve better performance on the Danish dataset, at the cost of slightly worse performance on the English dataset. Note that while doubling the training set, we did not change the model size. Had we done this, performance would likely improve on both languages.

	\subsubsection{Summary of Multiple 2-Speaker Separation Techniques}   
		\begin{table}[t]
			\caption{SDR (dB) and PESQ improvements for different separation methods on the WSJ0-2mix dataset without additional tracing (i.e., def. assign.).}
			\label{tab:WSJ0-2mix-summary}
			\centering
			\resizebox{0.6\textwidth}{!}{%
			\begin{tabular}{lccccc}
				\midrule \midrule
				Method & Config & \multicolumn{2}{c} {PESQ Imp.} & \multicolumn{2}{c} {SDR Imp.}\\  \cmidrule(l){3-4} \cmidrule(l){5-6}
				&                      		& CC & OC		&  CC 	& OC	\\
				\midrule
				Oracle NMF \cite{hershey_deep_2016}			& -	& -	& -		& 5.1  	& -     \\ 
				CASA \cite{hershey_deep_2016}   			& - & - & - 	& 2.9  	& 3.1   \\
				DPCL \cite{hershey_deep_2016} 				& - & - & -  	& 5.9  	& 5.8   \\
				DPCL+ \cite{chen_deep_2017}   				& - & -	& -		&  -  	& 9.1     \\ 
				DANet \cite{chen_deep_2017}   				& - & -	& -		&  -  	& 9.6     \\ 
				DANet$^\ddagger$ \cite{chen_deep_2017}  				& - & -	& -		&  -  	& 10.5     \\ 
				DPCL++ \cite{isik_single-channel_2016}   	& - & -	& -		&  -  	& 9.4     \\ 
				DPCL++$^\ddagger$ \cite{isik_single-channel_2016}   	& - & -	& -		&  -  	& \bf{10.8}     \\ 
				\midrule
				PIT-DNN     & 51\textbackslash51    				& 0.24 	&     0.23  	& 5.2 	& 5.2     \\ 
				PIT-CNN     & 51\textbackslash51   					& 0.52	&     0.50  	& 7.6 	& 7.6     \\ 
				uPIT-BLSTM   & PSM-ReLU    							& 0.66 	&     0.62  	& 9.4 	& 9.4     \\ 
				uPIT-BLSTM-ST& PSM-ReLU   							& {\bf{0.86}} & {\bf{0.82}} & {\bf{10.0}} & {10.0}  \\ 
				\midrule
				IRM  		& -     								& 2.15 	& 2.11 	& 12.4 	& 12.7     \\ 
				IPSM 		& - 									& 2.14 	& 2.10 	& 14.9 & 15.1 \\ 
				\midrule \midrule
				\multicolumn{6}{l} {\footnotesize $^\ddagger$ indicates curriculum training.}
			\end{tabular} }
		\end{table}

	Table~\ref{tab:WSJ0-2mix-summary} summarizes SDR (dB) and PESQ improvements for different separation methods on the WSJ0-2mix dataset. From the table we can observe that the models trained with PIT already achieve similar or better SDR than the original DPCL \cite{hershey_deep_2016}, respectively, with DNNs and CNNs. 
	Using the uPIT training criteria, we improve on PIT and achieve comparable performance with DPCL+, DPCL++ and DANet models\footnote{\cite{isik_single-channel_2016,chen_deep_2017} did not use the SDR measure from \cite{vincent_performance_2006}. Instead a related variant called scale-invariant SNR was used.} 
	reported in \cite{isik_single-channel_2016,chen_deep_2017}, which used curriculum training \cite{bengio_curriculum_2009}, and recurrent dropout \cite{gal_theoretically_2015}. 
	Note that, both uPIT and PIT models are much simpler than DANet, DPCL, DPCL+, and DPCL++, because uPIT and PIT models do not require any clustering step during inference or estimation of attractor points, as required by DANet.

	\subsection{Three-Talker Speech Separation} 
	In Fig.~\ref{fig:mse-uPIT-3sprk} we present the uPIT training progress as measured by MSE on the three-talker mixed speech training and validation sets WSJ0-3mix. We observe that similar to the two-talker scenario in Fig.~\ref{fig:mse-uPIT}, a low training MSE is achieved, although the validation MSE is slightly higher. A better balance between the training and validation MSEs may be achieved by hyperparameter tuning. We also observe that increasing the model size decreases both training and validation MSE, which is expected due to the more variability in the dataset.   
	
	\begin{figure}[h] 
		\centering
		\centerline{\includegraphics[trim={8mm 0mm 8mm 2mm},clip,width=0.75\linewidth]{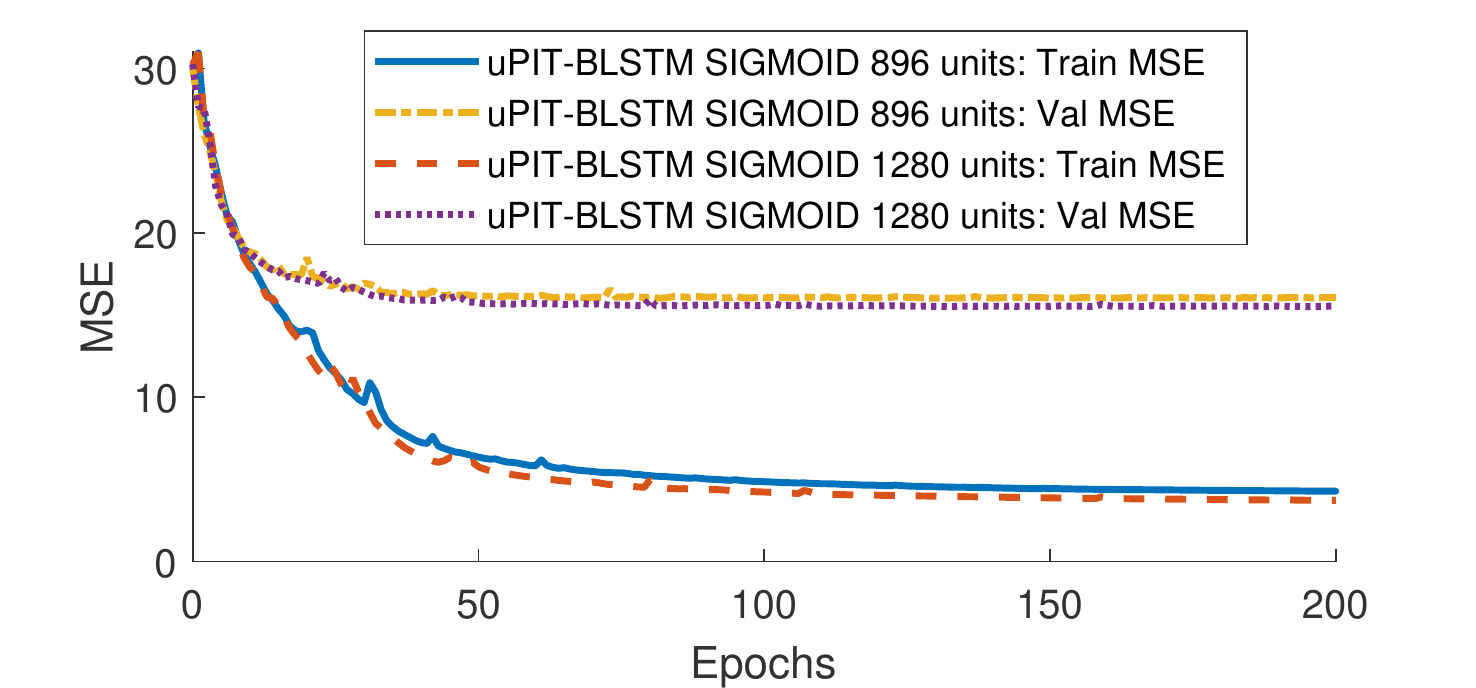}}
		\caption{MSE over epochs on the WSJ0-3mix NPSM training and validation sets wit uPIT.}\label{fig:mse-uPIT-3sprk}
	\end{figure}

	In Table~\ref{tab:WSJ0-3mix} we summarize the SDR improvement in dB from different uPIT separation configurations for three-talker mixed speech, in closed condition\;(CC) and open condition\;(OC).
	We observe that the basic uPIT-BLSTM model (896 units) compares favorably with DPCL++. 
	Furthermore, with additional units, further training and two-stage models (based on uPIT-BLSTM), uPIT achieves higher SDR than DPCL++ and similar SDR as DANet, without curriculum training, on this three-talker separation task.   

	\begin{table}[t]
		\caption{SDR improvements (dB) for different separation methods on the WSJ0-3mix dataset. $^\ddagger$ indicates curriculum training. }
		\label{tab:WSJ0-3mix}
		\centering
		\resizebox{0.75\textwidth}{!}{%
		\begin{tabular}{lcccccc}
			\midrule \midrule
			Method & \multirow{2}{*}{\makecell{ Units/\\ layer}}  & \multirow{2}{*}{\makecell{ Activation\\ function}}    & \multicolumn{2}{c} {Opt. Assign.} & \multicolumn{2}{c} {Def. Assign.} \\   \cmidrule(l){4-5} \cmidrule(l){6-7}
			&   &      							& CC & OC		&  CC 	& OC	\\
			\midrule
			Oracle NMF \cite{hershey_deep_2016} & - & - & 4.5 & - & - & - \\
			DPCL++$^\ddagger$ \cite{isik_single-channel_2016} 		 & - & - & - & - & - & 7.1 \\ 
			DANet \cite{chen_single_2017}   				& - & -	& -	 & - & - & 7.7     \\ 
			DANet$^\ddagger$ \cite{chen_deep_2017}   				& - & -	& -	 & - & - & \bf{8.8}     \\ 
			uPIT-BLSTM		& 896 	& Sigmoid  	   	& 10.0 & 9.9 & 7.4 & 7.2 \\ 
			uPIT-BLSTM		& 1280 	& Sigmoid  	   	& 10.1 & 10.0 & 7.5 & 7.4 \\ 
			uPIT-BLSTM-RD	& 1280 	& Sigmoid  	   	& 10.2 & 10.1 & 7.6 & 7.4 \\ 
			uPIT-BLSTM-ST	& 1280 	& Sigmoid  	   	& \bf{10.7} & \bf{10.6} & \bf{7.9} & 7.7 \\ 
			\midrule
			IRM		& - 	& -  	   	& 12.6	 & 12.8 & 12.6 & 12.8 \\
			IPSM 	& - 	& - 		& 15.1 	& 15.3 	& 15.1 & 15.3 \\
			\midrule \midrule
			\multicolumn{6}{l} {\footnotesize $^\ddagger$ indicates curriculum training.}
		\end{tabular}}
	\end{table}

	\subsection{Combined Two- and Three-Talker Speech Separation} 
	To illustrate the flexibility of uPIT, we summarize in Table~\ref{tab:WSJ0-2-3mix} the performance of the three-speaker uPIT-BLSTM, and uPIT-BLSTM-ST models (from Table~\ref{tab:WSJ0-3mix}), when they are trained and tested on both the WSJ0-2mix and WSJ0-3mix datasets, i.e. on both two- and three-speaker mixtures. 
	
	To be able to train the three-speaker models with the two-speaker WSJ0-2mix dataset, we extended WSJ0-2mix with a third "silent" channel. The silent channel consists of white Gaussian noise with an energy level 70 dB below the average energy level of the remaining two speakers in the mixture. 
	When we evaluated the model, we identified the two speaker-active output streams as the ones corresponding to the signals with the most energy. 
		
	We see from Table~\ref{tab:WSJ0-2-3mix} that uPIT-BLSTM achieves good, but slightly worse, performance compared to the corresponding two-speaker (Table~\ref{tab:WSJ0-2mix-summary}) and three-speaker (Table~\ref{tab:WSJ0-3mix}) models. 
	Surprisingly, the uPIT-BLSTM-ST model outperforms both the two-speaker (Table~\ref{tab:WSJ0-2mix-Further}) and three-speaker uPIT-BLSTM-ST (Table~\ref{tab:WSJ0-3mix}) models. These results indicate that a single model can handle a varying, and more importantly, unknown number of speakers, without compromising performance.    
	This is of great practical importance, since \emph{a priori} knowledge about the number of speakers is not needed at test time, as required by competing methods such as DPCL++ \cite{isik_single-channel_2016} and DANet \cite{chen_deep_2017,chen_single_2017}.  

	\begin{table}[t]
		\caption{SDR improvements (dB) for three-speaker models trained on both the WSJ0-2mix and WSJ0-3mix PSM datasets. }
		\label{tab:WSJ0-2-3mix}
		\centering
				\resizebox{0.5\textwidth}{!}{%
		\begin{tabular}{lcccc}
			\midrule \midrule
			Method     & \multicolumn{2}{c} {2 Spkr.} & \multicolumn{2}{c} {3 Spkr. } \\ \cmidrule(l){2-3} \cmidrule(l){4-5}
			& \multicolumn{2}{c} {Def. Assign.} & \multicolumn{2}{c} {Def. Assign.}	\\ \cmidrule(l){2-3} \cmidrule(l){4-5}
			& CC & OC		                           &  CC 	& OC	\\
			\midrule
			uPIT-BLSTM		  	   	& 9.4 & 9.3 & 7.2 & 7.1 \\ 
			uPIT-BLSTM-ST	  	   	& \bf{10.2} & \bf{10.1} & \bf{8.0} 	& \bf{7.8} \\ 
			\midrule
			IRM  					& 12.4 		& 12.7 		& 12.6 & 12.8 \\
			IPSM    				& 14.9 		& 15.1 		& 15.1 & 15.3 \\
			\midrule \midrule 
			\multicolumn{5}{l} {\footnotesize Both models have 1280 units per layer and ReLU outputs.}
		\end{tabular} }
	\end{table}
	
	During evaluation of the 3000 mixtures in the WSJ0-2mix test set, output stream one and two were the output streams with the most energy, i.e. the speaker-active output streams, in 2999 cases. Furthermore, output stream one and two had, on average, an energy level approximately 33 dB higher than the silent channel, indicating that the models successfully keep a constant permutation of the output masks throughout the test utterance. 
	%
	%
	\begin{figure}[ht!] 
		\centering
		\centerline{\includegraphics[trim={0mm 15mm 0mm 10mm},clip,width=0.8\linewidth]{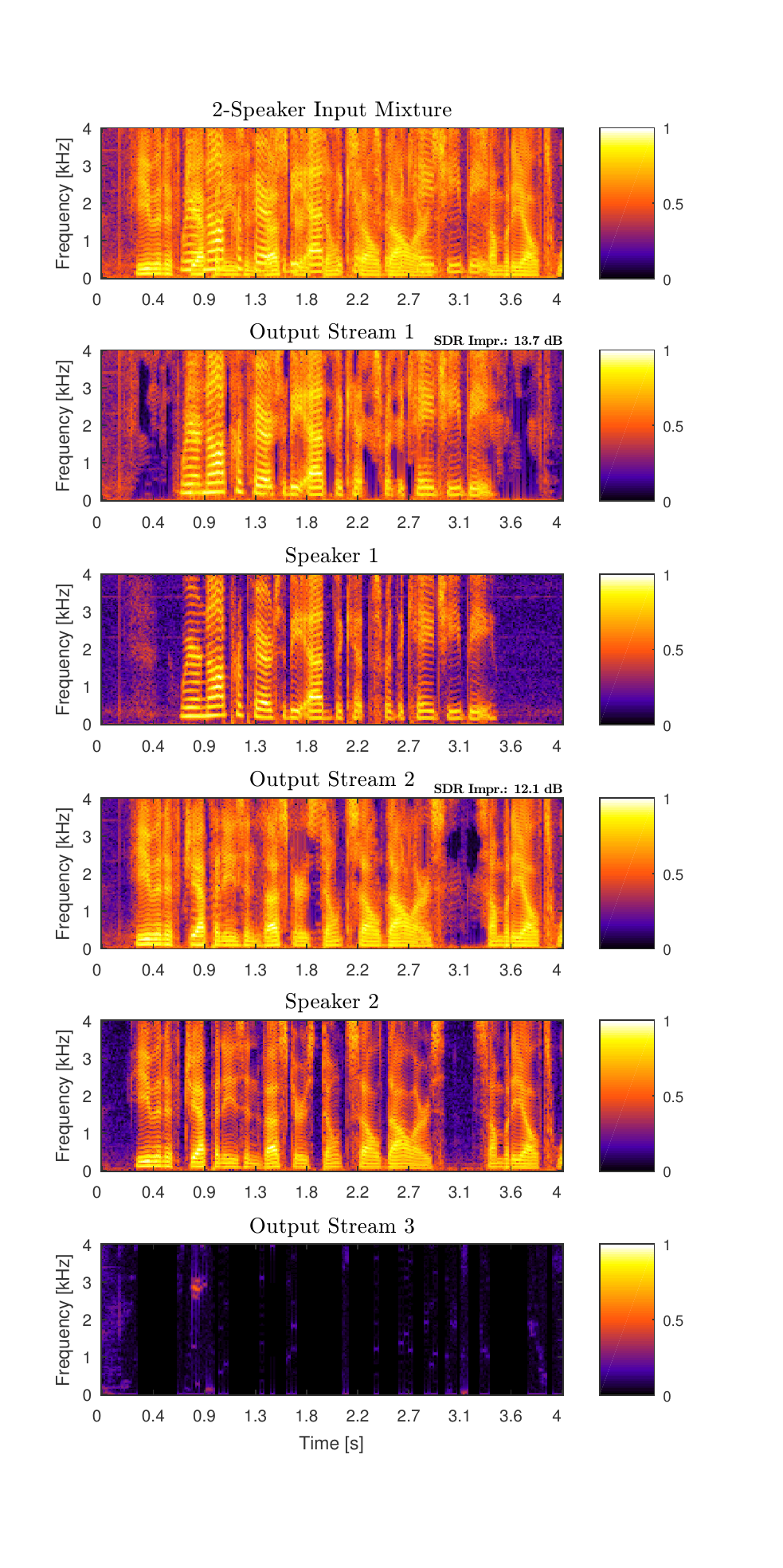}}
		\caption{Spectrograms showing how a three-speaker BLSTM model trained with uPIT can separate a two-speaker mixture while keeping a constant output-mask permutation. The energy in output stream three is $63$ dB lower than the energy in output stream one and two.}\label{fig:multSpkr}
	\end{figure}
	As an example, Fig.~\ref{fig:multSpkr} shows the spectrogram for a single two-speaker (male-vs-female) test case along with the spectrograms of the three output streams of the uPIT-BLSTM model, as well as the clean speech signals from each of the two speakers. 
	Clearly, output streams one and two contain the most energy and output stream three consists primarily of a low energy signal without any clear structure. 
	Furthermore, by comparing the spectrograms of the clean speech signals ("Speaker 1" and "Speaker 2" in Fig.~\ref{fig:multSpkr}) to the spectrogram of the corresponding output streams, it is observed that they share many similarities, which indicate that the model kept a constant output-mask permutation for the entire mixture and successfully separated the two speakers into two separate output streams.
	This is also supported by the SDR improvements, which for output stream one ("Speaker 1") is 13.7 dB, and for output stream two ("Speaker 2") is 12.1 dB. 

	\section{Conclusion and Discussion}\label{sec:conclusion}
	In this paper, we have introduced the utterance-level Permutation Invariant Training\;(uPIT) technique for speaker independent multi-talker speech separation. 
	We consider uPIT an interesting step towards solving the important cocktail party problem in a real-world setup, where the set of speakers is unknown during the training time.
	
	Our experiments on two- and three-talker mixed speech separation tasks indicate that uPIT can indeed effectively deal with the label permutation problem. These experiments show that bi-directional Long Short-Term Memory\;(LSTM) Recurrent Neural Networks\;(RNNs) perform better than uni-directional LSTMs and Phase Sensitive Masks\;(PSMs) are better training criteria than Amplitude Masks\;(AM). 
	Our results also suggest that the acoustic cues learned by the model are largely speaker and language independent since the models generalize well to unseen speakers and languages. 
	More importantly, our results indicate that uPIT trained models do not require \emph{a priori} knowledge about the number of speakers in the mixture. Specifically, we show that a single model can handle both two-speaker and three-speaker mixtures. This indicates that it might be possible to train a universal speech separation model using speech in various speaker, language and noise conditions.

	The proposed uPIT technique is algorithmically simpler yet performs on par with DPCL \cite{hershey_deep_2016,isik_single-channel_2016} and comparable to DANets \cite{chen_deep_2017,chen_single_2017}, both of which involve separate embedding and clustering stages during inference. 
	Since uPIT, as a training technique, can be easily integrated and combined with other advanced techniques such as complex-domain separation and multi-channel techniques, such as beamforming, uPIT has great potential for further improvement.

	\section*{Acknowledgment}
	We would like to thank Dr. John Hershey at MERL and Zhuo Chen at Columbia University for sharing the WSJ0-2mix and WSJ0-3mix datasets and for valuable discussions. We also thank Dr. Hakan Erdogan at Microsoft Research for discussions on PSM. 
	
    \pagebreak

{\small\bibliographystyle{bib/IEEEtran}\bibliography{bib/mybibD}}


  \cleardoublepage
  \setcounter{enumiii}{0}
  \setcounter{enumii}{0}
  \setcounter{enumiv}{0}
  \setcounter{enumi}{0}
  \setcounter{equation}{0}
  \setcounter{figure}{0}
  \setcounter{footnote}{0}
  \setcounter{mpfootnote}{0}
  \setcounter{paragraph}{0}
  \setcounter{parentequation}{0}
  \setcounter{part}{0}
  \setcounter{section}{0}
  \setcounter{subparagraph}{0}
  \setcounter{subsection}{0}
  \setcounter{subsubsection}{0}
  \setcounter{table}{0}
  \papertitlepage{%
  Joint Separation and Denoising of Noisy Multi-Talker Speech Using Recurrent Neural Networks and Permutation Invariant Training
}{paper:paperE}{%
  Morten Kolbæk, Dong Yu,  Zheng-Hua Tan, and Jesper Jensen
}{%
  The paper has been published in\\
\textit{Proceedings IEEE International Workshop on Machine Learning for Signal Processing}, pp. 1-6, September 2017.
}{%
  \noindent\copyright\ 2017 IEEE
}

\acresetall
\begin{abstract}
In this paper we propose to use utterance-level Permutation Invariant Training (uPIT) for speaker independent multi-talker speech separation and denoising, simultaneously. 
Specifically, we train deep bi-directional Long Short-Term Memory\;(LSTM) Recurrent Neural Networks\;(RNNs) using uPIT, for single-channel speaker independent multi-talker speech separation in multiple noisy conditions, including both synthetic and real-life noise signals. 
We focus our experiments on generalizability and noise robustness of models that rely on various types of \emph{a priori} knowledge e.g. in terms of noise type and number of simultaneous speakers.

We show that deep bi-directional LSTM RNNs trained using uPIT in noisy environments can improve the Signal-to-Distortion Ratio\;(SDR) as well as the Extended Short-Time Objective Intelligibility\;(ESTOI) measure, on the speaker independent multi-talker speech separation and denoising task, for various noise types and Signal-to-Noise Ratios\;(SNRs). 
Specifically, we first show that LSTM RNNs can achieve large SDR and ESTOI improvements, when evaluated using known noise types, and that a single model is capable of handling multiple noise types with only a slight decrease in performance. 
Furthermore, we show that a single LSTM RNN can handle both two-speaker and three-speaker noisy mixtures, without \emph{a priori} knowledge about the exact number of speakers. Finally, we show that LSTM RNNs trained using uPIT generalize well to noise types not seen during training.

\end{abstract}

\section{Introduction}\label{sec:intro5}
Focusing ones auditory attention towards a single speaker in a complex acoustic environment with multiple speakers and noise sources, is a task that  humans are extremely good at \cite{bronkhorst_cocktail_2000}. 
However, achieving similar performance with machines has so far not been possible \cite{mcdermott_cocktail_2009}, although it would be highly desirable for a vast range of applications, such as mobile communications, robotics, hearing aids, speaker verification systems, etc.   

Traditionally, speech denoising \cite{kolbaek_speech_2017,chen_long_2016,weninger_single-channel_2014,weninger_discriminatively_2014,erdogan_phase-sensitive_2015,chen_large-scale_2016} 
and multi-talker speech separation
\cite{du_speech_2014,yu_permutation_2017,hershey_deep_2016,isik_single-channel_2016,chen_deep_2017,weng_deep_2015,huang_joint_2015}  
have been considered as two separate tasks in the literature, although, for many applications both speech separation and denoising are desired. For example, in a human-machine interface the machine must be able to identify what is being said, and by who, before it can decide which signal to focus on, and consequently respond and act upon. 

The recent success of Deep Learning \cite{goodfellow_deep_2016} has revolutionized a large number of scientific fields, and is currently achieving state-of-the-art results on topics ranging from medical diagnosis \cite{gulshan_development_2016,esteva_dermatologist-level_2017} to Automatic Speech Recognition\;(ASR) \cite{xiong_achieving_2016,saon_english_2017}.  
Also the area of single-channel speech enhancement has seen improvement, with deep learning algorithms that have been reported to improve speech intelligibility for normal hearing, hearing impaired and cochlear implant users \cite{healy_algorithm_2015,chen_large-scale_2016,goehring_speech_2017,erdogan_deep_2017}. 
Speaker independent multi-talker speech separation, on the other hand, has so far not taken a similar leap forward, partly due to the long-lasting label permutation problem (further described in Section\;\ref{sec:train1}), which has prevented progress on deep learning based techniques for this task.

Recently, two technical directions have been proposed for speaker independent multi-talker speech separation; a clustering based approach \cite{hershey_deep_2016,isik_single-channel_2016,chen_deep_2017}, and a regression based approach \cite{yu_permutation_2017,kolbaek_multi-talker_2017-1}. 
The clustering based approaches include the Deep Clustering\;(DPCL) techniques \cite{hershey_deep_2016,isik_single-channel_2016} and the DANet technique \cite{chen_deep_2017}.  
The regression based approaches include the Permutation Invariant Training\;(PIT) technique \cite{yu_permutation_2017} and the utterance-level PIT\;(uPIT) technique \cite{kolbaek_multi-talker_2017-1}.   
The general idea behind the DPCL and DANet techniques is that the mixture signal can be represented in an embedding space, e.g. using Recurrent Neural Networks\;(RNNs), where the different source signals in the mixture form clusters. These clusters are then identified using a clustering technique, such as K-means. The clustering based techniques have shown impressive performance on two-speaker and three-speaker mixtures. 
The regression based PIT and uPIT techniques, which are described in detail in Section\;\ref{sec:train1} utilize a cost function that jointly optimizes the label assignment and regression error end-to-end, hence effectively solving the label permutation problem.

Both clustering based and regression based methods \cite{hershey_deep_2016,isik_single-channel_2016,chen_deep_2017,yu_permutation_2017,kolbaek_multi-talker_2017-1,erdogan_deep_2017} focus on ideal, noise-free training/testing conditions; i.e.\;situations where the mixtures contain clean speech only. For any practical application, background noise, e.g. due to interfering sound sources or non-ideal microphones must be expected. However, it is yet to be known how these techniques perform, when tested in noisy conditions that reflect a realistic usage scenario. 

In this paper we apply the recently proposed uPIT technique \cite{kolbaek_multi-talker_2017-1} for speaker independent multi-talker speech separation and denoising, simultaneously. 
Specifically, we train deep bi-directional Long Short-Term Memory\;(LSTM) RNNs using uPIT for speaker independent multi-talker speech separation in multiple noisy conditions, including both synthetic and real-life, known and unknown, noise signals at various Signal-to-Noise Ratios (SNRs). 

To the authors knowledge, this is the first attempt to perform speech separation and denoising simultaneously in a deep learning framework; hence, no competing baseline has been identified for this particular task.

\section{Source Separation Using Deep Learning}\label{sec:problem2}

The goal of single-channel speech separation is to separate a mixture of multiple speakers into the individual speakers using a single microphone recording. Similarly, single-channel speech denoising aims to extract a single target speech signal from a noisy single channel recording. 

Let $x_s[n]$, $n = 1,2, \dots, N$, $s = 1,2, \dots, S$ be the time domain source signal of length $N$
from source $s$ and let the 
observed mixture signal be defined as 
\begin{equation}
	y[n] = \sum_{s=1}^S x_s[n],
\label{eqMixed}
\end{equation}
where $x_1[n]$ is a speech signal and $x_s[n]$, $s = 2, \dots , S$ can be either speech or additive noise signals. 
Furthermore, let ${X}_s(i,f)$ and ${Y}(i,f)$, $i = 1,\dots,K$, $f=1,2,\dots,L$ be the $L$-point Short-Time discrete Fourier Transforms (STFT) of $x_s[n]$ and $y[n]$, respectively.
Also, let $\mathbf{x}_{s,i} = \left[ {{X}_s(i,1)} , {{X}_s(i,2)} , \dots, {{X}_s(i,\frac{L}{2}+1)} \right]^T$ $\in \mathbb{C}^{\frac{L}{2}+1}$ and  $\mathbf{y}_{i} = \left[ {{Y}(i,1)} , {{Y}(i,2)} , \dots, {{Y}(i,\frac{L}{2}+1)} \right]^T \in \mathbb{C}^{\frac{L}{2}+1}$ denote the single-sided STFT spectrum, at frame $i$, for sources $s=1,2, \dots ,S$ and the mixture signal, respectively. 

We define the magnitudes of the source signals and mixture signal as $A_s(i,f) \triangleq |{X}_s(i,f)|$ and $R(i,f)\triangleq |{Y}(i,f)|$, respectively, and their corresponding single-sided magnitude spectra as $\mathbf{a}_{s,i} = \left[ {{A_s}(i,1)} , \dots, {{A_s}(i,\frac{L}{2}+1)} \right]^T \in \mathbb{R}^{\frac{L}{2}+1}$ and $\mathbf{r}_{i} = \left[ {{R}(i,1)} ,\; {{R}(i,2)} ,\; \dots,\; {{R}(i,\frac{L}{2}+1)} \right]^T \in \mathbb{R}^{\frac{L}{2}+1}$.
For separating the mixture signal $\mathbf{y}_{i}$ into estimated target signal magnitudes $\mathbf{a}_{s,i}$, $s=1,2, \dots ,S$, we adopt the approach from \cite{kolbaek_multi-talker_2017-1} and estimate a set of masks ${M_s}(t,f)$, $s=1,2, \dots ,S$ using bi-directional LSTM RNNs. 

Let $\mathbf{m}_{s,i} = \left[ {{M_s}(i,1)} \;,\; {{M_s}(i,2)} \;,\; \dots,\; {{M_s}(i,\frac{L}{2}+1)} \right]^T \in \mathbb{R}^{\frac{L}{2}+1}$ be the ideal mask (to be defined in Sec.\;\ref{sec:mask1}) for speaker $s$ at frame $i$.
The masks $\mathbf{{m}}_{s,i}$, $s=1,2, \dots ,S$ are then used to extract the target signal magnitudes as $\mathbf{{a}}_{s,i} = \mathbf{{m}}_{s,i} \circ \mathbf{r}_i$, $s=1,2, \dots ,S$, $i=1,2, \dots ,K$ where $\circ$ is the element-wise product, i.e. the Hadamard product. 
Similarly, when the masks are estimated by a deep learning model we arrive at the estimated signal magnitudes as $\mathbf{\hat{a}}_{s,i} = \mathbf{\hat{m}}_{s,i} \circ \mathbf{r}_i$, $s=1,2, \dots ,S$, $i=1,2, \dots ,K$.
The overlap-and-add technique and the inverse discrete Fourier transform, using the phase of the mixture signal, is used for reconstructing $\mathbf{\hat{a}}_{s,i}$, $i=1,2, \dots ,K$ in the time domain.

\subsection{Mask Estimation and Loss functions}\label{sec:mask1}
A large number of training targets and loss functions have been proposed for masking based source separation \cite{erdogan_phase-sensitive_2015,wang_training_2014,erdogan_deep_2017}. Since the one reasonable goal is to have an accurate reconstruction, a loss function based on the reconstruction error instead of the mask estimation error is preferable \cite{erdogan_deep_2017}.  

In \cite{kolbaek_multi-talker_2017-1}, different such loss functions were investigated for speaker independent multi-talker speech separation and the best performing one was found to be the \ac{PSA} loss function \cite{erdogan_phase-sensitive_2015}, which for frame $i$ is given as  
\begin{equation}
\begin{aligned}
J_{i}^{PSA} = \;
& \sum_{s=1}^S \| \mathbf{\hat{m}}_{s,i}\circ\mathbf{{r}}_{i} - \mathbf{a}_{s,i}\cos(\phi_{s,i}) \|_2^2  \\
=\;& \sum_{s=1}^S \| \mathbf{\hat{a}}_{s,i} - \mathbf{a}_{s,i}\cos(\phi_{s,i}) \|_2^2,  \\
\end{aligned}
\label{eqLoss}
\end{equation}
where $\phi_{s,i} = \phi_{y,i} - \phi_{s,i} $ is the element-wise phase difference between the mixture $\mathbf{y}_i$ and the source $\mathbf{x}_{s,i}$ and $||\cdot||_2$ is the $\ell^2$-norm.   

In contrast to the classical squared error loss function, i.e. Eq.\;\eqref{eqLoss} without the cosine term, the PSA loss function accounts for some of the errors introduced by the noisy phase used in the reconstruction. 
When the PSA loss function is used for mask estimation, the actual mask estimated is the \ac{IPSF} \cite{erdogan_phase-sensitive_2015}, which due to the phase correction property is preferable over other commonly used masks such as the ideal ratio mask, or the ideal amplitude mask \cite{erdogan_deep_2017}.

\section{Permutation Invariant Training}\label{sec:train1}
Permutation Invariant Training\;(PIT) is a generalization of the traditional approach for training Deep Neural Networks\;(DNNs) for regression based source separation problems, such as speaker separation or denoising. 

For training a DNN based source separation model with $S$ output masks, $\mathbf{\hat{m}}_{s,i}$, $s=1, \dots ,S$, an MSE criterion is typically used and is computed between the true sources $\mathbf{a}_{s,i}$ and the estimated sources $\mathbf{\hat{a}}_{s,i} = \mathbf{\hat{m}}_{s,i} \circ \mathbf{r}_i$, $s = 1, \dots , S$, $i=1,\dots,K$. However, with multiple outputs, it is not trivial to pair the outputs with the correct targets.   
The commonly used approach for pairing a given output $\mathbf{\hat{a}}_{s,i}$ to a certain target $\mathbf{a}_{s,i}$ is to predefine the targets into an ordered list, such that output one is always paired with e.g. target one, i.e. $(\mathbf{a}_{1,i},\mathbf{\hat{a}}_{1,i})$, output two with target two $(\mathbf{a}_{2,i},\mathbf{\hat{a}}_{2,i})$, etc.

For tasks such as speech denoising with a single speaker in noise, or speech separation of known speakers \cite{huang_joint_2015}, simply predefining the ordering of the targets works well and the DNN can learn to correctly separate the sources and will provide the correct source at the output corresponding to the correct target. 
However, for mixtures containing similar signals, such as unknown equal energy male speakers, this standard training approach fails to converge \cite{yu_permutation_2017,weng_deep_2015,hershey_deep_2016}. 
Empirically, it is found that DNNs are likely to change permutation from one frame to another for highly similar sources. Hence, predefining the ordering of the targets, might not be the optimal solution, and clearly a bad solution for certain types of signals. This phenomenon, and the challenge of choosing the output-target permutation during training, is commonly known as the label permutation or ambiguity problem \cite{weng_deep_2015,hershey_deep_2016,chen_deep_2017,kolbaek_multi-talker_2017-1}. 

In \cite{yu_permutation_2017} a solution to the label permutation problem was proposed, where targets are provided as a set instead of an ordered list and the output-target permutation $\theta$, for a given frame, is defined as the permutation that minimizes the cost function in question (e.g. squared error) over all possible permutations $\mathcal{P}$. Following this approach combined with the PSA loss function, a permutation invariant training criterion and corresponding error $J_i^{PIT}$, for the $i^{th}$ frame, can be formulated as
\begin{equation}
\begin{aligned}
J_i^{PIT} = \;&  \underset{\theta\in \mathcal{P}}{\text{min}} 
& \sum_{s=1}^S \| \mathbf{\hat{a}}_{s,i} - \mathbf{a}_{\theta(s),i} \cos(\phi_{\theta(s),i}) \|_2^2.  \\
\end{aligned}
\label{eqPITseg}
\end{equation}

As shown in \cite{yu_permutation_2017}, Eq.\;\eqref{eqPITseg} effectively solves the label permutation problem. However, since PIT as defined in Eq.\;\eqref{eqPITseg} operates on frames, the DNN only learns to separate the input mixtures into sources at the frame level, and not the utterance level. 
In practice, this means that the mixture might be correctly separated, but the frames belonging to a particular speaker are not assigned the same output index throughout the utterance and without exact knowledge about the speaker-output permutation, it is very difficult to correctly reconstruct the separated sources.
In order to have the sources separated at the utterance-level, so that all frames from a particular output belong to the same source, additional speaker tracing or very large input-output contexts are needed \cite{yu_permutation_2017}.

\subsection{Utterance-Level Permutation Invariant Training } \label{subsec:pittrace}
In \cite{kolbaek_multi-talker_2017-1} an extension to PIT, known as utterance-level PIT\;(uPIT) was proposed for solving the speaker-output permutation problem. In uPIT, the output-target permutation $\theta$ is given as the permutation that gives the minimum squared error over all possible permutations for the entire utterance, instead of only a single frame. Formally, the utterance-level permutation used for training is found as   
\begin{equation}
\begin{aligned}
\theta^\ast = \underset{\theta\in \mathcal{P}}{\text{argmin}} \sum_{s=1}^S \sum_{i=1}^K  \lVert \hat{\mathbf{a}}_{s,i} -  \mathbf{a}_{\theta(s),i}\cos(\phi_{\theta(s),i}) \lVert_2^2,  
\end{aligned}
\label{eqPITutt2}
\end{equation}
and the permutation $\theta^\ast$ is then used \emph{for all} frames within the current utterance, hence an utterance-level loss $J_{\theta^\ast,i}^{uPIT}$ for the $i^{th}$ frame in a given utterance is defined as
\begin{equation}
\begin{aligned}
J_{\theta^\ast,i}^{uPIT} = \;& \sum_{s=1}^S \lVert \mathbf{\hat{a}}_{s,i} - \mathbf{a}_{\theta^\ast(s),i} \cos(\phi_{\theta^\ast(s),i}) \lVert_2^2.  \\
\end{aligned}
\label{eqPITutt3}
\end{equation}
Using the same permutation \emph{for all} frames in the entire utterance has the consequence that the smallest per-frame error will not always be used for training as with original PIT. Instead the smallest \emph{per-utterance} error will be used, which enforces the estimated sources to stay at the same DNN outputs for the entire utterance. Ideally, this means that each DNN output contains a single source. Finally, since the whole utterance is needed for computing the utterance-level permutation in Eq.\;\eqref{eqPITutt2}, RNNs are a natural choice of DNN model for this loss function.

\section{Experimental Design}\label{sec:expsetup}
To study the noise robustness of the uPIT technique, we have conducted several experiments with noise corrupted mixtures of multiple speakers. 
Since uPIT uses the noise-free source signals as training targets, a denoising capability is already present in the uPIT framework. By simply adding noise to the multi-speaker input mixture, a model trained with uPIT will not only learn to separate the sources but also to remove the noise.

\subsection{Noise-Free Multi-Talker Speech Mixtures} \label{subsec:speechdata}
We have used the noise-free two-speaker mixture (WSJ0-2mix) and three-speaker mixture (WSJ0-3mix)\footnote{Available at: http://www.merl.com/demos/deep-clustering} datasets for all experiments conducted in this paper. These datasets have been used in \cite{hershey_deep_2016,kolbaek_multi-talker_2017-1,yu_permutation_2017,isik_single-channel_2016}, which allows us to relate the performance of uPIT in noisy conditions with the performance in noise-free conditions. 
The feature representation is based on 129-dimensional STFT magnitude spectra, extracted from a 256 point STFT using a sampling frequency of 8 kHz, a hanning window size of 32 ms and a 16 ms frame shift.

The WSJ0-2mix dataset was derived from the WSJ0 corpus \cite{garofolo_csr-i_1993}. 
The WSJ0-2mix training set and validation set contain two-speaker mixtures generated by randomly selecting pairs of utterances from 49 male and 51 female speakers from the WSJ0 training set entitled si\_tr\_s. 
The two utterances are then mixed with a difference in active speech level \cite{noauthor_itu_1993} uniformly chosen between 0\;dB and 5\;dB. 
The training and validation sets consist of 20000 and 5000 mixtures, respectively, which is equivalent to approximately 30 hours of training data and 5 hours of validation data. The test set was similarly generated using utterances from 16 speakers from the WSJ0 validation set si\_dt\_05 and evaluation set si\_et\_05, and consists of 5000 mixtures or approximately 5 hours of data. That is, the speakers in the test set are different from the speakers in the training and validation sets. The WSJ0-3mix dataset was generated using a similar approach but contains mixtures of speech from three speakers.

Since we want a single RNN architecture that can handle both two-speaker and three-speaker mixtures, we have chosen a model architecture with three outputs. The specific architecture is described in detail in Sec.\;\ref{subsec:modelarch}. To ensure that the model can handle both two-speaker and three-speaker mixtures, the model must be trained on both scenarios, so we have combined the WSJ0-2mix and WSJ0-3mix datasets into a larger WSJ0-2+3mix dataset. To allow this fusion, we have extended the WSJ0-2mix dataset with a third "silent" speaker, such that the combined WSJ0-2+3mix dataset consists of only three speaker mixtures, but half of the mixtures contain three speakers, and the remaining half contain two speaker mixtures (and a "silent speaker"). To minimize the risk of numerical issues, e.g. in computing ideal masks, the third "silent" speaker consists of white Gaussian noise with an average energy level 70 dB below the average energy of the other two speakers in the mixture.  

\subsection{Noisy Multi-Talker Speech Mixtures} \label{subsec:noisedata}
To simulate noisy environments, we follow the common approach \cite{kolbaek_speech_2017} for generating noisy mixtures with additive noise and simply add the noise-free WSJ0-2+3mix mixture signal with a noise signal. To achieve a certain SNR the noise signal is scaled based on the active speech level of the noise-free mixture signal as per ITU P.56 \cite{noauthor_itu_1993}. 

To evaluate the robustness of the uPIT model against a stationary noise type, we use a synthetic Speech Shaped Noise\;(SSN) signal.  
The SSN noise signal is constructed by filtering a Gaussian white noise sequence through a $12^{th}$-order all-pole filter with coefficients found from linear predictive coding analysis of 100 randomly chosen TIMIT sentences \cite{garofolo_darpa_1993}.  

To evaluate the robustness against a highly non-stationary noise type we use a synthetic 6-speaker Babble\;(BBL) noise. 
The BBL noise signal is also based on TIMIT. The corpus, which consists of a total of 6300 spoken sentences, is randomly divided  into 6 groups of 1050 concatenated utterances. Each group is then normalized to unit energy and truncated to equal length followed by  addition of the six groups. This results in a BBL noise sequence with a duration of over 50 min. 

To evaluate the robustness against realistic noise types we use the street (STR), cafeteria\;(CAF), bus\;(BUS), and pedestrian\;(PED) noise signals from the CHiME3 dataset \cite{barker_third_2015}. These noise signals are real-life recordings in their respective environments.   

All six noise signals are divided into a 40 min.\;training sequence, a 5 min.\;validation sequence and a 5 min.\;test sequence. That is, the noise signals used for training and validation are different from the sequence used for testing.

\subsection{Model Architectures and Training} \label{subsec:modelarch}
For evaluating uPIT in noisy environments we have trained a total of seven bi-directional LSTM RNNs \cite{hochreiter_long_1997}, using the training conditions, i.e. datasets and noise types, presented in Table.\;\ref{tab:models}. 
\begin{table}
	\caption{Training conditions for different models.}
	\label{tab:models}
	\centering
	\setlength\tabcolsep{5pt} 
	\resizebox{0.6\columnwidth}{!}{
	\begin{tabular}{ccc}
		\midrule \midrule
		Model ID & \multicolumn{2}{l} {Dataset + Noise type (SNR: -5 dB -- 10 dB)}  \\ 
		\midrule
		LSTM1  & \multicolumn{2}{l} {WSJ0-2+3mix + SSN } 	       \\ 
		LSTM2   & \multicolumn{2}{l} {WSJ0-2+3mix + BBL }  	  \\ 
		LSTM3   & \multicolumn{2}{l} {WSJ0-2+3mix + STR }   	  \\ 
		LSTM4  & \multicolumn{2}{l} {WSJ0-2+3mix + CAF } 	    \\ 
		LSTM5  &\multicolumn{2}{l} {WSJ0-2+3mix + SSN + BBL + STR + CAF }    	  \\ 
		LSTM6  & \multicolumn{2}{l} {WSJ0-2mix + BBL }   	  \\ 
		LSTM7  & \multicolumn{2}{l} {WSJ0-3mix + BBL } \\ 
		\midrule \midrule
	\end{tabular}}
\end{table}
LSTM1-5 were trained on the WSJ0-2+3mix dataset, which contains a mix of both two-speaker and three-speaker mixtures. LSTM1-4 are noise type specific in the sense that they were trained using only a single noise type. LSTM5 was trained on all four noise types. LSTM6 and LSTM7 were trained using WSJ0-2mix and WSJ0-3mix datasets, respectively, and only a single noise type. 
LSTM5 will show the performance degradation, if any, when less \emph{a priori} knowledge about the noise types is available. Similarly, LSTM6-7 will show the potential performance improvement if the number of speakers in the mixture is known \emph{a priori}. 
Each mixture in the dataset was corrupted with noise at a specific SNR, uniformly chosen between -5\;dB and 10\;dB.   

Each model has three bi-directional LSTM layers, and a fully-connected output layer with ReLU \cite{goodfellow_deep_2016} activation functions. LSTM1-5 and LSTM7 have 1280 LSTM cells in each layer and LSTM6 has 896 cells, to be compliant with \cite{kolbaek_multi-talker_2017-1}. 
The input dimension is 129, i.e a single frame $\mathbf{r}_i$ and the output dimension is $3\times129 = 387$, i.e. $\mathbf{\hat{a}}_{s,i}$, $s=1,2,3$. We apply 50\% dropout \cite{goodfellow_deep_2016} between the LSTM layers, and the outputs from the forward and backward LSTMs, from one layer, are concatenated before they are used as input to the subsequent layer. 
LSTM6 has approximately $46\cdot10^6$ trainable parameters, and LSTM1-5 and 7 have approximately $94\cdot 10^6$ trainable parameters, which are found using stochastic gradient descent with gradients found by backpropagation.  
In all the experiments, the maximum number of epochs was set to 200 and the learning rates were set to $2 \cdot 10^{-5}$ per sample initially, and scaled down by $0.7$ when the training cost increased on the training set. The training was terminated when the learning rate got below $10^{-10}$. Each minibatch contains 8 randomly selected utterances.
All models are implemented using the Microsoft Cognitive Toolkit (CNTK) \cite{agarwal_introduction_2014}\footnote{Available at: \url{https://www.cntk.ai/}}.

\section{Experimental Results}\label{sec:exp2}
We evaluated the noise robustness of LSTM1-7 using the Signal to Distortion Ratio\;(SDR) \cite{vincent_performance_2006} and the Extended Short-Time Objective Intelligibility\;(ESTOI) measure \cite{jensen_algorithm_2016}.
The SDR is an often used performance metric for source separation and is defined in dB. 
The ESTOI measure estimates speech intelligibility and has been found to be highly correlated with human listening tests \cite{jensen_algorithm_2016}, especially for modulated maskers. The ESTOI measure is defined in the range $[-1, 1]$, and higher is better.   
When evaluating SDR and ESTOI, we choose the output-target permutation that maximizes the given performance metric. Furthermore, when evaluating two-speaker mixtures, we identify the silent speaker as the output with the least energy and then compute the performance metric based on the remaining two outputs.

\Cref{tab:sdr_ssn,tab:sdr_bbl,tab:sdr_str,tab:sdr_caf} summarize the SDR \emph{improvements} achieved by LSTM1-5 on two and three-speaker mixtures corrupted by SSN, BBL, STR, and CAF noise, respectively. The improvements are relative to the SDR of the noisy mixture without processing ("No Proc." in Tables). 
\Cref{tab:stoi_ssn,tab:stoi_bbl,tab:stoi_str,tab:stoi_caf} summarize ESTOI \emph{improvements} achieved by the same models in similar conditions. 
We evaluate the models at the challenging SNR of $-5$\;dB, as well as at $0$,\; $5$\;, and $20$\;dB. At an input SNR of $-5$\;dB, speech intelligibility, as estimated by ESTOI, is severely degraded, primarily due to the noise component, whereas speech intelligibility degradation at $20$\;dB is primarily caused by the competing talkers in the mixture itself.
As a reference, we also reported the IPSF performance, which uses oracle information and therefore serves as an upper performance bound on this particular task.

From \Cref{tab:sdr_ssn,tab:sdr_bbl,tab:sdr_str,tab:sdr_caf,tab:stoi_ssn,tab:stoi_bbl,tab:stoi_str,tab:stoi_caf} we see that all noise-type specific models, i.e. LSTM1-4, in general achieve large SDR and ESTOI improvements with an average improvement of $9.1$\;dB and $0.18$ for SDR and ESTOI, respectively, for two-speaker mixtures and $7.2$\;dB and $0.13$, respectively, for three-speaker mixtures. 
Furthermore, we see that LSTM5 performs only slightly worse than the noise type specific models, which is interesting, since LSTM5 and LSMT1-4 have all been trained with 60 hours of speech, but LSTM5 have only seen $15$ hours of each noise type, compared to $60$ hours for LSTM1-4. 
We also observe that the highly non-stationary BBL noise seems to be considerably harder than the three other noise types, which corresponds well with existing literature \cite{kolbaek_speech_2017,loizou_speech_2013,erkelens_minimum_2007}.       

\Cref{tab:sdr_bbl_spkr,tab:stoi_bbl_spkr} summarize the performance of LSTM6 and LSTM7. We observe that both models perform approximately similar to the noise-type-general LSTM5. More surprisingly, we see that LSTM2 consistently outperforms both LSTM6 and LSTM7, which corresponds well with a similar observation in the noise-free case in \cite{kolbaek_multi-talker_2017-1}. 
These results are of great importance, since they show that training a model on noisy three-speaker mixtures helps the model separating noisy two-speaker mixtures, and vice versa.

\Cref{tab:sdr_ped,tab:stoi_ped} summarize the performance of LSTM5, when evaluated using speech mixtures corrupted with the two unknown noise types, PED and BUS, i.e. noise types not included in the training set. We see that LSTM5 achieves large SDR and ESTOI improvements for both noise types, at almost all SNRs. More importantly, we observe that the scores are comparable with, and in some cases even exceed, the performance of LSTM5, when it was evaluated using known noise types as reported in \Cref{tab:sdr_ssn,tab:sdr_bbl,tab:sdr_str,tab:sdr_caf,tab:stoi_ssn,tab:stoi_bbl,tab:stoi_str,tab:stoi_caf}. These results indicate that LSTM5 is relatively robust against variations in the noise distribution.

In general, we observe SDR improvements for all models that are comparable in magnitude with the noise-free case \cite{yu_permutation_2017,kolbaek_multi-talker_2017-1,hershey_deep_2016,isik_single-channel_2016}. 
However, the SDR measure, as well as ESTOI, do not differentiate between distortions from other speakers (such as source to inference ratio from \cite{vincent_performance_2006}) and distortion from the noise source. This means that the trade-off between speech separation and noise-reduction is yet to be fully understood. We leave this topic for future research.

\begin{table}
	\caption{SDR improvements for LSTM1 and 5 tested on SSN.}
	\vspace{-2mm}
	\label{tab:sdr_ssn}
	\centering
	\setlength\tabcolsep{5pt} 
	\resizebox{0.7\columnwidth}{!}{%
		\begin{tabular}{ccccccccc}
			\midrule \midrule
			& \multicolumn{4}{c} {2-Speaker } & \multicolumn{4}{c} {3-Speaker} \\  \cmidrule(l){2-5} \cmidrule(l){6-9}
			\begin{tabular}[c]{@{}c@{}}SNR \\ {[dB]}\end{tabular} 	& 
			\begin{tabular}[c]{@{}c@{}}No \\ Proc.\end{tabular} 	& 
			\begin{tabular}[c]{@{}c@{}} IPSF \end{tabular} 			& 
			\begin{tabular}[c]{@{}c@{}}LSTM1\end{tabular} 		& 
			\begin{tabular}[c]{@{}c@{}}LSTM5\end{tabular} 		&   
			\begin{tabular}[c]{@{}c@{}}No \\ Proc.\end{tabular} 	& 
			\begin{tabular}[c]{@{}c@{}} IPSF \end{tabular} 			& 
			\begin{tabular}[c]{@{}c@{}}LSTM1\end{tabular} 		& 
			\begin{tabular}[c]{@{}c@{}}LSTM5\end{tabular} \\ 
			\midrule
						-5 & -8.8 & 15.9 & 9.6 & 9.4 & -10.3 & 16.6 & 8.0 & 7.8 \\ 
			0 & -5.1 & 14.5 & 9.1 & 9.0 & -7.0 & 15.2 & 7.6 & 7.4 \\ 
			5 & -2.4 & 13.9 & 8.6 & 8.4 & -4.8 & 14.6 & 7.0 & 6.9 \\ 
			20 & 0.0 & 14.8 & 8.7 & 8.8 & -3.0 & 15.1 & 6.6 & 6.7 \\ \midrule
			Avg. & -4.1 & 14.8 & 9.0 & 8.9 & -6.3 & 15.4 & 7.3 & 7.2 \\ \midrule \midrule
		\end{tabular}}
	\end{table}
	\begin{table}
	\caption{SDR improvements for LSTM2 and 5 tested on BBL.}
	\vspace{-2mm}
	\label{tab:sdr_bbl}
	\centering
	\setlength\tabcolsep{5pt} 
	\resizebox{0.7\columnwidth}{!}{%
		\begin{tabular}{ccccccccc}
			\midrule \midrule
			& \multicolumn{4}{c} {2-Speaker } & \multicolumn{4}{c} {3-Speaker} \\ \cmidrule(l){2-5} \cmidrule(l){6-9}
			\begin{tabular}[c]{@{}c@{}}SNR \\ {[dB]}\end{tabular} 	& 
			\begin{tabular}[c]{@{}c@{}}No \\ Proc.\end{tabular} 	& 
			\begin{tabular}[c]{@{}c@{}} IPSF \end{tabular} 			& 
			\begin{tabular}[c]{@{}c@{}}LSTM2\end{tabular} 		& 
			\begin{tabular}[c]{@{}c@{}}LSTM5\end{tabular} 		&   
			\begin{tabular}[c]{@{}c@{}}No \\ Proc.\end{tabular} 	& 
			\begin{tabular}[c]{@{}c@{}} IPSF \end{tabular} 			& 
			\begin{tabular}[c]{@{}c@{}}LSTM2\end{tabular} 		& 
			\begin{tabular}[c]{@{}c@{}}LSTM5\end{tabular} \\ 
			\midrule
						-5 & -8.9 & 17.2 & 6.0 & 5.4 & -10.4 & 17.8 & 4.4 & 3.8 \\ 
			0 & -5.1 & 15.4 & 8.1 & 7.6 & -7.1 & 16.0 & 6.3 & 5.8 \\ 
			5 & -2.4 & 14.5 & 8.5 & 8.1 & -4.8 & 15.1 & 6.7 & 6.5 \\ 
			20 & 0.0 & 14.8 & 9.0 & 8.8 & -3.0 & 15.2 & 6.8 & 6.7 \\ \midrule
			Avg. & -4.1 & 15.5 & 7.9 & 7.5 & -6.3 & 16.0 & 6.0 & 5.7 \\ \midrule \midrule
		\end{tabular}}
	\end{table}
	\begin{table}
	\caption{SDR improvements for LSTM3 and 5 tested on STR.}
	\vspace{-2mm}
	\label{tab:sdr_str}
	\centering
	\setlength\tabcolsep{5pt} 
	\resizebox{0.7\columnwidth}{!}{%
		\begin{tabular}{ccccccccc}
			\midrule \midrule
			& \multicolumn{4}{c} {2-Speaker } & \multicolumn{4}{c} {3-Speaker} \\ \cmidrule(l){2-5} \cmidrule(l){6-9}
			\begin{tabular}[c]{@{}c@{}}SNR \\ {[dB]}\end{tabular} 	& 
			\begin{tabular}[c]{@{}c@{}}No \\ Proc.\end{tabular} 	& 
			\begin{tabular}[c]{@{}c@{}} IPSF \end{tabular} 			& 
			\begin{tabular}[c]{@{}c@{}}LSTM3\end{tabular} 		& 
			\begin{tabular}[c]{@{}c@{}}LSTM5\end{tabular} 		&   
			\begin{tabular}[c]{@{}c@{}}No \\ Proc.\end{tabular} 	& 
			\begin{tabular}[c]{@{}c@{}} IPSF \end{tabular} 			& 
			\begin{tabular}[c]{@{}c@{}}LSTM3\end{tabular} 		& 
			\begin{tabular}[c]{@{}c@{}}LSTM5\end{tabular} \\ 
			\midrule
						-5 & -8.9 & 18.2 & 11.5 & 11.5 & -10.4 & 18.6 & 9.7 & 9.6 \\ 
			0 & -5.2 & 16.2 & 10.2 & 10.2 & -7.1 & 16.7 & 8.4 & 8.3 \\ 
			5 & -2.4 & 14.9 & 9.2 & 9.1 & -4.8 & 15.5 & 7.3 & 7.2 \\ 
			20 & 0.0 & 14.9 & 8.9 & 8.8 & -3.0 & 15.2 & 6.6 & 6.7 \\ \midrule
			Avg. & -4.1 & 16.1 & 9.9 & 9.9 & -6.3 & 16.5 & 8.0 & 7.9 \\ \midrule \midrule
		\end{tabular}}
	\end{table}
	\begin{table}
	\caption{SDR improvements for LSTM4 and 5 tested on CAF.}
	\vspace{-2mm}
	\label{tab:sdr_caf}
	\centering
	\setlength\tabcolsep{5pt} 
	\resizebox{0.7\columnwidth}{!}{%
		\begin{tabular}{ccccccccc}
			\midrule \midrule
			& \multicolumn{4}{c} {2-Speaker } & \multicolumn{4}{c} {3-Speaker} \\ \cmidrule(l){2-5} \cmidrule(l){6-9}
			\begin{tabular}[c]{@{}c@{}}SNR \\ {[dB]}\end{tabular} 	& 
			\begin{tabular}[c]{@{}c@{}}No \\ Proc.\end{tabular} 	& 
			\begin{tabular}[c]{@{}c@{}} IPSF \end{tabular} 			& 
			\begin{tabular}[c]{@{}c@{}}LSTM4\end{tabular} 		& 
			\begin{tabular}[c]{@{}c@{}}LSTM5\end{tabular} 		&   
			\begin{tabular}[c]{@{}c@{}}No \\ Proc.\end{tabular} 	& 
			\begin{tabular}[c]{@{}c@{}} IPSF \end{tabular} 			& 
			\begin{tabular}[c]{@{}c@{}}LSTM4\end{tabular} 		& 
			\begin{tabular}[c]{@{}c@{}}LSTM5\end{tabular} \\ 
			\midrule
						-5 & -8.9 & 18.2 & 10.0 & 9.9 & -10.4 & 18.6 & 8.4 & 8.2 \\ 
			0 & -5.1 & 16.3 & 9.7 & 9.5 & -7.1 & 16.8 & 7.9 & 7.7 \\ 
			5 & -2.4 & 15.1 & 9.0 & 8.9 & -4.8 & 15.6 & 7.1 & 6.9 \\ 
			20 & 0.0 & 14.8 & 8.8 & 8.8 & -3.0 & 15.2 & 6.7 & 6.6 \\ \midrule
			Avg. & -4.1 & 16.1 & 9.4 & 9.3 & -6.3 & 16.6 & 7.5 & 7.3 \\ \midrule \midrule
		\end{tabular}}
	\end{table}
	\begin{table}
	\caption{SDR improvements for LSTM6, 7 and 5 tested on BBL.}
	\vspace{-2mm}
	\label{tab:sdr_bbl_spkr}
	\centering
	\setlength\tabcolsep{5pt} 
	\resizebox{0.7\columnwidth}{!}{%
		\begin{tabular}{ccccccccc}
			\midrule \midrule
			& \multicolumn{4}{c} {2-Speaker } & \multicolumn{4}{c} {3-Speaker} \\ \cmidrule(l){2-5} \cmidrule(l){6-9}
			\begin{tabular}[c]{@{}c@{}}SNR \\ {[dB]}\end{tabular} 	& 
			\begin{tabular}[c]{@{}c@{}}No \\ Proc.\end{tabular} 	& 
			\begin{tabular}[c]{@{}c@{}} IPSF \end{tabular} 			& 
			\begin{tabular}[c]{@{}c@{}}LSTM6\end{tabular} 		& 
			\begin{tabular}[c]{@{}c@{}}LSTM5\end{tabular} 		&   
			\begin{tabular}[c]{@{}c@{}}No \\ Proc.\end{tabular} 	& 
			\begin{tabular}[c]{@{}c@{}} IPSF \end{tabular} 			& 
			\begin{tabular}[c]{@{}c@{}}LSTM7\end{tabular} 		& 
			\begin{tabular}[c]{@{}c@{}}LSTM5\end{tabular} \\ 
			\midrule
		    			-5 & -8.9 & 17.2 & 5.6 & 5.4 & -10.4 & 17.8 & 4.0 & 3.8 \\ 
			0 & -5.1 & 15.4 & 7.7 & 7.6 & -7.1 & 16.0 & 5.7 & 5.8 \\ 
			5 & -2.4 & 14.5 & 8.0 & 8.1 & -4.8 & 15.1 & 6.3 & 6.5 \\ 
			20 & 0.0 & 14.9 & 8.4 & 8.8 & -3.0 & 15.2 & 6.4 & 6.7 \\ \midrule
			Avg. & -4.1 & 15.5 & 7.4 & 7.5 & -6.3 & 16.0 & 5.6 & 5.7 \\ \midrule \midrule
		\end{tabular}}
\end{table}
\begin{table}
	\caption{SDR improvements for LSTM5 tested on BUS and PED.}
	\vspace{-2mm}
	\label{tab:sdr_ped}
	\centering
	\setlength\tabcolsep{5pt} 
	\resizebox{0.7\columnwidth}{!}{%
		\begin{tabular}{ccccccccccccc}
			\midrule \midrule
			& \multicolumn{6}{c} {2-Speaker } & \multicolumn{6}{c} {3-Speaker} \\ \cmidrule(l){2-7} \cmidrule(l){8-13}
			\begin{tabular}[c]{@{}c@{}}SNR \\ {[dB]}\end{tabular} 	& 
			\multicolumn{2}{c} {\begin{tabular}[c]{@{}c@{}}No \\ Proc.\end{tabular} }	& 
			\multicolumn{2}{c} {\begin{tabular}[c]{@{}c@{}} IPSF \end{tabular} 	}		& 
			\multicolumn{2}{c} {\begin{tabular}[c]{@{}c@{}}LSTM5\end{tabular} 	}	&   
			\multicolumn{2}{c} {\begin{tabular}[c]{@{}c@{}}No \\ Proc.\end{tabular}} 	& 
			\multicolumn{2}{c} {\begin{tabular}[c]{@{}c@{}} IPSF \end{tabular} 		}	& 
			\multicolumn{2}{c} {\begin{tabular}[c]{@{}c@{}}LSTM5\end{tabular}} \\ \cmidrule(l){2-7} \cmidrule(l){8-13}
			& BUS & PED & BUS & PED & BUS & PED & BUS & PED & BUS & PED & BUS & PED \\
			\midrule
						-5 & -9.0 & -8.9 & 19.6 & 16.7 & 11.7 & 7.3 & -10.5 & -10.4 & 19.9 & 17.4 & 9.7 & 5.7 \\ 
			0 & -5.2 & -5.2 & 17.3 & 14.9 & 10.7 & 7.8 & -7.2 & -7.1 & 17.6 & 15.7 & 8.5 & 6.3 \\ 
			5 & -2.4 & -2.4 & 15.7 & 14.1 & 9.5 & 7.9 & -4.8 & -4.8 & 16.1 & 14.8 & 7.4 & 6.3 \\ 
			20 & 0.0 & 0.0 & 14.9 & 14.8 & 8.8 & 8.7 & -3.0 & -3.0 & 15.2 & 15.2 & 6.7 & 6.7 \\ \midrule
			Avg. & -4.1 & -4.1 & 16.9 & 15.1 & 10.2 & 7.9 & -6.4 & -6.3 & 17.2 & 15.8 & 8.1 & 6.2 \\ \midrule \midrule
	\end{tabular}}
\end{table}

\begin{table}
	\caption{ESTOI improvements for LSTM1 and 5 tested on SSN.}
	\vspace{-2mm}
	\label{tab:stoi_ssn}
	\centering
	\setlength\tabcolsep{5pt} 
	\resizebox{0.7\columnwidth}{!}{%
		\begin{tabular}{ccccccccc}
			\midrule \midrule
			& \multicolumn{4}{c} {2-Speaker } & \multicolumn{4}{c} {3-Speaker} \\ \cmidrule(l){2-5} \cmidrule(l){6-9}
			\begin{tabular}[c]{@{}c@{}}SNR \\ {[dB]}\end{tabular} 	& 
			\begin{tabular}[c]{@{}c@{}}No \\ Proc.\end{tabular} 	& 
			\begin{tabular}[c]{@{}c@{}} IPSF \end{tabular} 			& 
			\begin{tabular}[c]{@{}c@{}}LSTM1\end{tabular} 		& 
			\begin{tabular}[c]{@{}c@{}}LSTM5\end{tabular} 		&   
			\begin{tabular}[c]{@{}c@{}}No \\ Proc.\end{tabular} 	& 
			\begin{tabular}[c]{@{}c@{}} IPSF \end{tabular} 			& 
			\begin{tabular}[c]{@{}c@{}}LSTM1\end{tabular} 		& 
			\begin{tabular}[c]{@{}c@{}}LSTM5\end{tabular} \\ 
			\midrule
						-5 & 0.18 & 0.65 & 0.17 & 0.16 & 0.14 & 0.69 & 0.10 & 0.09 \\ 
			0 & 0.29 & 0.58 & 0.23 & 0.22 & 0.22 & 0.63 & 0.15 & 0.14 \\ 
			5 & 0.39 & 0.50 & 0.23 & 0.22 & 0.29 & 0.58 & 0.17 & 0.16 \\ 
			20 & 0.54 & 0.39 & 0.17 & 0.18 & 0.38 & 0.53 & 0.15 & 0.15 \\ \midrule
			Avg. & 0.35 & 0.53 & 0.20 & 0.20 & 0.26 & 0.61 & 0.14 & 0.14 \\ \midrule \midrule
	\end{tabular}}
\end{table}
\begin{table}
	\caption{ESTOI improvements for LSTM2 and 5 tested on BBL.}
	\vspace{-2mm}	
	\label{tab:stoi_bbl}
	\centering
	\setlength\tabcolsep{5pt} 
	\resizebox{0.7\columnwidth}{!}{%
		\begin{tabular}{ccccccccc}
		\midrule \midrule
		& \multicolumn{4}{c} {2-Speaker } & \multicolumn{4}{c} {3-Speaker} \\ \cmidrule(l){2-5} \cmidrule(l){6-9}
		\begin{tabular}[c]{@{}c@{}}SNR \\ {[dB]}\end{tabular} 	& 
		\begin{tabular}[c]{@{}c@{}}No \\ Proc.\end{tabular} 	& 
		\begin{tabular}[c]{@{}c@{}} IPSF \end{tabular} 			& 
		\begin{tabular}[c]{@{}c@{}}LSTM2\end{tabular} 		& 
		\begin{tabular}[c]{@{}c@{}}LSTM5\end{tabular} 		&   
		\begin{tabular}[c]{@{}c@{}}No \\ Proc.\end{tabular} 	& 
		\begin{tabular}[c]{@{}c@{}} IPSF \end{tabular} 			& 
		\begin{tabular}[c]{@{}c@{}}LSTM2\end{tabular} 		& 
		\begin{tabular}[c]{@{}c@{}}LSTM5\end{tabular} \\ 
		\midrule
					-5 & 0.19 & 0.66 & 0.09 & 0.06 & 0.14 & 0.70 & 0.04 & 0.02 \\ 
			0 & 0.29 & 0.59 & 0.18 & 0.15 & 0.22 & 0.65 & 0.11 & 0.09 \\ 
			5 & 0.39 & 0.51 & 0.21 & 0.20 & 0.29 & 0.60 & 0.15 & 0.14 \\ 
			20 & 0.53 & 0.40 & 0.19 & 0.18 & 0.37 & 0.53 & 0.15 & 0.15 \\ \midrule
			Avg. & 0.35 & 0.54 & 0.17 & 0.15 & 0.26 & 0.62 & 0.11 & 0.10 \\ \midrule \midrule
	\end{tabular}}
\end{table}
\begin{table}
	\caption{ESTOI improvements for LSTM3 and 5 tested on STR.}
	\vspace{-2mm}	
	\label{tab:stoi_str}
	\centering
	\setlength\tabcolsep{5pt} 
	\resizebox{0.7\columnwidth}{!}{%
		\begin{tabular}{ccccccccc}
		\midrule \midrule
		& \multicolumn{4}{c} {2-Speaker } & \multicolumn{4}{c} {3-Speaker} \\ \cmidrule(l){2-5} \cmidrule(l){6-9}
		\begin{tabular}[c]{@{}c@{}}SNR \\ {[dB]}\end{tabular} 	& 
		\begin{tabular}[c]{@{}c@{}}No \\ Proc.\end{tabular} 	& 
		\begin{tabular}[c]{@{}c@{}} IPSF \end{tabular} 			& 
		\begin{tabular}[c]{@{}c@{}}LSTM3\end{tabular} 		& 
		\begin{tabular}[c]{@{}c@{}}LSTM5\end{tabular} 		&   
		\begin{tabular}[c]{@{}c@{}}No \\ Proc.\end{tabular} 	& 
		\begin{tabular}[c]{@{}c@{}} IPSF \end{tabular} 			& 
		\begin{tabular}[c]{@{}c@{}}LSTM3\end{tabular} 		& 
		\begin{tabular}[c]{@{}c@{}}LSTM5\end{tabular} \\ 
		\midrule
					-5 & 0.24 & 0.60 & 0.16 & 0.15 & 0.18 & 0.65 & 0.10 & 0.09 \\ 
			0 & 0.32 & 0.54 & 0.21 & 0.19 & 0.24 & 0.61 & 0.14 & 0.13 \\ 
			5 & 0.40 & 0.49 & 0.21 & 0.20 & 0.30 & 0.57 & 0.15 & 0.15 \\ 
			20 & 0.54 & 0.39 & 0.18 & 0.18 & 0.37 & 0.53 & 0.15 & 0.15 \\ \midrule
			Avg. & 0.38 & 0.51 & 0.19 & 0.18 & 0.27 & 0.59 & 0.14 & 0.13 \\ \midrule \midrule
	\end{tabular}}
\end{table}
\begin{table}
	\caption{ESTOI improvements for LSTM4 and 5 tested on CAF.}
	\vspace{-2mm}	
	\label{tab:stoi_caf}
	\centering
	\setlength\tabcolsep{5pt} 
	\resizebox{0.7\columnwidth}{!}{%
		\begin{tabular}{ccccccccc}
		\midrule \midrule
		& \multicolumn{4}{c} {2-Speaker } & \multicolumn{4}{c} {3-Speaker} \\ \cmidrule(l){2-5} \cmidrule(l){6-9}
		\begin{tabular}[c]{@{}c@{}}SNR \\ {[dB]}\end{tabular} 	& 
		\begin{tabular}[c]{@{}c@{}}No \\ Proc.\end{tabular} 	& 
		\begin{tabular}[c]{@{}c@{}} IPSF \end{tabular} 			& 
		\begin{tabular}[c]{@{}c@{}}LSTM4\end{tabular} 		& 
		\begin{tabular}[c]{@{}c@{}}LSTM5\end{tabular} 		&   
		\begin{tabular}[c]{@{}c@{}}No \\ Proc.\end{tabular} 	& 
		\begin{tabular}[c]{@{}c@{}} IPSF \end{tabular} 			& 
		\begin{tabular}[c]{@{}c@{}}LSTM4\end{tabular} 		& 
		\begin{tabular}[c]{@{}c@{}}LSTM5\end{tabular} \\ 
		\midrule
					-5 & 0.24 & 0.60 & 0.13 & 0.12 & 0.19 & 0.65 & 0.08 & 0.07 \\ 
			0 & 0.33 & 0.54 & 0.18 & 0.17 & 0.25 & 0.61 & 0.12 & 0.11 \\ 
			5 & 0.41 & 0.48 & 0.20 & 0.19 & 0.30 & 0.58 & 0.15 & 0.14 \\ 
			20 & 0.53 & 0.39 & 0.18 & 0.18 & 0.37 & 0.53 & 0.15 & 0.15 \\ \midrule
			Avg. & 0.38 & 0.50 & 0.17 & 0.17 & 0.28 & 0.59 & 0.12 & 0.12 \\ \midrule \midrule
	\end{tabular}}
\end{table}
\begin{table}
	\caption{ESTOI improvements for LSTM6, 7 and 5 tested on BBL.}
	\vspace{-2mm}
	\label{tab:stoi_bbl_spkr}
	\centering
	\setlength\tabcolsep{5pt} 
	\resizebox{0.7\columnwidth}{!}{%
		\begin{tabular}{ccccccccc}
		\midrule \midrule
		& \multicolumn{4}{c} {2-Speaker } & \multicolumn{4}{c} {3-Speaker} \\ \cmidrule(l){2-5} \cmidrule(l){6-9}
		\begin{tabular}[c]{@{}c@{}}SNR \\ {[dB]}\end{tabular} 	& 
		\begin{tabular}[c]{@{}c@{}}No \\ Proc.\end{tabular} 	& 
		\begin{tabular}[c]{@{}c@{}} IPSF \end{tabular} 			& 
		\begin{tabular}[c]{@{}c@{}}LSTM6\end{tabular} 		& 
		\begin{tabular}[c]{@{}c@{}}LSTM5\end{tabular} 		&   
		\begin{tabular}[c]{@{}c@{}}No \\ Proc.\end{tabular} 	& 
		\begin{tabular}[c]{@{}c@{}} IPSF \end{tabular} 			& 
		\begin{tabular}[c]{@{}c@{}}LSTM7\end{tabular} 		& 
		\begin{tabular}[c]{@{}c@{}}LSTM5\end{tabular} \\ 
		\midrule
					-5 & 0.20 & 0.66 & 0.07 & 0.06 & 0.14 & 0.69 & 0.02 & 0.02 \\ 
			0 & 0.30 & 0.59 & 0.16 & 0.16 & 0.22 & 0.65 & 0.08 & 0.09 \\ 
			5 & 0.39 & 0.52 & 0.20 & 0.20 & 0.29 & 0.60 & 0.13 & 0.14 \\ 
			20 & 0.54 & 0.40 & 0.17 & 0.19 & 0.38 & 0.53 & 0.14 & 0.15 \\ \midrule
			Avg. & 0.36 & 0.54 & 0.15 & 0.15 & 0.26 & 0.62 & 0.09 & 0.10 \\ \midrule \midrule
	\end{tabular}}
\end{table}
\begin{table}
	\caption{ESTOI improvements for LSTM5 tested on BUS and PED.}
	\vspace{-2mm}
	\label{tab:stoi_ped}
	\centering
	\setlength\tabcolsep{5pt} 
	\resizebox{0.7\columnwidth}{!}{%
		\begin{tabular}{ccccccccccccc}
			\midrule \midrule
			& \multicolumn{6}{c} {2-Speaker } & \multicolumn{6}{c} {3-Speaker} \\ \cmidrule(l){2-7} \cmidrule(l){8-13}
			\begin{tabular}[c]{@{}c@{}}SNR \\ {[dB]}\end{tabular} 	& 
			\multicolumn{2}{c} {\begin{tabular}[c]{@{}c@{}}No \\ Proc.\end{tabular} }	& 
			\multicolumn{2}{c} {\begin{tabular}[c]{@{}c@{}} IPSF \end{tabular} 	}		& 
			\multicolumn{2}{c} {\begin{tabular}[c]{@{}c@{}}LSTM5\end{tabular} 	}	&   
			\multicolumn{2}{c} {\begin{tabular}[c]{@{}c@{}}No \\ Proc.\end{tabular}} 	& 
			\multicolumn{2}{c} {\begin{tabular}[c]{@{}c@{}} IPSF \end{tabular} 		}	& 
			\multicolumn{2}{c} {\begin{tabular}[c]{@{}c@{}}LSTM5\end{tabular}} \\ \cmidrule(l){2-7} \cmidrule(l){8-13}
			& BUS & PED & BUS & PED & BUS & PED & BUS & PED & BUS & PED & BUS & PED \\
			\midrule
						-5 & 0.32 & 0.18 & 0.55 & 0.64 & 0.14 & 0.08 & 0.24 & 0.14 & 0.61 & 0.68 & 0.08 & 0.03 \\ 
			0 & 0.39 & 0.28 & 0.50 & 0.58 & 0.18 & 0.15 & 0.28 & 0.21 & 0.58 & 0.63 & 0.12 & 0.09 \\ 
			5 & 0.45 & 0.37 & 0.46 & 0.52 & 0.20 & 0.20 & 0.32 & 0.28 & 0.56 & 0.59 & 0.14 & 0.13 \\ 
			20 & 0.55 & 0.54 & 0.39 & 0.40 & 0.18 & 0.18 & 0.38 & 0.37 & 0.53 & 0.53 & 0.15 & 0.15 \\ \midrule
			Avg. & 0.43 & 0.34 & 0.47 & 0.54 & 0.17 & 0.15 & 0.31 & 0.25 & 0.57 & 0.61 & 0.12 & 0.10 \\ \midrule \midrule
	\end{tabular}}
\end{table}

\pagebreak
\section{Conclusion}\label{sec:conclusion2}
In this paper we have proposed utterance-level Permutation Invariant Training\;(uPIT) for speaker independent multi-talker speech separation and denoising.
Differently from prior works, that focus only on the ideal noise-free setting, we focus on the more realistic scenario of speech separation in noisy environments.  
Specifically, using the uPIT technique we have trained bi-directional Long Short-Term Memory\;(LSTM) Recurrent Neural Networks (RNNs), to separate two and three-speaker mixtures corrupted by multiple noise types at a wide range of Signal to Noise Ratios\;(SNRs).       

We show that bi-directional LSTM RNNs trained with uPIT are capable of improving both Signal to Distortion Ratio\;(SDR), as well as the Extended Short-Time Objective Intelligibility\;(ESTOI) measure for challenging noise types and SNRs. 
Specifically, we show that LSTM RNNs achieve large SDR and ESTOI improvements, when evaluated using noise types seen during training, and that a single model is capable of handling multiple noise types with only a slight decrease in performance. 
Furthermore, we show that a single LSTM RNN can handle both two-speaker and three-speaker noisy mixtures, without \emph{a priori} knowledge about the exact number of speakers. 
Finally, we show that LSTM RNNs trained using uPIT generalizes well to unknown noise types.

{\small\bibliographystyle{bib/IEEEtran}\bibliography{bib/mybibE}}


  \cleardoublepage
  \setcounter{enumiii}{0}
  \setcounter{enumii}{0}
  \setcounter{enumiv}{0}
  \setcounter{enumi}{0}
  \setcounter{equation}{0}
  \setcounter{figure}{0}
  \setcounter{footnote}{0}
  \setcounter{mpfootnote}{0}
  \setcounter{paragraph}{0}
  \setcounter{parentequation}{0}
  \setcounter{part}{0}
  \setcounter{section}{0}
  \setcounter{subparagraph}{0}
  \setcounter{subsection}{0}
  \setcounter{subsubsection}{0}
  \setcounter{table}{0}
  \papertitlepage{%
  Monaural Speech Enhancement Using Deep Neural Networks by Maximizing a Short-Time Objective Intelligibility Measure
}{paper:paperF}{%
  Morten Kolbæk, Zheng-Hua Tan, and Jesper Jensen
}{%
  The paper has been published in\\
  \textit{Proceedings IEEE International Conference on Acoustics, Speech, and Signal Processing}, pp. 5059-5063, April 2018.
}{%
  \noindent\copyright\ 2018 IEEE
}

\acresetall
\begin{abstract}
		
In this paper we propose a  Deep Neural Network\,(DNN) based \ac{SE} system that is designed to maximize an approximation of the Short-Time Objective Intelligibility\,(STOI) measure. 
We formalize an approximate-STOI cost function and derive analytical expressions for the gradients required for DNN training and show that these gradients have desirable properties when used together with gradient based optimization techniques.

We show through simulation experiments that the proposed SE system achieves large improvements in estimated speech intelligibility, when tested on matched and unmatched natural noise types, at multiple signal-to-noise ratios. 
Furthermore, we show that the SE system, when trained using an approximate-STOI cost function performs on par with a system trained with a mean squared error cost applied to short-time temporal envelopes.
Finally, we show that the proposed SE system performs on par with a traditional DNN based \acf{STSA} SE system in terms of estimated speech intelligibility.          
These results are important because they suggest that traditional DNN based STSA SE systems might be optimal in terms of estimated speech intelligibility.

\end{abstract}
\section{Introduction}
\label{sec:intro6}

Design and development of Speech Enhancement\,(SE) algorithms capable of improving speech quality and intelligibility has been a long-lasting goal in both academia and industry \cite{hendriks_dft-domain_2013,loizou_speech_2013}. 
Such algorithms are useful for a wide range of applications e.g. for mobile communications devices and hearing assistive devices \cite{hendriks_dft-domain_2013}.

Despite a large research effort for more than 30 years \cite{ephraim_speech_1984,loizou_speech_2013,hendriks_dft-domain_2013} modern single-microphone SE algorithms still perform unsatisfactorily in the complex acoustic environments, which users of e.g. hearing assistive devices are exposed to on a daily basis, e.g. traffic noise, cafeteria noise, or competing speakers.

Traditionally, SE algorithms have been divided into at least two groups; statistical-model based techniques and data-driven techniques.   
The first\linebreak group encompasses techniques such as spectral subtraction, the Wiener filter and the short-time spectral amplitude minimum mean squared error estimator \cite{ephraim_speech_1984,hendriks_dft-domain_2013,loizou_speech_2013}. 
These techniques make statistical assumptions about the probability distributions of the speech and noise signals, that enable them to suppress the noise dominated time-frequency regions of the noisy speech signal. 
In particularly, for stationary noise types this type of algorithms may perform well in terms of speech quality, but in general these techniques do not improve speech  intelligibility \cite{hu_comparative_2007,luts_multicenter_2010,jensen_spectral_2012}.         
The second group encompasses data-driven or machine learning techniques e.g. based on non-negative matrix factorization \cite{grais_single_2011}, support vector machines \cite{wang_towards_2013}, and Deep Neural Networks\,(DNNs) \cite{xu_regression_2015,healy_algorithm_2015}. 
These techniques make no statistical assumptions. Instead, they learn to suppress noise by observing a large number of representative pairs of noisy and noise-free speech signals in a supervised learning process. 
SE algorithms based on DNNs can, to some extent, improve speech intelligibility for hearing impaired and normal hearing people, in noisy conditions, if sufficient  \emph{a priori} knowledge is available e.g. the identity of the speaker or the noise type. \cite{chen_large-scale_2016,healy_algorithm_2017,kolbaek_speech_2017}.

Although the techniques mentioned above are fundamentally different, they typically share at least two common properties. First, they often aim to minimize a \acf{MSE} cost function, and secondly, they operate on short frames ($\approx$ 20 -- 30 ms ) in the Short-Time discrete Fourier Transform\,(STFT) domain \cite{hendriks_dft-domain_2013,loizou_speech_2013}. 
However, it is well known \cite{loizou_speech_2013,moore_introduction_2013} that the human auditory system has a non-linear frequency sensitivity, which is often approximated using e.g. a Gammatone or a one-third octave filter bank \cite{loizou_speech_2013}.
Furthermore, it is known that preservation of modulation frequencies below 7 Hz is critical for speech intelligibility \cite{elliott_modulation_2009,moore_introduction_2013}.      
This suggests that SE algorithms aimed at the human auditory system could benefit by incorporating such information.  
Numerous works exist, e.g. \cite{hu_perceptually_2003,ephraim_speech_1985,virag_single_1999,loizou_speech_2005,lightburn_sobm_2015,han_perceptual_2016,shivakumar_perception_2016,koizumi_dnn-based_2017,healy_algorithm_2015} and \cite[Sec. 2.2.3]{hendriks_dft-domain_2013} and the references therein, where SE algorithms have been designed with perceptual aspects in mind. 
However, although these algorithms do take some perceptual aspects into account, they do not directly optimize for speech intelligibility.

In this paper we propose an SE system that maximizes an objective speech intelligibility estimator. Specifically, we design a DNN based SE system that maximizes an approximation of the Short-Time Objective Intelligibility\,(STOI) \cite{taal_algorithm_2011} measure. The STOI measure has been found to be highly correlated with intelligibility as measured in human listening tests \cite{taal_algorithm_2011,loizou_speech_2013}.   
We derive analytical expressions for the required gradients used for the DNN weight updates during training and use these closed-form expressions to identify desirable properties of the approximate-STOI cost function. 
Finally, we study the potential performance gain between the proposed approximate-STOI cost function with a classical MSE cost function. 
We note that our goal is not to achieve state-of-the-art STOI improvements per se, but rather to study and compare the proposed approximate-STOI based SE system to existing DNN based enhancement schemes. 
Further improvement may straightforwardly be achieved with larger datasets and complex models like long short-term memory recurrent, or convolutional, neural networks \cite{goodfellow_deep_2016}.

\section{Speech Enhancement System}\label{sec:DNNSE}
In the following we introduce the approximate-STOI measure and we present the DNN framework used to maximize it. Finally, we discuss techniques used to reconstruct the enhanced and approximate-STOI optimal speech signal in the time-domain.

\subsection{Approximating Short-Time Objective Intelligibility}\label{sec:stoiApprox}
Let $x[n]$ be the $n^{th}$ sample of the clean time-domain speech signal and let a noisy observation $y[n]$ be defined as
\begin{equation}
	y[n] = x[n] + z[n], 
	\label{eq61}
\end{equation}
where $z[n]$ is an additive noise sample.   
Furthermore, let ${x}(k,m)$ and ${y}(k,m)$,  $k = 1,2,\dots, \frac{K}{2}+1$,  $m=1,2,\dots M, $ be the single-sided magnitude spectra of the $K$-point Short-Time discrete Fourier Transforms (STFT) of $x[n]$ and $y[n]$, respectively, where $M$ is the number of STFT frames.  
Also, let $\hat{x}(k,m)$ be an estimate of $x(k,m)$ obtained as $\hat{x}(k,m) = \hat{g}(k,m)y(k,m)$ where $\hat{g}(k,m)$ is an estimated gain value.
In this study we use a 10 kHz sample frequency and a 256 point STFT, i.e. $K=256$, with a Hann-window size of 256 samples (25.6 ms) and a 128 sample frame shift (12.8 ms).
Similarly to STOI \cite{taal_algorithm_2011}, we define a short-time temporal envelope vector of 
the $j^{th}$ one-third octave band for the clean speech signal as 
\begin{equation}
	\mathbf{x}_{j,m} = [ X_j( m-N+1 ), \; X_j( m-N+2 ), \dots , X_j( m ) ]^T,
\end{equation}
where 
\begin{equation}
	X_j( m ) = \sqrt{\sum_{k=k_1(j)}^{k_2(j)} x(k,m)^2},
	\label{eq62}
\end{equation}
and $k_1(j)$ and $k_2(j)$ denote the first and last STFT bin index of the $j^{th}$ one-third octave band, respectively. 
Similarly, we define $\mathbf{y}_{j,m}$ and $Y_j( m )$ for the noisy observation. 
Also, let $\hat{\mathbf{x}}_{j,m} = diag(\hat{\mathbf{g}}_{j,m})\mathbf{y}_{j,m}$ be the short-time temporal one-third octave band envelope vector of the enhanced speech signal, where $\hat{\mathbf{g}}_{j,m}$ is a gain vector defined in the $j^{th}$ one-third octave band and $diag(\hat{\mathbf{g}}_{j,m})$ is a diagonal matrix with the elements of $\hat{\mathbf{g}}_{j,m}$ on the main diagonal. 
We use $N=30$ such that the short-time temporal one-third octave band envelope vectors will span a duration of 384 ms, which ensures that important modulation frequencies are captured \cite{taal_algorithm_2011}.    
In total, $J=15$ one-third octave bands are used with the first band having a center frequency of 150 Hz and the last one of approximately 3.8 kHz. These frequencies are chosen such that they span the frequency range in which human speech normally lie \cite{taal_algorithm_2011}. 
For mathematical tractability, we discard the clipping step%
\footnote{It has been observed empirically, that omitting the clipping step most often does not affect the performance of STOI, e.g. \cite{lightburn_sobm_2015,jensen_algorithm_2016,andersen_predicting_2016,taal_matching_2012}.}%
, otherwise performed by STOI \cite{taal_algorithm_2011}, and define the approximated STOI measure as  
\begin{equation}
\mathcal{L} ( \mathbf{x}_{j,m},\hat{\mathbf{x}}_{j,m}) = \frac{\left({\mathbf{x}}_{j,m} - \mu_{{\mathbf{x}}_{j,m}}\right)^T  \left(\hat{\mathbf{x}}_{j,m} - \mu_{\hat{\mathbf{x}}_{j,m}}\right)}{ \left\lVert {\mathbf{x}}_{j,m} - \mu_{{\mathbf{x}}_{j,m}} \right\rVert  \; \left\lVert\hat{\mathbf{x}}_{j,m} - \mu_{\hat{\mathbf{x}}_{j,m}} \right\rVert },
\label{eq:stoicost}
\end{equation}
where $\left\lVert \cdot \right\rVert$ is the euclidean $\ell^2$-norm and $\mu_{{\mathbf{x}}_{j,m}}$ and $\mu_{\hat{\mathbf{x}}_{j,m}}$ are the sample means of $\mathbf{x}_{j,m}$ and $\hat{\mathbf{x}}_{j,m}$, respectively. 
Obviously, $\mathcal{L} ( \mathbf{x}_{j,m},\hat{\mathbf{x}}_{j,m})$ is simply the \ac{ELC}  between the vectors $\mathbf{x}_{j,m}$ and $\hat{\mathbf{x}}_{j,m}$.

\subsection{Maximizing Approximated STOI Using DNNs}\label{sec:stoiMax}
The approximated STOI measure given by Eq.\;\eqref{eq:stoicost} is defined in a one-third octave band domain and our goal is to find $\hat{\mathbf{x}}_{j,m} = diag(\hat{\mathbf{g}}_{j,m})\mathbf{y}_{j,m}$ such that Eq.\;\eqref{eq:stoicost} is maximized, i.e. finding an optimal gain vector $\hat{\mathbf{g}}_{j,m}$.   
In this study we estimate these optimal gains using DNNs. 
Specifically, we use Eq.\;\eqref{eq:stoicost} as a cost function and train multiple feed-forward DNNs, one for each one-third octave band, to estimate gain vectors $\hat{\mathbf{g}}_{j,m}$, such that the approximated STOI measure is maximized. For the remainder of this paragraph we omit the subscripts $j$ and $m$ for convenience.

Most modern deep learning toolkits, e.g. \ac{CNTK} \cite{agarwal_introduction_2014}, perform automatic differentiation, which allow one to train a DNN with a custom cost function, without the need of computing the gradients of the cost function explicitly \cite{goodfellow_deep_2016}. Nevertheless, when working with cost functions that have not yet been exhaustively studied, such as the approximated STOI measure, an analytic expression of the gradient can be valuable for studying important properties, such as gradient $\ell^2$-norm. 
It can be shown (details omitted due to space limitations) that the gradient of Eq.\;\eqref{eq:stoicost}, with respect to the desired signal vector $\hat{\mathbf{x}}$, is given by 
\begin{equation}
\nabla \mathcal{L} ( \mathbf{x},\hat{\mathbf{x}}) = \left[ \frac{\partial \mathcal{L} ( \mathbf{x},\hat{\mathbf{x}})}{\partial \hat{{x}}_{1}},
\frac{\partial \mathcal{L} ( \mathbf{x},\hat{\mathbf{x}})}{\partial \hat{{x}}_{2}}, \dots,
\frac{\partial \mathcal{L} ( \mathbf{x},\hat{\mathbf{x}})}{\partial \hat{{x}}_{N}}
\right]^T,
\label{eq:stoigrad}
\end{equation}
where
\begin{flalign}
\frac{\partial \mathcal{L} ( \mathbf{x},\hat{\mathbf{x}})}{\partial \hat{x}_{m}} = 
\frac{\mathcal{L} ( \mathbf{x},\hat{\mathbf{x}})  \left(x_m - \mu_{\mathbf{x}}\right)}{\left(\hat{\mathbf{x}} - \mu_{\hat{\mathbf{x}}}  \right)^T \left( \mathbf{x}  - \mu_{\mathbf{x}} \right)} -
\frac{\mathcal{L} ( \mathbf{x},\hat{\mathbf{x}}) \left(\hat{x}_m - \mu_{\hat{\mathbf{x}}}\right)}{\left(\hat{\mathbf{x}} - \mu_{\hat{\mathbf{x}}}  \right)^T\left(\hat{\mathbf{x}} - \mu_{\hat{\mathbf{x}}}  \right)}  ,
\label{eq:stoipart}
\end{flalign}
is the partial derivative of $\mathcal{L} ( \mathbf{x},\hat{\mathbf{x}})$ with respect to entry $m$ of $\hat{\mathbf{x}}$.   

Furthermore, it can be shown that the $\ell^2$-norm of the gradient as formulated by Eqs.\;\eqref{eq:stoigrad} and \eqref{eq:stoipart}, is given by
\begin{equation}
\left\lVert \nabla \mathcal{L} ( \mathbf{x},\hat{\mathbf{x}}) \right\rVert =  \sqrt{1-\mathcal{L} ( \mathbf{x},\hat{\mathbf{x}})^2} \left\lVert \hat{\mathbf{x}}  \right\rVert ^{-1},
\label{eq:stoigradnorm}
\end{equation}
which is shown in Fig.\;\ref{fig:stoigrad} as function of $\mathcal{L} ( \mathbf{x},\hat{\mathbf{x}})$ for the complete range $[-1,1]$, and for $\left\lVert \hat{\mathbf{x}}  \right\rVert = 1$. 
%
\begin{figure}[] 
	\centering
	\centerline{\includegraphics[width=0.8\linewidth]{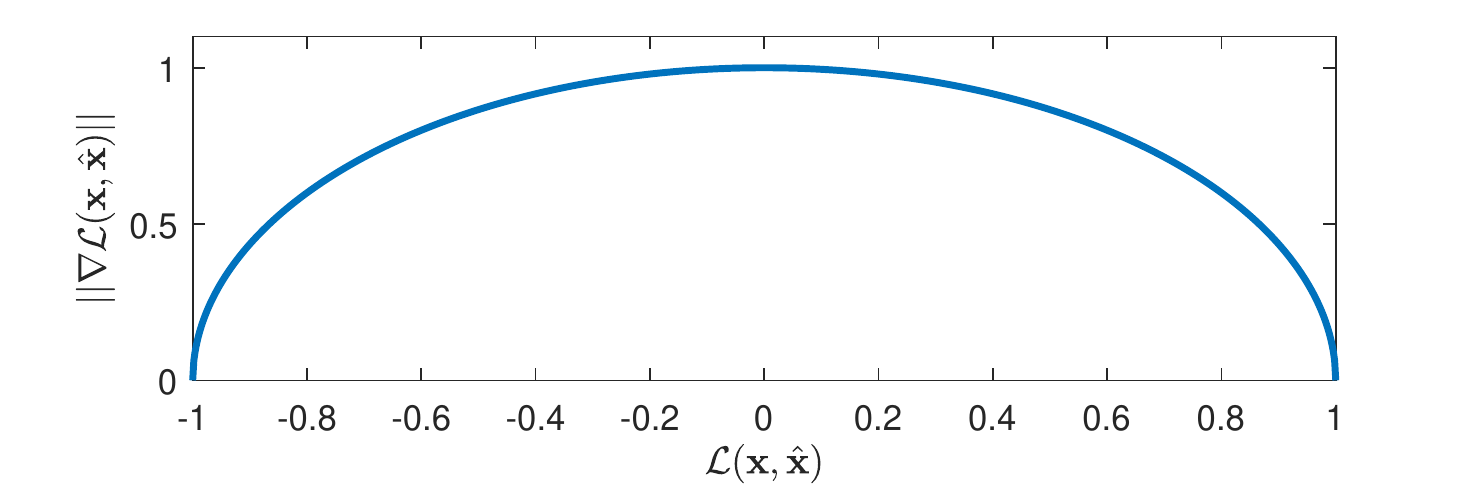}}
	\vspace{-3mm}
	\caption{$\ell^2$-norm of Eq.\;\eqref{eq:stoigrad} as function of cost function value.}
	\label{fig:stoigrad}
	\vspace{-5mm}
\end{figure}
We see from Fig.\,\ref{fig:stoigrad} that the $\ell^2$-norm of $\mathcal{L} ( \mathbf{x},\hat{\mathbf{x}})$ is a concave function with a global maximum at $\mathcal{L} ( \mathbf{x},\hat{\mathbf{x}}) = 0$ and is symmetric around zero. We also observe that $ \left\lVert\nabla \mathcal{L} ( \mathbf{x},\hat{\mathbf{x}}) \right\rVert$ is monotonically decreasing when $ \mathcal{L} (\mathbf{x},\hat{\mathbf{x}}) < 0$ and $ \mathcal{L} ( \mathbf{x},\hat{\mathbf{x}}) > 0$ with $ \left\lVert\nabla \mathcal{L} ( \mathbf{x},\hat{\mathbf{x}}) \right\rVert = 0$ when $\mathbf{x}$ and $\hat{\mathbf{x}}$ are either perfectly correlated or perfectly anti-correlated. 
Since $ \left\lVert\nabla \mathcal{L} ( \mathbf{x},\hat{\mathbf{x}}) \right\rVert$ is large when $\mathbf{x}$ and $\hat{\mathbf{x}}$ are uncorrelated and zero when perfectly correlated, and $ \left\lVert\nabla \mathcal{L} ( \mathbf{x},\hat{\mathbf{x}}) \right\rVert \neq 0$ otherwise, Eq.\;\eqref{eq:stoicost} is well suited as a cost function for gradient-based optimization techniques, such as Stochastic Gradient Descent\,(SGD) \cite{goodfellow_deep_2016}, since it guarantees non-zero step lengths for all inputs during optimization except at the optimal solution. In practice, to apply SGD we minimize $- \mathcal{L} ( \mathbf{x},\hat{\mathbf{x}})$.

\subsection{Reconstructing Approximate-STOI Optimal Speech}
When a gain vector $\hat{\mathbf{g}}_{j,m}$ has been estimated by a DNN, the enhanced speech envelope in the one-third octave band domain can be computed as $\hat{\mathbf{x}}_{j,m} = diag(\hat{\mathbf{g}}_{j,m})\mathbf{y}_{j,m}$. However, what we are really interested in is $\hat{x}(k,m)$, i.e. the estimated speech signal in the STFT domain, since $\hat{x}(k,m)$ can straightforwardly be transformed into the time-domain using the overlap-and-add technique \cite{loizou_speech_2013}.   
We therefore seek a mapping from the gain vector $\hat{\mathbf{g}}_{j,m}$ estimated in the one-third octave band domain, to the gain $\hat{g}(k,m)$, for a single STFT coefficient. 
To do so, let $\hat{g}_j(m)$ denote the gain value estimated by a DNN to be applied to the noisy one-third octave band amplitude in frame $m$.
We can then derive the relationship between the gain value $\hat{g}_j(m) \geq 0 $ in the one-third octave band, and the corresponding gain values $\hat{g}(k,m) \geq 0$ in the STFT domain as  
\begin{equation}
\hat{X}_j( m ) = \hat{g}_j(m) Y_j( m ) = \sqrt{\sum_{k=k_1(j)}^{k_2(j)-1} \left(\hat{g}(k,m)y(k,m) \right)^2}. 
\label{eq63} 
\end{equation}
One solution to Eq\;\eqref{eq63} is
\begin{equation}
\hat{g}_j(m) = \hat{g}(k,m) , \; k=k_1(j), \dots k_2(j)-1.
\label{eq64}
\end{equation}
Generally, the solution in Eq.\,\eqref{eq64}  is not unique; many choices of $\hat{g}(k,m)$ exist that give rise to the same estimated one-third octave band $\hat{X}_j( m )$ (and hence the same value of $\mathcal{L} ( \mathbf{x},\hat{\mathbf{x}})$). We choose, for convenience, a uniform gain across the STFT coefficients within a one-third octave band.
Since envelope estimates $\hat{X}_j( m )$ are computed for successive values of $m$, N estimates exist for each $\hat{X}_j( m )$, which are averaged during enhancement.     
When reconstructing the enhanced speech signal in the time domain, we use the overlap-and-add technique using the phase of the noisy STFT coefficients \cite{loizou_speech_2013}.

\section{Experimental Design}
\label{sec:expDesign}
To evaluate the performance of the approximate-STOI optimal DNN based SE system we have conducted series of experiments involving multiple matched and unmatched noise types at various SNRs.

\subsection{Noisy Speech Mixtures}
The clean speech signals used for training all models are from the \acf{WSJ0} corpus \cite{garofolo_csr-i_1993}. 
The utterances used for training and validation are generated by randomly selecting utterances from 44 male and 47 female speakers from the WSJ0 training set entitled si\_tr\_s. The training and validation sets consist of 20000 and 2000 utterances, respectively, which is equivalent to approximately 37 hours of training data and 4 hours of validation data.   
The test set is similarly generated using utterances from 16 speakers from the WSJ0 validation set si\_dt\_05 and evaluation set si\_et\_05, and consists of 1000 mixtures or approximately 2 hours of data, see \cite{kolbaek_supplemental_nodate} for further details. 
Notice, the speakers in the test set are different from the speakers in the validation and training sets. 
We use six different noise types: two synthetic signals and four noise signals recorded in real-life.  
The synthetic noise signals encompass a stationary Speech Shaped Noise\;(SSN) signal and a highly non-stationary 6-speaker Babble\;(BBL) noise. For real-life noise signals we use the street\;(STR), cafeteria\;(CAF), bus\;(BUS), and pedestrian\;(PED) noise signals from the CHiME3 dataset \cite{barker_third_2015}.    
The SSN noise signal is Gaussian white noise, shaped according to the long-term spectrum of the TIMIT corpus \cite{garofolo_darpa_1993}. Similarly, the BBL noise signal is constructed by mixing utterances from TIMIT. 
Further details on the design of the SSN and BBL noise signals can be found in \cite{kolbaek_speech_2017}. All noise signals are split into non-overlapping sequences with a 40 min.\;training sequence, a 5 min.\;validation sequence and a 5 min.\;test sequence, i.e. there is no overlap between the noise sequences used for training, validation and test.

\pagebreak
The noisy speech signals used for training and testing are constructed using Eq.\;\eqref{eq61}, where a clean speech signal $x[n]$ is added to a noise sequence $z[n]$ of equal length. To achieve a certain SNR, the noise signal is scaled based on the active speech level of the clean speech signal as per ITU P.56 \cite{itu_rec._1993}.  
The SNRs used for the training and validation sets are chosen uniformly from $[-5 , 10 ]$ dB. The SNR range is chosen to ensure that SNRs are included where intelligibility ranges from degraded to perfectly intelligible.  

\subsection{Model Architecture and Training}
To evaluate the performance of the proposed SE system a total of ten systems, identified as S0 -- S9, have been trained using different cost functions and noise types as presented in Table\;\ref{tab:models1}. 
\begin{table}
	\caption{Training conditions for different SE systems.}
	\label{tab:models1}
	\centering
	\setlength\tabcolsep{4pt} 
	\resizebox{0.75\columnwidth}{!}{
		\begin{tabular}{lcccccccccc}
			\midrule \midrule
			ID:    & S0  & S1  & S2  & S3  & S4  & S5  & S6  & S7  &S8   & S9   \\ 
			Cost:  & ELC & ELC & ELC & ELC & ELC & EMSE & EMSE & EMSE & EMSE & EMSE\\
			Noise: & SSN & BBL & CAF & STR & ALL & SSN & BBL & CAF & STR & ALL \\ 
			\midrule \midrule 
	\end{tabular}}
\end{table}
Five systems (S0--S4) have been trained using the ELC loss from Eq.\;\eqref{eq:stoicost} and five systems (S5--S9) have been trained using a standard MSE loss, denoted as \ac{EMSE}, since it operates on short-time temporal one-third octave band envelope vectors.
This is to investigate the potential performance difference between models trained with an approximate-STOI loss and models trained with the commonly used MSE loss.
Eight systems (S0--S3 and S5--S8) are trained as noise type specific systems, i.e. they are trained using only a single noise type. Two systems (S4 and S9) are trained as noise type general systems, i.e. they are trained on all noise types (Noise: "ALL" in Table\;\ref{tab:models1}). This is to investigate the performance drop, if any, when a single system is trained to handle multiple noise types.

Each DNN consists of three hidden layers with 512 units with ReLU activation functions and a sigmoid output layer. The DNNs are trained using SGD with the backpropagation technique and batch normalization \cite{goodfellow_deep_2016}. 
The DNNs are trained for a maximum of 200 epochs with a minibatch size of 256 randomly selected short-time temporal one-third octave band envelope vectors and the learning rates were set to $0.01$, and $5 \cdot 10^{-5}$ per sample initially, for S0--S4, and S5--S9, respectively. 
The learning rates were scaled down by $0.7$ when the training cost increased on the validation set. The training was terminated when the learning rate was below $10^{-10}$. The different learning rates for the systems trained with the ELC cost function and the systems trained with the EMSE cost functions were found from preliminary experiments.  
All models were implemented using CNTK \cite{agarwal_introduction_2014} and the script files needed to reproduce the reported results can be found in \cite{kolbaek_supplemental_nodate}. 

\section{Experimental Results}
\label{sec:expResults}
We have evaluated the performance of the ten systems based on their average ELC and STOI scores computed on the test set. 
The STOI score is computed using the enhanced and reconstructed time-domain speech signal, whereas the ELC score is computed using short-time one-third octave band temporal envelope vectors. 

\begin{table}[t]
	\caption{ELC results for S0 -- S9 tested with SSN, BBL, CAF, and STR}
	\label{tab:lincorr_combined}
	\centering
	\setlength\tabcolsep{5pt} 
	\resizebox{0.8\columnwidth}{!}{%
		\begin{tabular}{ccccccccccc}
			\midrule  \midrule 
			 & \multicolumn{5}{c} {SSN} & \multicolumn{5}{c} {BBL}  \\ \cmidrule(l){2-6} \cmidrule(l){7-11} 
			\begin{tabular}[c]{@{}c@{}}SNR \\ {[dB]}\end{tabular} 	& 
			UP.						&
			\begin{tabular}[c]{@{}c@{}}S0 \\ {\scriptsize (ELC)}\end{tabular} 		&
			\begin{tabular}[c]{@{}c@{}}S5 \\ {\scriptsize (EMSE)}\end{tabular} 		&
			\begin{tabular}[c]{@{}c@{}}S4 \\ {\scriptsize (ELC)}\end{tabular} 		&
			\begin{tabular}[c]{@{}c@{}}S9 \\ {\scriptsize (EMSE)}\end{tabular}  		&
			UP.						&
			\begin{tabular}[c]{@{}c@{}}S1 \\ {\scriptsize (ELC)}\end{tabular} 		&
			\begin{tabular}[c]{@{}c@{}}S6 \\ {\scriptsize (EMSE)}\end{tabular} 		&
			\begin{tabular}[c]{@{}c@{}}S4 \\ {\scriptsize (ELC)}\end{tabular} 		&
			\begin{tabular}[c]{@{}c@{}}S9 \\ {\scriptsize (EMSE)}\end{tabular}  			\\ \midrule
						-5 & 0.36 & 0.66 & 0.65 & 0.64 & 0.63 		& 0.34 & 0.50 & 0.51 & 0.48 & 0.48		\\ 
			0 & 0.52 & 0.77 & 0.76 & 0.75 & 0.74 		& 0.50 & 0.69 & 0.69 & 0.67 & 0.67		\\ 
			5 & 0.66 & 0.82 & 0.81 & 0.80 & 0.79 		& 0.64 & 0.78 & 0.77 & 0.77 & 0.77		\\ 
			Avg. & 0.51 & 0.75 & 0.74 & 0.73 & 0.72 	& 0.49 & 0.66 & 0.66 & 0.64 & 0.64		\\ 
			\midrule
		\end{tabular}}
	\begin{tabular}{c}
	\end{tabular}
\end{table}
\begin{table}[t]
	\centering
	\setlength\tabcolsep{5pt} 
	\resizebox{0.8\columnwidth}{!}{%
		\begin{tabular}{ccccccccccc}
			& \multicolumn{5}{c} {CAF} & \multicolumn{5}{c} {STR} \\ \cmidrule(l){2-6} \cmidrule(l){7-11}  
			\begin{tabular}[c]{@{}c@{}}SNR \\ {[dB]}\end{tabular} 	& 
			UP. 						&
			\begin{tabular}[c]{@{}c@{}}S2 \\ {\scriptsize (ELC)}\end{tabular} 		&
			\begin{tabular}[c]{@{}c@{}}S7 \\ {\scriptsize (EMSE)}\end{tabular} 		&
			\begin{tabular}[c]{@{}c@{}}S4 \\ {\scriptsize (ELC)}\end{tabular} 		&
			\begin{tabular}[c]{@{}c@{}}S9 \\ {\scriptsize (EMSE)}\end{tabular}  		&
			UP.						&
			\begin{tabular}[c]{@{}c@{}}S3 \\ {\scriptsize (ELC)}\end{tabular} 		&
			\begin{tabular}[c]{@{}c@{}}S8 \\ {\scriptsize (EMSE)}\end{tabular} 		&
			\begin{tabular}[c]{@{}c@{}}S4 \\ {\scriptsize (ELC)}\end{tabular} 		&
			\begin{tabular}[c]{@{}c@{}}S9 \\ {\scriptsize (EMSE)}\end{tabular}  		\\
			\midrule
						-5 		& 0.43 & 0.61 & 0.59 & 0.58 & 0.58		& 0.45 & 0.70 & 0.68 & 0.68 & 0.66\\ 
			0 		& 0.57 & 0.73 & 0.71 & 0.72 & 0.70		& 0.58 & 0.78 & 0.76 & 0.77 & 0.75\\ 
			5 		& 0.68 & 0.79 & 0.78 & 0.79 & 0.77		& 0.69 & 0.82 & 0.80 & 0.81 & 0.79\\ 
			Avg. 	& 0.56 & 0.71 & 0.69 & 0.70 & 0.68		& 0.57 & 0.77 & 0.75 & 0.75 & 0.73\\ 
			\midrule  \midrule
	\end{tabular}}
\end{table}
\begin{table}[t]
	\caption{STOI results for S0 -- S9 tested with SSN, BBL, CAF, and STR}
	\label{tab:stoi_combined}
	\centering
	\setlength\tabcolsep{5pt} 
	\resizebox{0.8\columnwidth}{!}{%
		\begin{tabular}{ccccccccccc}
			\midrule \midrule
			& \multicolumn{5}{c} {SSN} & \multicolumn{5}{c} {BBL}  \\ \cmidrule(l){2-6} \cmidrule(l){7-11}  
			\begin{tabular}[c]{@{}c@{}}SNR \\ {[dB]}\end{tabular} 	& 
			UP. 						&
			\begin{tabular}[c]{@{}c@{}}S0 \\ {\scriptsize (ELC)}\end{tabular} 		&
			\begin{tabular}[c]{@{}c@{}}S5 \\ {\scriptsize (EMSE)}\end{tabular} 		&
			\begin{tabular}[c]{@{}c@{}}S4 \\ {\scriptsize (ELC)}\end{tabular} 		&
			\begin{tabular}[c]{@{}c@{}}S9 \\ {\scriptsize (EMSE)}\end{tabular}  		&
			UP. 						&
			\begin{tabular}[c]{@{}c@{}}S1 \\ {\scriptsize (ELC)}\end{tabular} 		&
			\begin{tabular}[c]{@{}c@{}}S6 \\ {\scriptsize (EMSE)}\end{tabular} 		&
			\begin{tabular}[c]{@{}c@{}}S4 \\ {\scriptsize (ELC)}\end{tabular} 		&
			\begin{tabular}[c]{@{}c@{}}S9 \\ {\scriptsize (EMSE)}\end{tabular}  		\\ 
			\midrule
						-5 & 0.61 & 0.78 & 0.78 & 0.76 & 0.76 			& 0.59 & 0.66 & 0.67 & 0.65 & 0.65		\\ 
			0 & 0.74 & 0.88 & 0.88 & 0.87 & 0.87 			& 0.72 & 0.82 & 0.82 & 0.81 & 0.81		\\ 
			5 & 0.85 & 0.93 & 0.93 & 0.92 & 0.92 			& 0.83 & 0.90 & 0.90 & 0.89 & 0.90		\\ 
			Avg. & 0.73 & 0.86 & 0.86 & 0.85 & 0.85 		& 0.71 & 0.79 & 0.80 & 0.78 & 0.79		\\ 
			\midrule
	\end{tabular}}
\end{table}
\begin{table}[t]
	\centering
	\setlength\tabcolsep{5pt} 
	\resizebox{0.8\columnwidth}{!}{%
		\begin{tabular}{ccccccccccc}
			& \multicolumn{5}{c} {CAF} & \multicolumn{5}{c} {STR} \\ \cmidrule(l){2-6} \cmidrule(l){7-11}  
			\begin{tabular}[c]{@{}c@{}}SNR \\ {[dB]}\end{tabular} 	& 
			UP. 						&
			\begin{tabular}[c]{@{}c@{}}S2 \\ {\scriptsize (ELC)}\end{tabular} 		&
			\begin{tabular}[c]{@{}c@{}}S7 \\ {\scriptsize (EMSE)}\end{tabular} 		&
			\begin{tabular}[c]{@{}c@{}}S4 \\ {\scriptsize (ELC)}\end{tabular} 		&
			\begin{tabular}[c]{@{}c@{}}S9 \\ {\scriptsize (EMSE)}\end{tabular}  		&
			UP. 						&
			\begin{tabular}[c]{@{}c@{}}S3 \\ {\scriptsize (ELC)}\end{tabular} 		&
			\begin{tabular}[c]{@{}c@{}}S8 \\ {\scriptsize (EMSE)}\end{tabular} 		&
			\begin{tabular}[c]{@{}c@{}}S4 \\ {\scriptsize (ELC)}\end{tabular} 		&
			\begin{tabular}[c]{@{}c@{}}S9 \\ {\scriptsize (EMSE)}\end{tabular}  		\\
			\midrule
						-5 		& 0.67 & 0.76 & 0.76 & 0.75 & 0.75		& 0.68 & 0.81 & 0.82 & 0.80 & 0.80\\ 
			0 		& 0.78 & 0.86 & 0.86 & 0.85 & 0.86		& 0.78 & 0.88 & 0.89 & 0.88 & 0.88\\ 
			5 		& 0.87 & 0.91 & 0.92 & 0.91 & 0.92		& 0.87 & 0.92 & 0.93 & 0.92 & 0.92\\ 
			Avg. 	& 0.77 & 0.84 & 0.85 & 0.84 & 0.84		& 0.78 & 0.87 & 0.88 & 0.87 & 0.87\\ 
			\midrule  \midrule
	\end{tabular}}
\end{table}
\begin{table}[t]
	\caption{ELC and STOI for S4 and S9 tested with BUS and PED.}
	\label{tab:lincorr_stoi_bus_ped}
	\centering
	\resizebox{0.8\columnwidth}{!}{%
		\begin{tabular}{ccccccccccccc}
			\midrule \midrule
			& \multicolumn{6}{c} {ELC} & \multicolumn{6}{c} {STOI} \\ \cmidrule(l){2-7} \cmidrule(l){8-13}  
			& \multicolumn{3}{c} {BUS} & \multicolumn{3}{c} {PED} & \multicolumn{3}{c} {BUS} & \multicolumn{3}{c} {PED} \\ \cmidrule(l){2-4} \cmidrule(l){5-7} \cmidrule(l){8-10} \cmidrule(l){11-13}
			SNR & UP. & S4 & S9 &	UP.	& S4 & S9 &	UP. & S4 & S9 &	UP. & S4 & S9 \\ \midrule
						-5 & 0.56 & 0.71 & 0.68 & 0.35 & 0.55 & 0.53 				& 0.77 & 0.84 & 0.84 & 0.60 & 0.71 & 0.71 \\  
			0 & 0.66 & 0.79 & 0.76 & 0.50 & 0.70 & 0.68 				& 0.85 & 0.90 & 0.90 & 0.72 & 0.83 & 0.83 \\ 
			5 & 0.74 & 0.83 & 0.81 & 0.64 & 0.78 & 0.76 				& 0.91 & 0.94 & 0.94 & 0.83 & 0.90 & 0.90 \\ 
			Avg. & 0.65 & 0.78 & 0.75 & 0.50 & 0.68 & 0.66 				& 0.84 & 0.89 & 0.89 & 0.72 & 0.81 & 0.81 \\ 
			\midrule \midrule
	\end{tabular}}
\end{table}
\subsection{Matched and Unmatched Noise Type Experiments}\label{sec:noiseRees}
In \Cref{tab:lincorr_combined} we compare the ELC scores for the noise type specific systems trained using the ELC (S0--S4), and EMSE (S5--S8) cost functions, and tested in matched noise-type conditions (SSN, BBL, CAF, and STR) at an input SNR of -5, 0, and 5 dB. Results covering the SNR range from -10 to 20 dB can be found in \cite{kolbaek_supplemental_nodate}. 
All models achieve large improvements in ELC with an average improvement of approximately 0.15-0.20, for all SNRs and noise types, compared to the ELC score of the noisy, unprocessed signals (denoted UP. in \Cref{tab:lincorr_combined,tab:stoi_combined,tab:lincorr_stoi_bus_ped}). 
We also see that, as expected, models trained with the ELC cost function (S0--S4) in general achieve similar or slightly higher ELC scores compared to the models trained with EMSE (S5--S8). 
In \Cref{tab:stoi_combined} we report the STOI scores for the systems in \Cref{tab:lincorr_combined} tested in identical conditions.  
We see moderate to large improvements in STOI in all conditions with an average improvement from 0.07--0.13. 
We also observe that the systems trained with the EMSE cost function achieve similar improvement in STOI as the systems trained with the ELC cost function.
In \Cref{tab:lincorr_stoi_bus_ped}, the ELC and STOI scores for the noise type general systems (S4 and S9) tested with the unmatched BUS and PED noise types are summarized. 
We see average improvement in the order of 0.1--0.18 in terms of ELC score and 0.05 -- 0.09 in terms of STOI.
We also see the performance gap between the S4 system (trained with ELC cost function) is small compared to the S9 system (trained with  EMSE cost function) and that noise specific systems perform slightly better than the noise general one. 
The results in \Cref{tab:lincorr_combined,tab:stoi_combined,tab:lincorr_stoi_bus_ped} are interesting since they show roughly identical global behavior as measured by ELC and STOI for systems trained with the ELC and EMSE cost functions. 
\subsection{Gain Similarities Between ELC and EMSE Based Systems}
We now study to which extent ELC and EMSE based systems behave similarly on a more detailed level.  
Specifically, we compute correlation coefficients between the gain vectors produced by each of the two types of systems, for SSN, BBL, and STR noise types, and summarize them in \Cref{tab:lincorr}. 
In \Cref{tab:lincorr} we observe that high sample correlations ($>0.90$) are achieved for all noise types and both SNRs, which indicates that the gains produced by a system trained with the ELC cost function are quite similar to the gains produced by a system trained with the EMSE cost function, which supports the findings in Sec.\;\ref{sec:noiseRees}. 
Similar conclusions can be drawn for the remaining noise types (results omitted due to space limitations, see \cite{kolbaek_supplemental_nodate}).
\subsection{Approximate-STOI Optimal DNN vs. Classical SE DNN}
As a final study we compare the performance of an approximate-STOI optimal DNN based SE system with classical Short-Time Spectral Amplitude (STSA) DNN based enhancement systems that estimate $\hat{g}(k,m)$ directly for each STFT frame (see e.g. \cite{weninger_discriminatively_2014,kolbaek_speech_2016}).  
Similarly to S0--S9 these systems are three-layered feed-forward DNNs and use 30 STFT frames as input, but differently from S0--S9, they minimize the MSE between STFT magnitude spectra, i.e. across frequency.
The DNNs estimate five STFT frames per time-step and overlapping frames are averaged to construct the final gain.      
We have trained two of these classical systems, with 512 units and 4096 units, respectively, in each hidden layer, using the BBL noise corrupted training set. The results are presented in \Cref{tab:stddnn}.  

From \Cref{tab:stddnn} we see, for example, that such classical STSA-DNN based SE systems trained and tested with BBL noise achieve a maximum STOI score of 0.66 at an input SNR of -5 dB, which is equivalent to the STOI score of 0.66 achieved by S1 in \Cref{tab:stoi_combined}. We also see that the classical system performs on par with S1 at an input SNR of 5 dB SNR with a STOI score of 0.92 compared to 0.90 achieved by S1. 
Although surprising, this is an interesting result since it indicates that no improvement in STOI can be gained by a DNN based SE system that is designed to maximize an approximate-STOI measure using short-time temporal one-third octave band envelope vectors. 
The important implication of this is that traditional STSA-DNN based SE systems may be close to optimal from an estimated speech intelligibility perspective. 
\begin{table}
	\centering
		\parbox[t]{.45\linewidth}{
			\centering
			\setlength\tabcolsep{5pt} 
			\caption{Sample linear correlation between gain vectors.}
			\label{tab:lincorr}
			\resizebox{0.25\columnwidth}{!}{%
			\begin{tabular}{cccc}
				\midrule \midrule
				\begin{tabular}[c]{@{}c@{}}SNR \\ {[dB]}\end{tabular} 	& 
				\begin{tabular}[c]{@{}c@{}}SSN \\ \end{tabular} 	& 
				\begin{tabular}[c]{@{}c@{}}BBL \\ \end{tabular} 	& 
				\begin{tabular}[c]{@{}c@{}}STR \\ \end{tabular}  	\\ \midrule
				-5   & 0.93 & 0.91 & 0.92   \\ 
				5    & 0.94 & 0.96 & 0.92   \\ \midrule \midrule
		\end{tabular}} }
		\hspace{5mm}
		\parbox[t]{.45\linewidth}{
			\centering
			\setlength\tabcolsep{5pt} 
			\caption{STOI score for classical DNN, tested with BBL.}
			\label{tab:stddnn}
			\resizebox{0.25\columnwidth}{!}{%
			\begin{tabular}{cccc}
				\midrule \midrule
				\begin{tabular}[c]{@{}c@{}}SNR \\ {[dB]}\end{tabular} 	& 
				\begin{tabular}[c]{@{}c@{}}UP. \\ \end{tabular} 	& 
				\multicolumn{2}{c} {\begin{tabular}[c]{@{}c@{}}\# units \\ \midrule  512 \hspace{2.0mm}  4096 \hspace{-1.5mm} \end{tabular}  }	\\ \midrule
				-5   & 0.59 & 0.64 & 0.66  \\ 
				5   & 0.83 & 0.91 & 0.92  \\ \midrule \midrule
	\end{tabular}}}
\end{table}
\section{Conclusion}
\label{sec:con2}
In this paper we proposed a Speech Enhancement\,(SE) system based on Deep Neural Networks\,(DNNs) that optimizes an approximation of the Short-Time Objective Intelligibility\,(STOI) estimator. We proposed an approximate-STOI cost function and derived closed-form expressions for the required gradients.   
We showed that DNNs designed to maximize approximate-STOI, achieve large improvement in STOI when tested in matched and unmatched noise types at various SNRs.
We also showed that approximate-STOI optimal systems do not outperform systems that minimize a mean squared error cost.
Finally, we showed that approximate-STOI DNN based SE systems perform on par with classical DNN based SE systems.  
Our findings suggest that a potential speech intelligibility gain of approximate-STOI optimal systems over MSE based systems is modest at best.

{\small\bibliographystyle{bib/IEEEtran}\bibliography{bib/mybibF}}


  \cleardoublepage
  \setcounter{enumiii}{0}
  \setcounter{enumii}{0}
  \setcounter{enumiv}{0}
  \setcounter{enumi}{0}
  \setcounter{equation}{0}
  \setcounter{figure}{0}
  \setcounter{footnote}{0}
  \setcounter{mpfootnote}{0}
  \setcounter{paragraph}{0}
  \setcounter{parentequation}{0}
  \setcounter{part}{0}
  \setcounter{section}{0}
  \setcounter{subparagraph}{0}
  \setcounter{subsection}{0}
  \setcounter{subsubsection}{0}
  \setcounter{table}{0}
  \papertitlepage{%
  On the Relationship Between Short-Time Objective Intelligibility and Short-Time Spectral-Amplitude Mean-Square Error for Speech Enhancement
}{paper:paperG}{%
  Morten Kolbæk, Zheng-Hua Tan, and Jesper Jensen
}{%
  The paper has been published in\\
	\textit{IEEE/ACM Transactions on Audio, Speech, and Language Processing}, \\vol. 27, no. 2, pp. 283-295, February 2019.
}{%
  \noindent\copyright\ 2019 IEEE
}

\acresetall

\newcommand{\A}{\barbelow{{A}}}
\newcommand{\R}{\barbelow{{R}}}
\newcommand{\Ah}{\barbelow{{\hat{A}}}}
\newcommand{\Z}{\barbelow{{Z}}}
\newcommand{\Y}{\barbelow{{Y}}}

\newcommand{\sa}{\barbelow{{a}}}
\newcommand{\s}{\barbelow{{a}}}
\newcommand{\sr}{\barbelow{{r}}}
\newcommand{\sy}{\barbelow{{r}}}
\newcommand{\ah}{\barbelow{{\hat{a}}}}
\newcommand{\sh}{\barbelow{{\hat{a}}}}
\newcommand{\gh}{\barbelow{{\hat{g}}}}
\newcommand{\alfa}{\barbelow{{\alpha}}}
\newcommand{\bet}{\barbelow{{\beta}}}

\newcommand{\e}{\barbelow{{e}}}

\newcommand{\One}{\barbelow{{1}}}
\newcommand{\Zero}{\barbelow{{0}}}
\newcommand{\E}{\mathbb{E}}

\newcommand{\rhoAAh}{\rho \left(\A , \Ah \right)}
\newcommand{\rhoAAhR}{\rho \left(\A , \Ah\left( \R \right) \right)}
\newcommand{\rhoaahr}{\rho \left(\sa , \ah\left( \sr \right) \right)}
\newcommand{\rhoaa}{\rho \left(\sa , \ah \right)}

\newcommand{\far}{ f_{\A , \R} \left( \sa , \sr \right)}
\newcommand{\fagr}{ f_{\A \lvert \R} \left( \sa \lvert \sr \right)}
\newcommand{\fr}{ f_{\R} \left( \sr \right)}

\newcommand{\I}{ \Gamma \left( \sr \right) }

\newcommand{\AhR}{ \Ah\left( \R \right) }
\newcommand{\ahr}{ \ah\left( \sr \right) }

\newcommand{\Hidem}{ \barbelow{\barbelow{{H}}}}

\newcommand{\Agr}{ \A \lvert \sr} 

\newcommand{\Agri}{ {A \lvert r}_i}

\tikzset{%
	do path picture/.style={%
		path picture={%
			\pgfpointdiff{\pgfpointanchor{path picture bounding box}{south west}}%
			{\pgfpointanchor{path picture bounding box}{north east}}%
			\pgfgetlastxy\x\y%
			\tikzset{x=\x/2,y=\y/2}%
			#1
		}
	},
	cross/.style={do path picture={    
			\draw [line cap=round] (-1,-1) -- (1,1) (-1,1) -- (1,-1);
	}},
}

%
\title{On the Relationship Between Short-Time Objective Intelligibility and Short-Time Spectral-Amplitude Mean-Square Error for Speech Enhancement}

\begin{abstract}
The majority of \acf{DNN} based speech enhancement algorithms rely on the \acf{MSE} criterion of Short-Time Spectral Ampli-tudes\,(STSA), which has no apparent link to human perception, e.g. speech intelligibility. 
\acf{STOI}, a popular state-of-the-art speech intelligibility estimator, on the other hand, relies on linear correlation of speech temporal envelopes. This raises the question if a DNN training criterion based on \ac{ELC} can lead to improved speech intelligibility performance of DNN based speech enhancement algorithms compared to algorithms based on the STSA-MSE criterion.    	
In this paper we derive that, under certain general conditions, the STSA-MSE and ELC criteria are practically equivalent, and we provide empirical data to support our theoretical results.   
Furthermore, our experimental findings suggest that the standard STSA minimum-MSE estimator is near optimal, if the objective is to enhance noisy speech in a manner which is optimal with respect to the STOI speech intelligibility estimator.
\end{abstract}

\section{Introduction}
\label{sec:introG}
Despite the recent success of \acf{DNN} based speech enhancement algorithms \cite{erdogan_deep_2017,wang_deep_2017,wang_supervised_2017,kim_bitwise_2017,fakoor_reinforcement_2017}, it is yet unknown if these algorithms are optimal in terms of aspects related to human auditory perception, e.g. speech intelligibility, since existing algorithms do not directly optimize criteria designed with human auditory perception in mind.

Many current state-of-the-art DNN based speech enhancement algorithms use a \acf{MSE} training criterion \cite{chen_large-scale_2016,healy_algorithm_2017,kolbaek_speech_2017} on Short-Time Spectral Amplitudes\,(STSA).
This, however, might not be the optimal training criterion if the target is the human auditory system, and improvement in speech intelligibility or speech quality is the desired objective.  

It is well known that the frequency sensitivity of the human auditory system is non-linear ( e.g. \cite{schnupp_auditory_2011,moore_introduction_2013}) and, as a consequence, is often approximated in digital signal processing algorithms using e.g. a Gammatone filter bank \cite{patterson_complex_1992} or a one-third octave band filter bank \cite{taal_algorithm_2011}. It is also well known that preservation of modulation frequencies in the range 4-20 Hz are critical for speech intelligibility \cite{elliott_modulation_2009,schnupp_auditory_2011,drullman_effect_1994}.  
Therefore, it is natural to believe that, if prior knowledge about the human auditory system is incorporated into a speech enhancement algorithm, improvements in speech intelligibility or speech quality can be achieved \cite{lim_enhancement_1979}.

Indeed, numerous works exist that attempt to incorporate such knowledge (e.g.  \cite{healy_algorithm_2015,loizou_speech_2005,hendriks_dft-domain_2013,lightburn_sobm_2015,han_perceptual_2016,shivakumar_perception_2016,koizumi_dnn-based_2017,kolbaek_monaural_2018-1,zhao_perceptually_2018,zhang_training_2018,fu_end--end_2018} and references therein).
In \cite{healy_algorithm_2015} a transform-domain method based on a Gammatone filter bank was used, which incorporates a non-linear frequency resolution mimicking that of the human auditory system. 
In \cite{loizou_speech_2005} different perceptually motivated cost functions were used to derive STSA clean speech spectrum estimators in order to emphasize spectral peak information, account for auditory masking or penalize spectral over-attenuation. 
In \cite{han_perceptual_2016,shivakumar_perception_2016} similar goals were pursued, but instead of using classical statistically-based models, DNNs were used.
Finally, in \cite{koizumi_dnn-based_2017} a deep reinforcement learning technique was used to reward solutions that achieved a large score in terms of \ac{PESQ} \cite{rix_perceptual_2001}, a commonly used speech quality estimator.   

Although the works in e.g. \cite{loizou_speech_2005,healy_algorithm_2015,shivakumar_perception_2016,koizumi_dnn-based_2017} include knowledge about the human auditory system the techniques are not designed specifically to maximize speech intelligibility. 
While speech processing methods that improve speech intelligibility would be of vital importance for applications such as mobile communications, or hearing assistive devices, only very little research has been performed to understand if DNN-based speech enhancement systems can help improve speech intelligibility.
Very recent work \cite{kolbaek_monaural_2018-1,zhao_perceptually_2018,zhang_training_2018,fu_end--end_2018} has investigated if DNNs trained to maximize a state-of-the-art speech intelligibility estimator are capable of improving speech intelligibility as measured by the estimator \cite{kolbaek_monaural_2018-1,zhao_perceptually_2018,zhang_training_2018} or human listeners \cite{fu_end--end_2018}.     
Specifically, DNNs were trained to maximize the \acf{STOI} \cite{taal_algorithm_2011} estimator and were then compared, in terms of STOI, with DNNs trained to minimize the classical STSA-MSE criterion. Surprisingly, although all DNNs improved STOI, the DNNs trained to maximize STOI showed none or only very modest improvements in STOI compared to the DNNs trained with the classical STSA-MSE criterion \cite{kolbaek_monaural_2018-1,zhao_perceptually_2018,zhang_training_2018,fu_end--end_2018}.  

The STOI speech intelligibility estimator has proven to be able to quite accurately predict the intelligibility of noisy/processed speech in a large range of acoustic scenarios, including speech processed by mobile communication devices \cite{jorgensen_speech_2015}, ideal time-frequency weighted noisy speech \cite{taal_algorithm_2011}, noisy speech enhanced by single-microphone time-frequency weighting-based speech enhancement systems \cite{jensen_algorithm_2016,taal_algorithm_2011,jensen_speech_2014}, and speech processed by hearing assistive devices such as cochlear implants \cite{falk_objective_2015}. STOI has also been shown to be robust to variations in language types, including Danish \cite{taal_algorithm_2011}, Dutch \cite{jensen_speech_2014}, and Mandarin \cite{xia_evaluation_2012}. 
Finally, recent studies e.g. \cite{healy_algorithm_2017,chen_large-scale_2016} also show a good correspondence between STOI predictions of noisy speech enhanced by DNN-based speech enhancement systems, and speech intelligibility.  
As a consequence, STOI is currently the, perhaps, most commonly used speech intelligibility estimator for objectively evaluating the performance of speech enhancement systems \cite{healy_algorithm_2015,chen_large-scale_2016,healy_algorithm_2017,kolbaek_speech_2017}.  
Therefore, it is natural to believe that gains in speech intelligibility, as estimated by STOI, can be achieved by utilizing an optimality criterion based on STOI as opposed to the classical criterion based on STSA-MSE.  

In this paper we study the potential gain in speech intelligibility that can be achieved, if a DNN is designed to perform optimally with respect to the STOI speech intelligibility estimator. 
We derive that, under certain general conditions, maximizing an approximate-STOI criterion is equivalent to minimizing a STSA-MSE criterion. 
Furthermore, we present empirical data using simulation studies with DNNs applied to noisy speech signals, that support our theoretical results. 
Finally, we show theoretically under which conditions the equality between the approximate-STOI criterion and the STSA-MSE criterion holds for practical systems. 
Our results are in line with recent empirical work and might explain the somewhat surprising result in \cite{kolbaek_monaural_2018-1,zhao_perceptually_2018,zhang_training_2018,fu_end--end_2018}, where none or only very modest improvements in STOI were achieved with STOI optimal DNNs compared to MSE optimal DNNs.

\section{STFT-domain based Speech Enhancement}\label{sec:secse}
Fig.\;\ref{fig:sefig} shows a block-diagram of a classical gain-based speech enhancement system \cite{hendriks_dft-domain_2013,loizou_speech_2013}.
%
%
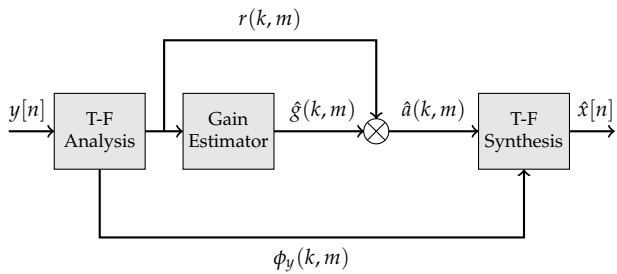
\begin{figure}
	\centering
	\vspace{2mm}
	\begin{tikzpicture}[baseline=(current bounding box.north)]
	\def\x{0.2}
	
	\draw [thick, ->] (0,2) -- (0.6,2);
	\node[above] at (0.25,2) {\footnotesize $y[n]$};
	\draw [thick, ->] (1.8,2) -- (2.3,2);
	\node[above] at (3.45,3.2) {\footnotesize $r(k,m)$};
	\draw [thick, ->]  (2.05,2) -- (2.05,3.2) -- (4.85,3.2) -- (4.85,2.16) ;
	\draw [thick, ->]  (1.2,1.5) -- (1.2,0.6) -- (6.8,0.6) -- (6.8,1.5) ;
	\node[below] at (4,0.6) {\footnotesize $\phi_y(k,m)$};
	\draw [thick, ->] (3.5,2) -- (4.69,2);
	\node[above] at (4.15,2.00) {\footnotesize $\hat{g}(k,m)$};
	
	\node [draw,circle,cross,minimum width=0.2 ] at (4.85,2){}; 
	
	\draw [thick, ->] (5.01,2) -- (6.2,2);
	\node[above] at (5.61,2.00) {\footnotesize $\hat{a}(k,m)$};
	
	\draw [thick, ->] (7.2,2) -- (8,2);
	\node[above] at (7.75,2) {\footnotesize $\hat{x}[n]$};

	\filldraw[fill=gray!20!white, draw=black] (0.6 , 1.5) rectangle (1.8 , 2.5) ;
	\node at (1.2,2) {\scriptsize {\begin{tabular}{c} T-F \\ Analysis \end{tabular}}};
	
	\filldraw[fill=gray!20!white, draw=black] (2.3 , 1.5) rectangle (3.5 , 2.5) ;
	\node at (2.9,2) {\scriptsize {\begin{tabular}{c} Gain \\ Estimator \end{tabular}}};
	
	\filldraw[fill=gray!20!white, draw=black] (6.2 , 1.5) rectangle (7.4 , 2.5) ;
	\node at (6.8,2) {\scriptsize {\begin{tabular}{c} T-F \\ Synthesis \end{tabular}}};

	\end{tikzpicture}
	\caption{Classical gain-based speech enhancement system. The noisy time-domain signal $y[n] = x[n] + v[n]$ is first decomposed into a \acf{T-F} representation $r(k,m)$ for time-frame $m$ and frequency index $k$. An estimator, e.g. a DNN, estimates a gain $\hat{g}(k,m)$ that is applied to the noisy short-term magnitude spectrum $r(k,m)$ to arrive at an enhanced signal magnitude  $\hat{a}(k,m) = \hat{g}(k,m)r(k,m)$. Finally, the enhanced time-domain signal $\hat{x}[n]$ is obtained from a T-F synthesis stage using the phase of the noisy signal $\phi_y(k,m)$.}
    \label{fig:sefig}
\end{figure}
Let $x[n]$ be the $n$th sample of the clean time-domain speech signal and let a noisy observation $y[n]$ be given by
\begin{equation}
y[n] = x[n] + v[n], 
\label{eq111}
\end{equation}
where $v[n]$ is a sample of additive noise.   
Furthermore, let $a(k,m)$ and $r(k,m)$,  $k = 1,\dots, \frac{K}{2}+1$,  $m=1,\dots M, $ denote the single-sided magnitude spectra of the $K$-point short-time discrete Fourier transform (STFT) of $x[n]$ and $y[n]$, respectively, where $M$ is the number of STFT frames.  
Also, let $\hat{a}(k,m)$ denote an estimate of $a(k,m)$ obtained as $\hat{a}(k,m) = \hat{g}(k,m)r(k,m)$. 
Here, $\hat{g}(k,m)$ is a scalar gain factor applied to the magnitude spectrum of the noisy speech $r(k,m)$ to arrive at an estimate $\hat{a}(k,m)$ of the clean speech magnitude spectrum $a(k,m)$. It is the goal of many STFT-based speech enhancement systems to find appropriate values for $\hat{g}(k,m)$ based on the available noisy signal $y[n]$.    
The gain factor $\hat{g}(k,m)$ is typically estimated using either statistical model-based methods such as classical STSA \acf{MMSE} estimators \cite{ephraim_speech_1984}, \cite{hendriks_dft-domain_2013,loizou_speech_2013}, or machine learning based techniques such as Gaussian mixture models \cite{kim_algorithm_2009}, support vector machines \cite{han_classification_2012}, or, more recently, DNNs \cite{healy_algorithm_2015,chen_large-scale_2016,healy_algorithm_2017,kolbaek_speech_2017}. 
For reconstructing the enhanced speech signal in the time domain, it is common practice to append the short-time phase spectrum of the noisy signal to the estimated short-time magnitude spectrum and then use the overlap-and-add technique \cite{allen_short_1977}, \cite{loizou_speech_2013}.

\section{Short-Time Objective Intelligibility (STOI)}\label{sec:secstoi}
In the following, we shortly review the STOI intelligibility estimator \cite{taal_algorithm_2011}. For further details we refer to \cite{taal_algorithm_2011}. 
Let the $j$th one-third octave band clean-speech amplitude, for time-frame $m$, be defined as 
\begin{equation}
a_j( m ) = \sqrt{\sum_{k=k_1(j)}^{k_2(j)} a(k,m)^2},
\label{eq22}
\end{equation}
where $k_1(j)$ and $k_2(j)$ denote the first and last STFT bin index, respectively, of the $j$th one-third octave band.
Furthermore, let a short-time temporal envelope vector that spans time-frames $m-N+1, \dots, m$, for the clean speech signal be defined as 
\begin{equation}
\s_{j,m} = [ a_j( m-N+1 ), \; a_j( m-N+2 ), \dots , a_j( m ) ]^T
\label{eq222}
\end{equation}
In a similar manner we define  $\sh_{j,m}$ and $\sy_{j,m}$ for the enhanced speech signal and the noisy observation, respectively. 

The parameter $N$ defines the length of the temporal envelope and for STOI $N=30$%
\footnote{With $N=30$, STOI is sensitive to temporal modulations of $2.6$ Hz and higher, which are frequencies important for speech intelligibility \cite{taal_algorithm_2011}.}, which for the STFT settings used in this study, as well as in \cite{taal_algorithm_2011}, corresponds to approximately $384$ ms.
Finally, the STOI speech intelligibility estimator for a pair of short-time temporal envelope vectors can then be approximated by the sample \acf{ELC} between the clean and enhanced envelope vectors $\s_{j,m}$ and $\sh_{j,m}$ given as 
\begin{equation}
\mathcal{L} ( \s_{j,m},\sh_{j,m}) = \frac{\left(\s_{j,m} - \mu_{\s_{j,m}}\right)^T  \left(\sh_{j,m} - \mu_{\sh_{j,m}}\right)}{ \left\lVert \s_{j,m} - \mu_{\s_{j,m}} \right\rVert  \; \left\lVert\sh_{j,m} - \mu_{\sh_{j,m}} \right\rVert },
\label{eq:stoicost}
\end{equation}
where $\left\lVert \cdot \right\rVert$ denotes the Euclidean $\ell^2$-norm and $\mu_{\s_{j,m}}$ and $\mu_{\sh_{j,m}}$ denote the sample means of $\s_{j,m}$ and $\sh_{j,m}$, respectively. 
Note that Eq.\;\eqref{eq:stoicost} is an approximation, since the clipping and normalization steps otherwise used in STOI, have been omitted. This has empirically been found not to have any significant effect on intelligibility prediction performance in most cases  \cite{taal_matching_2012,lightburn_sobm_2015,jensen_algorithm_2016,andersen_predicting_2016}. 
Furthermore, since the normalization step is applied for the entire vector $\sh_{j,m}$, the normalization procedure itself does not influence the final STOI score. Also, as clipping only occurs for time-frequency units for which the signal-to-distortion ratio (see Eq. (4) in \cite{taal_algorithm_2011}) is below $-15$ dB, clipping only occurs for a minority of the envelope vectors and approximating STOI with ELC is well valid, or even exact, in most cases, when evaluating speech signals at practical SNRs.

From $\mathcal{L} ( \s_{j,m},\sh_{j,m})$, the final STOI score for an entire speech signal is then defined as \cite{taal_algorithm_2011} the scalar, $-1 \leq d \leq 1$,    
\begin{equation}
d = \frac{1}{J(M-N+1)} \sum_{j=1}^{J} \sum_{m=N}^{M} \mathcal{L} ( \s_{j,m},\sh_{j,m}),
\label{eq:stoisum}
\end{equation}
where $J$ is the number of one-third octave bands and $M-N+1$ is the total number of short-time temporal envelope vectors.

Similarly to \cite{taal_algorithm_2011}, we use $J=15$ with a center frequency of the first one-third octave band at 150 Hz and the last at approximately 3.8 kHz to ensure a frequency range that covers the majority of the spectral information of human speech.
The STOI score in general has been shown to often have high correlation with listening tests involving human test subjects, i.e. the higher numerical value of Eq.\;\eqref{eq:stoisum}, the more intelligible is the speech signal.    

Since STOI, as approximated by Eq.\;\eqref{eq:stoisum}, is a sum of ELC values as given by Eq.\;\eqref{eq:stoicost}, maximizing Eq.\;\eqref{eq:stoicost} will also maximize the overall STOI score in Eq.\;\eqref{eq:stoisum}. 
As a consequence, in order to find an estimate $\hat{x}[n]$ of $x[n]$ so that STOI is maximized, one can focus on finding optimal estimates of the individual short-time temporal envelope vectors $\s_{j,m}$.      
Therefore, we define $\sh_{j,m} = \text{diag}(\gh_{j,m})\sy_{j,m}$ as the short-time temporal one-third octave band envelope vector of the enhanced speech signal, where $\gh_{j,m}$ is an estimated gain vector and $\text{diag}(\gh_{j,m})$ is a diagonal matrix with the elements of $\gh_{j,m}$ on the main diagonal.

\section{Envelope Linear Correlation Estimator}\label{sec:melcest}
We now introduce the approximate-STOI criterion in a stochastic context and derive the speech envelope estimator that maximizes it. We denote this estimator as the \ac{MMELC} estimator. 
Let $A_j(m)$ and $R_j(m)$ denote random variables representing a clean and a noisy, respectively, one-third octave band magnitude, for band $j$ and time frame $m$. Furthermore, let   
\begin{equation}
\A_j(m) = \left[ A_j(m-N+1), \, \dots \, A_j(m) \right]
\label{eq1}
\end{equation}
and
\begin{equation}
\R_j(m) = \left[ R_j(m-N+1), \, \dots \, R_j(m) \right]
\label{eq2}
\end{equation}
be the stack of these random variables in random envelope vectors. 
Finally, in a similar manner, let 
\begin{equation}
\Ah_j(m) = \left[ \hat{A}_j(m-N+1), \, \dots \, \hat{A}_j(m) \right], 
\label{eq3}
\end{equation}
be a random envelope vector representing an estimate of $\A_j(m)$.  
Now, the contribution of $\Ah_j(m)$ to speech intelligibility may be approximated
as the ELC between the envelope vectors $\A_j(m)$ and $\Ah_j(m)$. In the following, the indices $j$ and $m$ are omitted for convenience. 
Let $\One$ denote a vector of ones, and let $\barbelow{\mu}_{\A} = \frac{1}{N} \One^T \A \One$ be a vector, whose entries equal the sample mean of the entries in $\A$.
Let  $\barbelow{\mu}_{\Ah}$ be defined in a similar manner. 
Finally, let the ELC between $\A$ and $\Ah$, which is a random variable, be defined as
\begin{equation}
\rhoAAh \triangleq \frac{ \Big( \A - \barbelow{\mu}_{\A} \Big)^T  \Big( \Ah - \barbelow{\mu}_{\Ah} \Big) }{ \Big\lVert \A - \barbelow{\mu}_{\A} \Big\rVert \Big\lVert \Ah - \barbelow{\mu}_{\Ah} \Big\rVert },
\label{eq4}
\end{equation} 
and the expected ELC as
\begin{equation} 
\begin{split}
\Omega_{ELC}	& = \E_{\A,\R}  \left[ \rhoAAh \right]  \\
& = \int \int \rhoaa \far  d \sa \, d \sr \\
& = \int \underbrace{ \int \rhoaa \fagr \, d \sa \; }_{\I} \fr \, d \sr .
\end{split}
\label{eq5}
\end{equation}    
Here, $\ah$ is related to $\sr$ via a deterministic map, e.g. a DNN, and $\far$ denotes the joint \acf{PDF} of clean and noisy/processed one-third octave band envelope vectors. Furthermore, $\fagr$ and $\fr$ denote a conditional and marginal PDF, respectively.

An optimal estimator can be found by minimizing the Bayes risk \cite{kay_fundamentals_2010,loizou_speech_2013}, which is equivalent to maximizing Eq.\;\eqref{eq5}, hence arriving at the MMELC estimator, which we denote as $\ah_{MMELC}$.  
To do so, observe that for a particular noisy observation $\sr$ maximizing $\I$ maximizes Eq.\;\eqref{eq5}, since $\fr \geq 0 \; \forall \; \sr$. In other words, our goal is to maximize $\I$ for each and every $\sr$. Hence, for a particular observation, $\sr$, the MMELC estimate is given by
\begin{equation} 
\begin{split}
\ah_{MMELC} & = \arg\max_{\ah} \int \rhoaa \fagr \, d \sa \\
& = \arg\max_{\ah} \int \frac{ \big( \sa - \barbelow{\mu}_{\sa} \big)^T  \big( \ah - \barbelow{\mu}_{\ah} \big) }{ \big\lVert \sa - \barbelow{\mu}_{\sa} \big\rVert \big\lVert \ah - \barbelow{\mu}_{\ah} \big\rVert } \fagr \, d \sa  \\
& = \arg\max_{\ah} \underbrace{ \int \frac{ \big( \sa - \barbelow{\mu}_{\sa} \big)^T }{ \big\lVert \sa - \barbelow{\mu}_{\sa} \big\rVert }   \fagr \, d \sa }_{\E_{\A | \sr } \left[  \e(\A)^T \right] }   \underbrace{ \frac{\big( \ah - \barbelow{\mu}_{\ah} \big)}{\big\lVert \ah - \barbelow{\mu}_{\ah} \big\rVert} }_{\e(\ah)}  \\
& = \arg\max_{\ah} \; \E_{\A | \sr } \left[ \e(\A)^T \right] \e(\ah),  \\
\end{split}
\label{eq6}
\end{equation}     
where $\e(\cdot)$ is a function that normalizes its vector argument to zero sample mean and unit norm 
and where we used that for a given noisy observation $\sr$, $\ah$ is deterministic. 
Note that the solution to Eq.\;\eqref{eq6} is non-unique. For one given solution, say $\ah^\ast$, any affine transformation, $\delta \ah^\ast + \gamma \One \; \forall \; \delta,\gamma \, \in \mathcal{R} $, is also a solution, because any such transformation is undone by $\e(\cdot)$.
Hence, in the following we focus on finding one such particular solution, namely the zero sample mean, unit norm solution, i.e. the vector $\e(\ah)$ that maximizes the inner product with the vector $\E_{\A | \sr } \left[ \e(\A| \sr) \right]$. 
To do so, let $\alfa = \E_{\A | \sr } \left[ \e(\A| \sr) \right]$, and let $\e(\ah^\ast)$ denote the zero sample mean, unit norm vector that maximizes Eq.\;\eqref{eq6}. 
Then, using the method of Lagrange multipliers, it can be shown (see Appendix\;\ref{sec:lagran}) that the MMELC estimator is given by
\begin{equation} 
\begin{split}
\ah_{MMELC} &=  \e(\ah^\ast) \\
& = \frac{\big( \alfa - \barbelow{\mu}_{\alfa} \big)}{\big\lVert \alfa - \barbelow{\mu}_{\alfa} \big\rVert} \\
& = \frac{\alfa}{\Vert \alfa \Vert}, 
\end{split}
\label{eq7}
\end{equation}  
which is nothing more than the vector $\alfa$, normalized to unit norm. 
The fact that $\barbelow{\mu}_{\alfa} = \frac{1}{N}  \One^T \alpha \One = \Zero$ follows from Eq.\;\eqref{eq6}, where it is seen that $\alfa = \E_{\A | \sr } \left[ \e(\A| \sr) \right]$ is an expectation over vectors $ (\sa - \barbelow{\mu}_{\sa})  \big\lVert \sa - \barbelow{\mu}_{\sa} \big\rVert^{-1} $ whose sample mean is zero. 
By interpreting the expectation as an infinite linear combination of such vectors, it follows that $\barbelow{\mu}_{\alfa} = \Zero$.

\section{Relation to STSA-MMSE Estimators}\label{secrelation}
We now show that the MMELC estimator, Eq.\;\eqref{eq7}, is asymptotically equivalent to the one-third octave band STSA-MMSE estimator for large envelope lengths, i.e. as $N \to \infty$.
The STSA-MSE (e.g. \cite{ephraim_speech_1984}) is defined as
\begin{equation} 
\begin{split}
\Omega_{MSE}	& = \E_{\A,\R}  \left[ \left(\A-\Ah \right)^2 \right].  \\
\end{split}
\label{eq8}
\end{equation}  
It can be shown (e.g. \cite{ephraim_speech_1984,hendriks_dft-domain_2013,loizou_speech_2013}) that the optimal Bayesian estimator with respect to Eq.\;\eqref{eq8}, is the STSA-MMSE estimator given by the conditional mean defined as 
\begin{equation} 
\begin{split}
\ah_{MMSE} & =  \int \sa \; \fagr \, d \sa  \\
& = \E_{\A | \sr }  \left[ \A | \sr \right]. \\
\end{split}
\label{eq9}
\end{equation}  
To show that $\ah_{MMELC}$ is asymptotically equivalent to $\ah_{MMSE}$, let us introduce the idempotent, symmetric matrix
\begin{equation} 
\Hidem =  \barbelow{\barbelow{I}}_N - \frac{1}{N} \One\One^T,
\label{eq10}
\end{equation} 
where $\barbelow{\barbelow{I}}_N$ denotes the $N$-dimensional identity matrix. We can then rewrite the vector $\alfa$ as
\begin{equation} 
\begin{split}
\alfa & = \int \frac{ \big( \sa - \barbelow{\mu}_{\sa} \big) }{ \big\lVert \ah - \barbelow{\mu}_{\ah} \big\rVert }   \fagr \, d \sa   \\
& =  \int \frac{  \Hidem \sa }{ \big\lVert \Hidem \sa \big\rVert }   \fagr \, d \sa   \\
& = \E_{\A | \sr } \left[ \frac{  \Hidem \A | \sr }{ \big\lVert \Hidem \A | \sr \big\rVert }  \right] \\
& = \E_{\A | \sr } \left[ \frac{  \Z }{ \big\lVert \Z \big\rVert }  \right], \\
\end{split}
\label{eq11}
\end{equation} 
where $\A | \sr $ is a random vector, and we introduced the notation $\Z \triangleq \Hidem \A | \sr$.
We now employ the following conditional independence assumption
\begin{equation} 
\fagr = \prod_{j=1}^{N} f_{A_j | R_j = r_j} (a_j | r_j).
\label{eqz9}
\end{equation} 
This is a standard assumption in the area of speech enhancement, when operating in the STFT domain and has been the underlying assumption of a very large number of speech enhancement methods (see e.g. \cite{ephraim_speech_1984,ephraim_speech_1985,erkelens_minimum_2007,hendriks_dft-domain_2013,loizou_speech_2013} and references therein). 
The conditional independence assumption is, for example, valid, when speech and noise STFT coefficients may be assumed statistically independent across time and frequency and mutually independent \cite{mcaulay_speech_1980,ephraim_speech_1984,loizou_speech_2013}.

Using Kolmogorovs strong law of large numbers  \cite[pp. 67-68]{sen_large_1994} and the conditional independence assumption, it can be shown (see Appendix\;\ref{sec:zindepedent}) that asymptotically, as $N \to \infty$, the expectation in Eq.\;\eqref{eq11} factorizes as   
\begin{equation} 
\begin{split}
\lim_{N\to\infty} \;\; \alfa &= \lim\limits_{N \to \infty} \E_{\A | \sr } \left[ \frac{  1 }{ \big\lVert \Z \big\rVert }  \right]     \E_{\A | \sr }\left[ \Z  \right]. \\
\end{split}
\label{eq14}
\end{equation} 
Combining this result with Eq.\;\eqref{eq7} leads to
\begin{equation} 
\begin{split}
\lim_{N\to\infty} \ah_{MMELC} &=  \lim_{N\to\infty} \frac{  \alfa }{\big\lVert \alfa \big\rVert}\\
&= \lim_{N\to\infty}  \frac{   \E_{\A | \sr } \left[ \frac{  1 }{ \lVert \Z \rVert }  \right]     \E_{\A | \sr } \left[ \Z  \right]  }{ \Big\lVert \E_{\A | \sr } \left[ \frac{  1 }{ \lVert \Z \rVert }  \right]     \E_{\A | \sr } \left[ \Z  \right] \Big \rVert }\\
&=  \lim_{N\to\infty} \frac{   \E_{\A | \sr } \left[ \frac{  1 }{ \lVert \Z \rVert }  \right]     \E_{\A | \sr } \left[ \Z  \right]  }{ \E_{\A | \sr } \left[ \frac{  1 }{ \lVert \Z \rVert }  \right] \Big\lVert    \E_{\A | \sr } \left[ \Z  \right]   \Big \rVert   }\\
&= \lim_{N\to\infty} \frac{       \E_{\A | \sr } \left[ \Z  \right]  }{ \Big\lVert    \E_{\A | \sr } \left[ \Z  \right]   \Big \rVert   }.\\
\end{split}
\label{eq15}
\end{equation}  
Since Eq.\;\eqref{eq6} is invariant to affine transformations of its input arguments, we can scale  $\ah_{MMELC} $  with the scalar quantity $  \lVert \E_{\A | \sr } \left[  \Z   \right] \rVert  $ in Eq.\;\eqref{eq15} to arrive at 
\begin{equation} 
\begin{split}
\lim_{N\to\infty} \;\; \ah_{MMELC} &=  \E_{\A | \sr } \left[ \Z  \right].\\
\end{split}
\label{eq16}
\end{equation}  
Finally, as $N \to \infty $, the MMELC estimator  $\ah_{MMELC}$ is given by
\begin{equation} 
\begin{split}
\lim_{N\to\infty} \;\; \ah_{MMELC} &= \;  \E_{\A | \sr } \left[ \Z  \right] \\
&= \E_{\A | \sr } \big[   \Hidem \A | \sr  \big] \\
&=\E_{\A | \sr } \Bigg[ \left(  \barbelow{\barbelow{\mathbf{I}}}_N - \frac{1}{N} \One\One^T \right) \A | \sr  \Bigg] \\
&=\E_{\A | \sr } \bigg[  \A | \sr - \frac{1}{N} \One\One^T  \A | \sr  \bigg] \\
&=\E_{\A | \sr } \big[  \A | \sr \big]  - \frac{1}{N} \One\One^T  \E_{\A | \sr } \big[ \A | \sr  \big]  \\
&=\ah_{MMSE}   - \barbelow{\mu}_{\ah_{MMSE} }.\\
\end{split}
\label{eq17}
\end{equation} 
In words, the MMELC estimator, $\ah_{MMELC}$, is (asymptotically in $N$) an affine transformation of the STSA-MMSE estimator $\ah_{MMSE}$. 
In practice, this means that using the STSA-MMSE estimator leads to the same approximate-STOI criterion value as the estimator, $\ah_{MMELC}$, derived to maximize this criterion.  

In other words, applying the traditional STSA-MMSE estimator leads to maximum speech intelligibility as reflected by the approximate STOI estimator.
\section{Experimental Design}\label{sec:expdes}
We now investigate empirically the relationship between the MMELC estimator in Eq.\;\eqref{eq9} and the STSA-MMSE estimator in Eq.\;\eqref{eq6} using an experimental study.   
As defined in Eq.\;\eqref{eq6}, the MMELC estimator is the vector that maximizes the expectation of the ELC cost function given by Eq.\;\eqref{eq5}. 
This expectation, Eq.\;\eqref{eq5}, is defined via an integral of $\rhoaa$ for various realizations of $\sa$ and $\ah$, and weighted by the joint PDF $\far$. It is however, well known, that the integral may be approximated (arbitrarily well) as a sum of $\rhoaa$ terms, where realizations of $\sa$ and $\ah$ are drawn according to $\far$.      
This is similar to what a DNN approximates during a standard training process, where a gradient based optimization technique is used to minimize the cost on a representative training set \cite{goodfellow_deep_2016}. 
Therefore, training a DNN, e.g. using stochastic gradient ascent, to maximize Eq.\;\eqref{eq:stoicost} may be seen as an approximation of Eq.\;\eqref{eq6}, where the approximation becomes more accurate with increasing training set size. 

From the theoretical results presented in Sec.\;\ref{secrelation}, we would therefore expect that, for some sufficiently large $N$, one would obtain equality in an ELC sense, between a DNN trained to maximize an ELC cost function and one that is trained to minimize the classical STSA-MSE cost function.
To validate this expectation we follow the techniques formalized in Secs.\;\ref{sec:secse} and \ref{sec:secstoi} and train DNNs to estimate gain vectors, $\gh_{j,m}$, that we apply to noisy one-third octave band magnitude envelope signals $\sy_{j,m}$, to arrive at enhanced signals $\sh_{j,m}$. 

In principle, any supervised learning model would be applicable for these experiments but considering the universal function approximation capability of DNNs \cite{hornik_multilayer_1989}, this is our model of choice.
We use short-time temporal one-third octave band envelope vectors, as defined in Eq.\;\eqref{eq222}, and train multiple DNNs, one for each of the $J=15$ one-third octave bands, for various $N$, to investigate if for sufficiently large $N$, DNNs trained with a STSA-MSE cost function approach the ELC values of DNNs trained with a cost function based on ELC. 

We construct two types of enhancement systems, one type is trained using the STSA-MSE cost function, denoted as $\text{ES}_{MSE}$, and one that is trained using the ELC cost function denoted as $\text{ES}_{ELC}$. 
Each of the systems consists of $J=15$ DNNs, each estimating a gain vector $\gh_{j,m}$ for a particular one-third octave band directly from the STFT magnitudes of the noisy signal $r(k,m)$, with the input context given by $k = 1,\dots, \frac{K}{2}+1$,  $m-N+1 \dots, m $. This ensures that all DNNs have access to the same information for a particular value of $N$, as they all receive the same input data.  
Furthermore, we follow common practice (e.g. \cite{healy_algorithm_2015,healy_algorithm_2017,chen_large-scale_2016,kolbaek_monaural_2018-1}) and average overlapping estimated gain values, within a one-third octave band, during enhancement. We found during a preliminary study that this technique consistently lead to slightly larger STOI scores for both types of systems.

To compute the STFT coefficients for all signals we use a 10 kHz sample frequency and a $K=256$ point STFT with a Hann-window size of 256 samples (25.6 ms) and a 128 sample frame shift (12.8 ms). These coefficients are then used to compute one-third octave band envelopes for the clean and noisy signals using Eq.\;\eqref{eq222}.   
\subsection{Noise-free Speech Mixtures}
We have used the Wall Street Journal\,(WSJ0) speech corpus \cite{garofolo_csr-i_1993} as the clean speech data for both the training set, validation set, and test set. Specifically, the noise-free utterances used for training and validation are generated by randomly selecting utterances from 44 male and 47 female speakers from the WSJ0 training set entitled si\_tr\_s. In total 20000 utterances are used for the training set and 2000 are used for the validation set, which adds up to approximately 37 hours of training data and 4 hours of validation data.
For the test set, we have used a similar approach and sampled 1000 utterances among 16 speakers (10 males and 6 females) from the WSJ0 validation set si\_dt\_05 and evaluation set si\_et\_05, which is equivalent to approximately 2 hours of data, see \cite{kolbaek_supplemental_nodate-1} for further details.    
The speakers used in the training and validation sets are different than the speakers used for test, i.e. we test in a speaker independent setting. 
{\color{black} Finally, since WSJ0 utterances primarily include speech active regions we do not apply a VAD. This is motivated by the fact that noise-only regions are irrelevant for STOI, as these are discarded by an ideal VAD in the STOI front-end \cite{taal_algorithm_2011}.}
\subsection{Noise Types}\label{sec:nt}
To simulate a wide variety of sound scenes we have used six different noise types in our experiments: two synthetic noise signals and four natural noise signals, which are real-life recordings of naturally occurring sound scenes.  
For the two synthetic noise signals, we use a stationary \acf{SSN} signal and a highly non-stationary 6-speaker babble\;(BBL) noise.
For the naturally occurring noise signals, we use the street\;(STR), cafeteria\;(CAF), bus\;(BUS), and pedestrian\;(PED) noise signals from the CHiME3 dataset \cite{barker_third_2015}.    
The SSN noise signal is Gaussian white noise, spectrally shaped according to the long-term spectrum of the entire TIMIT speech corpus \cite{garofolo_darpa_1993}. Similarly, the BBL noise signal is constructed by mixing utterances from both genders from TIMIT. 
To ensure that all noise types are equally represented and with unique realizations in the training, validation and test sets, all six noise signals are split into non-overlapping segments such that 40 min.\;is used for training, 5 min.\;is used for validation and another 5 min. is used for test. 
\subsection{Noisy Speech Mixtures}
To construct the noisy speech signals used for training, we follow Eq.\;\eqref{eq111} and combine a noise-free training utterance $x[n]$ with a randomly selected noise sequence $v[n]$, of equal length, from the training noise signal. 
We scale the noise signal $v[n]$, to achieve a certain \acf{SNR}, according to the active speech level of $x[n]$ as defined by ITU P.56 \cite{itu_rec._1993}.      
For the training and validation sets, the SNRs are chosen uniformly from $[-5 , 10 ]$ dB to ensure that the intelligibility of the noisy speech mixtures $y[n]$ ranges from degraded to perfectly intelligible.    
\subsection{Model Architecture and Training}
The two types of enhancement systems, $\text{ES}_{ELC}$ and $\text{ES}_{MSE}$, each consist of 15 feed-forward DNNs. The DNNs in the $\text{ES}_{ELC}$ system are trained with the ELC cost function introduced in Eq.\;\eqref{eq:stoicost} and the DNNs in the $\text{ES}_{MSE}$ system are trained using the well-known STSA-MSE cost function given by   
\begin{equation}
\mathcal{J} ( \s,\sh) = \frac{1}{N} \left\lVert \s - \sh \right\lVert^2,
\label{eq:msecost}
\end{equation}
where the subscripts $j$ and $m$ are omitted for convenience.
We train both the $\text{ES}_{ELC}$ and $\text{ES}_{MSE}$ systems with 20000 training utterances and 2000 validation utterances and both data sets have been mixed uniformly with the SSN, BBL, CAF, and STR noise signals, which ensures that each noise type have been mixed with $25\%$ of the utterances in the training and validation sets. 
During test, we evaluate each system with one noise type at a time, i.e. each system is evaluated with 1000 noisy test utterances per noise type, and since BUS and PED are not included in the training and validation sets, these two noise signals serve as unmatched noise types, whereas SSN, BBL, CAF, and STR are matched noise types. This will allow us to study how the ELC optimal DNNs and STSA-MSE optimal DNNs generalize to unmatched noise types.

Each feed-forward DNN consists of three hidden layers with 512 units using ReLU activation functions. The $N$-dimensional output layer uses sigmoid functions which ensures that the output gain $\gh_{j,m}$ is confined between zero and one.
The DNNs are trained using stochastic gradient de-/ascent with the backpropagation technique and batch normalization \cite{goodfellow_deep_2016}. 
The DNNs are trained for a maximum of 200 epochs with a minibatch size of 256 randomly selected short-time temporal one-third octave band envelope vectors.

Since the $\text{ES}_{ELC}$ and $\text{ES}_{MSE}$ systems use different cost functions, they likely have different optimal learning rates. This is easily seen from the gradient norms of the two cost functions. 
It can be shown (details omitted due to space limitations) that the $\ell^2$-norm of the gradient of the ELC cost function in Eq.\;\eqref{eq:stoicost}, with respect to the desired signal vector $\sh$, is given by 
\begin{equation}
\left\lVert \nabla \mathcal{L} ( \s,\sh) \right\rVert =  \frac{ \sqrt{1-\mathcal{L} ( \s,\sh)^2} }{ \left\lVert \sh \right\rVert },
\label{eq:stoigradnorm}
\end{equation}
where the gradient $\nabla \mathcal{L} ( \s,\sh)$ is given by
\begin{equation}
\begin{split}
&\nabla \mathcal{L} \big( \s,\sh \big) = \\
& \left[ \frac{\partial \mathcal{L} \big( \s,\sh \big)}{\partial \hat{{a}}_{1}},
\frac{\partial \mathcal{L} \big( \s,\sh \big) }{\partial \hat{a}_{2}}, \dots,
\frac{\partial \mathcal{L} \big( \s,\sh \big) }{\partial \hat{a}_{N}}
\right]^T,
\label{eq:stoigrad}
\end{split}
\end{equation}
and
\begin{equation}
\begin{split}
&\frac{\partial \mathcal{L} \big( \s,\sh \big)}{\partial \hat{a}_{m}} = \\
&\frac{\mathcal{L} \big( \s,\sh \big)  \big( \s_m - \mu_{\s} \big)    }{  \big(\sh - \mu_{\sh}  \big)^T \big( \s  - \mu_{\s} \big) } -
\frac{\mathcal{L} \big( \s,\sh \big)  \big( \sh_m - \mu_{\sh} \big)    }{  \big(\sh - \mu_{\sh}  \big)^T \big( \sh - \mu_{\sh}  \big)   } . 
\label{eq:stoipart}
\end{split}
\end{equation}
is the partial derivative of $\mathcal{L} ( \s,\sh)$ with respect to entry $m$ of vector $\sh$.
Similarly, the gradient of the STSA-MSE cost function in Eq.\;\eqref{eq:msecost} is given by
\begin{equation}
\begin{split}
&\nabla \mathcal{J} \big( \s,\sh \big) = \frac{2}{N} \left(  \s - \sh \right), 
\label{eq:msegrad}
\end{split}
\end{equation}
such that
\begin{equation}
\begin{split}
\left\lVert  \nabla \mathcal{J} \big( \s,\sh \big) \right\rVert = \frac{2}{N} \left\lVert    \s - \sh   \right\rVert.
\label{eq:msegradnorm}
\end{split}
\end{equation}
Note, since $ \mathcal{L} ( \s,\sh) $ is invariant to the magnitude of $\left\lVert \sh \right\rVert$ (see Eq.\;\eqref{eq:stoicost}), and $\s$ and $N$ are constants during training, the gradient norm of the ELC cost function, Eq.\;\eqref{eq:stoigradnorm}, with respect to $\sh$, is inversely proportional to the gradient norm of the STSA-MSE cost function, Eq.\;\eqref{eq:msegradnorm}.
This suggests that the two cost functions have different optimal learning rates. 
This observation might partly explain why equality with respect to STOI between STOI optimal and STSA-MSE optimal DNNs were achieved in \cite{kolbaek_monaural_2018-1} but not in \cite{zhao_perceptually_2018,zhang_training_2018,fu_end--end_2018}, as \cite{kolbaek_monaural_2018-1} was the only study that explicitly stated that different learning rates for the two cost functions were used. 
In fact, in \cite{zhao_perceptually_2018,zhang_training_2018,fu_end--end_2018} the optimization method Adam \cite{kingma_adam:_2014} was used, and although Adam is an adaptive gradient method, it still has several critical hyper-parameters that can influence convergence \cite{wilson_marginal_2017}.

During a preliminary grid-search using the validation set corrupted with SSN at an SNR of 0 dB and $N=30$, we found learning rates of $0.01$ and $5 \cdot 10^{-5}$ per sample to be optimal for the $\text{ES}_{ELC}$ and $\text{ES}_{MSE}$ systems, respectively.  
During training, the cost on the validation set was evaluated for each epoch and the learning rates were scaled by $0.7$, if the cost increased compared to the cost for the previous epoch. The training was terminated, if the learning rate was below $10^{-10}$. 
We implemented the DNNs using CNTK \cite{agarwal_introduction_2014} and the scripts needed to reproduce the reported results can be found in \cite{kolbaek_supplemental_nodate-1}. 
Note, the goal of these experiments is not to achieve state-of-the-art enhancement performance. In fact, increasing the size of the dataset or DNNs might likely improve performance, although we have not reason to believe it will change the conclusion.

\section{Experimental Results}\label{sec:expres}
To study the relationship between $\text{ES}_{ELC}$ and $\text{ES}_{MSE}$ systems as function of $N$, we have trained multiple systems for various $N$. Specifically, a total of eight $\text{ES}_{ELC}$ systems and eight $\text{ES}_{MSE}$ systems have been trained with $N$ taking the values $N = \{4, 7, 15, 20, 30, 40, 50, 80\}$, which correspond to temporal envelope vectors with durations from approximately 50 to 1000 milliseconds.  
%
%
%
%
%
\begin{figure*}[ht] 
	\centering
	\centerline{\includegraphics[trim={57mm 20mm 46mm 17mm},clip,width=1.0\linewidth]{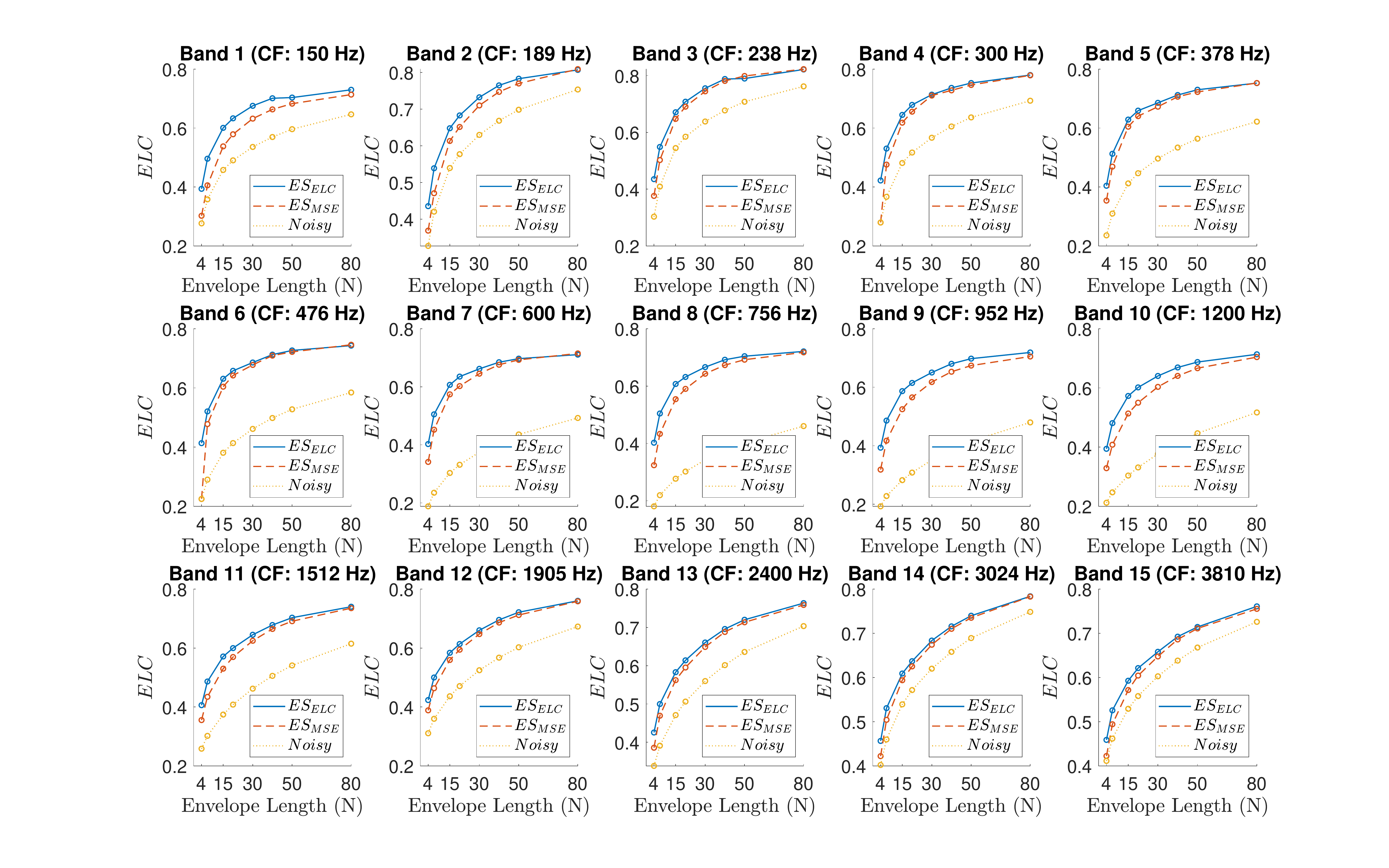}}
	\caption{ELC values for $\text{ES}_{ELC}$ and $\text{ES}_{MSE}$ systems trained using various envelope durations, $N$, and tested with corresponding values of $N$ using speech corrupted with BBL noise at an SNR of 0 dB. Each figure shows one out of $J=15$ one-third octave band DNNs (center frequency\,(CF) shown in parenthesis). It is seen that as $N \to 80$ the difference between the $\text{ES}_{ELC}$ and $\text{ES}_{MSE}$ DNNs, as measured by ELC, tends to zero. This is in line with the theoretical results of Sec.\;\ref{secrelation}.  }
	\label{fig:perband}
\end{figure*}
	\begin{figure*}[ht]
		\centering
		\begin{minipage}{.33\textwidth}
		\centering
		\centerline{\includegraphics[trim={0mm -11mm 9mm 0mm},clip,width=0.9\linewidth]{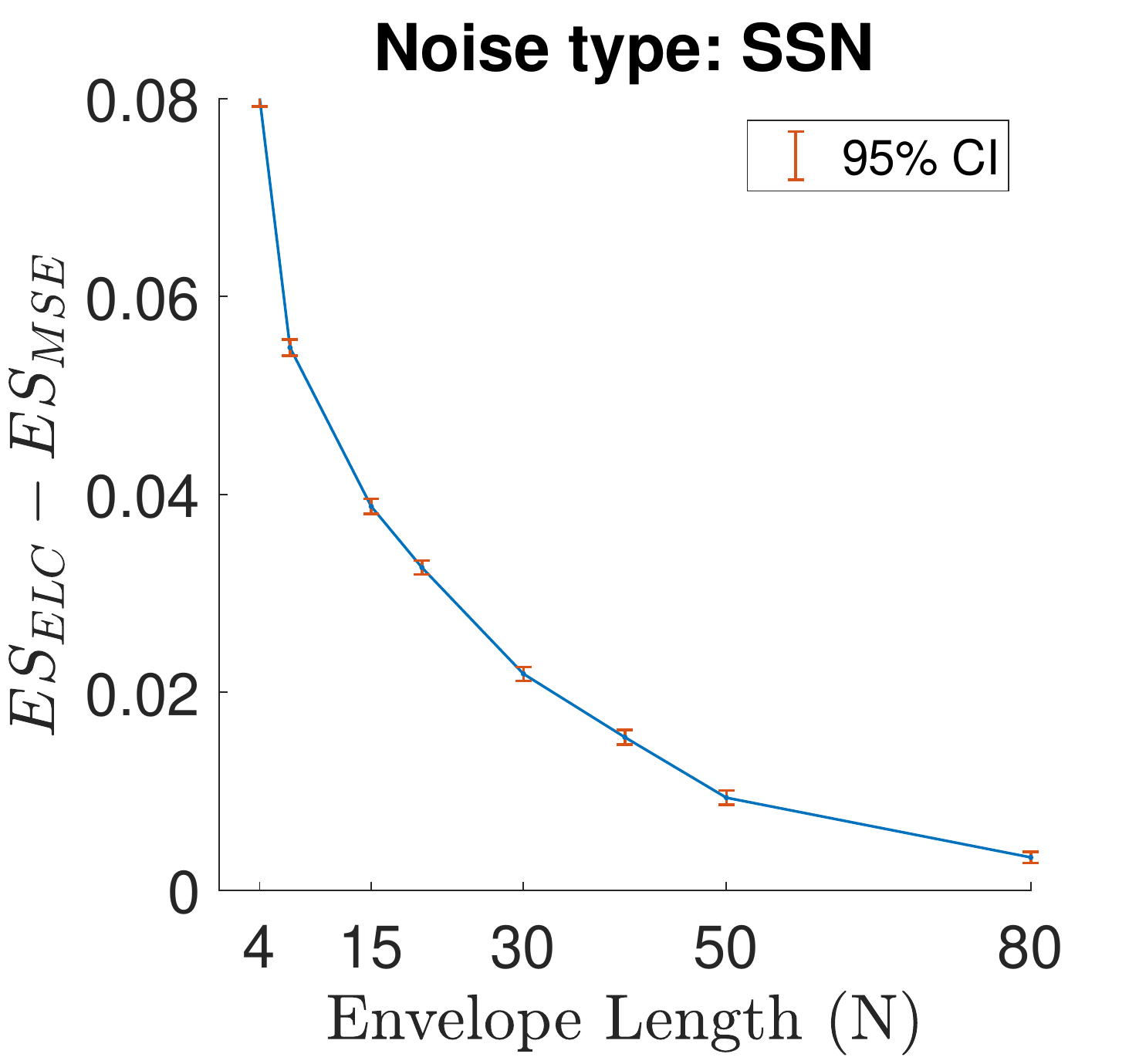}}
		\label{fig:test1}
	\end{minipage}%
	\begin{minipage}{.33\textwidth}
		\centering
		\centerline{\includegraphics[trim={0mm -11mm 9mm 0mm},clip,width=0.9\linewidth]{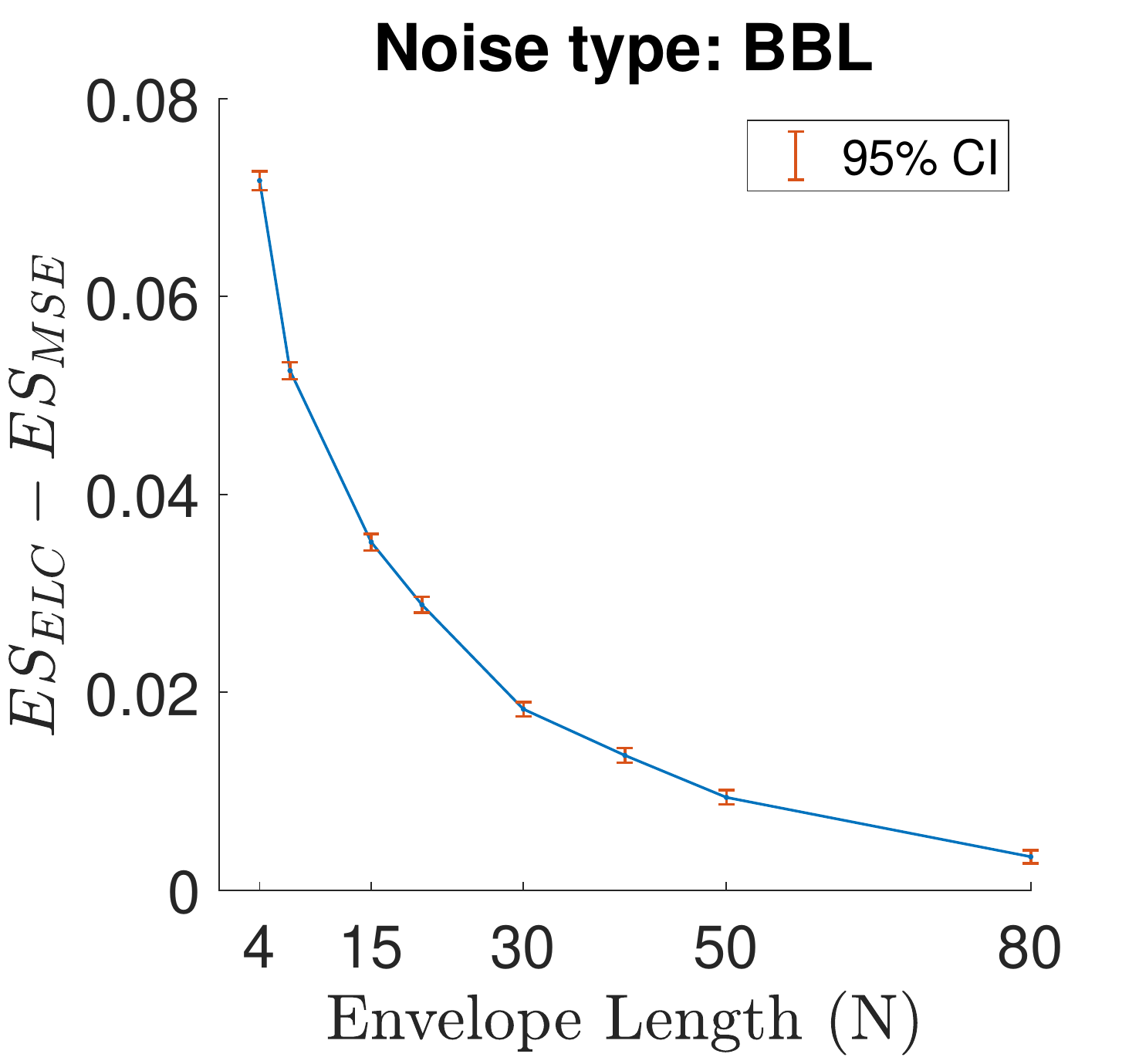}}
		\label{fig:test2}
	\end{minipage}
	\begin{minipage}{.33\textwidth}
		\centering
		\centerline{\includegraphics[trim={0mm -11mm 9mm 0mm},clip,width=0.9\linewidth]{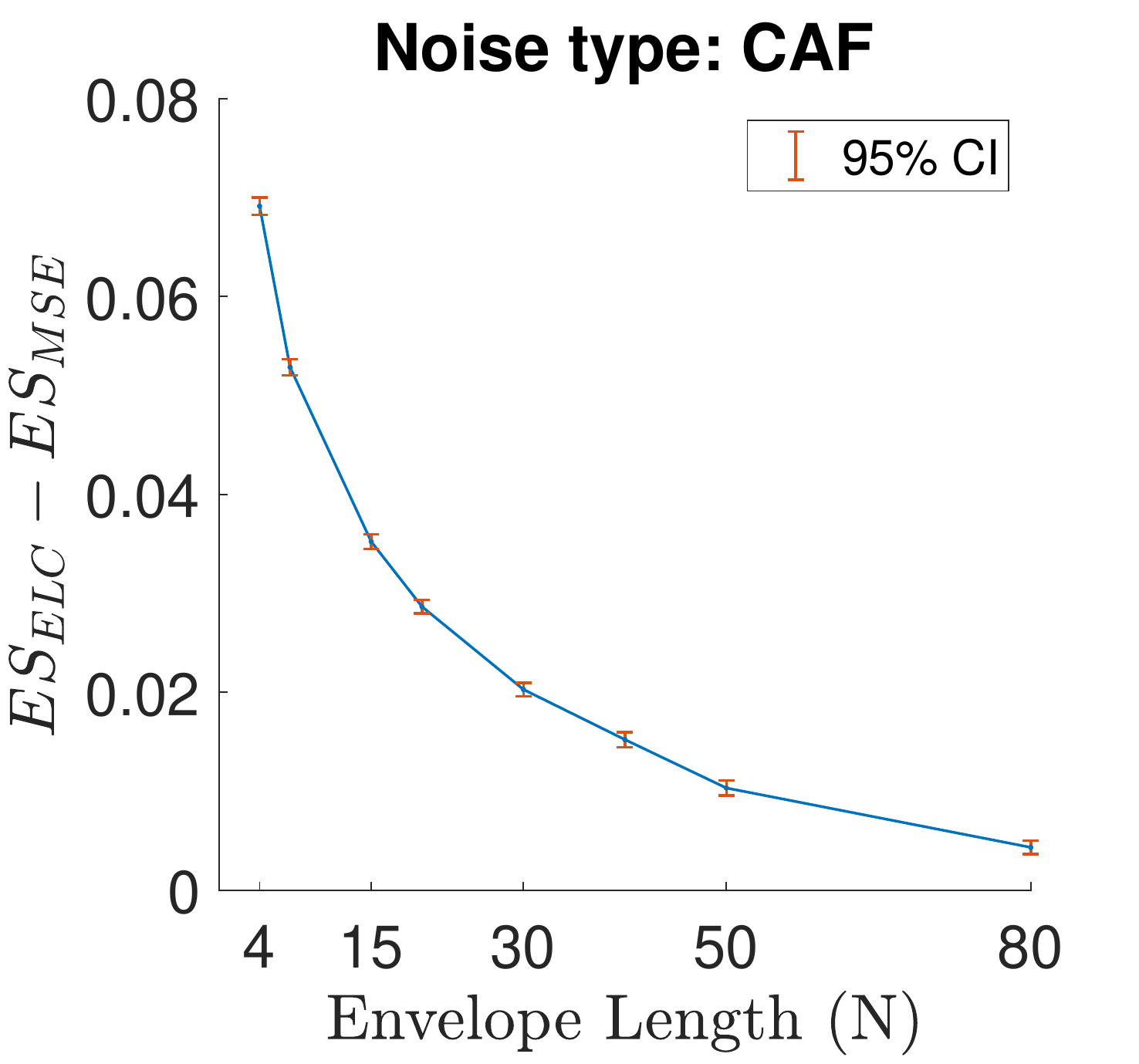}}
		\label{fig:test3}
	\end{minipage}
	\begin{minipage}{.33\textwidth}
		\centering
		\centerline{\includegraphics[trim={0mm 0mm 9mm 0mm},clip,width=0.9\linewidth]{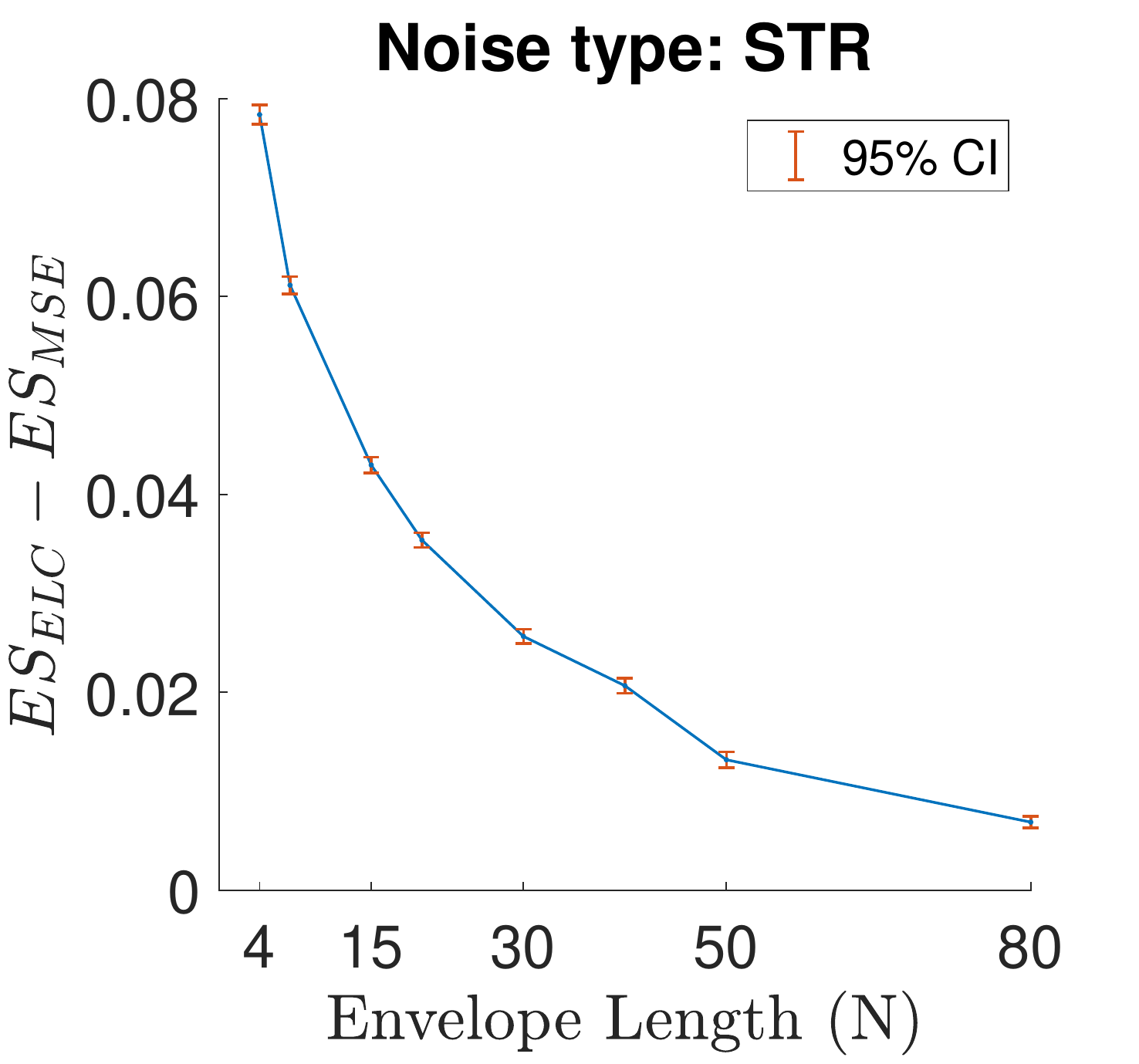}}
		\label{fig:test4}
	\end{minipage}%
	\begin{minipage}{.33\textwidth}
		\centering
		\centerline{\includegraphics[trim={0mm 0mm 9mm 0mm},clip,width=0.9\linewidth]{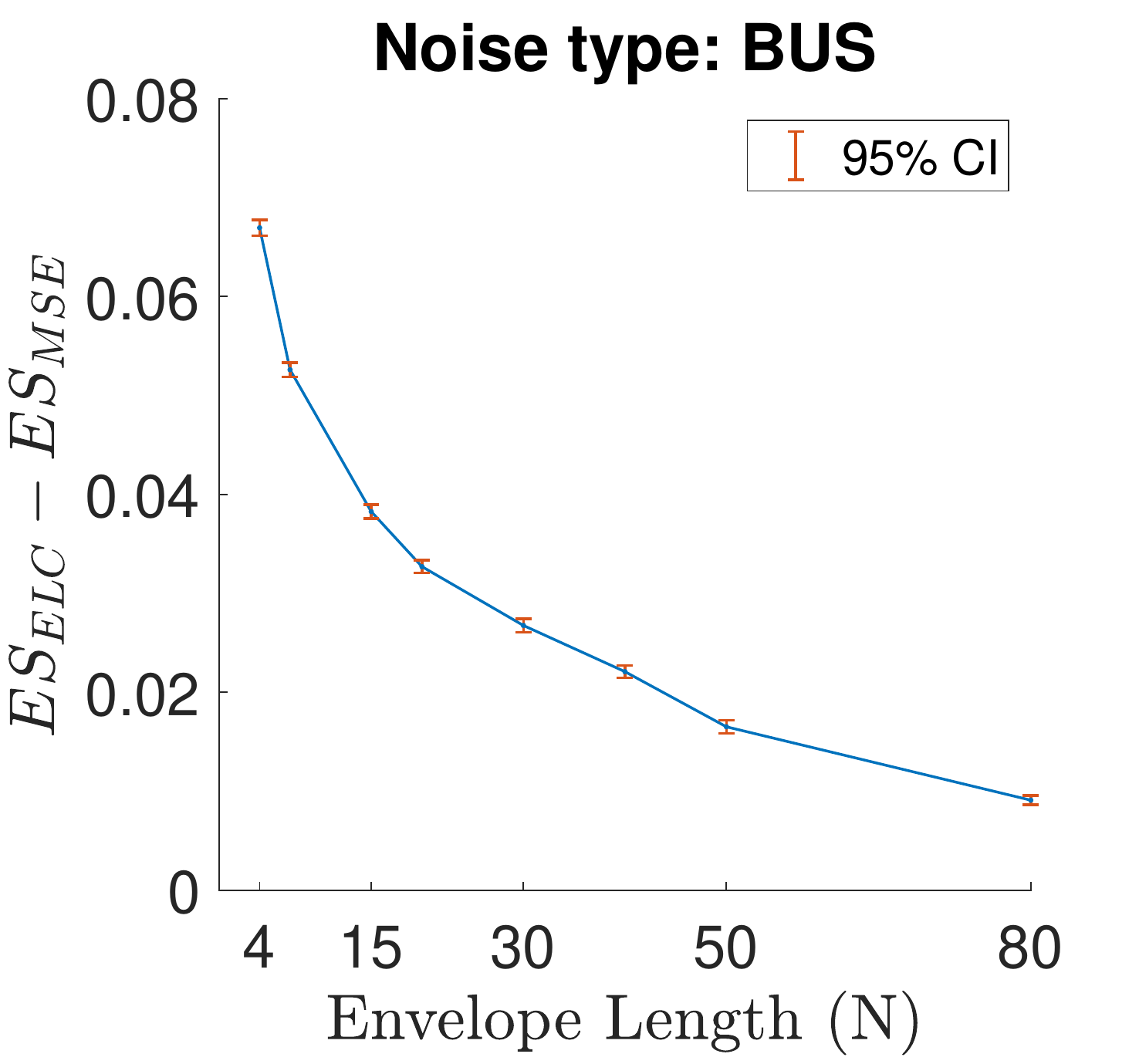}}
		\label{fig:test5}
	\end{minipage}
	\begin{minipage}{.33\textwidth}
		\centering
		\centerline{\includegraphics[trim={0mm 0mm 9mm 0mm},clip,width=0.9\linewidth]{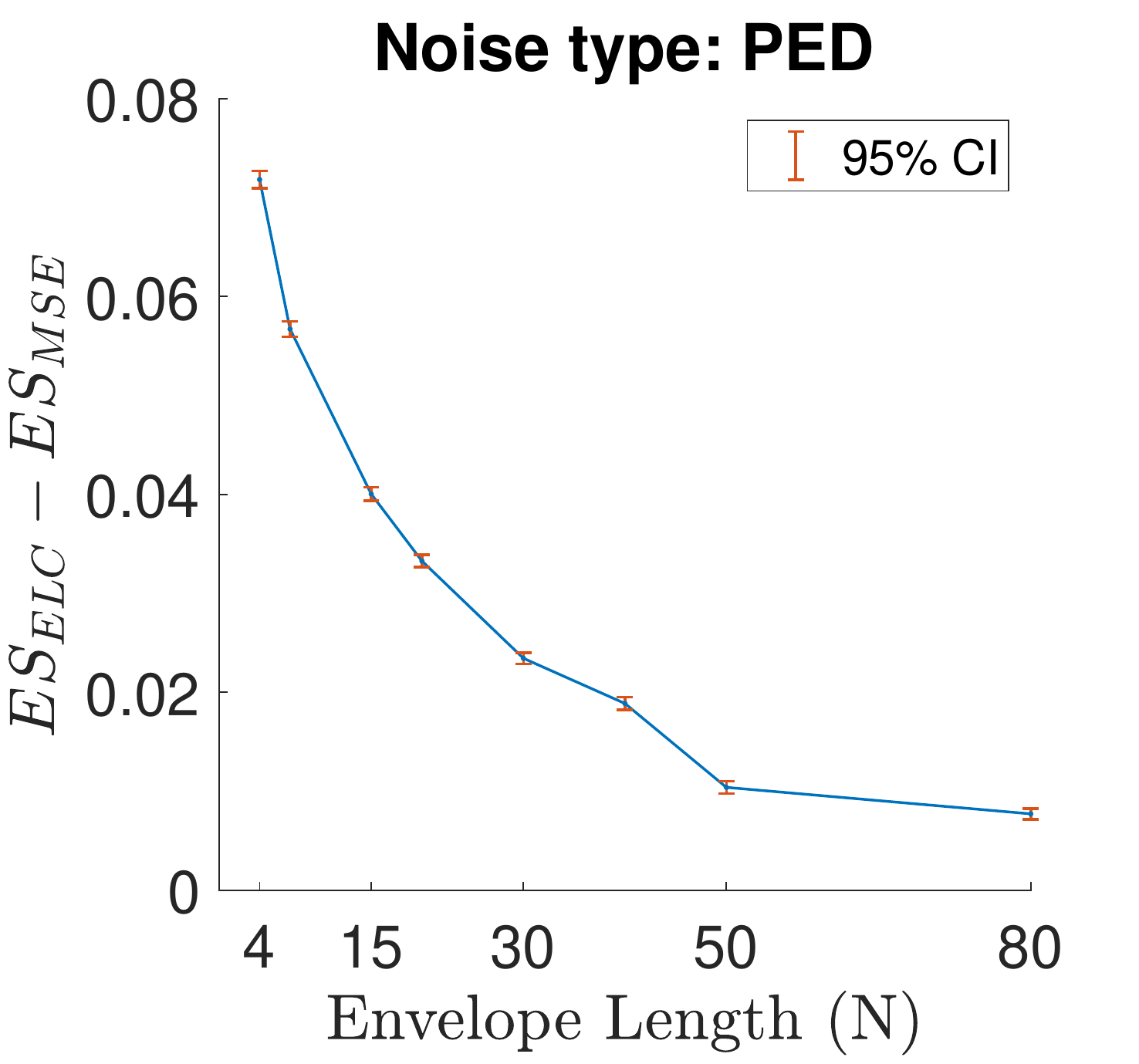}}
		\label{fig:test6}
	\end{minipage}
	\caption{Average ELC differences, as function of envelope durations $N$, between $\text{ES}_{ELC}$ and $\text{ES}_{MSE}$ systems, for different noise types. We observe a monotonic decreasing relationship between the average ELC difference and the envelope length and for $N=80$, the average ELC difference between the $\text{ES}_{ELC}$ and $\text{ES}_{MSE}$ systems is close to zero. This is in line with the theoretical results of Sec.\;\ref{secrelation}.}
	\label{fig:pernoisetype}
\end{figure*}
\subsection{Comparing One-third Octave Bands}
In Fig.\;\ref{fig:perband} we present the ELC scores, as function of envelope duration $N$, for each of the $J=15$ one-third octave band DNNs in the $\text{ES}_{ELC}$ and $\text{ES}_{MSE}$ systems. All DNNs are tested using speech corrupted with BBL noise at an SNR of 0 dB. 
First, we observe that both systems manage to improve the ELC score considerably, when compared to the ELC score of the noisy speech signals, i.e. both systems enhance the noisy speech, which is in line with known results \cite{kolbaek_speech_2017}. 
Furthermore, we can observe that the DNNs trained with the ELC cost function, i.e. the $\text{ES}_{ELC}$ systems, in general achieve higher, or similar, ELC scores than the DNNs trained with the STSA-MSE cost function, i.e. the $\text{ES}_{MSE}$ systems. 
This is an important observation, since it verifies that DNNs trained to maximize ELC indeed achieve the highest, or similar, ELC scores compared to DNNs trained to optimize a different cost function, STSA-MSE in this case.    
Finally, and most importantly, we observe that the difference in ELC score between the $\text{ES}_{ELC}$ and $\text{ES}_{MSE}$ DNNs generally decrease with increasing $N$. For $N=80$ the ELC score of the $\text{ES}_{ELC}$ and $\text{ES}_{MSE}$ DNNs practically coincide. 
\subsection{Comparing ELC across Noise Types}
In Fig.\;\ref{fig:pernoisetype} we present the ELC score difference, as function of envelope duration $N$, for $\text{ES}_{ELC}$ and $\text{ES}_{MSE}$ systems, when tested using speech material corrupted with various noise types at an SNR of 0 dB. 
Specifically, we compute the difference in ELC score for each pair of one-third octave band DNNs in the $\text{ES}_{ELC}$ and $\text{ES}_{MSE}$ systems, and then compute the average ELC difference as function of envelope duration $N$. We do this for all the 1000 test utterances and for each of the six noise types introduced in Sec\;\ref{sec:nt}: SSN, BBL, CAF, STR, BUS, and PED. 
Finally, we compute the $95\%$ confidence interval\,(CI) on the mean ELC difference.

From Fig.\;\ref{fig:pernoisetype} we observe that the average ELC difference, i.e. $\text{ES}_{ELC} - \text{ES}_{MSE}$, appears to be monotonically decreasing with respect to the duration of the envelope $N$. 
Furthermore, we observe that the average ELC difference approaches zero as the duration of the envelope $N$ increases, and similarly to Fig.\;\ref{fig:perband}, for $N=80$, the difference between the $\text{ES}_{ELC}$ and $\text{ES}_{MSE}$ systems is close to zero.  
Finally, we observe that the $95\%$ confidence intervals are relatively narrow for all envelope durations and noise types, which indicate that our test set is sufficiently large to provide accurate estimates of the true mean ELC difference.        
Similarly to Fig.\;\ref{fig:perband}, the results in Fig.\;\ref{fig:pernoisetype} support the theoretical results of Sec.\;\ref{secrelation}. 
Additionally, the results in Fig.\;\ref{fig:pernoisetype} show consistency across multiple noise types, which suggests that the theory in practice applies for various noise type distributions.

\subsection{Comparing STOI across Noise Types}
We now investigate if the global behavior observed for approximate-STOI, i.e. ELC, in Fig.\;\ref{fig:pernoisetype} also applies for real STOI.  
To do this, we reconstruct the test signals used for Fig.\;\ref{fig:pernoisetype} in the time domain. We follow the technique proposed in \cite{kolbaek_monaural_2018-1}, where a uniform gain across STFT coefficients within a one-third octave band is used before an inverse DFT is applied using the phase of the noisy signal.   
In Table\,\ref{tab:stoi_scorr1} we present the STOI scores for $\text{ES}_{ELC}$ and $\text{ES}_{MSE}$ systems, as a function of $N$, when tested using speech material corrupted with different noise types at an SNR of 0 dB.
Note that these test signals are similar to the test signals used for Fig.\;\ref{fig:pernoisetype} except that we now evaluate them according to STOI and not ELC. 

From Table\,\ref{tab:stoi_scorr1} we observe that the average STOI difference between the $\text{ES}_{ELC}$ and $\text{ES}_{MSE}$ systems is maximum for $N=4$, but quickly tends to zero as $N$ increases and for $N\ge15$, the STOI difference is practically zero, i.e. $\leq 0.01$. 
{\color{black}
Also, we observe that the gap in STOI between the $\text{ES}_{ELC}$ and $\text{ES}_{MSE}$ systems closes faster at a lower value of $N$ in Table\;\ref{tab:stoi_scorr1} compared to Fig.\;\ref{fig:pernoisetype}.
We believe this is due to the transformation of the, potentially "invalid", sequences of (e.g. \cite{nawab_signal_1983,griffin_signal_1984}) modified magnitude spectra, when reconstructing enhanced time-domain signals, whose intelligibility is estimated by STOI in  Table\;\ref{tab:stoi_scorr1}. 
Therefore, STOI in Table\,\ref{tab:stoi_scorr1} might be computed based on slightly different magnitude spectra compared to the magnitude spectra used for computing the ELC scores in Fig.\;\ref{fig:pernoisetype}. 
Furthermore, we observe that the $\text{ES}_{MSE}$ achieve slightly higher STOI scores than the $\text{ES}_{ELC}$ systems for $N=4$, which might be due to sub-optimal learning rates as the ones actually used during training of the systems at, e.g. $N=4$,  were found based on a grid-search using systems with $N=30$ (see Sec. VI.D). More importantly, the maximum improvement in STOI is achieved for $N=\{15,20,30\}$, where both systems achieve similar STOI scores.} 
Finally, while the theoretical results of Sec.\;\ref{secrelation} show that approximate-STOI performance of $\ah_{MMELC}$ and $\ah_{MMSE}$ is identical, asymptotically, for $N \to \infty$, the empirical results in Table\,\ref{tab:stoi_scorr1} suggest that $N\ge15$ is sufficient for practical equality to hold for DNN based speech enhancement systems. 
\begin{table}
	\caption{{\color{black} STOI scores as function of $N$ for $\text{ES}_{ELC}$ and $\text{ES}_{MSE}$ systems tested using different noise types at an SNR of 0 dB.}}
	\label{tab:stoi_scorr1}
	\centering
	\setlength\tabcolsep{5pt} 
	\resizebox{0.7\columnwidth}{!}{%
		{\color{black}\begin{tabular}{cccccccccc}
				\midrule \midrule
				$N:$ & & $4$ & $7$ & $15$ & $20$ & $30$ & $40$ & $50$ & $80$ \\
				\midrule
				\multirow{ 2}{*}{SSN:} & ELC :& 0.81 & 0.85 & 0.88 & 0.88 & 0.87 & 0.86 & 0.85 & 0.84 \\
				& MSE :& 0.84 & 0.87 & 0.87 & 0.87 & 0.87 & 0.86 & 0.85 & 0.84 \\ [6pt]
				\multirow{ 2}{*}{BBL:} & ELC :& 0.77 & 0.80 & 0.82 & 0.82 & 0.81 & 0.80 & 0.80 & 0.78 \\
				& MSE :& 0.79 & 0.82 & 0.82 & 0.82 & 0.81 & 0.80 & 0.80 & 0.78 \\ [6pt]
				\multirow{ 2}{*}{CAF:} & ELC :& 0.82 & 0.85 & 0.87 & 0.87 & 0.86 & 0.85 & 0.84 & 0.83 \\
				& MSE :& 0.85 & 0.87 & 0.87 & 0.87 & 0.86 & 0.85 & 0.85 & 0.84 \\ [6pt]
				\multirow{ 2}{*}{STR:} & ELC :& 0.83 & 0.86 & 0.88 & 0.89 & 0.88 & 0.87 & 0.87 & 0.85 \\
				& MSE :& 0.86 & 0.88 & 0.88 & 0.88 & 0.88 & 0.87 & 0.87 & 0.85 \\ [6pt]
				\multirow{ 2}{*}{PED:} & ELC :& 0.77 & 0.81 & 0.83 & 0.83 & 0.83 & 0.82 & 0.81 & 0.80 \\
				& MSE :& 0.80 & 0.82 & 0.83 & 0.83 & 0.82 & 0.82 & 0.81 & 0.80 \\ [6pt]
				\multirow{ 2}{*}{BUS:} & ELC :& 0.87 & 0.89 & 0.90 & 0.91 & 0.90 & 0.89 & 0.89 & 0.89 \\
				& MSE :& 0.89 & 0.90 & 0.90 & 0.90 & 0.90 & 0.90 & 0.89 & 0.89 \\ [3pt] 
				\midrule \midrule 
	\end{tabular}}}
\end{table}
\pagebreak

\subsection{Comparing Gain-Values}
{\color{black}
Figures\;\ref{fig:perband} and \ref{fig:pernoisetype}, and Table\,\ref{tab:stoi_scorr1} show that $\text{ES}_{ELC}$ systems achieve approximately the same ELC and STOI values as $\text{ES}_{MSE}$ systems and that the ELC and STOI difference between the two types of systems approach zero as $N$ becomes large. These empirical results are in line with the theoretical results in Sec.\;\ref{secrelation}. }
However, the results in Sec.\;\ref{secrelation} predict that not only do $\text{ES}_{ELC}$, and $\text{ES}_{MSE}$ systems produce identical ELC scores, they also predict that the systems are, in fact, essentially identical, i.e. up to an affine transformation. Hence, in this section, we compare how the systems actually operate. 
Specifically, we compare the gains estimated by $\text{ES}_{ELC}$ systems with gains estimated by $\text{ES}_{MSE}$ systems.      

In Fig.\;\ref{fig:gaincorr} we present scatter plots, one for each one-third octave band for pairs of gains estimated by $\text{ES}_{ELC}$ and $\text{ES}_{MSE}$ systems tested with BBL noise at an SNR of 5 dB. Each scatter plot consists of 10000 pairs of gains acquired by sampling 10 gain-pairs randomly and uniformly distributed from each of the 1000 test utterances. In Fig.\;\ref{fig:gaincorr}, yellow indicates high density of gain-pairs and dark blue indicates low density.   
{\color{black}
From Fig.\;\ref{fig:gaincorr} it is seen that a correlation no smaller than $0.88$ is achieved for all 15 one-third octave bands. The highest correlation of $r=0.98$ is achieved by bands $5$ to $7$ and the lowest is $r=0.88$ achieved by band $2$ followed by band $1$ with $r=0.89$. 
}
It is also seen that a large number of gain values are either zero, or one, as one would expect due to the sparse nature of speech in the T-F domain. 
However, although a strong correlation is observed for all bands, the gain-pairs are slightly more scattered at the first few bands than for the remaining bands. 
This might be explained simply by the fact that low one-third octave bands correspond to single STFT bins, whereas higher one-third octave bands are sums of a large number of STFT bins. 
This, in turn, may have the consequence that for finite $N$ $(N=30)$, Kolmogorovs strong law of large numbers (see Appendix.\;\ref{sec:zindepedent}) is better valid at higher frequencies than at lower frequencies (so that gain vectors produced by one system is closer to an affine transformation of gain vectors produced by the other system).
In fact, if we compute $r_1$ for models trained with $N=50$, we get $r_1 = 0.93$, i.e. increased correlation between the gain vectors produced by the two systems.  
Finally, in Table.\;\ref{tab:lincorr} we present average correlation coefficients and we observe correlation coefficients  $\geq 0.87$ for all, both matched and unmatched, noise types, at multiple SNRs.
%
%
\begin{figure*}[ht] 
	\centering
	\centerline{\includegraphics[trim={23mm 5mm 23mm 5mm},clip,width=1.0\linewidth]{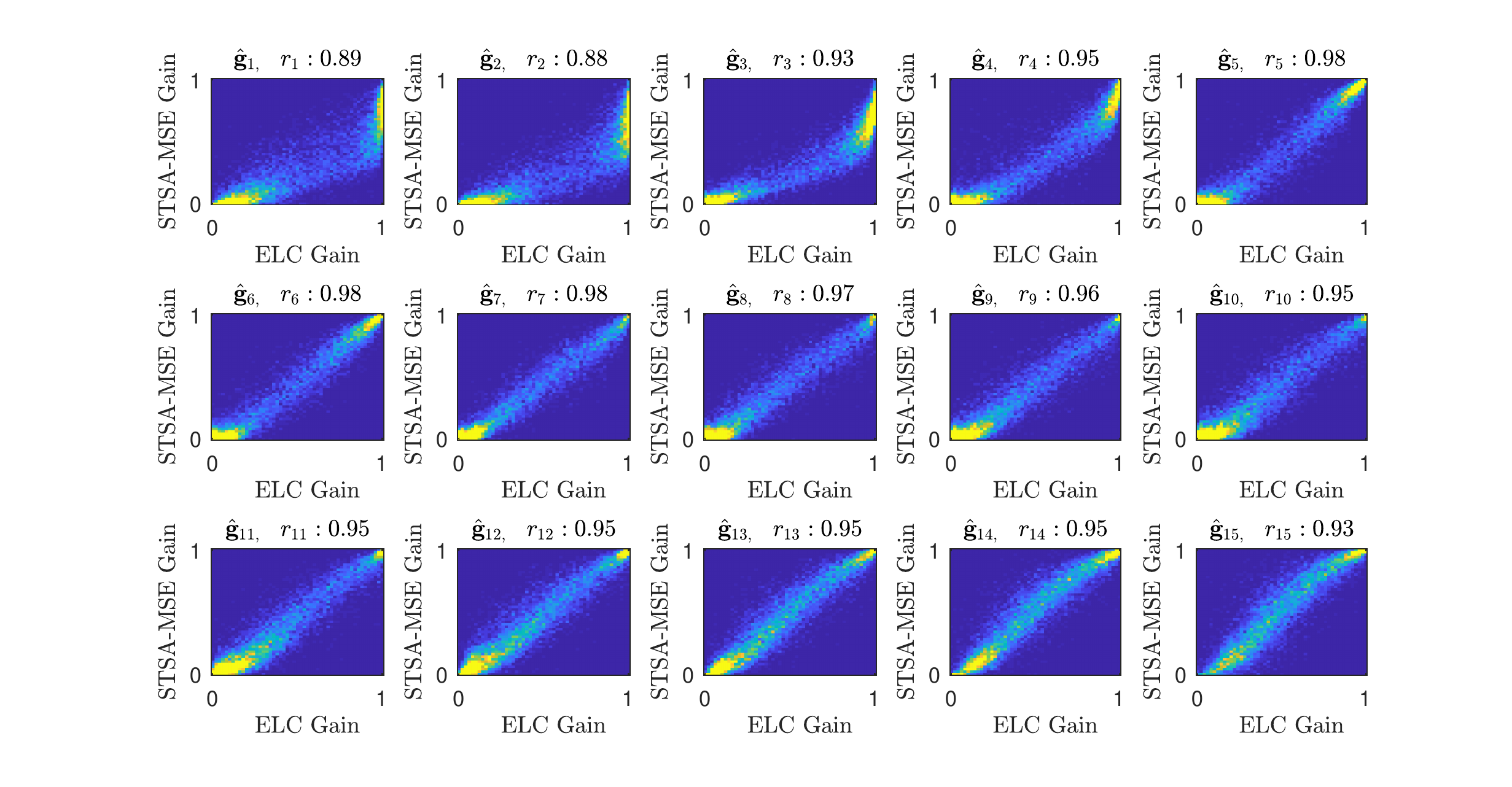}}
	\caption{Scatter plots based on gain values from $\text{ES}_{ELC}$ and $\text{ES}_{MSE}$ systems with an envelope length of $N=30$. Dark blue indicate low density and bright yellow indicate high density. The systems are tested with BBL noise corrupted speech at an SNR of 5 dB. Each figure shows one of 15 ($\hat{\mathbf{g}}_1, \hat{\mathbf{g}}_2, \dots, \hat{\mathbf{g}}_{15}$) one-third octave bands. A correlation no smaller than $0.88$ is achieved for all one-third octave bands, which indicates that the $\text{ES}_{ELC}$ and $\text{ES}_{MSE}$ systems estimate fairly similar gain vectors. }
	\label{fig:gaincorr}
\end{figure*}
\begin{table}
\caption{Sample correlations between gains from $\text{ES}_{ELC}$ and $\text{ES}_{MSE}$ systems with $N=30$. See Fig.\;\ref{fig:gaincorr} for per band correlations.}
\label{tab:lincorr}
\centering
\setlength\tabcolsep{5pt} 
\resizebox{0.55\columnwidth}{!}{%
	\begin{tabular}{ccccccc}
		\toprule
		\begin{tabular}[c]{@{}c@{}}SNR \\ {[dB]}\end{tabular} 	& 
		\begin{tabular}[c]{@{}c@{}}SSN \\\end{tabular} 	& 
		\begin{tabular}[c]{@{}c@{}}BBL \\\end{tabular} 	& 
		\begin{tabular}[c]{@{}c@{}}CAF \\\end{tabular} 	&
		\begin{tabular}[c]{@{}c@{}}STR \\\end{tabular} 	&
		\begin{tabular}[c]{@{}c@{}}BUS \\\end{tabular} 	&
		\begin{tabular}[c]{@{}c@{}}PED \\\end{tabular} 	\\
		\midrule
		-5   & 0.94 & 0.87 & 0.89 & 0.93  & 0.87 & 0.90\\ 
		0    & 0.94 & 0.92 & 0.92 & 0.93  & 0.88 & 0.92\\ 
		5    & 0.95 & 0.95 & 0.93 & 0.93  & 0.90 & 0.92\\ 
		10   & 0.95 & 0.95 & 0.92 & 0.92  & 0.91 & 0.93\\ 
		\midrule \midrule
\end{tabular}}
\end{table}

\section{Conclusion}\label{sec:con}
This study is motivated by the fact that most estimators used for speech enhancement, being either data-driven models, e.g. \acfp{DNN}, or statistical model-based techniques such as the Short-Time Spectral Amplitude Minimum Mean-Squared Error\,(STSA-MMSE) estimator, use the STSA Mean-Squared Error\,(MSE) cost function as a performance indicator.  
Short-Time Objective Intelligibility\,(STOI), a state-of-the-art speech intelligibility estimator, on the other hand, rely on the Envelope Linear Correlation\,(ELC) of speech temporal envelopes. 
Since the primary goal of many speech enhancement systems is to improve speech intelligibility, it raises the question if estimators can benefit from an ELC cost function. 

In this paper we derive the Maximum Mean Envelope Linear Correlation\,(MMELC) estimator and study its relationship to the well-known STSA-MMSE estimator.    
We show theoretically that the MMELC estimator, under a commonly used conditional independence assumption, is asymptotically equivalent to the STSA-MMSE estimator.
Furthermore, we demonstrate experimentally that this relationship also holds for DNN based speech enhancement systems, when the DNNs are trained to either maximize ELC or minimize MSE and the systems are evaluated using both ELC and STOI.
Finally, our experimental findings suggest, that applying the traditional STSA-MMSE estimator on noisy speech signals in practice leads to essentially maximum speech intelligibility as reflected by the STOI speech intelligibility estimator.	 
%
%


%

\begin{subappendices}

\pagebreak
\section{Maximizing a Constrained Inner Product}\label{sec:lagran}
This appendix derives an expression for the zero-mean, unit-norm vector $\e(\ah)$, which maximizes the inner product with the vector $\E_{\A | \sr } \left[ \e(\A| \sr) \right]$.
For notational convenience, let $\alfa = \E_{\A | \sr } \left[ \e(\A| \sr) \right]$, and $\bet = \e(\ah)$.  The constrained optimization problem from Eq.\;\eqref{eq6} is then defined as  
\begin{equation}
\begin{aligned}
& \underset{\bet}{\text{maximize}}
& & \alfa^T \bet \\
& \text{subject to} 
& & \bet^T \One = 0, \\
& & &\bet^T\bet = 1.
\end{aligned}
\label{eq:eqLag1}
\end{equation}
The vector $\bet^\ast$ that solves Eq.\;\eqref{eq:eqLag1} can be found using the method of Lagrange multipliers \cite{boyd_convex_2004}. Introducing two scalar Lagrange multipliers, $\lambda_1$ and $\lambda_2$, for the two equality constraints, the Lagrangian is given by\footnote{We solve the equivalent problem that minimizes $-\alfa^T \bet$.}    
\begin{equation}
\mathcal{L}(\bet,\lambda_1,\lambda_2) = -\alfa^T \bet + \lambda_1 \bet^T \One + \lambda_2(\bet^T\bet - 1).
\label{eqLag2}
\end{equation}
Setting the partial derivatives $\frac{\partial \mathcal{L}}{\partial \bet}$ equal to zero 
\begin{equation}
\begin{aligned}
\frac{\partial \mathcal{L}}{\partial \bet} & = -\alfa + \lambda_1 \One + 2\lambda_2\bet = \barbelow{0}, \\
\end{aligned}
\label{eqLag30}
\end{equation}
and solving for $\bet$, we arrive at 
\begin{equation}
\begin{aligned}
\bet & = \frac{\alfa - \lambda_1 \One}{2\lambda_2}.
\end{aligned}
\label{eqLag3}
\end{equation}
Using the same approach for $\frac{\partial \mathcal{L}}{\partial \lambda_1}$ and $\frac{\partial \mathcal{L}}{\partial \lambda_2}$, substituting in Eq.\;\eqref{eqLag3} and solving for $\lambda_1$, and $\lambda_2$ such that the two constraints are fulfilled, we find   
\begin{equation}
\begin{aligned}
\lambda_1 & =   \frac{1}{N} \alfa^T \One =  {\mu}_{\alfa},\\
\end{aligned}
\label{eqLag40}
\end{equation}
and
\begin{equation}
\begin{aligned}
\lambda_2 & =  \frac{\lVert \alfa - \barbelow{\mu}_{\alfa} \One \rVert }{2}.
\end{aligned}
\label{eqLag4}
\end{equation}
Inserting $\lambda_1$ and $\lambda_2$ into Eq.\;\eqref{eqLag3} results in 
\begin{equation}
\begin{aligned}
\bet^\ast & = \frac{ \alfa - \barbelow{\mu}_{\alfa} \One  }{\lVert \alfa - \barbelow{\mu}_{\alfa} \One \rVert},
\end{aligned}
\label{eqLag5}
\end{equation}
which is simply the vector $\alfa$, normalized to zero sample mean and unit norm.

\section{Factorization of Expectation}\label{sec:zindepedent}
This appendix shows that the expectation in Eq.\;\eqref{eq11} factorizes into the product of expectations in Eq.\;\eqref{eq14},  asymptotically as $N \to \infty$. 
Let
\begin{equation} 
\Y \triangleq \A | \sr,
\label{eqz1}
\end{equation}
and
\begin{equation} 
\Hidem \triangleq  \barbelow{\barbelow{I}}_N - \frac{1}{N} \One\One^T,
\label{eqz2}
\end{equation}
so that
\begin{equation} 
\Z = \Hidem \Y,
\label{eqz3}
\end{equation}
where $\barbelow{\barbelow{I}}_N$ denotes the $N$-dimensional identity matrix and $\A | \sr $ is a random vector distributed according to the conditional probability density function $\fagr$.
A specific element $Z_i$, of $\Z$ is then given by
\begin{equation} 
\begin{split}
Z_i &= \barbelow{{h}}_i^T \Y \\
& =   S_i - \frac{1}{N} \One^T \Y,  \\
\end{split}
\label{eqz4}
\end{equation} 
where $\barbelow{{h}}_i$ is the $i$th column of matrix $\Hidem$. 

We now define the covariance between $Z_i$ and $1/\lVert \Z \rVert$ as
\begin{equation} 
\begin{split}
\text{cov}(Z_i , \frac{1}{\big\lVert \Z \big\rVert} ) &\triangleq \E \left[ \bigg( Z_i - \E \left[ Z_i \right] \bigg) \bigg( \frac{1}{ \big\lVert \Z \big\rVert } - \E \left[ \frac{1}{ \big\lVert \Z \big\rVert } \right] \bigg)    \right]  \\
& = \E \left[ \frac{  Z_i }{ \big\lVert \Z \big\rVert }  \right] - \E \left[ Z_i \right] \E \left[ \frac{1}{\big\lVert \Z \big\rVert } \right].  \\
\end{split}
\label{eqz5}
\end{equation} 
We can rewrite the factors on the right-hand side of Eq.\;\eqref{eqz5} as follows
\begin{equation} 
\begin{split}
\E \left[ Z_i \right] &= \E \left[ \barbelow{\mathbf{h}}_i^T \Y  \right]  \\
& = \E \left[ S_i - \frac{1}{N} \One^T \Y  \right]  \\
& = \E \left[ S_i  \right] - \frac{1}{N} \One^T  \E \left[ \Y  \right]  \\
& = \E \left[ S_i  \right] - \frac{1}{N} \sum_{j=1}^{N}  \E \left[ S_j  \right],  \\
\end{split}
\label{eqz6}
\end{equation} 
\begin{equation} 
\begin{split}
\E \left[ \frac{1}{\big\lVert \Z \big\rVert } \right] &= \E \left[ \frac{1}{ \sqrt{ \Y^T \Hidem \Hidem^T \Y} }  \right]   \\
& =  \E \left[ \frac{1}{ \sqrt{ \Y^T \Hidem \Y} }  \right]   \\
& =  \E \left[ \frac{1}{ \sqrt{ \Y^T\Y  - \frac{1}{N} \Y^T \One \One^T \Y} }  \right]   \\
& =  \E \left[ \frac{1}{ \sqrt{ \sum_{j=1}^{N}S_j^2  - \frac{1}{N} \left( \sum_{j=1}^{N}S_j \right)^2 } }  \right]   \\
& =  \E \left[ \frac{\sqrt{\frac{1}{N}}}{ \sqrt{ \frac{1}{N} \sum_{j=1}^{N}S_j^2  - \left( \frac{1}{N}  \sum_{j=1}^{N}S_j \right)^2 } }  \right],   \\
\end{split}
\label{eqz7}
\end{equation} 
and
\begin{equation} 
\begin{split}
\E \left[ \frac{  Z_i }{ \big\lVert \Z \big\rVert }  \right] &= 	\E \left[ \frac{  \sqrt{ \frac{1}{N}}  \left( S_i - \frac{1}{N} \sum_{j=1}^{N}S_j \right)}{ \sqrt{ \frac{1}{N} \sum_{j=1}^{N}S_j^2  - \left( \frac{1}{N} \sum_{j=1}^{N}S_j \right)^2 } }  \right].   \\
\end{split}
\label{eqz8}
\end{equation} 

In Eqs.\;\eqref{eqz6}, \eqref{eqz7} and \eqref{eqz8} two different sums of random variables occur, 
\begin{equation} 
\begin{split}
\frac{1}{N}  \sum_{j=1}^{N}S_j, \\
\end{split}
\label{eqz10}
\end{equation}     
and
\begin{equation} 
\begin{split}
\frac{1}{N} \sum_{j=1}^{N}S_j^2.  \\
\end{split}
\label{eqz11}
\end{equation} 
Since, by assumption, Eq.\;\eqref{eqz9}, $S_j\;\forall \;j$ are independent random variables with finite variances\footnote{Assuming a finite variance of $S_j$ is motivated by the fact that $S_j$ model speech signals, which always take finite values due to both physical and physiological limitations of sound and speech production systems, respectively.}%
, according to Kolmogorovs strong law of large numbers \cite{sen_large_1994}, the sums given by Eqs.\;\eqref{eqz10} and \eqref{eqz11} will converge (almost surely, i.e. with probability\;($\Pr$) one) to their average means $\mu_S = \frac{1}{N}\sum_{j=1}^{N}\E[S_j]$, and $\mu_{S^2} = \frac{1}{N}\sum_{j=1}^{N}\E[S_j^2]$, respectively, as $N \to \infty$. Formally, we can express this as  
\begin{equation} 
\begin{split}
\Pr &\left( \lim\limits_{N \to \infty} \frac{1}{N}  \sum_{j=1}^{N}S_j = \mu_{S} \right) = 1, \\
\end{split}
\label{eqz12}
\end{equation}     
and
\begin{equation} 
\begin{split}%
\Pr &\left( \lim\limits_{N \to \infty} \frac{1}{N}  \sum_{j=1}^{N}S_j^2 = \mu_{S^2} \right) = 1. \\
\end{split}
\label{eqz13}
\end{equation}     

By substituting Eqs.\;\eqref{eqz12}, and \eqref{eqz13} into Eqs.\;\eqref{eqz6}, \eqref{eqz7} and \eqref{eqz8}, we arrive at
\begin{equation} 
\begin{split}
\lim\limits_{N \to \infty} \E \left[ Z_i \right] &= \E \left[ S_i  \right] - \mu_S,  \\
\end{split}
\label{eqz14}
\end{equation} 
\begin{equation} 
\begin{split}
\lim\limits_{N \to \infty} \E \left[ \frac{1}{\big\lVert \Z \big\rVert } \right] &= \frac{\lim\limits_{N \to \infty} \sqrt{\frac{1}{N}}}{ \sqrt{ \mu_{S^2}  - \mu_S^2 } }     ,    \\
\end{split}
\label{eqz15}
\end{equation} 
and
\begin{equation} 
\begin{split}
\lim\limits_{N \to \infty} \E \left[ \frac{  Z_i }{ \big\lVert \Z \big\rVert }  \right] &=   \left(\E \left[ S_i  \right] - \mu_S \right) \frac{\lim\limits_{N \to \infty} \sqrt{\frac{1}{N}}  }{ \sqrt{ \mu_{S^2}  - \mu_S^2 } }  \\
& = \lim\limits_{N \to \infty} \E \left[ Z_i  \right]   \E \left[ \frac{  1 }{ \big\lVert \Z \big\rVert }  \right]  ,   \\
\end{split}
\label{eqz16}
\end{equation} 
where the last line follows from Eq.\;\eqref{eqz14} and \eqref{eqz15}.
In words, as $N \to \infty$, the covariance between $Z_i$ and $1/\lVert \Z \rVert$ tends to zero and, consequently, the expectation in Eq.\;\eqref{eq11} factorizes into the product of expectations in Eq.\;\eqref{eq14}.

\end{subappendices}

{\small\bibliographystyle{bib/IEEEtran}\bibliography{bib/mybibG}}

\end{document}